\documentclass[useAMS,usenatbib]{mnras}

\usepackage{pdflscape}
\usepackage{epstopdf}
\usepackage{graphicx}
\usepackage{textcomp}
\usepackage{courier}
\usepackage{caption}
\usepackage{amssymb,amsmath,amstext}
\usepackage{url}
\usepackage{deluxetable}
\usepackage{gensymb}
\usepackage{appendix}
\usepackage{longtable}
\usepackage{booktabs}
\usepackage{orcidlink}

\usepackage{lipsum}

\usepackage{epsfig}\usepackage{epsf}
\usepackage{soul}

\title[Type IIn Spectropolarimetry]{Multi-Epoch Spectropolarimetry for a Sample of Type IIn Supernovae: Persistent Asymmetry in Dusty Circumstellar Material}
\author[Bilinski et. al.]{Christopher Bilinski$^{1}$\thanks{E-mail:
cgbilinsk@gmail.com} \orcidlink{0000-0002-8826-3571}, Nathan Smith$^{1}$ \orcidlink{0000-0001-5510-2424}, G. Grant Williams$^{2}$ \orcidlink{0000-0002-3452-0560},  \newauthor Paul S. Smith$^{1}$ \orcidlink{0000-0002-5083-3663}, 
Douglas C. Leonard$^{3}$, Jennifer L. Hoffman$^{4}$ \orcidlink{0000-0003-1495-2275}, Jennifer Andrews$^{1,5}$,  \newauthor Peter Milne$^{1}$\\
$^{1}$Steward Observatory, University of Arizona, 933 N. Cherry Avenue, Tucson AZ 85719, USA\\
$^{2}$MMT Observatory, Tucson, AZ 85721-0065, USA\\
$^{3}$Department of Astronomy, San Diego State University, San Diego, CA 92812, USA\\
$^{4}$Department of Physics \& Astronomy, University of Denver, 2112 East Wesley Avenue, Denver, CO 80208, USA \\
$^{5}$Gemini Observatory, 670 North A‘ohoku Place, Hilo, HI 96720-2700, USA}

\begin{document}
\date{Accepted 0000. Received 0000; in original form 0000}

\pagerange{\pageref{firstpage}--\pageref{lastpage}} \pubyear{2023}

\maketitle 

\label{firstpage}
 
\begin{abstract}
We present multi-epoch spectropolarimetry and spectra for a sample of 14 Type~IIn supernovae (SNe IIn). We find that after correcting for likely interstellar polarization, SNe IIn commonly show intrinsic continuum polarization of 1--3\% at the time of peak optical luminosity, although a few show weaker or negligible polarization.  While some SNe IIn have even stronger polarization at early times, their polarization tends to drop smoothly over several hundred days after peak.  We find a tendency for the intrinsic polarization to be stronger at bluer wavelengths, especially at early times.  While polarization from an electron scattering region is expected to be grey, scattering of SN light by dusty circumstellar material (CSM) may induce such a wavelength-dependent polarization.  For most SNe IIn, changes in polarization degree and wavelength dependence are not accompanied by changes in the position angle, requiring that asymmetric pre-SN mass loss had a persistent geometry.  While 2$-$3\% polarization is typical,  about 30\% of SNe IIn have very low or undetected polarization.  Under the simplifying assumption that all SN IIn progenitors have axisymmetric CSM (i.e. disk/torus/bipolar), then the distribution of polarization values we observe is consistent with similarly asymmetric CSM seen from a distribution of random viewing angles.  This asymmetry has very important implications for understanding the origin of pre-SN mass loss in SNe IIn, suggesting that it was shaped by binary interaction.
\end{abstract}


\section{Introduction}
\label{sec:Int}
The environment into which a supernova (SN) explodes may substantially influence what we observe, and that surrounding environment is the product of immediate pre-SN mass loss. While a range of densities are inferred around various types of SNe, the highest density circumstellar material (CSM) is inferred for Type IIn events (SNe~IIn), which show strong narrow emission lines in their spectra.  The origin of this dense CSM remains uncertain, but it cannot be produced by any normal winds observed in massive stars \citep{2017hsn..book..403S}.  Luminous blue variables (LBVs) such as $\eta$ Car -- which shows substantial axisymmetric CSM in the Homunculus Nebula \citep{1949Obs....69...31T,1950ApJ...111..408G} -- are often suggested as progenitors of SNe IIn \citep{2005ASPC..332..302S,2006ApJ...645L..45S,2007ApJ...666.1116S,2011MNRAS.412.1522S,2014ARA&A..52..487S,2007ApJ...656..372G,2008A&A...483L..47T}.  Studying the geometry of the SN environment at the time of death may help connect progenitor types to various SNe, since some scenarios for mass loss (binary interaction, mergers, rapid rotation) are expected to produce strong asymmetry in the CSM. 

SNe IIn are thought to be the result of fast SN ejecta colliding with dense CSM that was expelled just prior to the death of the progenitor (see \citealt{2014ARA&A..52..487S} or \citealt{2017hsn..book..403S} for reviews).  This slow, pre-shock CSM is seen spectroscopically as narrow (100-500\,km\,$\mathrm{s^{-1}}$) and intermediate-width (1000-3000\,km\,$\mathrm{s^{-1}}$) Balmer-series emission lines that dominate the optical spectrum, and the shock interacting with CSM generally causes enhanced luminosity with a smooth blue continuum at early times \citep{1990MNRAS.244..269S,1997ARA&A..35..309F,2017hsn..book..403S}.  Although CSM interaction produces most of the luminosity for typical SNe~IIn, the underlying SN ejecta may also be detected as broad emission and absorption features at later times.

A common tool used in studying the shape of SNe IIn is that of spectroscopy.  Spectroscopic line profiles can reveal expansion asymmetries along our line of sight, and how they evolve with time.  Indeed, numerous observations imply asphericity in SNe IIn in both their CSM and SN ejecta (for example, SN~1988Z: \citealt{1994MNRAS.268..173C}; SN~1995N: \citealt{2002ApJ...572..350F}; SN~1997eg: \citealt{2008ApJ...688.1186H}; SN~1998S: \citealt{2000ApJ...536..239L,2001ApJ...550.1030W,2005ApJ...622..991F,2012MNRAS.424.2659M}; SN~2005ip: \citealt{2009ApJ...695.1334S,2014ApJ...780..184K}; SN~2006jd: \citealt{2012ApJ...756..173S}; SN~2006tf: \citealt{2008ApJ...686..467S};  SN~2009ip: \citealt{2014MNRAS.442.1166M,2017MNRAS.470.1491R}; SN~2010jl: \citealt{2012AJ....143...17S,2014ApJ...797..118F}; PTF11iqb: \citealt{2015MNRAS.449.1876S}; SN~2012ab: \citealt{2018MNRAS.475.1104B}; SN~2013L: \citealt{2017MNRAS.471.4047A}; SN~2014ab: \citealt{2020MNRAS.498.3835B}; iPTF14hls: \citealt{2018MNRAS.477...74A}; SN~2017hcc: \citealt{2020MNRAS.499.3544S}).

A less commonly used tool, though powerful for constraining the geometry of the continuum photosphere, is that of spectropolarimetry.  Several studies have been published using spectropolarimetry to study the nature of SNe II-P (see \citealt{2001ApJ...553..861L,2006Natur.440..505L,2010ApJ...713.1363C,2021A&A...651A..19D}).  Generally, in SNe II the polarization signal is initially small, increases throughout the photospheric phase, reaches a maximum at the beginning of the nebular phase, and then drops as the inverse of time squared. In SNe II-P, the emission arises from freely expanding SN ejecta, and so polarization points to asymmetry in the ejecta, and therefore, asymmetry in the explosion itself.  In SNe IIn, on the other hand, emission arises from the CSM, the shock interaction between the ejecta and CSM, and possibly the freely expanding SN ejecta.  As such, observed polarization in SNe~IIn may trace the geometry of the CSM more than that of the SN explosion, making the interpretation complicated.  Spectropolarimetric results have been published for only a handful of SNe~IIn (SN~1997eg: \citealt{2008ApJ...688.1186H}; SN~1998S: \citealt{2000ApJ...536..239L}; SN~2006tf: \citealt{2008ApJ...686..467S}; SN~2010jl: \citealt{2011A&A...527L...6P}; SN~2009ip: \citealt{2014MNRAS.442.1166M,2017MNRAS.470.1491R}; SN~2012ab: \citealt{2018MNRAS.475.1104B}; SN~2013fs: \citealt{2018MNRAS.476.1497B};  SN 2014ab: \citealt{2020MNRAS.498.3835B}; SN~2017hcc: \citealt{2019MNRAS.488.3089K}).  The majority of SNe~IIn with published spectropolarimetric data exhibit continuum polarization signals on the order of 1--3\%.  These data do not yet provide a unified picture, however, as the timing of this polarization signal, variations in the position angle, diverse polarization changes across line profiles, and uncertain interstellar polarization (ISP) estimates make the overall picture complicated.  It is difficult in the case of SNe IIn to disentangle the underlying SN ejecta and their geometry from that of the bright and spatially more extended CSM interaction regions.  Unlike in normal core-collapse SNe, where the polarization arises from  centrosymmetric scattering in homologously expanding ejecta, in SNe IIn, the main source of luminosity (the shock running into CSM) is itself spatially extended and likely asymmetric.  Furthermore, at least one SN IIn (SN~2014ab) exhibits relatively low levels of continuum polarization but shows signs of significant asymmetry spectroscopically, suggesting that the viewing angle may substantially impact the continuum polarization measurements of SNe~IIn \citep{2020MNRAS.498.3835B}.  If this is the case, then more SNe~IIn with relatively low intrinsic continuum polarization are likely to have been detected, but perhaps the data remain unpublished because of a bias toward publishing significantly polarized events.

A further dilemma in SN IIn polarization studies is the mismatch between the very few models that have been created for these particular type of objects and the observations.  \citet{2015MNRAS.449.4304D,2016MNRAS.458.1253V,2020A&A...642A.214K,Williamsinprep} obtain polarization signals of up to $\sim$ 2\% when modelling various geometries with both symmetric and asymmetric SN ejecta and diverse surrounding environments.  These models do not account for the polarization signals seen for recent objects such as SN~2017hcc with polarization as high as $\sim$ 6\%.

We present spectropolarimetric data obtained for a sample of 14 SNe IIn: SN~2010jl, SN~2011cc, PTF11iqb, SN~2011ht, SN~2012ab, SN~2009ip, SN~2014ab, Master OT J044212.20+230616.7 (M04421), ASASSN-14il, SN~2015da, SN~2015bh, PS15cwt, SN~2017gas, and SN~2017hcc, obtained over the course of $\sim 8$\,yr by the Supernova Spectropolarimetry (SNSPOL) project.\footnote{\url{http://grb.mmto.arizona.edu/~ggwilli/snspol/}} 
 Some objects have only 1 epoch of spectropolarimetry (SN~2011cc, SN~2011ht, PTF11iqb, M04421, and PS15cwt), while others have multiple epochs (as many as 11 epochs in the case of SN~2010jl).  Although the population of SNe IIn has proven itself to be polarimetrically diverse, we attempt to form a more unified picture for SNe IIn using this large spectropolarimetric data sample.

\section{Observations}
\label{sec:Obs}
\subsection{Photometry}
We reference photometry for objects available on the Open SN catalog \citep{2017ApJ...835...64G} or from the Super-LOTIS (Livermore Optical Transient Imaging System; \citealt{2008AIPC.1000..535W}) telescope to constrain the approximate date of peak absolute magnitude for all of the SNe~IIn we study herein.  Table \ref{tab:peakdates} shows the peak dates and peak absolute magnitudes (including the filter they were taken with) for each of our objects.  We use these peak dates as a reference point for every object throughout the paper.  In general, the SNe IIn within our sample show a wide diversity of peak absolute magnitudes and light curve durations.

\begin{table}
\caption{Peak Dates and Magnitudes}
\label{tab:peakdates}
\begin{tabular}{ccccc}
  \hline
SN Name & MJD & Abs. Mag. & Band & Ref. \\
\hline \hline
SN~2010jl & 55487.62 & -19.9 & V & a \\
SN~2011cc & 55757.32 & -19.0 & R & b \\
PTF11iqb & 55776.46 & -18.6 & R & c \\
SN~2011ht & 55888.37 & -17.5 & V & d \\
SN~2012ab & 55984.28 & -19.5 & R & e \\
SN~2009ip & 56207.22 & -18.5 & R & f \\
SN~2014ab & 56669.48\tablenotemark{a} & -19.5 & V & g \\
M04421 & 56954.50 & -20.1 & R & SL \\
ASASSN-14il & 56971.70 & -21.5 & V & SL \\
SN~2015da & 57162.20 & -20.8 & R & SL \\
SN~2015bh & 57166.32 & -17.9 & R & h \\
PS15cwt & 57237.51\tablenotemark{a} & -15.8 & i & PS \\
SN~2017gas & 57996.20 & -21.4 & R & SL \\
SN~2017hcc & 58091.10 & -21.4 & R & SL \\
\hline
\tablenotetext{a}{This peak date is also either the discovery date or very close to the discovery date for the object, implying that the estimate of the peak date is highly uncertain and was likely sometime before this date.}
\end{tabular}
\\Sources: SL (Super-LOTIS), PS (Pan-STARRS1) a \citep{2011ApJ...730...34S}, b \citep{2014ApJ...789..104O}, c \citep{2015MNRAS.449.1876S}, d \citep{2014ApSS.354...89B}, e \citep{2018MNRAS.475.1104B}, f \citep{2013MNRAS.430.1801M}, g \citep{2020MNRAS.498.3835B}, h \citep{2016arXiv160900731G}.
\end{table}

\subsection{Spectropolarimetry}
\label{sec:Obs_SPOL}
Our spectropolarimetric observations were obtained using the CCD Imaging/Spectropolarimeter \citep[SPOL; ][]{1992ApJ...398L..57S} on the 61'' Kuiper, 90'' Bok, and 6.5\,m MMT telescopes.  We enumerate all of our observations in detail in Table \ref{tab:Obs_SPOL}.  Observations often spanned multiple nights within a single observing run, so we combined these data into a single epoch to improve the signal-to-noise ratio in the data.

\begin{table}
\caption[Spectropolarimetric Observations]{Spectropolarimetric observations taken with the 61'' Kuiper, 90'' Bok, and 6.5\,m MMT telescopes.  Days are measured relative to the date of the peak observed magnitude for each SN.  Ap. indicates the slit width size in arcseconds for the aperture used.  Multiple slit widths are listed if multiple images were taken on the same day using different slit widths for the different exposures.  Exp. indicates exposure times for each full $Q$ or $U$ sequence at every waveplate position, so the total exposure time on the target is twice this value.}
\label{tab:Obs_SPOL}
\begin{tabular}{ccccc}
\hline \vspace{-2mm}\\
Start Time & MJD & Day & Ap.  & Exp. \\
(UTC) &  &  & ($^{\prime\prime}$) & (s)\vspace{2mm}\\
\hline \hline \vspace{-1mm}\\

\hline \vspace{-2mm}\\ \multicolumn{5}{c}{SN~2010jl\tablenotemark{a}, Epoch 1, Day 25, Kuiper} \\ \vspace{-2mm} \\
\hline
2010 Nov 10.46 & 55510.46 & 23 & 4.1 & 720 \\ 
2010 Nov 11.46 & 55511.46 & 24 & 4.1 & 480 \\ 
2010 Nov 12.46 & 55512.46 & 25 & 4.1 & 240 \\ 
2010 Nov 14.44 & 55514.44 & 27 & 4.1 & 720 \\ 
2010 Nov 15.43 & 55515.43 & 28 & 4.1 & 480 \\ \vspace{-2mm} \\

\hline \vspace{-2mm}\\ \multicolumn{5}{c}{SN~2010jl, Epoch 2, Day 45, Kuiper} \\ \vspace{-2mm} \\
\hline
2010 Dec 1.45 & 55531.45 & 44 & 4.1 & 720 \\
2010 Dec 2.44 & 55532.44 & 45 & 4.1 & 720 \\
2010 Dec 3.44 & 55533.44 & 46 & 4.1 & 800 \\ \vspace{-2mm} \\ 

\hline \vspace{-2mm}\\ \multicolumn{5}{c}{SN~2010jl, Epoch 3, Day 76, Bok} \\ \vspace{-2mm} \\
\hline
2011 Jan 2.40 & 55563.40 & 76 & 4.1 & 480 \\
2011 Jan 3.39 & 55564.39 & 77 & 4.1 & 960 \\ \vspace{-2mm} \\

\hline \vspace{-2mm}\\ \multicolumn{5}{c}{SN~2010jl, Epoch 4, Day 109, Bok} \\ \vspace{-2mm} \\
\hline
2011 Feb 2.35 & 55594.35 & 107 & 4.1 & 720 \\ 
2011 Feb 5.32 & 55597.32 & 110 & 4.1 & 720 \\ 
2011 Feb 6.33 & 55598.33 & 111 & 4.1 & 720 \\ \vspace{-2mm} \\

\hline \vspace{-2mm}\\ \multicolumn{5}{c}{SN~2010jl, Epoch 5, Day 137, Bok} \\ \vspace{-2mm} \\
\hline
2011 Mar 2.31 & 55622.31 & 135 & 4.1 & 720 \\
2011 Mar 4.27 & 55624.27 & 137 & 4.1 & 720 \\
2011 Mar 6.28 & 55626.28 & 139 & 4.1 & 720 \\ \vspace{-2mm} \\

\hline \vspace{-2mm}\\ \multicolumn{5}{c}{SN~2010jl, Epoch 6, Day 168, Kuiper} \\ \vspace{-2mm} \\
\hline
2011 Apr 4.23 & 55655.23 & 168 & 4.1 & 960 \\ \vspace{-2mm} \\

\hline \vspace{-2mm}\\ \multicolumn{5}{c}{SN~2010jl, Epoch 7, Day 221, Bok} \\ \vspace{-2mm} \\
\hline
2011 May 27.19 & 55708.19 & 221 & 4.1 & 720 \\ \vspace{-2mm} \\

\hline \vspace{-2mm}\\ \multicolumn{5}{c}{SN~2010jl, Epoch 8, Day 239, Bok} \\ \vspace{-2mm} \\
\hline
2011 Jun 14.18 & 55726.18 & 239 & 4.1 & 720 \\ 
2011 Jun 15.17 & 55727.17 & 240 & 4.1 & 720 \\ \hline\vspace{-2mm} \\
\tablenotetext{a}{All epochs of SN~2010jl data also appear in a more detailed capacity in a forthcoming paper by \citet{Williamsinprep}.}
\end{tabular}
\end{table}

\begin{table}\ContinuedFloat
\caption{continued}
\label{tab:Obs_SPOL2}
\begin{tabular}{ccccc}
\hline \vspace{-2mm}\\
Start Time & MJD & Day & Ap.  & Exp. \\
(UTC) &  &  & ($^{\prime\prime}$) & (s)\vspace{2mm}\\
\hline \hline \vspace{-1mm}\\

\hline \vspace{-2mm}\\ \multicolumn{5}{c}{SN~2010jl, Epoch 9, Day 465, Bok} \\ \vspace{-2mm} \\
\hline
2012 Jan 23.38 & 55949.38 & 462 & 3.0 & 960 \\ 
2012 Jan 25.30 & 55951.30 & 464 & 3.0 & 960 \\ 
2012 Jan 27.35 & 55953.35 & 466 & 3.0 & 960 \\ 
2012 Jan 29.31 & 55955.31 & 468 & 3.0 & 960 \\ \vspace{-2mm} \\

\hline \vspace{-2mm}\\ \multicolumn{5}{c}{SN~2010jl, Epoch 10, Day 488, Bok} \\ \vspace{-2mm} \\
\hline
2012 Feb 13.28 & 55970.28 & 483 & 3.0 & 960 \\ 
2012 Feb 18.35 & 55975.35 & 488 & 3.0 & 960 \\ 
2012 Feb 19.33 & 55976.33 & 489 & 3.0 & 960 \\ 
2012 Feb 20.39 & 55977.39 & 490 & 3.0 & 960 \\ 
2012 Feb 21.29 & 55978.29 & 491 & 3.0 & 960 \\ \vspace{-2mm} \\

\hline \vspace{-2mm}\\ \multicolumn{5}{c}{SN~2010jl, Epoch 11, Day 546, MMT} \\ \vspace{-2mm} \\
\hline
2012 Apr 16.13 & 56033.13 & 546 & 1.9 & 1440 \\ \vspace{-2mm} \\

\hline \vspace{-2mm}\\ \multicolumn{5}{c}{SN~2011cc, Epoch 1, Day 63, Bok} \\ \vspace{-2mm} \\
\hline
2011 Sep 16.22 & 	55820.22 & 63 & 3.0 & 720 \\ \vspace{-2mm} \\

\hline \vspace{-2mm}\\ \multicolumn{5}{c}{PTF11iqb, Epoch 1, Day 176, Bok} \\ \vspace{-2mm} \\
\hline
2012 Jan 26.13 & 55952.13 & 176 & 4.1 & 800 \\\vspace{-2mm} \\

\hline \vspace{-2mm}\\ \multicolumn{5}{c}{SN~2011ht, Epoch 1, Day 64, Bok} \\ \vspace{-2mm} \\
\hline
2012 Jan 23.45 & 55949.45 & 61 & 3.0 & 960 \\
2012 Jan 25.38 & 55951.38 & 63 & 3.0 & 960 \\
2012 Jan 27.41 & 55953.41 & 65 & 3.0 & 960 \\
2012 Jan 29.33 & 55955.33 & 67 & 3.0 & 960 \\ \vspace{-2mm} \\

\hline \vspace{-2mm}\\ \multicolumn{5}{c}{SN~2012ab, Epoch 1, Day 25, Bok} \\ \vspace{-2mm} \\
\hline
2012 March 23.41 & 56009.41 & 25 & 3.0 & 1200 \\ \vspace{-2mm} \\

\hline \vspace{-2mm}\\ \multicolumn{5}{c}{SN~2012ab, Epoch 2, Day 49, MMT} \\ \vspace{-2mm} \\
\hline
2012 Apr 16.38 & 56033.38 & 49 & 1.9 & 1440 \\\vspace{-2mm} \\

\hline \vspace{-2mm}\\ \multicolumn{5}{c}{SN~2009ip, Epoch 1, Day -14, MMT} \\ \vspace{-2mm} \\
\hline
2012 Sep 21.22 & 56191.22 & -16 & 1.9 & 960 \\
2012 Sep 23.23 & 56193.23 & -14 & 1.1 & 960 \\
2012 Sep 24.19 & 56194.19 & -13 & 1.5 & 1920 \\ \vspace{-2mm} \\

\hline \vspace{-2mm}\\ \multicolumn{5}{c}{SN~2009ip, Epoch 2, Day 7, Kuiper} \\ \vspace{-2mm} \\
\hline
2012 Oct 11.18 & 56211.18 & 4 & 5.1 & 1920 \\ 
2012 Oct 14.16 & 56214.16 & 7 & 5.1 & 960 \\
2012 Oct 16.16 & 56216.16 & 9 & 5.1 & 1920 \\ \hline \vspace{-2mm} \\
\end{tabular}
\end{table}

\begin{table}\ContinuedFloat
\caption{continued}
\label{tab:Obs_SPOL3}
\begin{tabular}{ccccc}
\hline \vspace{-2mm}\\
Start Time & MJD & Day & Ap.  & Exp. \\
(UTC) &  &  & ($^{\prime\prime}$) & (s)\vspace{2mm}\\
\hline \hline \vspace{-1mm}\\

\hline \vspace{-2mm}\\ \multicolumn{5}{c}{SN~2009ip, Epoch 3, Day 37, Kuiper} \\ \vspace{-2mm} \\
\hline
2012 Nov 12.13 & 56243.13 & 36 & 4.1 & 1920 \\ 
2012 Nov 14.12 & 56245.12 & 38 & 4.1 & 1920 \\ \vspace{-2mm} \\

\hline \vspace{-2mm}\\ \multicolumn{5}{c}{SN~2009ip, Epoch 4, Day 60, Bok} \\ \vspace{-2mm} \\
\hline
2012 Dec 05.07 & 56266.07 & 59 & 4.1 & 1600 \\ 
2012 Dec 06.08 & 56267.08 & 60 & 4.1 & 1600 \\
2012 Dec 07.09 & 56268.09 & 61 & 4.1 & 960 \\ \vspace{-2mm} \\

\hline \vspace{-2mm}\\ \multicolumn{5}{c}{SN~2014ab, Epoch 1, Day 77, Bok} \\ \vspace{-2mm} \\
\hline
2014 Mar 29.37 & 56745.37 & 76 & 4.1 & 1920 \\
2014 Mar 30.41 & 56746.41 & 77 & 4.1 & 960 \\
2014 Mar 31.38 & 56747.38 & 78 & 4.1 & 960 \\
2014 Apr 02.34 & 56749.34 & 80 & 4.1 & 960 \\\vspace{-2mm} \\

\hline \vspace{-2mm}\\ \multicolumn{5}{c}{SN~2014ab, Epoch 2, Day 99, MMT} \\ \vspace{-2mm} \\
\hline
2014 Apr 20.42 & 56767.42 & 98 & 1.5 & 720 \\
2014 Apr 21.42 & 56768.42 & 99 & 1.5 & 960 \\\vspace{-2mm} \\

\hline \vspace{-2mm}\\ \multicolumn{5}{c}{SN~2014ab, Epoch 3, Day 106, Kuiper} \\ \vspace{-2mm} \\
\hline
2014 Apr 26.40 & 56773.40 & 104 & 5.1 & 960 \\
2014 Apr 27.39 & 56774.39 & 105 & 4.1 & 960 \\
2014 Apr 28.41 & 56775.41 & 106 & 5.1 & 960 \\
2014 May 01.31 & 56778.31 & 109 & 5.1 & 1920 \\\vspace{-2mm} \\

\hline \vspace{-2mm}\\ \multicolumn{5}{c}{SN~2014ab, Epoch 4, Day 132, Bok} \\ \vspace{-2mm} \\
\hline
2014 May 23.33 & 56800.33 & 131 & 4.1 & 1920 \\
2014 May 24.31 & 56801.31 & 132 & 4.1 & 960 \\
2014 May 26.35 & 56803.35 & 134 & 4.1 & 960 \\\vspace{-2mm} \\

\hline \vspace{-2mm}\\ \multicolumn{5}{c}{SN~2014ab, Epoch 5, Day 162, Kuiper} \\ \vspace{-2mm} \\
\hline
2014 June 22.24 & 56830.24 & 161 & 4.1 & 1920 \\
2014 June 23.24 & 56831.24 & 162 & 4.1 & 1920 \\
2014 June 24.27 & 56832.27 & 163 & 4.1 & 960 \\
2014 June 26.25 & 56834.25 & 165 & 4.1 & 960 \\\vspace{-2mm} \\

\hline \vspace{-2mm}\\ \multicolumn{5}{c}{M04421, Epoch 1, Day 118, Kuiper} \\ \vspace{-2mm} \\
\hline
2015 Feb 19.19 & 57072.19 & 118 & 4.1 & 2880 \\\vspace{-2mm} \\

\hline \vspace{-2mm}\\ \multicolumn{5}{c}{ASASSN-14il, Epoch 1, Day -15, Kuiper} \\ \vspace{-2mm} \\
\hline
2014 Oct 25.22 & 56955.22 & -16 & 4.1 & 1920 \\
2014 Oct 28.22 & 56958.22 & -13 & 4.1 & 960  \\ \hline \vspace{-2mm} \\

\end{tabular}
\end{table}

\begin{table}\ContinuedFloat
\caption{continued}
\label{tab:Obs_SPOL4}
\begin{tabular}{ccccc}
\hline \vspace{-2mm}\\
Start Time & MJD & Day & Ap.  & Exp. \\
(UTC) &  &  & ($^{\prime\prime}$) & (s)\vspace{2mm}\\
\hline \hline \vspace{-1mm}\\

\hline \vspace{-2mm}\\ \multicolumn{5}{c}{ASASSN-14il, Epoch 2, Day 17, Kuiper} \\ \vspace{-2mm} \\
\hline
2014 Nov 27.17 & 56988.17 & 16 & 4.1 & 1920 \\
2014 Nov 28.18 & 56989.18 & 17 & 4.1 & 1920 \\\vspace{-2mm} \\

\hline \vspace{-2mm}\\ \multicolumn{5}{c}{ASASSN-14il, Epoch 3, Day 73, Bok} \\ \vspace{-2mm} \\
\hline
2015 Jan 22.13 & 57044.13 & 72 & 5.1 & 960 \\
2015 Jan 23.13 & 57045.13 & 73 & 5.1 & 960 \\\vspace{-2mm} \\

\hline \vspace{-2mm}\\ \multicolumn{5}{c}{SN~2015da, Epoch 1, Day -55, Kuiper} \\ \vspace{-2mm} \\
\hline
2015 Mar 25.37 & 57106.37 & -56 & 4.1 & 960 \\
2015 Mar 26.37 & 57107.37 & -55 & 4.1 & 2880 \\\vspace{-2mm} \\

\hline \vspace{-2mm}\\ \multicolumn{5}{c}{SN~2015da, Epoch 2, Day -33, Kuiper} \\ \vspace{-2mm} \\
\hline
2015 Apr 15.33 & 57127.33 & -35 & 4.1 & 1920 \\
2015 Apr 18.39 & 57130.39 & -32 & 4.1 & 960 \\
2015 Apr 19.39 & 57131.39 & -31 & 4.1 & 960 \\\vspace{-2mm} \\

\hline \vspace{-2mm}\\ \multicolumn{5}{c}{SN~2015da, Epoch 3, Day -22, MMT} \\ \vspace{-2mm} \\
\hline
2015 Apr 27.31 & 57139.31 & -23 & 1.9 & 2400 \\
2015 Apr 29.36 & 57141.36 & -21 & 2.8 & 1920 \\\vspace{-2mm} \\

\hline \vspace{-2mm}\\ \multicolumn{5}{c}{SN~2015da, Epoch 4, Day 4, Bok} \\ \vspace{-2mm} \\
\hline
2015 May 22.33 & 57164.33 & 2 & 4.1 & 960 \\
2015 May 24.35 & 57166.35 & 4 & 4.1 & 1920 \\
2015 May 25.34 & 57167.34 & 5 & 4.1 & 960 \\\vspace{-2mm} \\

\hline \vspace{-2mm}\\ \multicolumn{5}{c}{SN~2015bh, Epoch 1, Day -3, Bok} \\ \vspace{-2mm} \\
\hline
2015 May 19.20 & 57161.20 & -5 & 4.1 & 960 \\
2015 May 20.18 & 57162.18 & -4 & 4.1 & 960 \\
2015 May 22.23 & 57164.23 & -2 & 4.1 & 960 \\
2015 May 23.20 & 57165.20 & -1 & 4.1 & 960 \\\vspace{-2mm} \\

\hline \vspace{-2mm}\\ \multicolumn{5}{c}{SN~2015bh, Epoch 2, Day 18, MMT} \\ \vspace{-2mm} \\
\hline
2015 Jun 11.15 & 57184.15 & 18 & 1.5 & 960 \\
2015 Jun 12.15 & 57185.15 & 19 & 1.5 & 960 \\\vspace{-2mm} \\

\hline \vspace{-2mm}\\ \multicolumn{5}{c}{SN~2015bh, Epoch 3, Day 23, Kuiper} \\ \vspace{-2mm} \\
\hline
2015 Jun 14.18 & 57187.18 & 21 & 4.1 & 800 \\
2015 Jun 15.17 & 57188.17 & 22 & 4.1 & 960 \\
2015 Jun 16.18 & 57189.18 & 23 & 4.1 & 800 \\
2015 Jun 18.18 & 57191.18 & 25 & 4.1 & 720 \\
2015 Jun 19.18 & 57192.18 & 26 & 4.1 & 560 \\\hline \vspace{-2mm} \\
\end{tabular}
\end{table}

\begin{table}\ContinuedFloat
\caption{continued}
\label{tab:Obs_SPOL5}
\begin{tabular}{ccccc}
\hline \vspace{-2mm}\\
Start Time & MJD & Day & Ap.  & Exp. \\
(UTC) &  &  & ($^{\prime\prime}$) & (s)\vspace{2mm}\\
\hline \hline \vspace{-1mm}\\

\hline \vspace{-2mm}\\ \multicolumn{5}{c}{PS15cwt, Epoch 1, Day 74, Bok} \\ \vspace{-2mm} \\
\hline
2015 Oct 16.31 & 57311.31 & 74 & 4.1 & 3840 \\\vspace{-2mm} \\

\hline \vspace{-2mm}\\ \multicolumn{5}{c}{SN~2017gas, Epoch 1, Day 0, MMT} \\ \vspace{-2mm} \\
\hline
2017 Aug 30.21 & 57995.21 & -1 & 1.5 & 2400 \\
2017 Aug 31.22 & 57996.22 & 0 & 1.5 & 2400 \\\vspace{-2mm} \\

\hline \vspace{-2mm}\\ \multicolumn{5}{c}{SN~2017gas, Epoch 2, Day 20, Kuiper} \\ \vspace{-2mm} \\
\hline
2017 Sep 19.21 & 58015.21 & 19 & 4.1 & 1920 \\
2017 Sep 20.23 & 58016.23 & 20 & 4.1 & 1920 \\\vspace{-2mm} \\

\hline \vspace{-2mm}\\ \multicolumn{5}{c}{SN~2017gas, Epoch 3, Day 49, Kuiper} \\ \vspace{-2mm} \\
\hline
2017 Oct 17.20 & 58043.20 & 47 & 4.1 & 1920 \\
2017 Oct 21.16 & 58047.16 & 51 & 4.1 & 1920 \\\vspace{-2mm} \\

\hline \vspace{-2mm}\\ \multicolumn{5}{c}{SN~2017gas, Epoch 4, Day 81, Kuiper} \\ \vspace{-2mm} \\
\hline
2017 Nov 19.16 & 58076.16 & 80 & 4.1 & 960 \\
2017 Nov 20.11 & 58077.11 & 81 & 4.1 & 1920 \\
2017 Nov 22.10 & 58079.10 & 82 & 4.1 & 960 \\\vspace{-2mm} \\

\hline \vspace{-2mm}\\ \multicolumn{5}{c}{SN~2017gas, Epoch 5, Day 113, MMT} \\ \vspace{-2mm} \\
\hline
2017 Dec 22.08 & 58109.08 & 113 & 2.8 & 1920 \\\vspace{-2mm} \\

\hline \vspace{-2mm}\\ \multicolumn{5}{c}{SN~2017hcc, Epoch 1, Day -45, Kuiper} \\ \vspace{-2mm} \\
\hline
2017 Oct 17.31 & 58043.31 & -48 & 4.1 & 1600 \\
2017 Oct 20.23 & 58046.23 & -45 & 4.1 & 960 \\
2017 Oct 21.31 & 58047.31 & -44 & 5.1 & 1600 \\
2017 Oct 22.26 & 58048.26 & -43 & 5.1 & 1600 \\\vspace{-2mm} \\

\hline \vspace{-2mm}\\ \multicolumn{5}{c}{SN~2017hcc, Epoch 2, Day -15, Kuiper} \\ \vspace{-2mm} \\
\hline
2017 Nov 15.23 & 58072.23 & -19 & 5.1 & 1600 \\
2017 Nov 16.12 & 58073.12 & -18 & 5.1 & 1600 \\
2017 Nov 21.13 & 58078.13 & -13 & 4.1,5.1 & 4640 \\
2017 Nov 22.20 & 58079.20 & -12 & 4.1 & 1920 \\\vspace{-2mm} \\

\hline \vspace{-2mm}\\ \multicolumn{5}{c}{SN~2017hcc, Epoch 3, Day 9, Kuiper} \\ \vspace{-2mm} \\
\hline
2017 Dec 08.23 & 58095.23 & 4 & 7.6 & 960 \\
2017 Dec 09.12 & 58096.12 & 5 & 5.1 & 1600 \\
2017 Dec 15.19 & 58102.19 & 11 & 7.6 & 1600 \\
2017 Dec 16.12 & 58103.12 & 12 & 5.1 & 1920 \\ \hline \vspace{-2mm} \\
\end{tabular}
\end{table}

\begin{table}\ContinuedFloat
\caption{continued}
\label{tab:Obs_SPOL6}
\begin{tabular}{ccccc}
\hline \vspace{-2mm}\\
Start Time & MJD & Day & Ap.  & Exp. \\
(UTC) &  &  & ($^{\prime\prime}$) & (s)\vspace{2mm}\\
\hline \hline \vspace{-1mm}\\
\hline \vspace{-2mm}\\ \multicolumn{5}{c}{SN~2017hcc, Epoch 4, Day 17, MMT} \\ \vspace{-2mm} \\
\hline
2017 Dec 21.08 & 58108.08 & 17 & 2.8 & 960 \\
2017 Dec 22.11 & 58109.11 & 18 & 6.5 & 960 \\\vspace{-2mm} \\

\hline \vspace{-2mm}\\ \multicolumn{5}{c}{SN~2017hcc, Epoch 5, Day 41, Kuiper} \\ \vspace{-2mm} \\
\hline
2018 Jan 12.16 & 58130.16 & 39 & 7.6 & 480 \\
2018 Jan 13.13 & 58131.13 & 40 & 5.1 & 960 \\
2018 Jan 14.14 & 58132.14 & 41 & 7.6 & 960 \\
2018 Jan 15.15 & 58133.15 & 42 & 4.1 & 800 \\
2018 Jan 16.15 & 58134.15 & 43 & 5.1 & 800 \\\vspace{-2mm} \\

\hline \vspace{-2mm}\\ \multicolumn{5}{c}{SN~2017hcc, Epoch 6, Day 48, Bok} \\ \vspace{-2mm} \\
\hline
2018 Jan 19.14 & 58137.14 & 46 & 5.1 & 800 \\
2018 Jan 20.08 & 58138.08 & 47 & 4.1,7.6 & 1280 \\
2018 Jan 22.12 & 58140.12 & 49 & 7.6 & 960 \\
2018 Jan 23.14 & 58141.14 & 50 & 5.1 & 360 \\\vspace{-2mm} \\

\hline \vspace{-2mm}\\ \multicolumn{5}{c}{SN~2017hcc, Epoch 7, Day 328, Bok} \\ \vspace{-2mm} \\
\hline
2018 Oct 28.20 & 58419.20 & 328 & 3.0 & 4800 \\ \hline \vspace{-2mm} \\
\end{tabular}
\end{table}

All spectropolarimetric observations were obtained using a rotatable semi-achromatic half-wave plate to modulate incident polarization and a Wollaston prism in the collimated beam to separate the orthogonally polarized spectra onto a thinned antireflection-coated pixel $800\times1{,}200$ SITe CCD.  In order to account for detector quantum efficiency differences and pixel-to-pixel variations, we took a series of four separate exposures that sample a total of 16 different orientations of the waveplate.  Although only four waveplate positions are necessary to isolate the ordinary and extraordinary beams into Stokes $Q_1, Q_2, U_1$, and $U_2$, which accounts for variations in detector efficiency, we use a redundant set of 16 orientations to minimize instrumental variation associated with waveplate orientation.  The data are then combined using the prescription in \citet{1988MillerSPOL}.

We used the 964 lines mm$^{-1}$ grating blazed at $14.6\degree$ (4639\,{\AA}) on the MMT telescope.  In this configuration, we obtain a slit demagnification of 0.76, a dispersion of 2.62\,{\AA} per pixel, and a spectral coverage of $3140$\,{\AA}.  We used 600\,lines\,mm$^{-1}$ grating blazed at $11.35\degree$ (5819\,{\AA}) on the Kuiper and Bok telescopes.  In this configuration, we obtain a slit demagnification of 0.81, a dispersion of 4.14\,{\AA} per pixel, and a spectral coverage of $4970$\,{\AA}.  A variety of slit widths were used at each telescope, depending on weather conditions.  These exact settings can be found in Table \ref{tab:Obs_SPOL} for each specific observation.  A typical slit width (4.1$^{\prime\prime}$) at the Bok and Kuiper thus provides spectral resolution of $\sim$26 {\AA}, while a typical slit width (1.5$^{\prime\prime}$) at the MMT provides spectral resolution of $\sim$16 {\AA}.  Our analysis is restricted to a wavelength range of 4400--7000\,{\AA} to avoid spurious detections and fluctuations at the edge of our detector.  Observations at the MMT were made at the parallactic angle except in cases where this would result in significant background contamination.  Observations made at the Bok and Kuiper telescopes relied on another program using SPOL to observe active galactic nuclei, in which the rotation angle was fixed, so observations were made without regard to the parallactic angle.

A number of polarized stars (Hiltner 960, VI~Cyg~12, BD +64 106, BD +59 389, HD~245310, and HD~155528) were used to calibrate the position angle \citep{1992AJ....104.1563S}.  We found the discrepancy between the measured and the expected position angle to be $<0.2\degree$ between multiple polarimetric standard stars.  We also observed a number of unpolarized standard stars (BD +29 4211, G191B2B, and HD~212311) to verify that we had low instrumental polarization (typically $<0.1\%$) for each set of observations \citep{1990AJ.....99.1621O}.  The data were then also flux calibrated using the unpolarized standard stars.

Spectropolarimetric data reduction was performed using $IRAF$\footnote{IRAF is distributed by the National Optical Astronomy Observatory, which is operated by the Association of Universities for Research in Astronomy, Inc., under cooperative agreement with the National Science Foundation.}.  Specifically, each observation was bias subtracted, flat fielded, and wavelength calibrated (typically using He, Ne, and Ar lamp spectra). We used a fully-polarizing Nicol prism placed in the beam above the slit to correct for the efficiency of the waveplate as a function of wavelength.  When binning our data, we bin the data in $q$ and $u$ weighted by photon count first (though flux-weighting provides nearly identical results), then compute derivative properties, such as the polarization or position angle.  Throughout the paper we use two continuum wavelength bins:  5100--5700~{\AA} and 6000--6300~{\AA}. 

\subsection{Non-Polarization Spectroscopy}
In order to better constrain the ISP by using interstellar Na~\textsc{i}~D absorption line equivalent widths, we also obtained higher resolution spectra than SPOL provides.  We obtained moderate-resolution  ($R\sim 4000$) spectra using the 1200 lines mm$^{-1}$ grating in the Blue Channel (BC) spectrograph mounted on the MMT. We obtained these spectra for SN~2011cc, PTF11iqb, SN~2011ht, SN~2009ip, M04421, ASASSN-14il, PS15cwt, and SN~2017gas at times while the SNe were still bright.  All spectra were taken with the long slit at the parallactic angle.  Standard spectral reduction procedures were followed for all of the spectra.  As discussed in more detail in \S~\ref{IIn:sec:Ext}, these spectra are used for the purpose of estimating Na~\textsc{i}~D absorption line equivalent widths since these objects did not have previously determined values in the literature (or had such low estimates that previous authors chose to neglect the implied host-galaxy reddening).

\section{Background Information on Targets}
\label{sec:IndObj}
We list basic parameters for each of the SNe IIn in our sample in Table \ref{tab:basicsummary} and discuss them in more detail below.  Additionally, we summarize key published results for each object.

\begin{table}
\centering	
\caption{Basic Parameters for our SNe IIn Sample}	
\label{tab:basicsummary}
\hspace*{-1.2cm}
\begin{tabular}{lcccc}	
  \hline	
SN Name & RA & Dec & Total $A_V$ & Dist. \\
 & (J2000) &  (J2000) & & (Mpc) \\	
\hline \hline	
SN~2010jl   & $09^\mathrm{h}42^\mathrm{m}53^\mathrm{s}.33$  & $+09\degree29^{\prime}41^{\prime\prime}.8$  & 0.168 & 48.8  \\	
SN~2011cc   & $16^\mathrm{h}33^\mathrm{m}49^\mathrm{s}.44$  & $+39\degree15^{\prime}48^{\prime\prime}.7$  & 0.246 & 137.5 \\	
PTF11iqb    & $00^\mathrm{h}34^\mathrm{m}04^\mathrm{s}.84$  & $-09\degree42^{\prime}17^{\prime\prime}.9$  & 0.147 & 50.2  \\	
SN~2011ht   & $10^\mathrm{h}08^\mathrm{m}10^\mathrm{s}.58$  & $+51\degree50^{\prime}57^{\prime\prime}.1$  & 0.118 & 20.4  \\	
SN~2012ab   & $12^\mathrm{h}22^\mathrm{m}47^\mathrm{s}.63$  & $+05\degree36^{\prime}24^{\prime\prime}.83$ & 0.243 & 82.3  \\	
SN~2009ip   & $22^\mathrm{h}23^\mathrm{m}08^\mathrm{s}.26$  & $-28\degree56^{\prime}52^{\prime\prime}.4$  & 0.100 & 25.8  \\	
SN~2014ab   & $13^\mathrm{h}48^\mathrm{m}06^\mathrm{s}.05$  & $+07\degree23^{\prime}16^{\prime\prime}.12$ & 0.259 & 104.4 \\	
M04421      & $04^\mathrm{h}42^\mathrm{m}12^\mathrm{s}.20$  & $+23\degree06^{\prime}16^{\prime\prime}.7$  & 1.091 & 71.5 \\	
ASASSN-14il & $00^\mathrm{h}45^\mathrm{m}32^\mathrm{s}.55$  & $-14\degree15^{\prime}34^{\prime\prime}.6$  & 1.466 & 88.5  \\	
SN~2015da   & $13^\mathrm{h}52^\mathrm{m}24^\mathrm{s}.11$  & $+39\degree41^{\prime}28^{\prime\prime}.6$  & 3.046 & 37.0  \\	
SN~2015bh   & $09^\mathrm{h}09^\mathrm{m}35^\mathrm{s}.12$  & $+33\degree07^{\prime}21^{\prime\prime}.3$  & 0.713 & 31.3  \\	
PS15cwt     & $02^\mathrm{h}33^\mathrm{m}16^\mathrm{s}.24$  & $+19\degree15^{\prime}25^{\prime\prime}.2$  & 0.281 & 60.36 \\	
SN~2017gas  & $20^\mathrm{h}17^\mathrm{m}11^\mathrm{s}.320$ & $+58\degree12^{\prime}08^{\prime\prime}.00$ & 2.462 & 54.5  \\	
SN~2017hcc  & $00^\mathrm{h}03^\mathrm{m}50^\mathrm{s}.58$  & $-11\degree28^{\prime}28^{\prime\prime}.78$ & 0.141 & 73    \\	
\hline	
\end{tabular}	
\\A detailed summary of each of these basic parameters and the sources from which they are derived is discussed in \S~\ref{sec:IndObj}.
\end{table}

\subsubsection{SN~2010jl}	
\label{IIn:sec:Res:2010jl}	
  SN~2010jl was discovered by the Puckett Observatory Supernova Search on 2010 Nov. 3.52 (UT dates are used in this paper) at an unfiltered apparent magnitude of 13.5 \citep{2010CBET.2532....1N}.  SN~2010jl is located near the galaxy UGC 5189A (redshift $z = 0.010697$; \citealt{1999PASP..111..438F}).  We adopt a Milky Way extinction along the line of sight of $A_V=0.075\,\mathrm{mag}$ ($E_{B-V} = 0.024 \,\mathrm{mag};\,$\citealt{2011ApJ...737..103S}) and a redshift-based distance of $48.8\pm 3.5\,\mathrm{Mpc}$ from the NASA/IPAC Extragalactic Database\footnote{The NASA/IPAC Extragalactic Database (NED) is operated by the Jet Propulsion Laboratory, California Institute of Technology, under contract with the National Aeronautics and Space Administration (NASA; \url{http://ned.ipac.caltech.edu}).}[assuming H$_0=73\,\mathrm{km\,s^{-1}\,Mpc^{-1}}$ \citep{2005ApJ...627..579R} and taking into account influences from the Virgo cluster, the Great Attractor, and the Shapley supercluster, as we do for all of our targets].  We use a host-galaxy reddening of $A_V= 0.093\,\mathrm{mag}$ ($E_{B-V} = 0.030 \,\mathrm{mag}$), taken from \citet{2011A&A...527L...6P} as shown in Table \ref{tab:NaIDResults}.  The total extinction (host-galaxy and Milky Way) for SN~2010jl is $A_V = 0.168$.	

SN~2010jl shows many similarities to the bipolar geometry of $\eta$ Car--it likely arose from a luminous blue variable (LBV) detonating as a SN into a dense bipolar CSM \citep{2011ApJ...732...63S,2012AJ....143...17S,2014ApJ...797..118F}.  \citet{2011ApJ...732...63S} identified a candidate massive ($>30 M_{\odot}$) progenitor to SN~2010jl in archival \textit{Hubble Space Telescope} imaging, consistent with the LBV progenitor scenario.  However, with a more precise position from post-exposion {\it HST} imaging, \citet{2017ApJ...836..222F} find that this source is somewhat offset from the SN position and instead suggest that the progenitor was fully obscured.  A diversity of interpretations of the dust properties of SN~2010jl have emerged, some claiming it has pre-existing dust \citep{2011AJ....142...45A}, some invoking post-shock dust formation \citep{2012AJ....143...17S,2013ApJ...776....5M,2014Natur.511..326G}, and some positing no dust \citep{2012AJ....144..131Z,2014ApJ...797..118F}.  Spectropolarimetric data obtained by \citet{2011A&A...527L...6P,2019BAAA...61...90Q} show continuum polarization at $\sim$ 1.7-2\% with strong line depolarization, suggesting very low levels of ISP ($<$0.3\%), substantial asphericity, and a line forming region external to the photosphere. 

\subsubsection{SN~2011cc}	
SN~2011cc was discovered by the Lick Observatory Supernova Search on 2011 Mar. 17.52 at an unfiltered apparent magnitude of 17.7 \citep{2011CBET.2712....1M}.  SN~2011cc is located in the galaxy IC 4612 (redshift $z = 0.031895$; \citealt{2002AJ....124.1266R}).  We adopt a Milky Way extinction along the line of sight of $A_V= 0.028\,\mathrm{mag}$ ($E_{B-V} = 0.0090 \,\mathrm{mag};\,$ \citealt{2011ApJ...737..103S}) and a redshift-based distance of $137.5\pm 9.6\,\mathrm{Mpc}$ from the NASA/IPAC Extragalactic Database.  We estimate a host-galaxy reddening of $A_V= 0.22\,\mathrm{mag}$ ($E_{B-V} = 0.070 \,\mathrm{mag}$) from Na~\textsc{i}~D absorption line equivalent widths (see \S~\ref{IIn:sec:Ext} for a detailed discussion on how we estimate this) in spectra taken on day $-$66 with the BC on the MMT as shown in Table \ref{tab:NaIDResults}.  Only brief Astronomer's Telegrams discovering and then identifying SN~2011cc as a SN IIn have been published so far \citep{2011CBET.2712....2F}.  The total extinction (host-galaxy and Milky Way) for SN~2011cc is $A_V = 0.246$.	

\subsubsection{PTF11iqb}	
PTF11iqb was discovered by the Palomar Transient Factory on 2011 Jul. 23.41 at an unfiltered apparent magnitude of 16.8 \citep{2011ATel.3510....1P}. PTF11iqb is located in the galaxy NGC 151 (redshift $z = 0.012499$; \citealt{2016A&A...595A.118V}).  We adopt a Milky Way extinction along the line of sight of $A_V=0.088\,\mathrm{mag}$ ($E_{B-V} = 0.028 \,\mathrm{mag};\,$ \citealt{2011ApJ...737..103S}) and a redshift-based distance of $50.2\pm 3.5\,\mathrm{Mpc}$ from the NASA/IPAC Extragalactic Database.  We estimate a host-galaxy reddening of $A_V= 0.06\,\mathrm{mag}$ ($E_{B-V} = 0.019 \,\mathrm{mag}$) from the Na~\textsc{i}~D absorption line equivalent widths in spectra taken on day 57 with the BC on the MMT as shown in Table \ref{tab:NaIDResults}.  The total extinction (host-galaxy and Milky Way) for PTF11iqb is $A_V = 0.147$.	

PTF11iqb was spectroscopically very similar to SN~1998S \citep{2015MNRAS.449.1876S}.  Although it initially appeared as a SNe IIn, it quickly transformed into something more like a Type II-L or Type II-P SN, but with additional evidence of interaction again at late times \citep{2015MNRAS.449.1876S}.  The progenitor for this object may have been a cool giant with an extended envelope, and the early spectra showed Wolf-Rayet-like features indicative of dense slow CSM heated by a shock \citep{2015MNRAS.449.1876S}.  PTF11iqb showed extremely asymmetric line profiles at late times after $\sim$100 days, and \citet{2015MNRAS.449.1876S} proposed that the progenitor was surrounded by an inner disk that was overrun by the SN photosphere.  The expanding photosphere engulfed the disk and temporarily masked signs of CSM interaction, which were then revealed again at late times as the SN photosphere receded.

\subsubsection{SN~2011ht}	
 SN~2011ht was discovered by T. Boles on Sep. 29.182 at an unfiltered apparent magnitude of 17.0 \citep{2011CBET.2851....2P} in the galaxy UGC 5460 (redshift $z = 0.003646$; \citealt{1991rc3..book.....D}).  We adopt a Milky Way extinction along the line of sight of $A_V=0.029\,\mathrm{mag}$ ($E_{B-V} = 0.0094 \,\mathrm{mag};\,$ \citealt{2011ApJ...737..103S}) and a redshift-based distance of $20.4\pm 1.4\,\mathrm{Mpc}$ from the NASA/IPAC Extragalactic Database.  We estimate a host-galaxy reddening of $A_V= 0.09\,\mathrm{mag}$ ($E_{B-V} = 0.029 \,\mathrm{mag}$) from the Na~\textsc{i}~D1 absorption line equivalent width in spectra taken on day 57 with the BC on the MMT as shown in Table \ref{tab:NaIDResults}.  The total extinction (host-galaxy and Milky Way) for SN~2011ht is $A_V = 0.118$.

Some initial studies of SN~2011ht suggested that it may have been a SN impostor \citep{2011CBET.2851....2P}.  However, more extensive studies \citep{2011CBET.2903....1P,2013MNRAS.431.2599M} and UV observations \citep{2012ApJ...751...92R} suggested instead that SN~2011ht was indeed a core-collapse SN IIn, though subluminous due to a low ${}^{56}$Ni yield.  SN~2011ht serves as a prototype for a subclass of objects known as Type IIn-P that could arise from electron capture SNe or massive stars experiencing fallback of the SN ejecta \citep{2013MNRAS.431.2599M}.  SNe IIn with a low ${}^{56}$Ni yield might be similar to the event that originated the Crab Nebula \citep{2013MNRAS.434..102S}.  A progenitor outburst was detected at the location of SN~2011ht between 287 and 170 days prior to the discovery date, further supporting the idea that the later explosion likely ran into previously ejected CSM \citep{2013ApJ...779L...8F}.

\subsubsection{SN~2012ab}	
 SN~2012ab was discovered by the Robotic Optical Transient Search Experiment on Jan. 31.35 at an unfiltered apparent magnitude of 15.8 \citep{2012CBET.3022....1V}.  SN~2012ab is located in the galaxy 2MASX J12224762+0536247 (redshift $z = 0.018$; \citealt{2014ApJS..210....9B}).  We adopt a Milky Way extinction along the line of sight of $A_V=0.057\,\mathrm{mag}$ ($E_{B-V} = 0.018 \,\mathrm{mag};\,$ \citealt{2011ApJ...737..103S}) and a redshift-based distance of $82.3\pm 5.8\,\mathrm{Mpc}$ from the NASA/IPAC Extragalactic Database.  We use a host-galaxy reddening of $A_V= 0.19\,\mathrm{mag}$ ($E_{B-V} = 0.060 \,\mathrm{mag}$), taken from \citet{2018MNRAS.475.1104B} as shown in Table \ref{tab:NaIDResults}.  The total extinction (host-galaxy and Milky Way) for SN~2012ab is $A_V = 0.243$.	

Spectroscopy of SN~2012ab suggests that the SN ejecta interact mostly with bluesifted CSM on the near side of the SN at early times, but then transition to having increased shock interaction with redshifted CSM on the far side of the SN at later times \citep{2018MNRAS.475.1104B,2020MNRAS.499..129G}.  Spectropolarimetry of SN~2012ab shows an initial polarization of 1.7\% at early times that rises to 3.5\% about 24 days later, which is around the same time that the receding CSM interaction began.	

\subsubsection{SN~2009ip}	
\label{IIn:sec:Res:2009ip}	
After already being known as a SN impostor transient since 2009 \citep{2010AJ....139.1451S}, SN~2009ip was then discovered in yet another outburst on 2012 Jul. 24 by the Catalina Real-Time Transient Survey SN Hunt project, marking the start if its final rebrightening event \citep{2012ATel.4334....1D}.  SN~2009ip is located in the galaxy NGC 7259 (redshift $z = 0.005944$; \citealt{2006MNRAS.371.1855W}).  We adopt a Milky Way extinction along the line of sight of $A_V=0.054\,\mathrm{mag}$ ($E_{B-V} = 0.017 \,\mathrm{mag};\,$ \citealt{2011ApJ...737..103S}) and a redshift-based distance of $25.8\pm 1.8\,\mathrm{Mpc}$ from the NASA/IPAC Extragalactic Database.  We estimate a host-galaxy reddening of $A_V= 0.05\,\mathrm{mag}$ ($E_{B-V} = 0.015 \,\mathrm{mag}$) from the Na~\textsc{i}~D2 absorption line equivalent width in spectra taken on day 8 with the BC on the MMT as shown in Table \ref{tab:NaIDResults}.  The total extinction (host-galaxy and Milky Way) for SN~2009ip is $A_V = 0.100$.	

SN~2009ip is a unique SN in that it was studied extensively before explosion.  The initial event from 2009 was quickly categorized as an outburst from an LBV showing variability in the prior decade \citep{2010AJ....139.1451S}.  A detection in archival HST images also revealed the presence of a quiescent progenitor star \citep{2010AJ....139.1451S,2011ApJ...732...32F}.  Then, in 2012, SN~2009ip garnered much more attention when it resurfaced with two connected brigthening events \citep{2012ATel.4334....1D,2012ATel.4412....1S,2012ATel.4423....1B,2013ApJ...763L..27P}.  Although the terminal nature of these rebrightening events was contested \citep{2013ApJ...767....1P}, SN~2009ip has since faded to levels  below that of the progenitor, confirming that it was a true core-collapse SN \citep{smith22}.  
\citet{2013MNRAS.434.2721S} found evidence for pre-SN CSM dust in early near-infrared spectroscopy of SN~2009ip, while comparing the observed evolution of the light curve and spectra to models suggested that SN~2009ip was the initially faint explosion of a blue supergiant much like SN~1987A, except with much stronger CSM interaction at peak \citep{2014MNRAS.438.1191S}.

The polarization of SN~2009ip has been studied in detail.  \citet{2014MNRAS.442.1166M} measured a $V$-band polarization of $\sim$ 0.9\% at a position angle of $\theta \sim 166\degree$ during the 2012a event, transitioning to a polarization of $\sim$ 1.7\% at a position angle of $\theta \sim 72\degree$ during the 2012b event, and then fading thereafter with further changes in the position angle.  The evolution for SN~2009ip was interpreted to have arisen from an initially prolate explosion seen in the 2012a event colliding with an oblate CSM distribution during the 2012b event \citep{2014MNRAS.442.1166M}.  \citet{2017MNRAS.470.1491R} looked at the spectropolarimetric evolution of specific line features observed for SN~2009ip and found that an inclined disk-like CSM best explained the absorption features along with evolution of the position angle seen in SN~2009ip.	

\subsubsection{SN~2014ab}	
SN~2014ab was discovered by the Catalina Sky Survey on 2014 Mar. 9.43 at an apparent $V$-band magnitude of 16.4 ($M_{V} = -19.0$\,mag) \citep{2014CBET.3826....1H}.  SN~2014ab is located in the galaxy  VV~306c (redshift $z = 0.023203$; \citealt{1959VV....C......0V,1999PASP..111..438F}).  We adopt a Milky Way extinction along the line of sight of $A_V=0.083\,\mathrm{mag}$ ($E_{B-V} = 0.027 \,\mathrm{mag};\,$ \citealt{2011ApJ...737..103S}) and a redshift-based distance of $104.4\pm 7.3\,\mathrm{Mpc}$ from the NASA/IPAC Extragalactic Database.  We use a host-galaxy reddening of $A_V= 0.18\,\mathrm{mag}$ ($E_{B-V} = 0.057 \,\mathrm{mag}$), taken from \citet{2018MNRAS.475.1104B} as shown in Table \ref{tab:NaIDResults}.  The total extinction (host-galaxy and Milky Way) for SN~2014ab is $A_V = 0.259$.

SN~2014ab was found to exhibit many spectral properties similar to that of SN~2010jl \citep{2020A&A...641A.148M,2020MNRAS.498.3835B}.   In particular, spectra of SN~2014ab showed blueshifted intermediate-width components indicative of either an optically thick CSM occulting the far side, obscuration by large dust grains, or inherent asymmetry along our line of sight.  \citet{2020A&A...641A.148M} found evidence of pre-existing dust within the CSM around SN~2014ab.  Spectropolarimetric data presented in \citet{2020MNRAS.498.3835B}, which are also presented in this work, reveal small levels of instrinsic polarization for SN~2014ab, suggesting a mostly symmetric photosphere in the plane of the sky.

\subsubsection{M04421}	
M04421 was discovered by the MASTER Global Robotic Net on 2014 Sep. 20.81259 at an unfiltered apparent magnitude 15.4 \citep{2014ATel.6484....1T}.  We estimate a redshift of $z = 0.01717$ from the narrow component of H$\alpha$ emission detected in day -9 spectra taken with the BC on the MMT.  We adopt a Milky Way extinction along the line of sight of $A_V=1.091\,\mathrm{mag}$ ($E_{B-V} = 0.352 \,\mathrm{mag};\,$ \citealt{2011ApJ...737..103S}).  Since the estimated redshift is very different from that of the claimed host galaxy, 2MASX J04421256+2306209, we instead estimate a redshift-based distance of $71.5\,\mathrm{Mpc}$ using our estimate of the redshift.  We estimate a host-galaxy reddening of $A_V < 0.10\,\mathrm{mag}$ ($E_{B-V} < 0.032 \,\mathrm{mag}$) from an upper limit on the Na~\textsc{i}~D absorption line equivalent widths in spectra taken on day -9 with the BC on the MMT as shown in Table \ref{tab:NaIDResults}.  Although we do not correct the data using this reddening estimate, we do use it to set a rough limit on the ISP inferred from Na~\textsc{i}~D as discussed in \S~\ref{IIn:sec:Ext}.  The total extinction (host-galaxy and Milky Way) for M04421 is $A_V = 1.091$.  Only brief Astronomer's Telegrams discovering and then identifying M04421 as a SN IIn have been published so far \citep{2014ATel.6484....1T,2014ATel.6487....1S}.	

\subsubsection{ASASSN-14il}	
ASASSN-14il was discovered by the All Sky Automated Survey for SuperNovae on 2014 Oct. 1.11 at an apparent $V$-band magnitude of 16.5 \citep{2014ATel.6525....1B}.  ASASSN-14il is located in the galaxy 2MASX J00453260-1415328 (redshift $z = 0.021989$; \citealt{2009MNRAS.399..683J}).  We adopt a Milky Way extinction along the line of sight of $A_V=0.061\,\mathrm{mag}$ ($E_{B-V} = 0.020 \,\mathrm{mag};\,$ \citealt{2011ApJ...737..103S}) and a redshift-based distance of $88.5\pm 6.2\,\mathrm{Mpc}$ from the NASA/IPAC Extragalactic Database.  We estimate a host-galaxy reddening of $A_V= 1.40\,\mathrm{mag}$ ($E_{B-V} = 0.453 \,\mathrm{mag}$) from the Na~\textsc{i}~D absorption line equivalent widths in spectra taken on day -26 with the BC on the MMT as shown in Table \ref{tab:NaIDResults}.  The total extinction (host-galaxy and Milky Way) for ASASSN-14il is $A_V = 1.466$.  Only brief Astronomer's Telegrams discovering and then identifying ASASSN-14il as a SN IIn have been published so far \citep{2014ATel.6525....1B,2014ATel.6536....1C}, but a more in-depth study of this super-luminous SN~IIn is forthcoming in \citet{Dickinsoninprep}.

\subsubsection{SN~2015da}	
SN~2015da was discovered by the Xingming Sky Survey on 2015 Jan. 9.89694 at an unfiltered apparent magnitude of 16.9\footnote{\url{http://www.cbat.eps.harvard.edu/unconf/followups/J13522411+3941286.html}}.  SN~2015da is located near the galaxy NGC 5337 (redshift $z = 0.007222$; \citealt{1999PASP..111..438F}).  We adopt a Milky Way extinction along the line of sight of $A_V=0.039\,\mathrm{mag}$ ($E_{B-V} = 0.013 \,\mathrm{mag};\,$ \citealt{2011ApJ...737..103S}) and a redshift-based distance of $37.0\pm 2.6\,\mathrm{Mpc}$ from the NASA/IPAC Extragalactic Database.  We use a host-galaxy reddening of $A_V= 3.01\,\mathrm{mag}$ ($E_{B-V} = 0.97 \,\mathrm{mag}$), taken from \citet{2020A&A...635A..39T} as shown in Table \ref{tab:NaIDResults}.  The total extinction (host-galaxy and Milky Way) for SN~2015da is $A_V = 3.046$.	
\citet{2020A&A...635A..39T} observe narrow Balmer lines indicative of SN ejecta interacting with CSM continuously over the course of 4 years.

\subsubsection{SN~2015bh}	
SN~2015bh was discovered by the Catalina Real-Time Transient Survey on 2015 Feb. 7.39 and imaged within a day at an apparent $V$-band magnitude of 19.9\footnote{\url{http://www.cbat.eps.harvard.edu/iau/cbet/004200/CBET004229.txt}}.  SN~2015bh is located in the galaxy NGC 2770 (redshift $z = 0.006494$; \citealt{2005ApJS..160..149S}).  We adopt a Milky Way extinction along the line of sight of $A_V=0.062\,\mathrm{mag}$ ($E_{B-V} = 0.020 \,\mathrm{mag};\,$ \citealt{2011ApJ...737..103S}) and a redshift-based distance of $31.3\pm 2.2\,\mathrm{Mpc}$ from the NASA/IPAC Extragalactic Database.  We use a host-galaxy reddening of $A_V= 0.65\,\mathrm{mag}$ ($E_{B-V} = 0.21 \,\mathrm{mag}$), taken from \citet{2017A&A...599A.129T} as shown in Table \ref{tab:NaIDResults}.  The total extinction (host-galaxy and Milky Way) for SN~2015bh is $A_V = 0.713$.	

SN~2015bh shows many similarities to SN~2009ip.  Initially, SN~2015bh was classified as an SN impostor.  Spectroscopic studies showed that SN~2015bh began to interact with CSM not long after its first brightening event in 2015 \citep{2016ApJ...824....6O,2016arXiv160900731G,2016MNRAS.463.3894E,2018A&A...617A.115B}.  The initial brightening in 2015 may have been an actual faint SN core collapse event with the second brightening being due to the onset of CSM interaction \citep{2016MNRAS.463.3894E}, much as is hypothesized for SN~2009ip  \citep{2014MNRAS.438.1191S}.  Pre-explosion observations reveal an LBV undergoing outbursts over the last $\sim$ 20\,yr prior to explosion \citep{2016ApJ...824....6O,2016MNRAS.463.3894E,2017A&A...599A.129T}.

\subsubsection{PS15cwt}	
Other aliases for PS15cwt include PSN J02331624+1915252 and CSS150920:023316+191525 \citep{2017ApJ...835...64G}.  PS15cwt was discovered by Puckett et al. on 2015 Aug. 21.4002 at an unfiltered apparent magnitude of 16.2\footnote{\url{http://www.cbat.eps.harvard.edu/unconf/followups/J02331624+1915252.html}}.  \citet{2015ATel.7955....1S} estimate a redshift of $z = 0.0135$ from early spectra.  We adopt a Milky Way extinction along the line of sight of $A_V=0.281\,\mathrm{mag}$ ($E_{B-V} = 0.091 \,\mathrm{mag};\,$ \citealt{2011ApJ...737..103S}) and a redshift-based distance of $60.36\,\mathrm{Mpc}$.  We estimate a host-galaxy reddening of $A_V < 0.08\,\mathrm{mag}$ ($E_{B-V} < 0.026 \,\mathrm{mag}$) from an upper limit on the Na~\textsc{i}~D absorption line equivalent widths in spectra taken on day 73 with SPOL on the Bok telescope as shown in Table \ref{tab:NaIDResults}.  Although we do not correct the data using this reddening estimate, we do use it to set a rough limit on the ISP inferred from Na~\textsc{i}~D as discussed in \S~\ref{IIn:sec:Ext}.  The total extinction (host-galaxy and Milky Way) for PS15cwt is $A_V = 0.281$.  Only brief Astronomer's Telegrams discovering and then identifying PS15cwt as a SN IIn have been published so far \citep{2015ATel.7955....1S}.	

\subsubsection{SN~2017gas}	
SN~2017gas was discovered by the All Sky Automated Survey for SuperNovae on 2017 Aug. 10.41 at an apparent $V$-band magnitude of 16.0 \citep{2017ATel10652....1B}. SN~2017gas is located in the galaxy 2MASX J20171114+5812094 (redshift $z = 0.01$; \citealt{2017ATel10669....1B}).  We estimate a redshift of $z = 0.0106$ from the narrow component of H$\alpha$ emission detected in day 39 spectra taken with the BC on the MMT.  We adopt a Milky Way extinction along the line of sight of $A_V=1.066\,\mathrm{mag}$ ($E_{B-V} = 0.344 \,\mathrm{mag};\,$ \citealt{2011ApJ...737..103S}) and a redshift-based distance of $54.5\pm 3.8\,\mathrm{Mpc}$ from the NASA/IPAC Extragalactic Database.  We estimate a host-galaxy reddening of $A_V= 1.40\,\mathrm{mag}$ ($E_{B-V} = 0.450 \,\mathrm{mag}$) from the Na~\textsc{i}~D absorption line equivalent widths in spectra taken on day 39 with the BC on the MMT as shown in Table \ref{tab:NaIDResults}.  The total extinction (host-galaxy and Milky Way) for SN~2017gas is $A_V = 2.462$.  Only brief Astronomer's Telegrams discovering and then identifying SN~2017gas as a SN IIn have been published so far \citep{2017ATel10652....1B,2017ATel10669....1B}.

\subsubsection{SN~2017hcc}
\label{IIn:sec:Res:2017hcc}
SN~2017hcc was discovered by the Asteroid Terrestrial-impact Last Alert System on 2017 Oct. 2.38 at an orange filter apparent magnitude of 17.44 \citep{2017TNSTR1070....1T}.  SN~2017hcc is located in an anonymous galaxy (redshift $z = 0.0173$; \citealt{2017ATel11015....1N,2017RNAAS...1...28P}).  We adopt a Milky Way extinction along the line of sight of $A_V=0.091\,\mathrm{mag}$ ($E_{B-V} = 0.029 \,\mathrm{mag};\,$ \citealt{2011ApJ...737..103S}) and a redshift-based distance of $73\,\mathrm{Mpc}$ \citep{2017RNAAS...1...28P}.  We use a host-galaxy reddening of $A_V= 0.050\,\mathrm{mag}$ ($E_{B-V} = 0.016 \,\mathrm{mag}$), taken from \citet{2020MNRAS.499.3544S} as shown in Table \ref{tab:NaIDResults}.  The total extinction (host-galaxy and Milky Way) for SN~2017hcc is $A_V = 0.141$.

SN~2017hcc is of particular interest because it shattered records for polarization measurements of all types of SNe, not just SNe IIn.  \citet{2017ATel10911....1M} measured an integrated $V$-band continuum polarization of 4.84\%.  They also estimated a low contribution from the ISP based on a high Galactic latitude, small extinction in both the Milky Way and the host galaxy, and strong line depolarization in the core of H$\alpha$ and H$\beta$.  \citet{2019MNRAS.488.3089K} also find low host-galaxy extinction and measure a decline of the intrinsic polarization of SN~2017hcc of $\sim$ 3.5\% over $\sim$ 2 months.  Both studies suggest an origin of the continuum polarization in a region with significant asymmetry, such as a toroidal or disk-like CSM \citep{2017ATel10911....1M,2019MNRAS.488.3089K}.  \citet{2017ATel10911....1M} suggest that unpolarized line emission arises in the photoionized pre-shock CSM, consistent with narrow line components seen at early times in spectra reported by \citet{2020MNRAS.499.3544S}.  \citet{2020MNRAS.499.3544S} also found evidence of dust formation in the post-shock shell and within the SN ejecta for SN~2017hcc. The very high early polarization and the polarization evolution will be discussed in more detail in a separate paper \citep{Mauerhaninprep}.

\section{Results}
\label{IIn:sec:Res}
\subsection{Extinction and Reddening}
\label{IIn:sec:Ext}
When available, we reference past studies on individual SNe to obtain an estimate of the host-galaxy extinction along our line of sight.  Individual results are mentioned in the sections corresponding to that object within \S~\ref{sec:IndObj} and are summarized in Table \ref{tab:NaIDResults}.  We assume a total to selective absorption ratio of $R_{V} = 3.1$ \citep{1994ApJ...422..158O}, though this may not be true everywhere within the Milky Way Galaxy, nor in other host galaxies.  Reddening estimates often utilize the correlation found between the narrow Na~\textsc{i}~D absorption lines $\lambda\lambda$5890 (D2), 5896 (D1) and the interstellar dust extinction along a particular line of sight, though the correlation requires that the lines are not saturated and not blended in moderate-resolution spectra \citep{2012MNRAS.426.1465P}.  A number of studies have examined the correlation between the equivalent width of the Na~\textsc{i}~D doublet absorption lines and interstellar extinction \citep{1994AJ....107.1022R,1997A&A...318..269M,2003fthp.conf..200T,2012MNRAS.426.1465P}.  We use the relations provided by \citet{2012MNRAS.426.1465P} in all of our estimations in \S~\ref{sec:IndObj}.  \citet{2013ApJ...779...38P} found that the dust-extinction values estimated from the Na~\textsc{i}~D doublet absorption for one-fourth of their sample of SNe~Ia was stronger than expected when compared to those derived from SN colour.  

\begin{figure}
\centering
\includegraphics[width=0.5\textwidth,height=1\textheight,keepaspectratio,clip=true,trim=0cm 0cm 0cm 0cm]{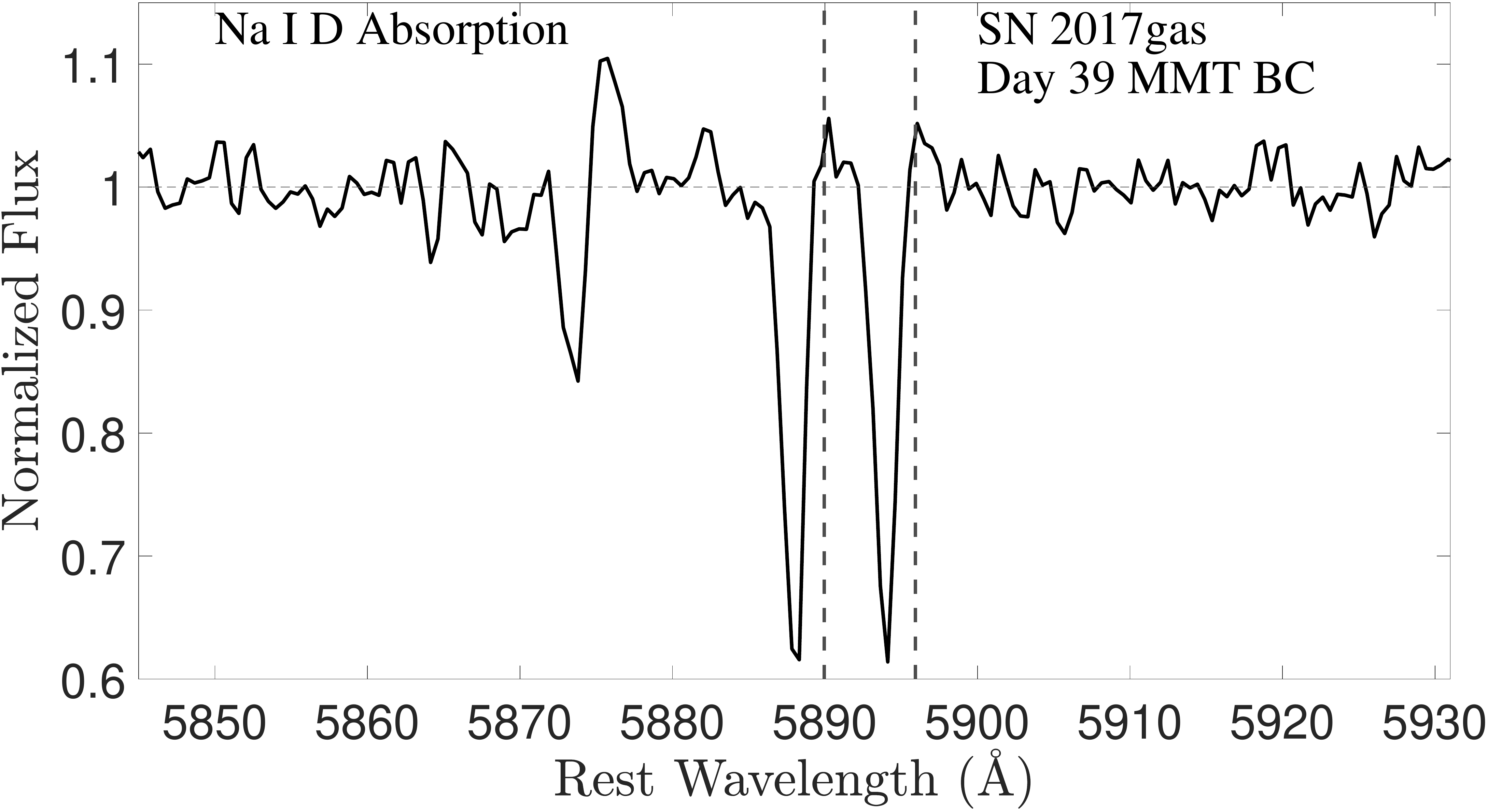}
\caption[Host-galaxy Na~\textsc{i}~D absorption for SN~2017gas]{A BC spectrum ($R\sim 4000$) of SN~2017gas taken on day 39 at the MMT showing the Na~\textsc{i}~D absorption doublet that we use to constrain host-galaxy extinction.  Since the spectrum has been redshift-corrected using the average redshift of the host galaxy and then aligned to the narrow component of H$\alpha$ emission, the observed Na~\textsc{i}~D absorption lines are offset slightly ($\sim 94\pm24 $\,km\,$\mathrm{s^{-1}}$) from their rest wavelengths due to host-galaxy rotation.}
\label{fig:IIn:NaID}
\end{figure}

For several objects, either no literature estimate of the host-galaxy extinction existed (SN~2011cc, PTF11iqb, M04421, ASASSN-14il, PS15cwt, and SN~2017gas) or the host-galaxy extinction was deemed from Na~\textsc{i}~D upper limits to be low enough that it could be neglected (SN~2009ip: \citealt{2014MNRAS.442.1166M} and SN~2011ht: \citealt{2012ApJ...751...92R}).  In the cases that moderate-resolution spectra were available from the BC on the MMT (SN~2011cc, SN~2011ht, PTF11iqb, SN~2009ip,  M04421, ASASSN-14il, and SN~2017gas), we measured the equivalent width of the Na~\textsc{i}~D absorption lines in order to estimate the dust-extinction values using the relations described above.  When absorption was not detectable, we used the 1$\sigma$ noise level to derive an upper limit to the strength of the Na~\textsc{i}~D absorption doublet.  Figure \ref{fig:IIn:NaID} shows one of our moderate-resolution spectra used to estimate the equivalent width of the Na~\textsc{i}~D absorption lines.  No moderate-resolution spectra were available for PS15cwt, so we attempted to estimate the host-galaxy extinction using our relatively low-resolution Bok SPOL data.  Since no absorption lines were clearly present in this data, we instead set an upper limit to the equivalent width of the Na~\textsc{i}~D absorption doublet.  All of these spectroscopic observations are detailed in Table \ref{tab:NaIDResults}.

\begin{table}
\caption[Host-galaxy Extinction]{Estimates of host-galaxy extinction using Na~\textsc{i}~D line absorption equivalent widths.  Day is measured relative to peak.}
\label{tab:NaIDResults}
\begin{tabular}{lcccc}
  \hline
SN Name & Tel./Instr. & Day & $A_{V}$\tablenotemark{a} & Source\tablenotemark{b} \\
\hline \hline
SN~2010jl & TNG/SARG & -19 & 0.09 & D2,a \\
SN~2011cc & MMT/BC & -66 & 0.22 & D1 \& D2 \\
PTF11iqb & MMT/BC & 57 & 0.06 & D1 \& D2 \\
SN~2011ht & MMT/BC & 57 & 0.09 & D1 \\
SN~2012ab & HET/LRS & 3 & 0.19 & b \\
SN~2009ip & MMT/BC & 8 & 0.05 & D2 \\
SN~2014ab & VLT/X-shooter & 70 & 0.18 & D1 \& D2,c \\
M04421 & MMT/BC & -9 & $<0.10$ & Noise Limit \\
ASASSN-14il & MMT/BC & -26 & 1.40 & D1 \& D2 \\
SN~2015da & Keck/DEIMOS & 86 & 0.97 & d \\
SN~2015bh & GTC/OSIRIS & -11 & 0.21 & D1 \& D2,e \\
PS15cwt & Bok/SPOL & 73 & $<0.08$ & Noise Limit \\
SN~2017gas & MMT/BC & 39 & 1.40 & D1 \& D2 \\
SN~2017hcc & Mag/MIKE & 40 & 0.05 & D1 \& D2,f \\
\hline
\end{tabular}
\tablenotetext{a}{If the source of $A_{V}$ is from another paper that only quoted $E_{B-V}$, we convert it using $A_{V} = 3.1E_{B-V}$ \citep{1994ApJ...422..158O}.}
\tablenotetext{b}{Sources: a \citep{2011A&A...527L...6P}, b \citep{2018MNRAS.475.1104B}, c \citep{2020MNRAS.498.3835B}, d \citep{2020A&A...635A..39T}, e \citep{2017A&A...599A.129T}, f \citep{2020MNRAS.499.3544S}.}
\end{table}

\subsection{Spectropolarimetry Parameters}
\label{IIn:sec:Res:Specpol}
Our spectropolarimetric analysis is performed primarily using the linear Stokes parameters, $q=Q/I$ and $u=U/I$, which are rotated 45$\degree$ with respect to each other, allowing us to decompose the polarization signal into orthogonal components in position angle space. Typically, one can combine the Stokes parameters to obtain the polarization level, $p = \sqrt{q^2 + u^2}$, and the position angle on the sky, $\theta = (1/2)\, \mathrm{tan}^{-1} (u/q)$.  However, since the definition of the polarization makes it a positive-definite value, it may seem artificially high in cases where we have a low signal-to-noise ratio because fluctuations will raise the mean polarization level significantly.  In this section, we discuss alternatives for the traditional definition of polarization.

Given the issue of the positively-based nature of the traditional definition of polarization when the signal-to-noise ratio is low, we instead consider a few alternative formulations to describe the polarization.  First, the debiased polarization \citep{1978ApJ...220L..67S}:

\begin{equation}
p_{db} = \pm\sqrt{\lvert q^{2} + u^{2} - (\sigma^{2}_{q} + \sigma^{2}_{u})\rvert}
\end{equation}

\noindent where $\sigma_q$ and $\sigma_u$ are the statistical errors in the measurements of $q$ and $u$, respectively, and the sign of $p_{db}$ is chosen to match the sign of $[q^2 + u^2 - (\sigma^2_q + \sigma^2_u)]$.  Next, we consider the optimal polarization \citep{1997ApJ...476L..27W}:

\begin{equation}
p_{opt} = p - \frac{\sigma^{2}_{p}}{p}
\end{equation}

\noindent where $\sigma_p$ is the statistical error on the polarization propagated from $\sigma_q$ and $\sigma_u$.  Lastly, we consider the rotated stokes parameters (RSP; \citealt{1993ApJ...402..249T}):

\begin{equation}
\begin{aligned}
q_{RSP} = q\, \mathrm{cos}(2\theta_{\rm smooth}) + u\, \mathrm{sin}(2\theta_{\rm smooth}) \\
u_{RSP} = -q\,\mathrm{sin}(2\theta_{\rm smooth}) + u\, \mathrm{cos}(2\theta_{\rm smooth})
\end{aligned}
\end{equation}

\noindent where $\theta_{smooth}$ is chosen such that the majority of the polarization signal is rotated onto $q_{RSP}$ \citep{2001ApJ...553..861L}.  As discussed in detail in \citet{2001ApJ...553..861L}, each estimation of polarization has its limitations, but the traditional definition of polarization and the debiased definition perform worse than the optimal polarization at estimating the true polarization in a simulation of low signal-to-noise ratio polarization spectra across emission lines.  In particular, the traditional polarization fails to detect an unpolarized line feature in the simulation, as is expected when the signal-to-noise ratio drops low.  The debiased polarization contains a double-peaked probability distribution and results in negative polarization spikes when the signal-to-noise ratio is low \citep{1988MillerSPOL}.  While the optimal polarization is formally undefined for values where $p_{trad} < \sigma_{p_{trad}}$, this formulation still results in less negative polarization spikes than the debiased polarization if this restriction on the formal definition is neglected (as we choose to do throughout our paper) and also matches the true polarization signal in the simulation of \citet{2001ApJ...553..861L} more accurately.   

\begin{figure*}
\centering
\includegraphics[width=1.0\textwidth,height=1\textheight,keepaspectratio,clip=true,trim=0cm 0cm 0cm 0cm]{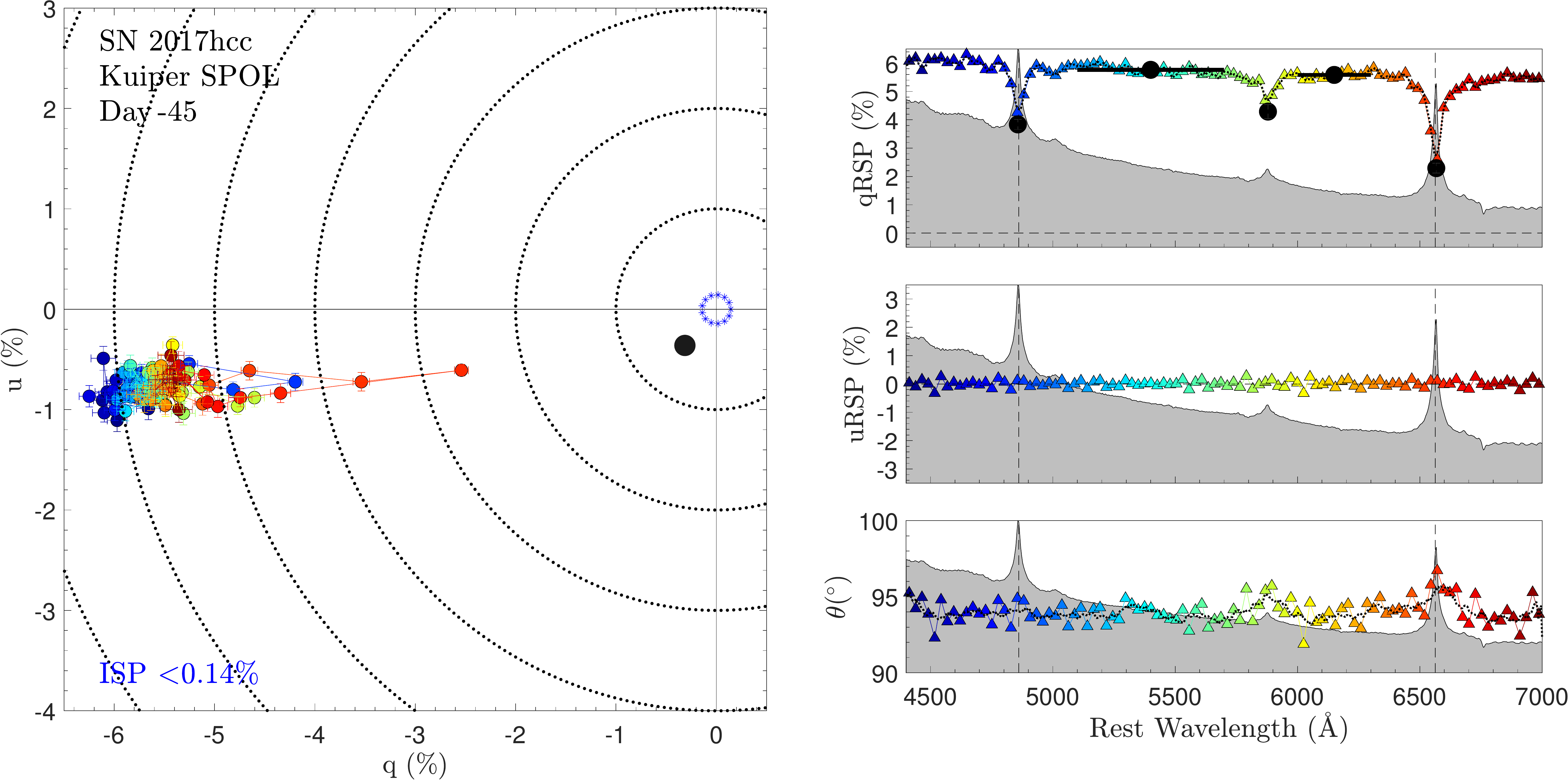}
\caption[$q-u$ plot for SN~2017hcc on day -45 without ISP correction]{A $q-u$ plot showing the spectropolarimetric data for SN~2017hcc on day -45 without ISP correction.  On the right we also show $q_{RSP}$, $u_{RSP}$, and the position angle $\theta$.  The dotted black line in the $q$RSP plot indicates the smoothed optimal polarization.  The dotted black line in the position angle plot indicates the smoothed position angle, $\theta_{smooth}$.  The solid black points in the $q_{RSP}$ plot indicate the optimal polarization values measured across the continuum regions designated by the black horizontal bars.  If an estimate of the polarization was made at the wavelength of a narrow-component of an emission line, we show this estimate with another solid black point at the location of the line (H$\alpha$, H$\beta$, or He~\textsc{i} $\lambda$5876).  Shaded regions show a scaled flux spectrum from the same day.  Vertical dashed lines are used to indicate the wavelengths of H$\alpha$ and H$\beta$.  Horizontal dashed lines are included in the $q_{RSP}$ and $u_{RSP}$ plots for clarity.  Black dotted circles demark each integer value of polarization in the $q-u$ plot for clarity.  Colours, bins, and error bars in the $q-u$ plot on the left correspond to those on the right, with the colors mapped to the wavelength axis labeled on the right.  We estimate an ISP magnitude $<0.14$ (shown as a circle of blue asterisks) based on Na~\textsc{i}~D absorption-line measurements (see \S~\ref{IIn:sec:Ext}).  We mark our estimate of the actual ISP derived from depolarization of H$\alpha$ narrow emission with a black circle in the $q-u$ plot (see \S~\ref{sec:ISPdepol}).  The data are grouped into $\sim$28~{\AA} bins.} 
\label{fig:QU_example}
\end{figure*}

\begin{figure}
\centering
\includegraphics[width=0.47\textwidth,height=1\textheight,keepaspectratio,clip=true,trim=0cm 0cm 0cm 0cm]{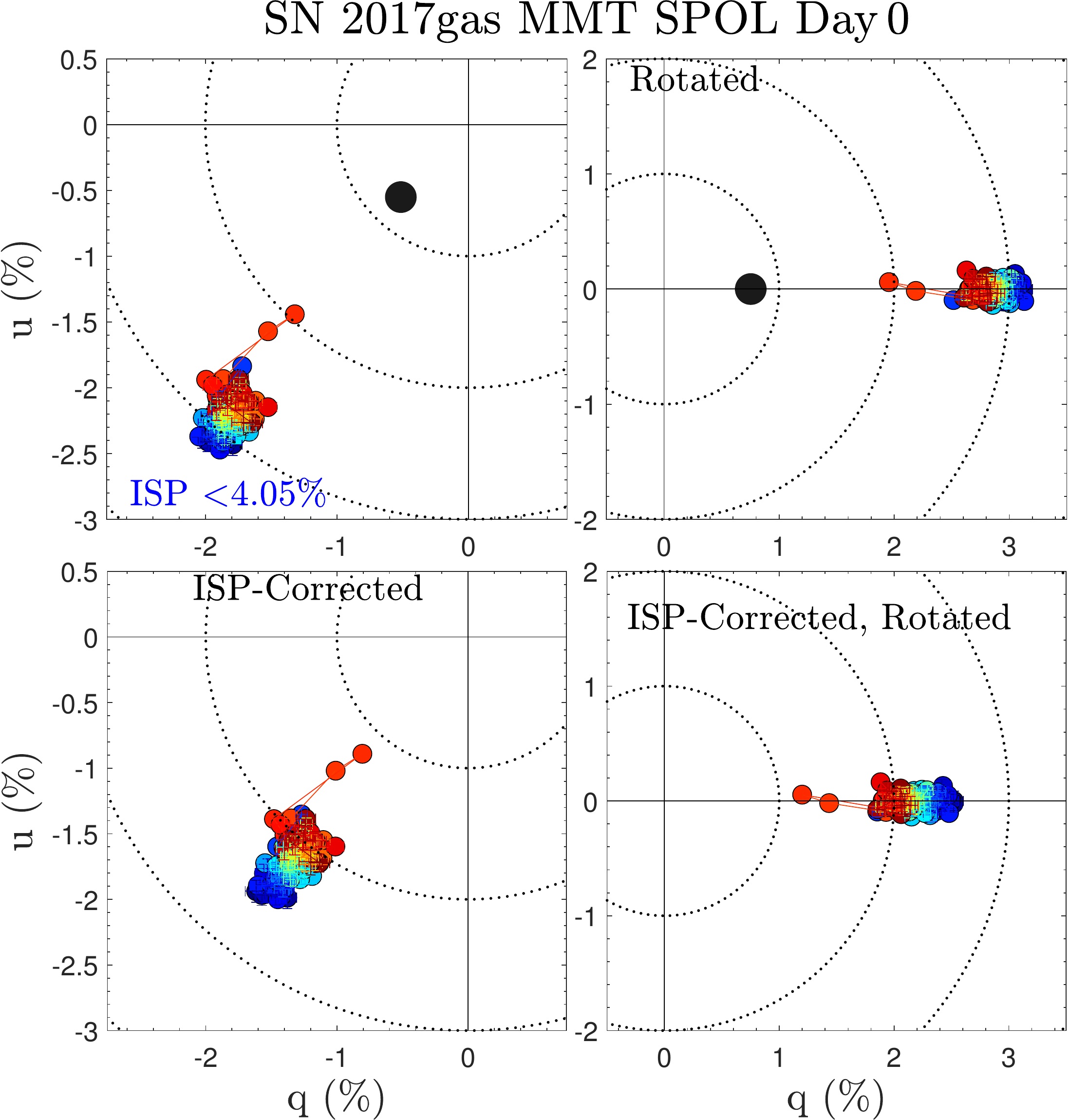}
\caption[$q-u$ plots of SN~2017gas on day 0 with and without rotation and ISP-correction]{A $q-u$ plot showing the spectropolarimetric data for SN~2017gas on day 0.  The solid black circle represents the estimate of the ISP at the wavelength of H$\alpha$.  The top left panel shows the raw $q-u$ data after correcting only for reddening and redshift.  The top right panel shows the data rotated according to the RSP prescription, discussed in \S~\ref{IIn:sec:Res:Specpol}.  The bottom left panel shows the $q-u$ data after correcting for the estimate of the ISP at all wavelengths.  The bottom right panel shows the $q-u$ data after correcting for the estimate of the ISP at all wavelengths and then rotated according to the RSP prescription.}
\label{fig:RSPISPcomp}
\end{figure}

For these reasons, when determining the average polarization over a large bandwidth (see \citealt{2001ApJ...553..861L} for a detailed discussion of the advantages of $p_{opt}$ when binning data over large wavelengths), we take the photon-count weighted average of our data and then compute $p_{opt}$.  We also make use of $p_{opt}$ when comparing our maximal polarization spectra in Figure \ref{fig:maximalpol} because $qRSP$ might fail to represent changes in polarization across lines accurately.  However, when studying the evolution of spectral features in one object from epoch to epoch, we prefer to study the RSP because they yield the most accurate representation of the true polarization when the signal-to-noise ratio is low, as is often the case in our later epochs.  In the case that we are studying the evolution of the polarization of one object epoch by epoch, we are also able to inspect $uRSP$ at the same time, as seen in all of our individual object spectropolarimetry figures in Appendix \ref{App:A}, so the choice of inspecting the RSP does not risk overlooking polarization signal that was rotated out of $qRSP$ across line changes.  We show an example of one of such figures with $q$, $u$, $qRSP$, $uRSP$, and $\theta$ in Figure \ref{fig:QU_example}.  We also show a comparison of a $q-u$ plot with that of a $qRSP-uRSP$ plot in Figure \ref{fig:RSPISPcomp} to illustrate what the process of generating the RSP looks like.

\subsection{Choosing the Interstellar Polarization}
In order to study the polarization signal intrinsic to our targets, we first must deal with the complicated issue of the ISP.  After the light from our target leaves its location (with some intrinsic polarization signature), it must then pass through the interstellar medium of its host galaxy that lies along our line of sight, which can impart changes to its polarization signal due to magnetically aligned dust grains.  Changes to the polarization signal may also be imparted by dust grains along the line of sight within the Milky Way.  Since each change to the polarization signal acts as a vectorial change in the $q-u$ plane, it can be difficult to pin down separate contributions to the ISP, but estimating the bulk ISP effect is more tractable.  We first attempt to use relations discovered by \citet{1975ApJ...196..261S} relating extinction to an upper limit on ISP in an attempt to constrain the influence of the ISP on our intrinsic polarization signals.  Since this only sets an upper limit on the magnitude of the ISP with no constraint on the position angle of the ISP in the $q-u$ plane, we further attempt to estimate the ISP from depolarization at the wavelengths of strong H$\alpha$ emission seen in some of our targets.  Each of these approaches is explained in further detail in the following sections.  


\subsubsection{ISP constraint based on reddening measurements}
\label{sec:ISPred}
\citet{1975ApJ...196..261S} suggest that an upper limit on the ISP can be set from the reddening along the line of sight according to:
 
\begin{equation}
    p\leq 9E_{(B-V)}
\end{equation}

\noindent where $E_{(B-V)}$ is given in magnitudes and $p$ is the per cent polarization.  Estimates of the reddening along various lines of sight within the Milky Way are available \citep{2011ApJ...737..103S}.  In order to estimate the reddening along the line of sight within the host galaxy for our targets, we use Na~\textsc{i}~D relations discussed in \S~\ref{IIn:sec:Ext}.  Combining both of these estimates of the reddening, we then place an upper limit on the ISP for each target, as shown by a blue circle of asterisks in all of our $q-u$ plots in Appendix \ref{App:A}.  Keep in mind that this is an upper limit, not a statement of equality.  Many of the targets in the sample within \citet{1975ApJ...196..261S} found the ISP to be far below the relation that they use for an upper limit.  Additionally, this relation assumes dust properties in the host galaxies are similar to those of the Milky Way, which may not be the case \citep{2000ApJ...536..239L,2016ApJ...828...24P}. 

In many cases, the ISP upper limit inferred from Na~\textsc{i}~D is not very restrictive, corresponding to ISP levels larger than the signals we detect.  We reiterate that the ISP polarization degree inferred from $E_{(B-V)}$ (which is, in turn, inferred from Na~\textsc{i}~D) is only an upper limit; the original empirical relation is derived from an upper threshold, not a fit to a correlation.  Physically, it is an upper limit because multiple interstellar medium (ISM) clouds along the line of sight might produce Na~\textsc{i}~D absorption and dust reddening that add together, but the magnetically aligned dust grains in these multiple clouds might not have the same orientation, and so their induced polarization may therefore cancel.  Thus, when the Na~\textsc{i}~D equivalent width and reddening are very low, this upper limit is useful, but when the ISM reddening has a higher value, the upper limit is not meaningful.

\begin{figure*}
\centering
\includegraphics[width=0.92\textwidth,height=1\textheight,keepaspectratio,clip=true,trim=0cm 0cm 0cm 0cm]{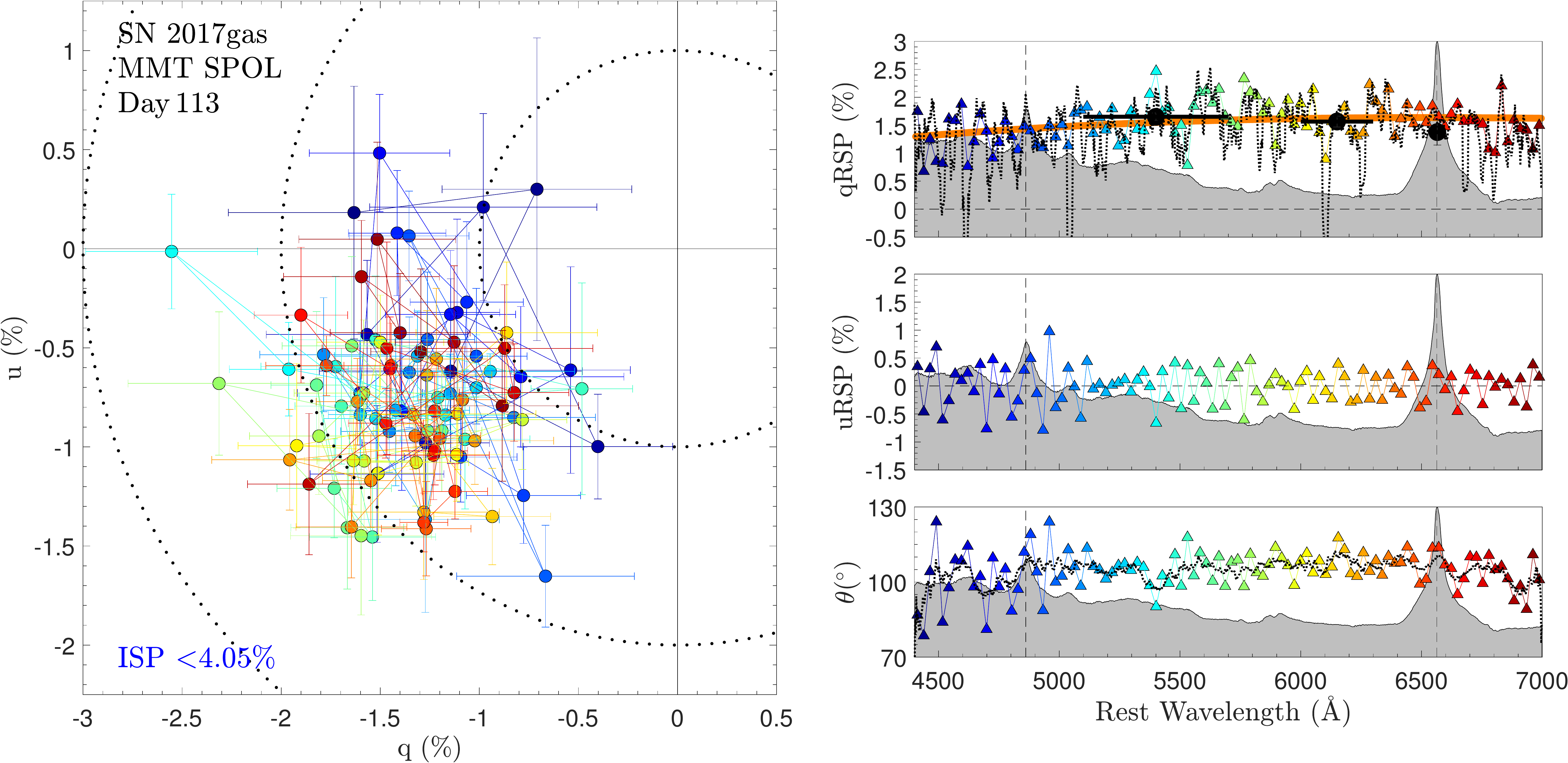}
\caption[$q-u$ plot for SN~2017gas on day 113 with Serkowski fit]{Same as Figure \ref{fig:QU_example}, but for SN~2017gas observed on day 113 after peak with the MMT.  We also overplot the Serkowski law fit to the data (see \S~\ref{sec:ISPdepol} for detailed discussion of the fitting process) in orange in the $q_{RSP}$ panel, with $\lambda_{max} = 6521${\AA}.}
\label{fig:SerkowskiTemplate}
\end{figure*}

\begin{table}
\caption[Serkowski ISP Fits]{Fits to the $q$ and $u$ values at the location of depolarized H$\alpha$ emission lines using $\lambda_{max} = 6521$ {\AA} (see \S~\ref{sec:ISPdepol} for a detailed discussion on the fitting process).  The scale factor was used to make the estimate more consistent with the ISP constraint from reddening, as discussed in \S~\ref{IIn:sec:Ext}.}
\label{tab:ISPfits}
\begin{tabular}{ccccc}
  \hline
SN Name & Day & $q_{max}$ & $u_{max}$ & Scale Factor \\
\hline \hline
SN~2010jl & 25 & 0.2246 & -0.2245 & 0.85 \\
SN~2012ab & 76 & 0.2975 & -0.1064 & 0.50 \\
SN~2009ip & 7 & -0.2655 & -1.3944 & 0.50 \\
SN~2015bh & 18 & 0.0612 & 0.6324 & 0.32 \\
SN~2017gas & 0 & -1.0284 & -1.1023 & 0.75 \\
SN~2017hcc & -45 & -0.6264 & -0.7213 & 0.50 \\
\hline
\end{tabular}
\end{table}

\subsubsection{ISP estimate from depolarization in strong emission lines}
\label{sec:ISPdepol}

For a handful of our objects (SN~2010jl, SN~2009ip, SN~2012ab, SN~2015bh, SN~2017gas, and SN~2017hcc), strong depolarization of H$\alpha$ is seen early on.  Because recombination line emission that comes from regions outside the electron scattering photosphere can be assumed to reach us without scattering, we can treat its light as unpolarized.  Since the line flux dominates that of the continuum, we expect that the overall polarization signal in the H$\alpha$ line will approach that of the ISP (although we note that an external polarizing source such as CSM dust can scatter line emission even if it is external to the electron scattering photosphere).  The narrow component of the H$\alpha$ line is not resolved in our spectropolarimetric data, so we attempt to estimate the $q_{ISP}$ and $u_{ISP}$ values based on the three pixels closest to the narrow H$\alpha$ emission-line peak.  We estimate the continuum flux by fitting the continuum level on either side of the H$\alpha$ emission line.  

We then assume that the continuum at the center of the narrow H$\alpha$ line has similiar polarization to that measured in the 6000--6300{\AA} regions (although there is evidence for a wavelength-dependent polarization in our data, the wavelength dependence is not steep enough to compromise this assumption).  After removing the polarization from this continuum flux, we then assume that the remaining polarization signal is associated with unpolarized H$\alpha$ emission-line flux, and thus is a reasonable estimate of the ISP at the wavelength of the narrow component of H$\alpha$ emission.  The more intrinsically polarized the H$\alpha$ lines are, or the more polarization that external CSM dust contributes to the overall signal, the worse this correction of the ISP becomes.

In order to estimate the ISP as a function of wavelength, we fit a Serkowski law \citep{1975ApJ...196..261S}:
\begin{equation}
    p(\lambda) = p_{\mathrm{max}}e^{-K\mathrm{ln}^2\frac{\lambda_{\mathrm{max}}}{\lambda}}
\end{equation}
where $\lambda$ is the wavelength in {\AA}, $p_{max}$ is the maximum polarization across all wavelengths, and $K$ is the Serkowski parameter, through the $q_{depol}$ and $u_{depol}$ values we estimated from the narrow H$\alpha$ emission line.  Specifically, we minimize the errors on a simultaneous fit to

\begin{equation}
\begin{aligned}
    q_{\mathrm{ISP}}(\lambda) = p_{\mathrm{ISP}}(\lambda) \mathrm{cos}(2\theta_{\mathrm{ISP}}) \\   
    u_{\mathrm{ISP}}(\lambda) = p_{\mathrm{ISP}}(\lambda) \mathrm{sin}(2\theta_{\mathrm{ISP}}) \\
    p_{\mathrm{max,ISP}} = q_{\mathrm{max,ISP}} \mathrm{cos}(2\theta_{\mathrm{ISP}}) + u_{\mathrm{max,ISP}} \mathrm{sin}(2\theta_{\mathrm{ISP}}) \\
    \theta_{\mathrm{ISP}} = \frac{1}{2}\mathrm{tan}^{-1}(\frac{u_{max}}{q_{max}})
\end{aligned}
\end{equation}

\noindent (where $\theta$ is the position angle) in order to obtain $\lambda_{\mathrm{max}}$, $q_{\mathrm{max}}$, and $u_{\mathrm{max}}$.  A similar procedure was previously performed in \citet{2021A&A...651A..19D}, where the ISP curve was fit to a number of depolarized lines in one epoch.  \citet{2018A&A...615A..42C} found that the $K-\lambda_{\mathrm{max}}$ relation is an instrinsic property of polarization in the ISM, with $K_{\mathrm{ISP}} = -1.13 + 0.000405\lambda_{\mathrm{max,ISP}}$, so we use this relation.  

In order to estimate $q_{max}$ and $u_{max}$ for each of our objects exhibiting strong depolarization, we first estimate a reasonable value for $\lambda_{max}$ from a late-time (day 113) measurement of SN~2017gas when the continuum is still bright but the polarization signal has significantly faded.  We also prefer this epoch of SN~2017gas spectropolarimetry since reddening constraints (see \S~\ref{sec:ISPred}) suggest that extensive ISP may be contributing to SN~2017gas.  Additionally, since this estimate of the ISP may be convoluted by distant CSM dust as is discussed in greater detail in \S~\ref{sec:Dis:wavedepend}, we chose an epoch that did not show polarization stronger at blue wavelengths than red ones.  Although this may bias our choice of $\lambda_{max}$ towards redder wavelengths, we continued with this route because it placed our estimate of $\lambda_{max}$ closer to those found in past literature estimates of ISP dependence on wavelength ($\lambda_{max} \sim 5500$ {\AA}: \citealt{2012JQSRT.113.2334V}, though \citealt{2015A&A...577A..53P,2019MNRAS.483.3636C} have found bluer values for $\lambda_{max}$ in the ISP) and it allowed us to more confidently avoid fitting distant CSM dust polarization that we find very likely in our targets.  

The epoch 5 (day 113) data for SN~2017gas with the ISP fit overplotted on the $q_{RSP}$ is shown in Figure \ref{fig:SerkowskiTemplate}.  The best-fit value for $\lambda_{\mathrm{max}} = 6521$ \, {\AA}.  Using this $\lambda_{\mathrm{max}}$ value from our fit to the late-time SN~2017gas data along with $q_{depol}$ and $u_{depol}$ from the estimate of the ISP at the wavelength of the narrow component of H$\alpha$ emission for each of our objects that show strong depolarization, we estimate a wavelength-dependent ISP for each target with significant H$\alpha$ depolarization.  However, the assumption that the ISP along the line of sight to each of the SNe~IIn in our sample is well-represented by a similar $\lambda_{max}$ value has limitations--$\lambda_{max}$ values for MW dust vary with position on the sky.  A summary of the $q_{max}$ and $u_{max}$ values that resulted from our fits is shown in Table \ref{tab:ISPfits}.

\begin{figure}
\centering
\includegraphics[width=0.5\textwidth,height=1\textheight,keepaspectratio,clip=true,trim=0cm 0cm 0cm 0cm]{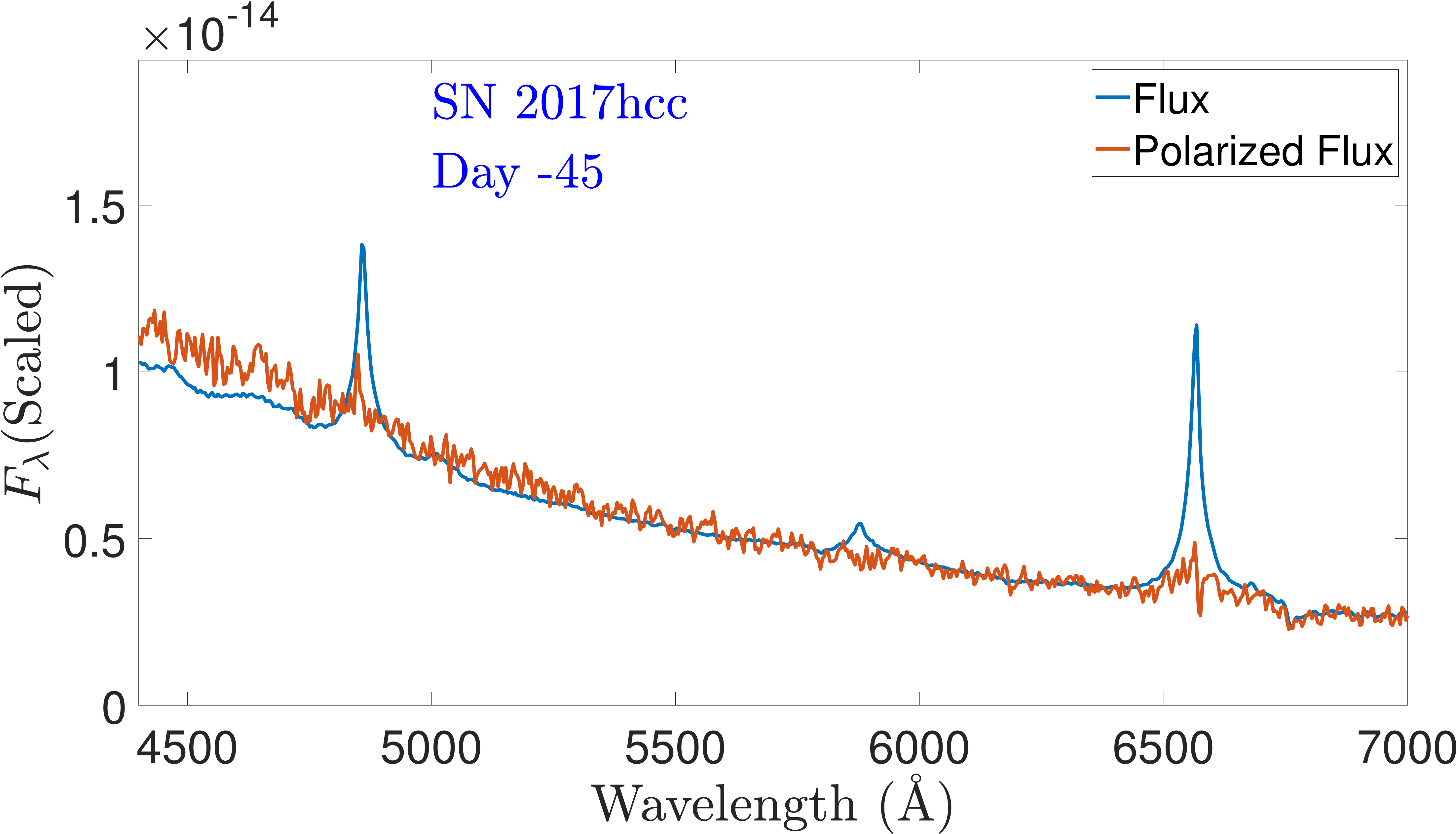}
\includegraphics[width=0.5\textwidth,height=1\textheight,keepaspectratio,clip=true,trim=0cm 0cm 0cm 0cm]{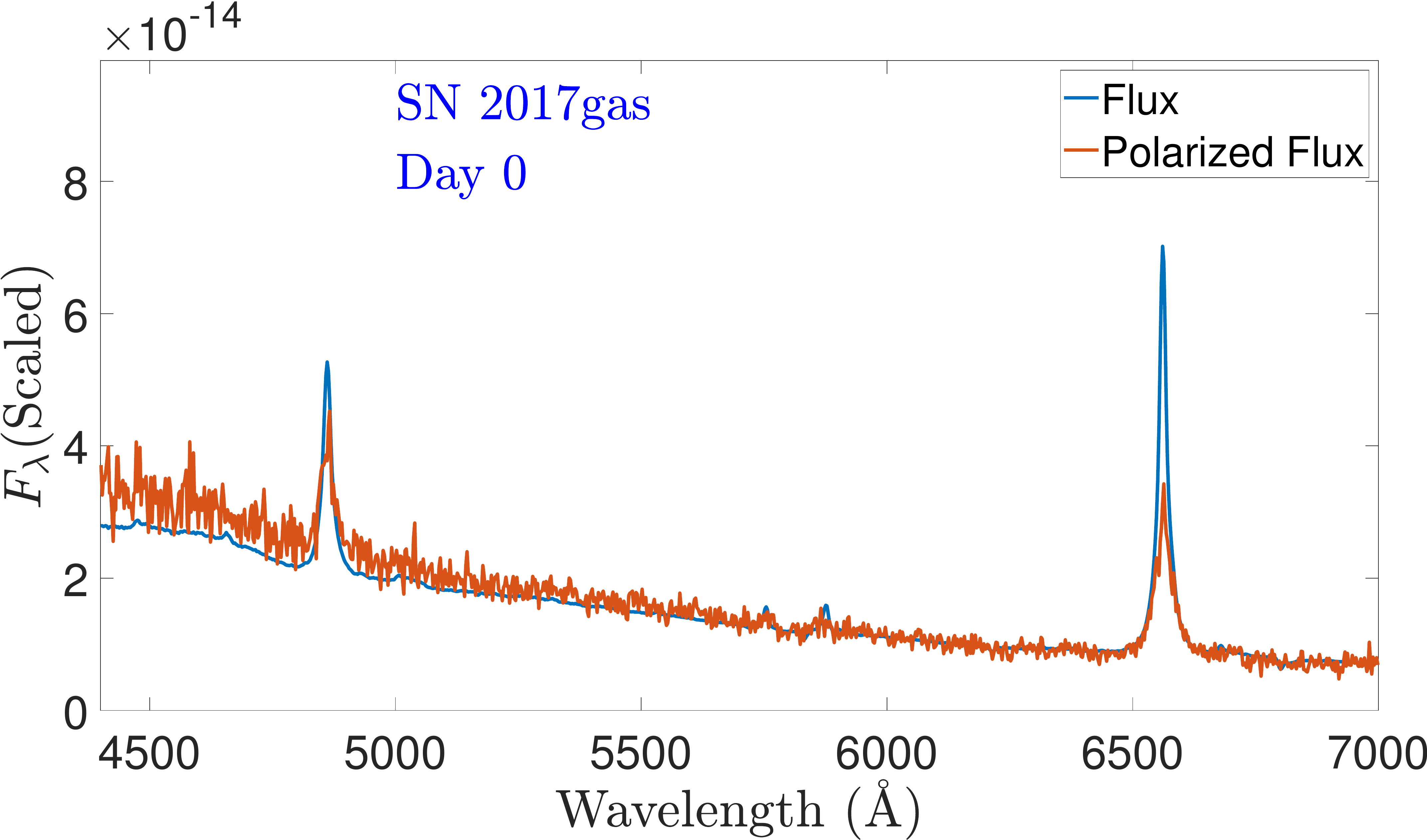}
\caption[Flux and polarized flux for SN~2017hcc and SN~2017gas]{{\it Top panel:} Flux (blue) and polarized flux (red, polarization times flux) for SN~2017hcc taken on day -45.  The polarized flux is scaled to match the flux at the continuum region 6000--6300~{\AA}.  If line emission were completely depolarized, we would expect to see the polarized flux follow the continuum of the flux.  However, the polarized flux increases slightly at the location of broad H$\alpha$ emission, suggesting that a portion of this line is polarized.  {\it Bottom panel:} The same for SN~2017gas on day 0, showing much more significant polarization within the H$\alpha$ emission line.}
\label{fig:polarizedflux}
\end{figure}

The above estimate of the wavelength-dependent ISP assumes that the H$\alpha$ narrow-line flux is completely depolarized.  However, careful inspection of the polarized flux across the H$\alpha$ emission lines reveals that the lines are not completely depolarized.  Figure \ref{fig:polarizedflux} shows the first epoch of spectropolarimetry for SN~2017hcc (day -45) and SN~2017gas (day 0).  SN~2017hcc only shows small levels of polarization in the line flux across the broad component, while SN~2017gas shows significant line polarization across the entire line.  The polarized flux would trace the continuum flux if the line emission were completely unpolarized.  However, even polarized lines could cause the polarized flux to trace the continuum flux if the lines are polarized at a similar magnitude as the continuum, but at a different angle (this would be seen as a rotation in the $q-u$ plane, but not a change in the magnitude of the polarization).  In our case, both polarized flux signals show increases at wavelengths of emission lines compared to the continuum, which cannot arise from unpolarized emission lines.  Additionally, it is worth noting that polarization is relatively stronger at blue wavelengths, which we discuss in more detail in \S~\ref{IIn:sec:Res:wavedep}.

Because the line emission is likely polarized to some extent in many of our targets (especially at early times when the optical depth is highest in the CSM interaction region such that the line forming region is beneath the electron scattering photosphere), our estimates of the depolarization correction only go a portion of the way from the continuum polarization to the true ISP location in the $q-u$ plane.\footnote{In principle, the line emission could be polarized opposite to the continuum in the $q-u$ plane, so that when it is combined vectorially it actually causes our estimates to overestimate the displacement between the continuum polarization and the ISP in the $q-u$ plane.  However, since the H$\alpha$ emission-line profiles at early times when strong depolarization is seen often exhibit broad Lorentzian profiles (see \S~\ref{IIn:sec:Res:halphaprofile} for more discussion on the H$\alpha$ emission line profiles), we expect that the broad emission-line flux, which appears polarized, originates in a region coincident with the electron-scattering continuum photosphere (when electron scattering dominates, it imparts a symmetric broadening around zero velocity with a Lorenztian shape; \citealt{2001MNRAS.326.1448C,2017hsn..book..403S}).}  Given the generally Lorentzian shape in early-time H$\alpha$ emission lines for our targets and the lack of significant changes in the position angle across emission lines, we find it most likely that the emission-line region shares a geometry with that of the continuum photosphere \citep{2015MNRAS.449.4304D}.  Thus, our estimate of the displacement between the continuum polarization and the ISP derived from depolarization is likely an underestimate of the true offset.  

When our constraint on the ISP from reddening suggests an ISP value further from the continuum than that estimated from depolarized H$\alpha$ emission, we adjust the wavelength-dependent ISP estimate from depolarization to be closer to being consistent with the reddening constraint.  In cases where the reddening constraint is less stringent on the ISP (keep in mind that this constraint just sets an upper limit on the magnitude of the ISP, not an actual estimate of it), we instead adjust the ISP estimate from depolarization by a factor of 0.5 to reflect the possibility that a significant fraction of the line emission is polarized.  We list the scale factors used for our sample in Table \ref{tab:ISPfits}.  Larger data sets or a more in-depth study of line polarization at early times in SNe IIn may lead to a better estimate of this scale factor. 

We then correct the $q$ and $u$ data for each SN that exhibits strong depolarization using this wavelength-dependent ISP estimate.  As an example, we compare the spectropolarimetric data for SN~2017gas with the ISP signal included and removed side-by-side in Figure \ref{fig:RSPISPcomp}.  In all cases where we have adjusted our data using an estimate of the wavelength-dependent ISP, the variation of the polarization signal from epoch to epoch is significantly greater than the ISP estimate.  This implies that even if our estimate of the ISP was made incorrectly, our targets still exhibit significant intrinsic polarization through the changes in their polarization signal from epoch to epoch.

\begin{figure*}
\centering
\includegraphics[width=0.7\textwidth,height=1\textheight,keepaspectratio,clip=true,trim=0cm 0cm 0cm 0cm]{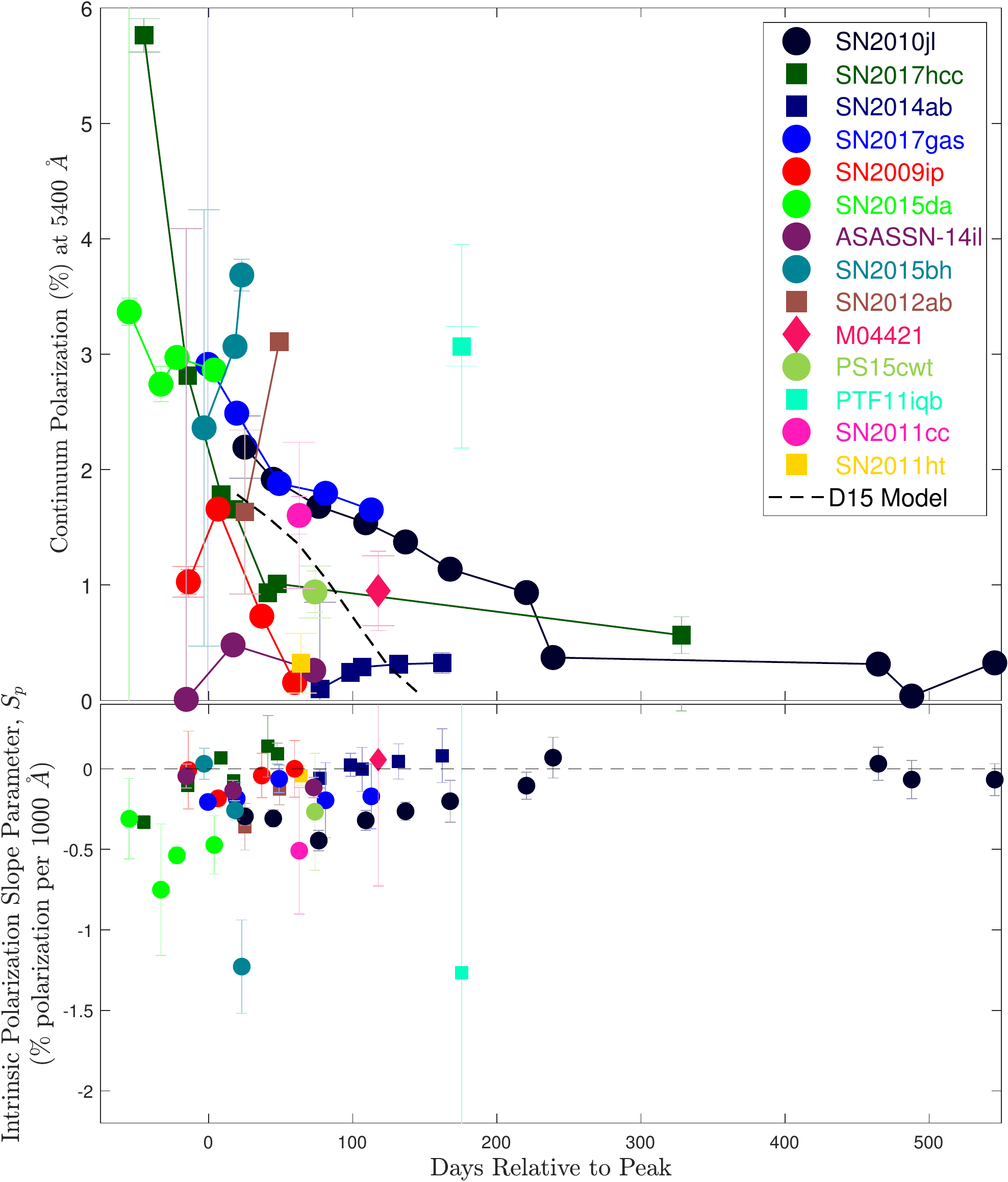}
\caption[Continuum polarization and instrinsic polarization slope parameter evolution]{{\it Top panel:} Continuum polarization measurements across 5100--5700~{\AA} relative to the peak date for our entire SNe IIn sample.  Statistical error bars are faded and have short end caps.  Error bars for ISP constraints from Na~\textsc{i}~D are labeled only on the first epoch for each target with  larger end caps than the statistical error bars for clarity.  Although the ISP affects data in every epoch, it would shift all of the data along the same vector in the $q-u$ plane, having little effect on changes in the polarization signal over time. {\it Bottom panel:} $S_{p}$ values relative to peak date for our entire SNe IIn sample.  $S_{p}$ is the intrinsic polarization slope parameter, which measures the wavelength dependence of the polarization signal (see \S~\ref{IIn:sec:Res:wavedep}).  An $S_{p}$ of -1 specifies an average change in polarization across the continuum of the spectrum of -1\% for every 1000~{\AA}, indicating a more strongly polarized continuum at bluer wavelengths.}
\label{fig:contpol_slope}
\end{figure*}

\begin{figure*}
\centering
\includegraphics[width=0.7\textwidth,height=1\textheight,keepaspectratio,clip=true,trim=0cm 0cm 0cm 0cm]{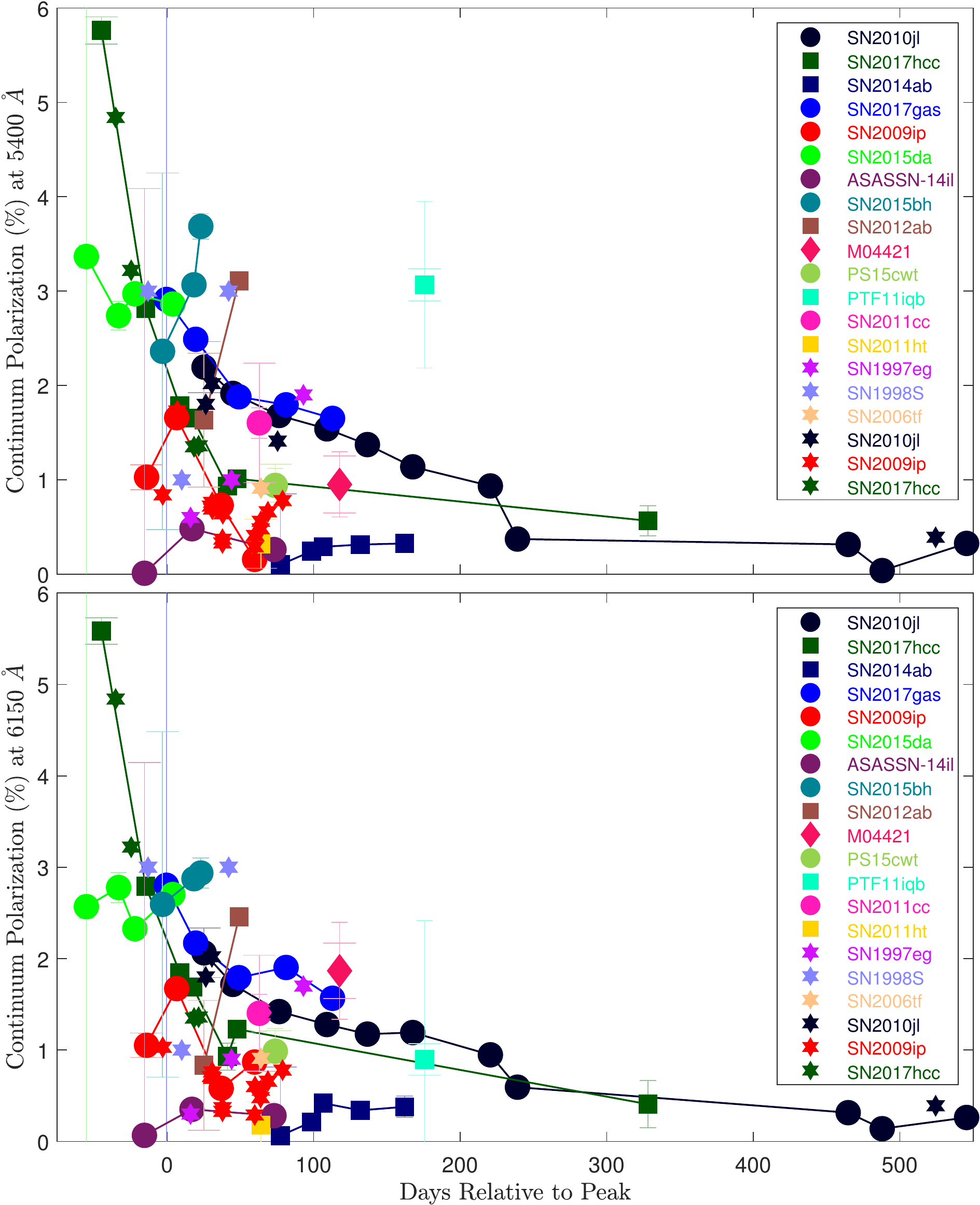}
\caption[Continuum polarization evolution across 5100--5700~{\AA} and 6000--6300~{\AA}]{Same as in Figure \ref{fig:contpol_slope}, but with previously published spectropolarimetric data included and marked by hexagonal stars.  {\it Top panel:} Polarization measurements across the 5100--5700~{\AA} bin.  {\it Bottom panel:} Polarization measurements across the 6000--6300~{\AA} bin.}
\label{fig:contpol12_lit}
\end{figure*}

\subsection{Intrinsic Spectropolarimetry}
\label{IIn:sec:Res:intrspol}
Our spectropolarimetric results, after correcting the data using our estimates of the wavelength-dependent ISP for any objects with clear depolarization in H$\alpha$ emission (as discussed in detail in \S~\ref{sec:ISPdepol}), are enumerated in Table \ref{tab:contpolval}.  Individual $q-u$ figures for every SN IIn in our sample are also included in Appendix \ref{App:A}, showing the data both before and after wavelength-dependent ISP correction if it was implemented.

\begin{table*}
\caption[ISP-corrected Continuum Polarization Measurements]{ISP-corrected continuum polarization measurements for our sample of SNe IIn.  Day is measured relative to the observed peak shown in Table \ref{tab:peakdates} and often includes data from many nights that have been combined into a single epoch, as detailed in Table \ref{tab:Obs_SPOL}.  $p_{opt}$ is the optimal polarization, which alleviates the positive-definite bias of traditional polarization definitions, as discussed in \S~\ref{IIn:sec:Res:Specpol}.  ``5400" indicates data binned across 5100-5700~{\AA}, while ``6150" indicates data binned across 6000-6300~{\AA}.  $\theta$ is the position angle.}
\label{tab:contpolval}
\begin{tabular}{cccccccccc}
  \hline
Name & Day & $p_{opt,5400}(\sigma)$ & $\theta_{5400}(\sigma)$ & $q_{5400}(\sigma)$ & $u_{5400}(\sigma)$ & $p_{opt,6150}(\sigma)$ & $\theta_{6150}(\sigma)$ & $q_{6150}(\sigma)$ & $u_{6150}(\sigma)$ \\
\hline \hline
SN2010jl & 25 & 2.19(0.03) & 130.9(0.4) & -0.31(0.03) & -2.17(0.03) & 2.06(0.04) & 132(0.5) & -0.21(0.04) & -2.05(0.04) \\
SN2010jl & 45 & 1.92(0.02) & 130.3(0.4) & -0.31(0.03) & -1.89(0.02) & 1.72(0.04) & 131(0.6) & -0.21(0.03) & -1.71(0.04) \\
SN2010jl & 76 & 1.68(0.02) & 132(0.4) & -0.18(0.02) & -1.67(0.02) & 1.42(0.04) & 130(0.7) & -0.25(0.03) & -1.4(0.04) \\
SN2010jl & 109 & 1.54(0.02) & 132.7(0.4) & -0.13(0.02) & -1.54(0.02) & 1.28(0.03) & 132(0.7) & -0.14(0.03) & -1.27(0.03) \\
SN2010jl & 137 & 1.37(0.02) & 131.3(0.5) & -0.18(0.02) & -1.36(0.02) & 1.17(0.03) & 129(0.8) & -0.24(0.03) & -1.15(0.03) \\
SN2010jl & 168 & 1.14(0.05) & 133.9(1.3) & -0.04(0.05) & -1.14(0.05) & 1.19(0.07) & 131(1.6) & -0.18(0.07) & -1.18(0.07) \\
SN2010jl & 221 & 0.93(0.03) & 124.9(1) & -0.32(0.03) & -0.88(0.03) & 0.95(0.05) & 124(1.4) & -0.36(0.05) & -0.88(0.05) \\
SN2010jl & 239 & 0.37(0.07) & 151.8(4.7) & 0.21(0.06) & -0.32(0.07) & 0.59(0.07) & 133(4.5) & -0.05(0.09) & -0.6(0.07) \\
SN2010jl & 465 & 0.31(0.05) & 135.4(4.1) & 0(0.05) & -0.32(0.05) & 0.32(0.07) & 123(5.7) & -0.13(0.07) & -0.3(0.07) \\
SN2010jl & 488 & 0.04(0.05) & 25.2(20.1) & 0.05(0.05) & 0.06(0.05) & 0.14(0.07) & 156(11.9) & 0.11(0.07) & -0.13(0.07) \\
SN2010jl & 546 & 0.33(0.05) & 173.4(3.9) & 0.32(0.05) & -0.08(0.05) & 0.26(0.07) & 167(7.1) & 0.25(0.08) & -0.13(0.07) \\ \hline
SN2011cc & 63 & 1.6(0.16) & 125.3(3.2) & -0.54(0.19) & -1.53(0.16) & 1.4(0.21) & 124(4.2) & -0.56(0.21) & -1.32(0.21) \\ \hline
PTF11iqb & 176 & 3.07(0.88) & 0.6(7.7) & 3.3(0.88) & 0.07(0.89) & 0.9(1.52) & 177(22) & 2.02(1.52) & -0.2(1.56) \\ \hline
SN2011ht & 64 & 0.32(0.03) & 78.5(2.2) & -0.3(0.03) & 0.13(0.02) & 0.18(0.03) & 58(5.4) & -0.08(0.03) & 0.17(0.03) \\ \hline
SN2012ab & 25 & 1.63(0.05) & 126.2(0.8) & -0.49(0.05) & -1.56(0.05) & 0.83(0.06) & 129(2.7) & -0.18(0.08) & -0.82(0.06) \\
SN2012ab & 49 & 3.11(0.02) & 118.6(0.2) & -1.68(0.02) & -2.61(0.02) & 2.46(0.04) & 118(0.4) & -1.41(0.04) & -2.02(0.04) \\ \hline
SN2009ip & -14 & 1.03(0.11) & 176(4.3) & 1.03(0.11) & -0.14(0.16) & 1.05(0.13) & 171(5.3) & 1.02(0.12) & -0.32(0.21) \\
SN2009ip & 7 & 1.66(0.02) & 71.6(0.3) & -1.33(0.02) & 0.99(0.02) & 1.67(0.03) & 73(0.5) & -1.4(0.03) & 0.91(0.03) \\
SN2009ip & 37 & 0.73(0.06) & 60.5(2.2) & -0.38(0.06) & 0.63(0.06) & 0.58(0.08) & 48(4) & -0.07(0.08) & 0.59(0.09) \\
SN2009ip & 60 & 0.16(0.08) & 105.8(12.3) & -0.16(0.07) & -0.1(0.08) & 0.87(0.09) & 120(2.8) & -0.44(0.09) & -0.76(0.09) \\ \hline
SN2014ab & 77 & 0.1(0.03) & 13.1(7.2) & 0.1(0.03) & 0.05(0.03) & 0.06(0.04) & 13(13.3) & 0.07(0.04) & 0.04(0.04) \\
SN2014ab & 99 & 0.24(0.03) & 73(3.4) & -0.2(0.03) & 0.14(0.03) & 0.21(0.04) & 72(5.5) & -0.18(0.04) & 0.13(0.04) \\
SN2014ab & 106 & 0.29(0.07) & 68.1(6.3) & -0.22(0.07) & 0.21(0.07) & 0.42(0.09) & 44(6.3) & 0.02(0.1) & 0.44(0.09) \\
SN2014ab & 132 & 0.31(0.05) & 65.1(4.8) & -0.21(0.05) & 0.25(0.05) & 0.34(0.08) & 54(5.4) & -0.11(0.07) & 0.34(0.08) \\
SN2014ab & 162 & 0.32(0.09) & 45.4(7.2) & 0(0.09) & 0.35(0.09) & 0.38(0.12) & 88(8.5) & -0.41(0.12) & 0.02(0.12) \\ \hline
M04421 & 118 & 0.96(0.34) & 90.3(9.8) & -1.07(0.34) & -0.01(0.37) & 1.87(0.53) & 32(7.7) & 0.89(0.54) & 1.8(0.53) \\ \hline
ASASSN-14il & -15 & 0.01(0.03) & 146.1(24.7) & 0.01(0.03) & -0.04(0.03) & 0.07(0.05) & 157(15.3) & 0.06(0.04) & -0.07(0.06) \\
ASASSN-14il & 17 & 0.48(0.03) & 161.3(2) & 0.38(0.03) & -0.29(0.03) & 0.35(0.05) & 162(3.7) & 0.29(0.04) & -0.21(0.05) \\
ASASSN-14il & 73 & 0.26(0.03) & 158.1(3.4) & 0.19(0.03) & -0.18(0.03) & 0.28(0.04) & 170(4.4) & 0.27(0.04) & -0.1(0.04) \\ \hline
SN2015da & -55 & 3.37(0.12) & 149.7(1) & 1.65(0.12) & -2.94(0.12) & 2.57(0.13) & 149(1.4) & 1.25(0.12) & -2.25(0.13) \\
SN2015da & -33 & 2.74(0.15) & 151(1.6) & 1.46(0.15) & -2.33(0.15) & 2.78(0.16) & 150(1.6) & 1.41(0.15) & -2.4(0.17) \\
SN2015da & -22 & 2.97(0.02) & 148.9(0.2) & 1.39(0.02) & -2.63(0.02) & 2.33(0.03) & 149(0.4) & 1.09(0.03) & -2.06(0.03) \\
SN2015da & 4 & 2.86(0.08) & 149(0.8) & 1.34(0.08) & -2.53(0.08) & 2.7(0.09) & 151(0.9) & 1.41(0.09) & -2.31(0.08) \\ \hline
SN2015bh & -3 & 2.36(0.04) & 100.7(0.5) & -2.2(0.04) & -0.87(0.04) & 2.59(0.07) & 100(0.7) & -2.43(0.07) & -0.91(0.06) \\
SN2015bh & 18 & 3.07(0.03) & 108(0.3) & -2.48(0.03) & -1.81(0.03) & 2.87(0.05) & 109(0.4) & -2.27(0.05) & -1.77(0.04) \\
SN2015bh & 23 & 3.69(0.14) & 109.7(1.1) & -2.85(0.14) & -2.35(0.14) & 2.94(0.17) & 108(1.6) & -2.39(0.17) & -1.73(0.16) \\ \hline
PS15cwt & 74 & 0.94(0.18) & 119.3(5.5) & -0.51(0.19) & -0.83(0.17) & 0.99(0.25) & 126(7.1) & -0.33(0.26) & -0.99(0.25) \\ \hline
SN2017gas & 0 & 2.91(0.01) & 115.5(0.1) & -1.83(0.01) & -2.27(0.01) & 2.8(0.02) & 116(0.2) & -1.75(0.01) & -2.19(0.02) \\
SN2017gas & 20 & 2.49(0.05) & 115.8(0.6) & -1.55(0.05) & -1.95(0.05) & 2.17(0.07) & 115(0.9) & -1.4(0.07) & -1.66(0.07) \\
SN2017gas & 49 & 1.88(0.05) & 113.3(0.7) & -1.29(0.05) & -1.37(0.05) & 1.79(0.06) & 113(1) & -1.23(0.07) & -1.3(0.06) \\
SN2017gas & 81 & 1.79(0.07) & 110.3(1.2) & -1.36(0.07) & -1.17(0.07) & 1.9(0.1) & 109(1.4) & -1.52(0.1) & -1.15(0.09) \\
SN2017gas & 113 & 1.65(0.07) & 104.1(1.2) & -1.46(0.07) & -0.78(0.07) & 1.57(0.1) & 107(1.7) & -1.3(0.1) & -0.88(0.09) \\ \hline
SN2017hcc & -45 & 5.76(0.02) & 93.8(0.1) & -5.71(0.02) & -0.77(0.02) & 5.58(0.03) & 94(0.1) & -5.54(0.03) & -0.7(0.03) \\
SN2017hcc & -15 & 2.82(0.02) & 98.8(0.2) & -2.68(0.02) & -0.85(0.01) & 2.79(0.03) & 99(0.3) & -2.67(0.03) & -0.82(0.02) \\
SN2017hcc & 9 & 1.78(0.02) & 102(0.3) & -1.63(0.02) & -0.72(0.02) & 1.84(0.02) & 103(0.3) & -1.66(0.02) & -0.8(0.02) \\
SN2017hcc & 17 & 1.65(0.01) & 103.9(0.2) & -1.46(0.01) & -0.77(0.01) & 1.69(0.02) & 105(0.4) & -1.48(0.02) & -0.82(0.02) \\
SN2017hcc & 41 & 0.93(0.08) & 96(2.6) & -0.92(0.08) & -0.2(0.09) & 0.93(0.15) & 114(4.5) & -0.64(0.15) & -0.71(0.14) \\
SN2017hcc & 48 & 1.01(0.02) & 103.3(0.6) & -0.9(0.02) & -0.45(0.02) & 1.23(0.03) & 103(0.8) & -1.11(0.03) & -0.53(0.03) \\
SN2017hcc & 328 & 0.57(0.16) & 172.9(6.9) & 0.59(0.16) & -0.15(0.15) & 0.41(0.26) & 131(14.6) & -0.08(0.27) & -0.53(0.26) \\
\hline
\end{tabular}
\end{table*}

The temporal evolution of the continuum polarization for our sample of SNe IIn is shown in the top panel of Figure \ref{fig:contpol_slope}, where we have aligned the SNe relative to their times of peak brightness (in $R$-, $V$-, or $i$-bands; see Table \ref{tab:peakdates}).  This shows the per cent polarization when binned across 5100--5700~{\AA}, which is the main bin size we use to discuss these results throughout the paper (the numbers are slightly different in the 6000--6300~{\AA} bin, but this does not affect our conclusions).  The range of polarization degree and the rise/decline rates are diverse.  The statisical errors (shown with smaller endcap sizes compared to ISP errors in Figure \ref{fig:contpol_slope}) are generally quite small because we have combined the polarization signal over large wavelength bins.  Although we have removed an estimate of the wavelength-dependent ISP for objects that contained strong depolarization of H$\alpha$ in at least one epoch (SN~2010jl, SN~2012ab, SN~2009ip, SN~2015bh, SN~2017gas, and SN~2017hcc), we still show an uncertainty due to the magnitude of the ISP estimated from reddening on the first data point (using a larger endcap size than for the statistical errors) for each target to reflect this potential uncertainty.  

The highest polarization signals observed (3$-$6\%) are only seen at times near peak brightness or before ($<50$ days).  SN~2017hcc shows the strongest drop in polarization from 5.76\% on day $-$45 to 2.82\% on day $-$15.  Between days 0 and 113, SN~2017gas drops from 2.91\% to 1.65\%, roughly a 0.011\% per day decline.  SN~2010jl also exhibits a gradual decline from days 25 to 239 of 2.19\% to 0.37\%, similar to that of SN~2017gas, and flattens out thereafter.  The polarization signal generally drops steadily over time for the majority of the targets in our sample with several epochs of spectropolarimetry (SN~2010jl, SN~2009ip, SN~2015da, SN~2017hcc, and SN~2017gas). There are, however, some cases where the continuum polarization instead increases over a short time period (before declining at later times, when data are available).  This can be seen clearly in the early data for SN~2012ab, SN~2009ip, ASASSN-14il, and SN~2015bh.  We interpret these various changes and trends in continuum polarization in detail in \S~\ref{sec:Dis:contpolevo}.

We also show the continuum polarization measured across 6000--6300~{\AA} side by side with that measured across 5100--5700~{\AA} in Figure \ref{fig:contpol12_lit}.  For the most part, the trends across both wavelength bins match each other quite well.  However, increases in the polarization that were seen in some objects (SN~2009ip, ASASSN-14il, and SN~2015bh) are generally less pronounced in the 6000--6300~{\AA} bin than they are in the 5100--5700~{\AA} bin.  This is a hint that the polarization has some possible wavelength dependence that is not expected for pure electron scattering, as discussed more below in \S~\ref{sec:Dis:wavedepend}.

\begin{figure*}
\centering
\includegraphics[width=0.65\textwidth,height=1\textheight,keepaspectratio,clip=true,trim=0cm 0cm 0cm 0cm]{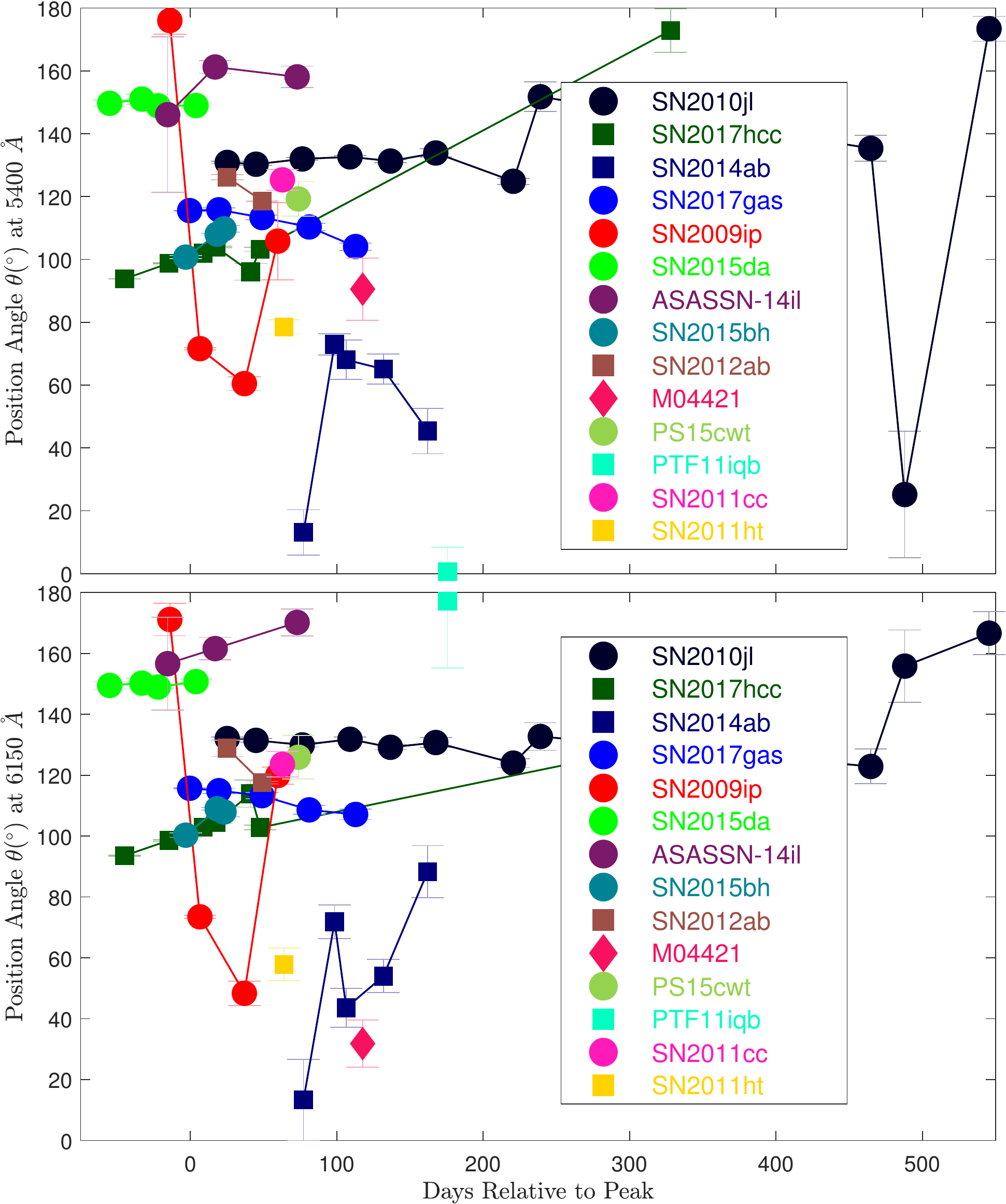}
\caption[Position angle evolution across 5100--5700~{\AA} and 6000--6300~{\AA}]{{\it Top panel:} Position angle measurements across 5100--5700~{\AA} relative to the peak date for our entire SNe IIn sample.  {\it Bottom panel:} Position angle measurements across 6000--6300~{\AA} relative to the peak date for our entire SNe IIn sample.}
\label{fig:theta12}
\end{figure*}

Figure ~\ref{fig:theta12} compares the evolution of the position angle $\theta$ with time when binned across the two different bandpasses we chose.  The position angle remains roughly constant in both bands for most of our targets.  However, there are some pronounced exceptions to this trend in SN~2009ip and SN~2014ab.  On day $-$14, SN~2009ip exhibits a position angle of $176\degree$ in the bin centered on 5400 {\AA}, progressing to $71.6\degree$ on day 7, and continuing to evolve past $60.5\degree$ on day 37 all the way back to $105.8\degree$ on day 60 (this evolution in $\theta$ was discussed in detail previously by \citealt{2014MNRAS.442.1166M}).  SN~2014ab exhibits a position angle of $13.1\degree$ on day 77, but changes to $73\degree$ by day 99 with a gradual decrease half of the way back towards the initial value thereafter.  In some cases, the significant change in the position angle can be attributed to the polarization being very low and nearly centered on the origin by this time, so slight changes in the polarization can result in large changes in the polarization angle.  We discuss the implications for these changes in more detail in \S~\ref{sec:Dis:contpolevo}.

\subsection{H$\alpha$ Emission Line Evolution}
\label{IIn:sec:Res:halphaprofile}
\begin{figure*}
\centering
\includegraphics[width=0.65\textwidth,height=1\textheight,keepaspectratio,clip=true,trim=0cm 0cm 0cm 0cm]{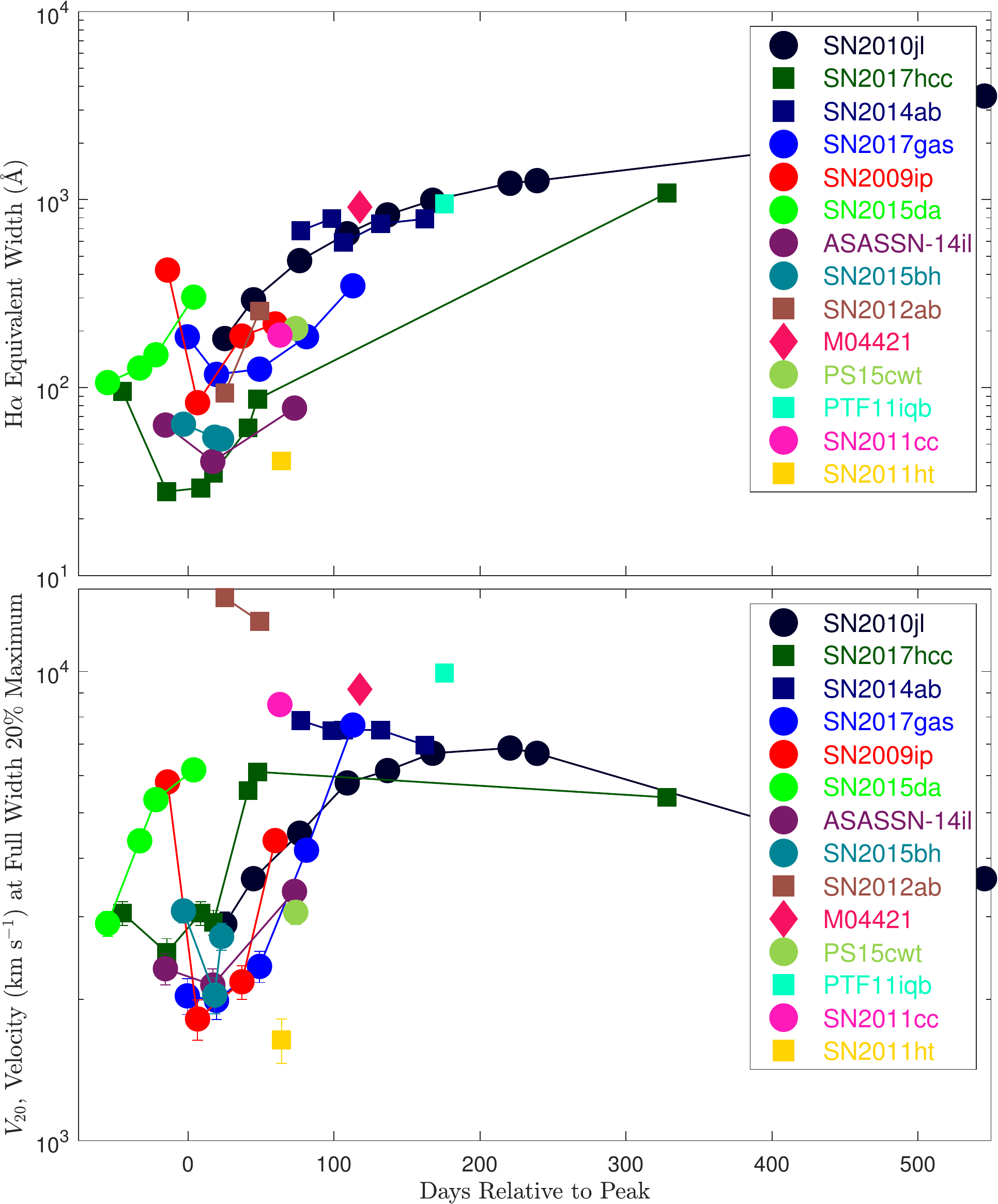}
\caption[Evolution of H$\alpha$ equivalent width and velocity at full width 20\% maximum]{{\it Top panel:} H$\alpha$ equivalent width relative to the peak date for our entire SNe IIn sample.  We have cut off the late-time data that shows very little continuum flux and thus strong H$\alpha$ equivalent widths for clarity in the early-time data we are most interested in.  {\it Bottom panel:} Velocity of the full-width-at-20\%-maximum relative to the peak date for our entire SNe IIn sample.}
\label{fig:halphaEQWandFW20}
\end{figure*}

We measured the H$\alpha$ equivalent widths and the full velocity widths at 20\%  maximum ($V_{20}$) for each of our targets at each epoch (we chose 20\% maximum instead of half maximum in order to better sample the broad wings of the H$\alpha$ line).  The results are shown in Figure \ref{fig:halphaEQWandFW20}.  

The H$\alpha$ equivalent width generally increases over time past peak for all of our targets, as is typical for SNe IIn \citep{2014MNRAS.438.1191S}. There are, however, a few notable cases in which it also decreases with time leading up to the time of peak.  In particular, between the first two epochs of data for SN~2017hcc, SN~2017gas, and SN~2009ip, we see a significant drop in the H$\alpha$ equivalent width, which has been noted previously for some SNe IIn that are discovered well before their peak luminosity phase (e.g. SN~2009ip: \citealt{2014MNRAS.438.1191S} and SN~2006gy: \citealt{2010ApJ...709..856S}).

The evolution of $V_{20}$ is similar to that of the H$\alpha$ equivalent width.  For most of our objects, $V_{20}$ for H$\alpha$ increases with time, except for the same few outliers as mentioned for H$\alpha$ equivalent width.  We discuss implications for the general trend of increasing H$\alpha$ equivalent width and $V_{20}$ (and exceptions) in detail in \S~\ref{sec:Dis:contpolevo}.  

Though it is beyond the scope of this paper, it is worth noting that the H$\alpha$ emission line sometimes has a more complex evolution through time across different velocity components of the line profile.  In particular, we note that the position angle in some of our spectropolarimetric results changes across the Balmer-series lines (most prominently for SN~2010jl when the continuum polarization has significantly faded; \citealt{Williamsinprep}).   A more in-depth study into changes across individual emission lines could discover more about the implied geometrical differences between the line-forming region and the electron-scattering continuum photosphere.

\subsection{Wavelength-Dependent Polarization}
\label{IIn:sec:Res:wavedep}
Polarization arising from an electron scattering photosphere where Thomson scattering dominates is expected to be wavelength independent. While the polarized flux (shown in Figure \ref{fig:polarizedflux}) is of course wavelength dependent, the polarization degree ($p$) should be flat for electron scattering.  However, the polarization signal for many of our objects shows a general trend of stronger polarization at bluer wavelengths.  In order to quantify this trend, we fit a line through the continuum in the polarization data, excluding notable emission and absorption line regions, to estimate a slope parameter.  The intrinsic polarization slope parameter, $S_{p}$, for all of our data is shown in the bottom panel of Figure \ref{fig:contpol_slope}.  A negative value for $S_{p}$ indicates that the polarization skews upward at blue wavelengths, while a positive value indicates that the polarization skews upward at red wavelengths.  The evolution of this slope parameter does not follow the same trend for each of our targets.  In some cases (SN~2009ip, SN~2014ab, SN~2017hcc) it becomes increasingly positive as the polarization signal decreases and the object fades.  In other cases (SN~2010jl, ASASSN-14il, SN~2015da), the slope parameter becomes more negative at first and then reverses back toward null or positive values.  In  other cases (SN~2017gas), the slope parameter begins strongly negative, becomes more positive for a few epochs, and then decreases again.  Lastly, in some cases the slope parameter quickly becomes more negative (SN~2015bh) or more positive (SN~2012ab) without any late-time data to constrain how it eventually evolves from this initial trend.  We discuss the implications of this wavelength-dependent polarization on the SN environments in \S~\ref{sec:Dis:wavedepend}.

\begin{figure}
\centering
\includegraphics[width=0.5\textwidth,height=1\textheight,keepaspectratio,clip=true,trim=0cm 0cm 0cm 0cm]{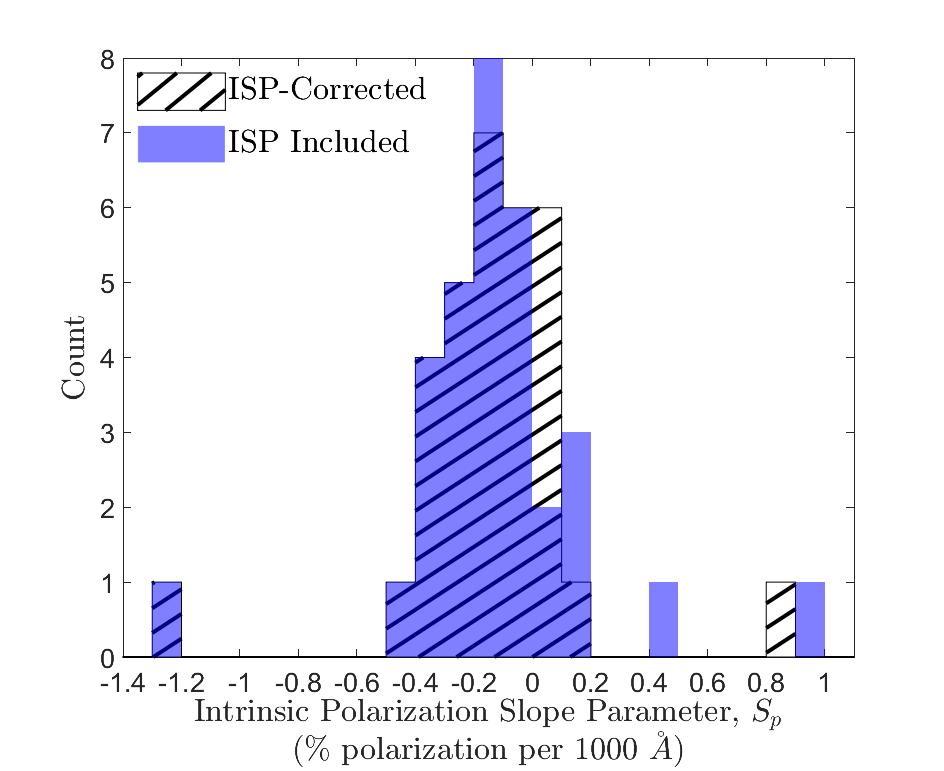}
\caption[Histogram of the intrinsic polarization slope parameter with and without ISP correction]{A histogram showing $S_{p}$ for our sample of SNe IIn that exhibit strong line depolarization with ISP correction (dashed region) and without (blue region).  Although $S_{p}$ becomes more negative in general after ISP correction, ISP correction does not significantly affect the claim that $S_{p}$ is predominantly negative across our sample.}
\label{fig:HistSlopeISP}
\end{figure}

To test whether our ISP correction had contributed significantly to this polarization slope, we fit a line to the polarization signal for each epoch of our data without any ISP correction as well.  We show a comparison of the slope parameters before and after ISP correction in Figure \ref{fig:HistSlopeISP}.  Although ISP correction does push the slope parameters to slightly more negative values in general (which is reasonable given the ISP estimates peaking at $\lambda_{\mathrm{max}}$), even the uncorrected data have slopes that are skewed to the blue.  We perform a two-sample Kolmogorov-Smirnov test on the two populations which returns a p-value of 0.951, suggesting that they do indeed arise from similar populations.  Therefore, our ISP correction does not introduce a strong bias to artificially produce an inherently blue slope.

\section{Discussion}
\subsection{Prior Spectropolarimetry from the Literature}
\label{dis:prior}
In this section we summarize the current state of spectropolarimetric observations of SNe IIn.  There has been no previous study of a sample of multiple SNe IIn as presented here, but there are several polarization studies of individual~SNe IIn that have been published (SN~1997eg: \citealt{2008ApJ...688.1186H}; SN~1998S: \citealt{2000ApJ...536..239L}; SN~2006tf: \citealt{2008ApJ...686..467S}; SN~2010jl: \citealt{2011A&A...527L...6P}; SN~2009ip: \citealt{2014MNRAS.442.1166M,2017MNRAS.470.1491R}; SN~2012ab: \citealt{2018MNRAS.475.1104B}; SN~2013fs: \citealt{2018MNRAS.476.1497B}; SN 2014ab: \citealt{2020MNRAS.498.3835B} SN~2017hcc: \citealt{2019MNRAS.488.3089K}).  Note that four of these (SN~2009ip, SN 2010jl, SN~2012ab, and SN~2014ab) are also included in this study.  We list the published values for the continuum polarization for these SNe~IIn in Table \ref{tab:externalspol}, and we show their continuum polarization values alongside the ones from our sample in Figure \ref{fig:contpol12_lit}.  Since the continuum polarization was often integrated over a variety of bandwidths in the past literature, we chose the bandwidths that most resembled ours for the purposes of Figure \ref{fig:contpol12_lit} and Table \ref{tab:externalspol}.  Most SNe IIn with published spectropolarimetric results show high continuum polarization values in the range $\sim$ 1--3\%, suggesting significantly aspherical shapes for their electron scattering regions.  Additionally, whenever spectropolarimetry on an object within our sample already had previous data published, we found good agreement between the past data and  our results.  We summarize the past results and their implications for each SNe IIn target with previously published spectropolarimetric data individually below.

\begin{table}
\caption[Prior Published Spectropolarimetric Values for SNe IIn]{Prior published spectropolarimetric values for SNe IIn.  Some of these polarization estimates take into account an estimate of the ISP, which can be very uncertain, especially when there are very few epochs of data.}
\label{tab:externalspol}
\begin{tabular}{cccccc}
  \hline
Name & Day & $P(\sigma)$ & $\theta(\sigma)$ & $\lambda$ range & Ref. \\
 & & $(\%)$ & $(\deg)$ & (\AA) \\
\hline \hline
SN~1997eg & 16 & 0.6(0.1) & 37(2) & 5200-5500 & a \\
SN~1997eg & 16 & 0.3(0.1) & 23(2) & 6100-6200 & a \\
SN~1997eg & 44 & 1.0(0.1) & 41(2) & 5200-5500 & a \\
SN~1997eg & 44 & 0.9(0.1) & 35(2) & 6100-6200 & a \\
SN~1997eg & 93 & 1.9(0.1) & 36(2) & 5200-5500 & a \\
SN~1997eg & 93 & 1.7(0.1) & 33(2) & 6100-6200 & a \\
SN~1998S & -13 & 3 & 135 & 4300-6800 & b \\
SN~1998S & 10 & 1 & - & 4500-7500 & c \\
SN~1998S & 42 & 3 & - & 4500-7500 & c \\
SN~2006tf & 64 & 0.91(0.03) & 135.4(0.8) & 5050-5950 & d \\
SN~2010jl & 31 & 2.02(0.05) & 132.1(1.0) & 5000-5600 & e \\
SN~2010jl & 26 & 1.80(0.01) & 135.8 & 4500-8000 & f \\
SN~2010jl & 75 & 1.41(0.01) & 137.6 & 4500-8000 & f \\
SN~2010jl & 524 & 0.39(0.01) & 19.3 & 4500-8000 & f \\
SN~2009ip & -3 & 1.03(0.10) & - & 6100-6200 & g \\
SN~2009ip & -3 & 0.84(0.12) & 81(4) & 5050-5950 & g \\
SN~2009ip & 7 & 1.65(0.05) & - & 6100-6200 & g \\
SN~2009ip & 7 & 1.73(0.05) & 72(1) & 5050-5950 & g \\
SN~2009ip & 30 & 0.71(0.07) & - & 6100-6200 & g \\
SN~2009ip & 30 & 0.72(0.06) & 49(3) & 5050-5950 & g \\
SN~2009ip & 31 & 0.70(0.05) & - & 6100-6200 & g \\
SN~2009ip & 31 & 0.69(0.03) & 47(2) & 5050-5950 & g \\
SN~2009ip & 38 & 0.64(0.09) & 66(6) & 5050-5950 & g \\
SN~2009ip & 60 & 0.60(0.05) & - & 6100-6200 & g \\
SN~2009ip & 60 & 0.41(0.04) & 122(2) & 5050-5950 & g \\
SN~2009ip & 64 & 0.56(0.06) & 132(3) & 6100-6200 & g \\
SN~2009ip & 31 & 0.76(0.21) & 47(8) & 5800-7200 & h\tablenotemark{a} \\
SN~2009ip & 31 & 0.69(0.20) & 41(8) & 5800-7200 & h\tablenotemark{a} \\
SN~2009ip & 38 & 0.37(0.30) & 57(21) & 5800-7200 & h\tablenotemark{a} \\
SN~2009ip & 38 & 0.33(0.26) & 63(20) & 5800-7200 & h\tablenotemark{a} \\
SN~2009ip & 60 & 0.29(0.24) & 100(15) & 5800-7200 & h \\
SN~2009ip & 64 & 0.48(0.39) & 119(22) & 5800-7200 & h \\
SN~2009ip & 69 & 0.66(0.55) & 89(21) & 5800-7200 & h \\
SN~2009ip & 79 & 0.78(1.07) & 149(39) & 5800-7200 & h \\
SN~2017hcc & -35 & 4.84(0.02) & 96.5(0.1) & 5070-5950 &  i \\
SN~2017hcc & -25 & 3.22(0.13) & 95(1.4) & 5890-7270 &  j \\
SN~2017hcc & 18 & 1.35(0.20) & 101.4(4.7) & 5890-7270 &  j \\
SN~2017hcc & 22 & 1.36(0.10) & 100.8(2.2) & 5070-5950 &  j \\
\hline
\end{tabular}
List of sources: a \citep{2008ApJ...688.1186H}, b \citep{2000ApJ...536..239L}, c \citep{2001ApJ...550.1030W}, d \citep{2008ApJ...686..467S}, e \citep{2011A&A...527L...6P}, f \citep{2019BAAA...61...90Q}, g \citep{2014MNRAS.442.1166M}, h \citep{2017MNRAS.470.1491R}, i \citep{2017ATel10911....1M}, j \citep{2019MNRAS.488.3089K}.
\tablenotetext{a}{Two different gratings were used for observations on this date.  We cite both values as separate observations here.  The first observation listed was taken with the 300V resolution grating, while the following observation was taken with the 1200R grating.}
\end{table}

\subsubsection{SN~1997eg}
Detailed multi-epoch spectropolarimetry of SN~1997eg was interpreted as aspherical SN ejecta misaligned with a surrounding CSM disk \citep{2008ApJ...688.1186H}.  These authors favored a toroidal shell model as was proposed by \citet{2003ApJ...593..788K} because of the double-peaked intermediate-width H$\alpha$ line profile and the observed loops in the $q-u$ plane across the strong polarized emission lines in  SN~1997eg.  The choice of the ISP for SN~1997eg is particularly important, as the ISP chosen by \citet{2008ApJ...688.1186H} suggests a continuum polarization that increased steadily with time (they also discuss alternate choices).

\subsubsection{SN~1998S}
Spectropolarimetry of SN~1998S showed strong linear polarization ($\sim$3\%) indicative of an equatorial CSM disk, similar to that seen for SN~1987A, but much denser and closer to the SN \citep{2000ApJ...536..239L,2001ApJ...550.1030W}.  The high continuum polarization measurement in this scenario is contingent once again on an estimate of the ISP, in which \citet{2000ApJ...536..239L} favor a model with mostly unpolarized broad lines due to negligible deviations from the continuum seen in the polarized flux.  This results in an interpretation where SN~1998S underwent CSM interaction early on (strong polarization in epoch 1), after which the SN ejecta engulfed the closest CSM region (weak polarization in epoch 2), and then eventually ran into another disk of CSM (strong polarization in epoch 3) with which interaction persisted for hundreds of days \citep{2000ApJ...536..239L}.

\subsubsection{SN~2006tf}
Although only one epoch of spectropolarimetric data exists for SN~2006tf, it shows relatively strong continuum polarization ($\sim$ 1\%) with mild depolarization across many emission-line features \citep{2008ApJ...686..467S}.  Without an accurate estimate of the ISP it is difficult to assess the intrinsic continuum polarization, but the relatively weak depolarization in emission lines suggests a somewhat less intrinsically polarized SN IIn than other SNe IIn studied spectropolarimetrically prior to SN~2006tf \citep{2008ApJ...686..467S}.  Note that, like SN~2010jl, SN~2006tf was a super-luminous SN IIn.

\subsubsection{SN~2013fs}
Although SN~2013fs was measured spectropolarimetrically, the polarization signal was only constrained by an upper limit of $<1$\% at all epochs after subtracting off an ISP assumed from the second epoch \citep{2018MNRAS.476.1497B}.  Additionally, although SN~2013fs initially appeared as a SNe IIn much like PTF11iqb, it transitioned to a SNe II-P or II-L as the fleeting SNe IIn signatures faded after a short time \citep{2018MNRAS.476.1497B}.  For these reasons, we do not show these polarization constraints in Figure \ref{fig:contpol12_lit}.

\subsubsection{SN~2010jl, SN~2009ip, and SN~2017hcc}
These three objects have previously been studied in the literature but they are also a part of our sample of SNe IIn.  Please see \S~\ref{IIn:sec:Res:2010jl}, \ref{IIn:sec:Res:2009ip}, and \ref{IIn:sec:Res:2017hcc} for a detailed discussion of their key features.

\subsection{Peak Polarization}
\label{dis:peakpol}
Until recently, the highest polarization signals measured for SNe IIn were seen in SN~2010jl ($\sim 2\%$, \citealt{2011A&A...527L...6P}) and SN~1998S ($\sim 3\%$, \citealt{2000ApJ...536..239L}, though this continuum polarization estimate for SN~1998S is contingent on an ISP measurement that is somewhat uncertain, implying that it may never have been detected at an intrinsic polarization of 3\%).  Recently, however, SN~2017hcc was detected with broadband polarization measurements extending to 4.84\% on day $-$35 \citep{2019MNRAS.488.3089K}.  We measure the highest instrinsic polarization ever recorded for a SN in SN~2017hcc on day $-$45 with a continuum polarization measurement of 5.76\% (see also \citealt{Mauerhaninprep}) Additionally, our sample contains a number of other SNe IIn that exhibit continuum polarization measurements around the $2-3\%$ range.

\begin{figure*}
\centering
\includegraphics[width=0.85\textwidth,height=1\textheight,keepaspectratio,clip=true,trim=0cm 0cm 0cm 0cm]{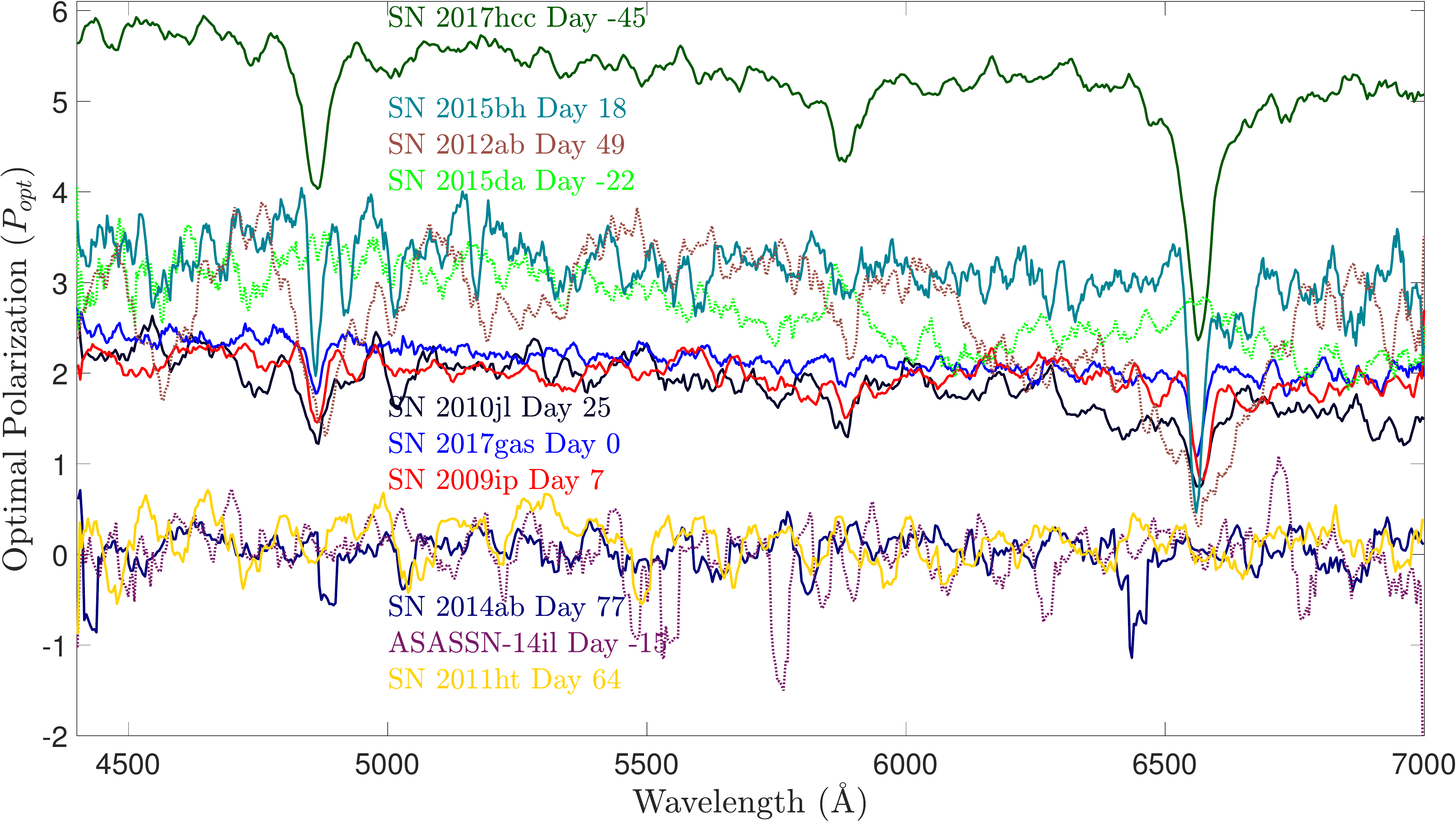}
\caption[Peak polarization spectra]{A comparison of the peak observed polarization spectra (unbinned) for most of our SNe IIn sample.  In cases where the polarization signal is low, we instead use high signal-to-noise ratio data temporally close to the peak date (see \S~\ref{dis:peakpol} for a detailed discussion on how we choose the spectra plotted here and why the observed gap is likely due to sample size).  Targets with spectropolarimetry that do not meet our noise cutoff are excluded from this plot for clarity.}
\label{fig:maximalpol}
\end{figure*}

We show a comparison of the peak polarization spectrum for each of our targets with high signal-to-noise ratio in Figure \ref{fig:maximalpol}.  For targets with an average optimal polarization degree above 1\% and a standard deviation in the optimal polarization below 1.25\%, we selected the spectrum with the highest intrinsic optimal polarization.  If the target was not observed to have an optimal polarization above 1\%, we instead plot the spectrum closest to the date of peak magnitude that meets our standard deviation cutoff in the continuum of $<1.25\%$.  There is a dearth of SNe with continuum polarization between $\sim 0\%$ to $\sim 2\%$, but with the small number of objects, it is uncertain whether this gap in intrinsic polarization is real or simply due to stochastic sampling.  If we had caught some of our more highly polarized targets at later times, we might have measured their peak polarization spectra in this region.  Thus, we suggest that the gaps in the peak polarization plot shown in Figure \ref{fig:maximalpol} may be due to sample size.  

Of particular interest here is the peak polarization spectrum of SN~2017hcc, because it reaches almost $6\%$ on day $-$45.  Early models of SN polarization that focused on SN~1987A predicted signals of up to 4\% for an oblate ellipsoid with a fattening of $E = 0.2$, where the fattening is the axis ratio of the elliptical density distribution of the envelope \citep{1991A&A...246..481H}.  Later work done on modeling SNe specifically with interaction \citep{2015MNRAS.449.4304D,2016MNRAS.458.1253V,2020A&A...642A.214K} only predicted polarization signals of up to about 2\%.  The various models considered included both symmetric and asymmetric SN ejecta interacting with a CSM disk, colliding wind shells in binary stars, bipolar lobes similar to those of the Homunculus nebula in $\eta$ Car \citep{2006ApJ...644.1151S},  and a relic disk similar to that considered for SN~1997eg \citep{2008ApJ...688.1186H}.  One key limitation of these models is perhaps that the full 3-D geometry is not modeled.  In general, if the CSM interaction region has a toroidal geometry, the SN ejecta  will progress more rapidly into the polar regions where the CSM is less dense, making the photosphere take on a prolate geometry and potentially engulf the disk (see \citealt{2015MNRAS.449.1876S}).  This can result in a complicated flux source geometry that the models fail to accurately depict when photons are deposited into the interaction regions at various angles in a 3-D consideration.

Models are generally unable to reach polarization as high as 6\% at early times. Perhaps this is because higher optical depths in the CSM that would be required to reach higher polarization degrees would also result in multiple scatterings that wash out the polarization signal from most viewing angles \citep{1996ApJ...461..828W}.  Additionally, the orthogonal geometries of the interaction region and the photosphere end up competing, producing polarization signals that may significantly cancel each other in the models (L. Dessart, private communication).  As discussed below in \S~\ref{sec:Dis:wavedepend}, however, an alternative way to easily achieve high polarization levels is with scattering by CSM dust, which is also consistent with the wavelength dependence that we observe.

\subsection{Continuum Polarization Evolution Through Time}
\label{sec:Dis:contpolevo}
In \S~\ref{IIn:sec:Res:intrspol} we summarized the observed temporal evolution of the continuum polarization for our sample of SNe IIn.  The strongest polarization signals in our sample exceed previously published values for other SNe IIn as discussed in \S~\ref{dis:peakpol}.  The temporal evolution we measure has proven to be diverse.  There are some emerging trends, however, such as steadily declining continuum polarization at late times.

\subsubsection{Declining Polarization at Late Times}
In the most strongly polarized objects for which we also have multiple epochs of spectropolarimetry at late times (SN~2010jl, SN~2017gas, SN~2017hcc), a simple drop in optical depth due to lower densities at larger radii may help explain most of the steady drop in polarization.  From epoch 2 onward for SN~2017gas and SN~2017hcc, as well as from epoch 1 for SN~2010jl, we generally see a drop in continuum polarization matched with an increase in the H$\alpha$ equivalent width and an increase in $V_{20}$.  In general, the H$\alpha$ equivalent width in SNe IIn increases as the continuum optical depth drops and the continuum luminosity fades \citep{2010ApJ...709..856S,2014MNRAS.438.1191S,2017hsn..book..403S}, whereas broad lines from the fast SN ejecta are typically exposed at late times as the optical depth of the CSM interaction region drops \citep{2017hsn..book..403S,2020MNRAS.499.3544S}. Figure \ref{fig:halphaEQWandFW20} shows that $V_{20}$ typically increases to 6000-10000\,km\,$\mathrm{s^{-1}}$ after day 100, consistent with the emergence of the fast SN ejecta.

In the case of other SNe II, we often see the polarization signal increase as the object enters the nebular phase, revealing the central mechanism of the explosion \citep{2001ApJ...553..861L,2006Natur.440..505L,2010ApJ...713.1363C,2021A&A...651A..19D}.  However, this is not seen in our sample of SNe IIn, likely because the CSM interaction regions are overwhelming the central SN ejecta while they remain bright.  After the CSM interaction has faded sufficiently to reveal the central SN ejecta (though in many cases for SNe~IIn, CSM interaction still dominates the brightness at late times, thus preventing a nebular phase from truly occurring) the SN ejecta are likely no longer bright enough to produce polarization comparable to that seen at nebular times in SNe II.

Although the drop in polarization for SN~2010jl, SN~2017gas, and SN~2017hcc can generally be attributed to decreasing optical depths, the changes that occur between the first two epochs of spectropolarimetry of SN~2107gas and SN~2017hcc defy this explanation.  The continuum polarization drops precipitously, matched instead with a decrease in $V_{20}$ and a decrease in the H$\alpha$ equivalent width.  Thus, the rapid drop in continuum polarization seen in SN~2017gas and SN~2017hcc between their first two epochs of data may be due to real geometrical changes in the photosphere, increased multiple scattering within the CSM interaction region, or decreased contribution to the luminosity from a light echo originating in CSM dust.  An increase in multiple scattering can wash out the polarization signal (\citealt{2016ApJ...817...32K}, though see \citealt{2003ApJ...598..572H} for a discussion on multiple scattering and viewing angle).  As the optical depth in the continuum likely increased during these epochs, multiple scattering may have begun to play a bigger role.  This may be more important in the case of SN~2017gas where the line emission was more significantly polarized at early times than that of SN~2017hcc, suggesting a line-emission region beneath the electron scattering photosphere.  Lastly, in both cases (though much more pronunced in SN~2017hcc), $S_{p}$ becomes less negative between the first two epochs.  This is consistent with early CSM dust producing a strong wavelength-dependent polarization in the first epoch, but then being obliterated before the 2nd epoch.

\subsubsection{Increasing Polarization at Early Times}
Sometimes the continuum polarization increases rather suddenly at early times near peak in our sample of SNe IIn.  Since SNe IIn are thought to be the result of SN ejecta interacting with CSM regions that were produced within years or decades prior to the death of the progenitor \citep{2014ARA&A..52..487S,2017hsn..book..403S}, it would make sense that multiple CSM shells could exist at a variety of distances from the SN.  As the SN ejecta expand and reaches new CSM shells, this interaction could cause the polarization signal to suddenly increase if the newly overtaken shells are asymmetric.  

If the external CSM shells have different geometries than each other or the underlying SN ejecta, this might not only result in a change in the magnitude of the polarization, but also the polarization angle, as is seen in the case of SN~2009ip.  Indeed, \citet{2014MNRAS.442.1166M} found that the change in polarization paired with the change in position angle is likely due to a mismatch between the initial SN ejecta photosphere occulted by a disk during the 2012a event and the later interaction with the disk that turns on during the 2012b event.  The other SNe IIn for which we observe a sudden increase in the polarization (SN~2012ab and SN~2015bh) do not show a large change in the position angle as in the case of SN~2009ip.  Their increase in polarization without a change in the position angle does suggest that any new CSM interaction that began later was still aligned with the previous interaction geometry.  This, in turn, suggests that multiple CSM regions may exist around SNe IIn that are aligned along the same axis.

For SN~2012ab, the sudden increase in polarization is paired with only a small change in position angle, suggesting instead a distant CSM region that contains a geometry aligned with that of the earlier source of polarization. Considering the spectral evolution of the line profiles, \citet{2018MNRAS.475.1104B} found that SN~2012ab's rise in polarization is likely due to interaction beginning on the far side of the SN, which had an axis of symmetry similar to that of the early-time interaction observed on the near side of the SN.

\begin{figure}
\centering
\includegraphics[width=0.5\textwidth,height=1\textheight,keepaspectratio,clip=true,trim=0cm 0cm 0cm 0cm]{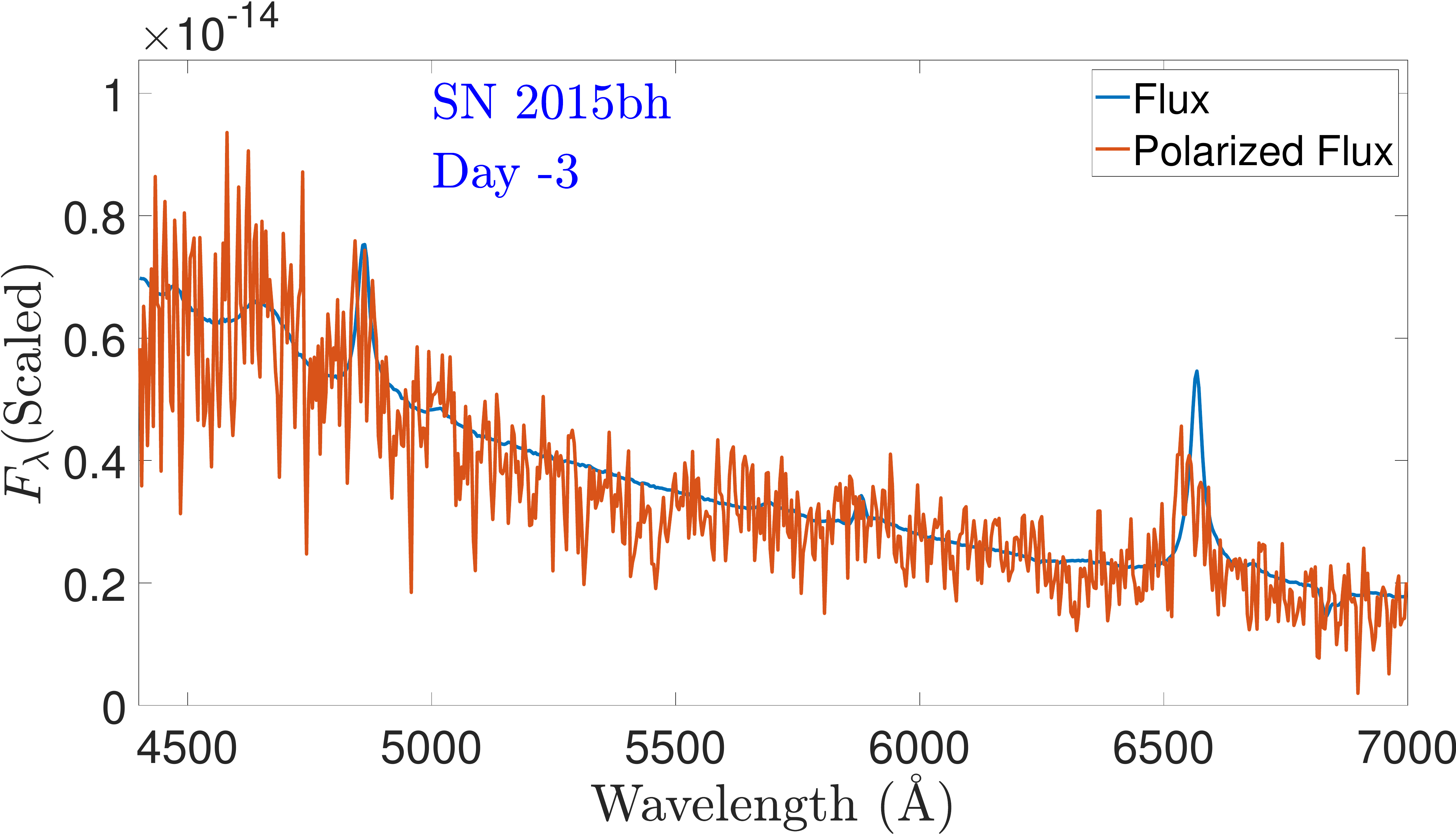}
\includegraphics[width=0.5\textwidth,height=1\textheight,keepaspectratio,clip=true,trim=0cm 0cm 0cm 0cm]{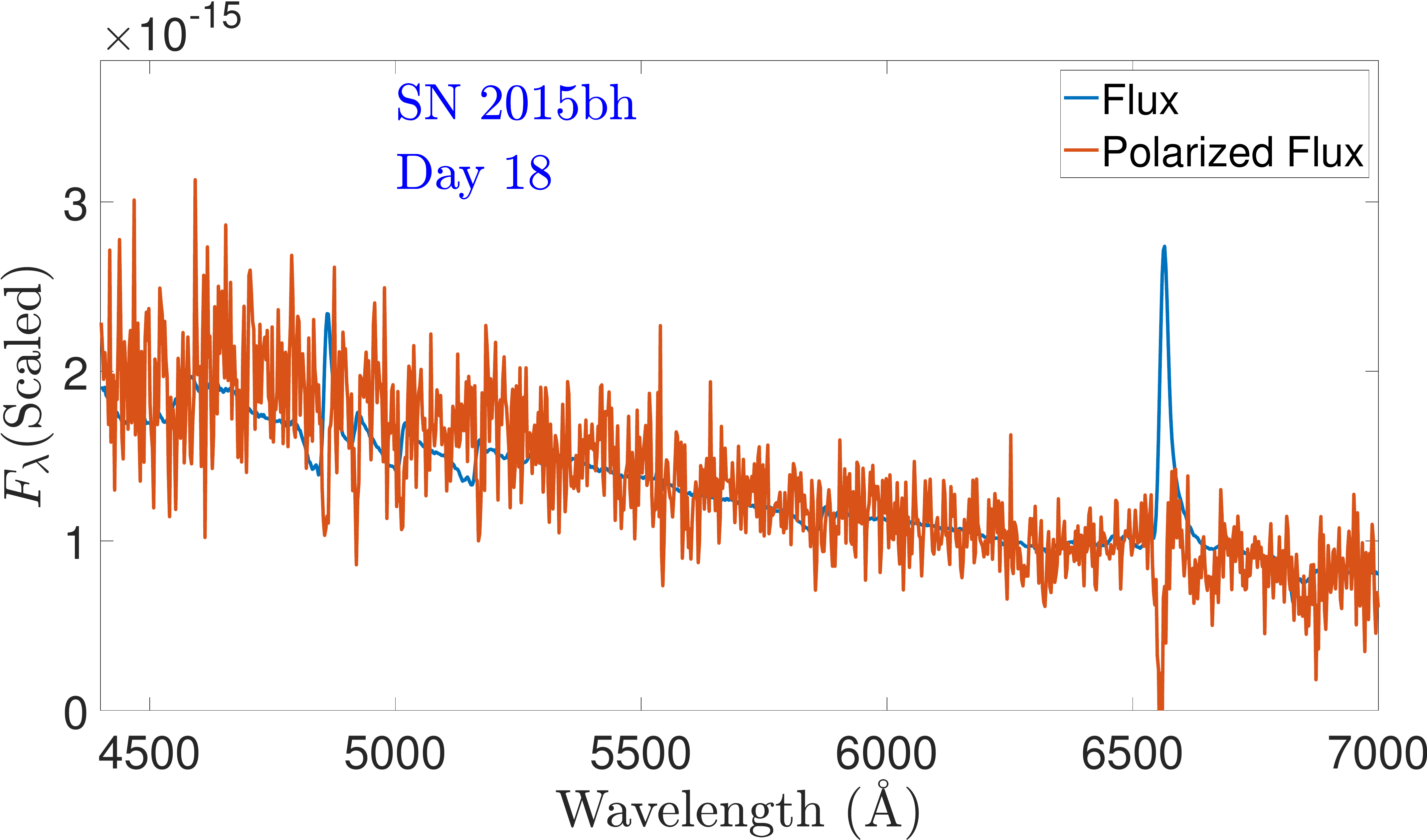}
\caption[Flux and polarized flux for SN~2015bh on days -3 and 18]{Same as Figure \ref{fig:polarizedflux} but for SN~2015bh on days -3 and 18, showing a transition between an initially nearly fully polarized H$\alpha$ emission line to one that shows negligible line polarization.}
\label{fig:bh}
\end{figure}

In the case of SN~2015bh, the rise in polarization is paired with a slight but significant change in the position angle, paired with a sudden onset of much stronger depolarization seen in Balmer emission lines.  This is likely due to the onset of CSM interaction with a geometric footprint similar to that of the photosphere arising at earlier times.  We expect that either the CSM interaction that began by day 18 for SN~2015bh has begun powering the photoionization of a new region of more distant CSM or the earlier line-emitting region has proceeded beyond the electron scattering photosphere, so that less of the line emission is polarized.  Figure \ref{fig:bh} shows the polarized flux for SN~2015bh between days $-$3 and 18, where initially the line emission shows significant polarization, but this line polarization decreases dramatically by day 18.  This confirms that by day 18 SN~2015bh has a new source of depolarizing line emission, either due to a new photoionized CSM shell or to the pre-existing line emission region now being external to the electron scattering photosphere.

A similar case arises for ASASSN-14il where the polarization increases initially and then remains constant, but in this case we do not observe strong line depolarization.  We suggest two possible scenarios in this case.  Either ASASSN-14il was initially polarized on day $-$15 and has become unpolarized by day 17 (with the remaining polarization signal seen on days 17-73 arising due to a strong ISP), or perhaps new CSM interaction began between days $-$15 and 17 and persisted until at least day 73 without any strong depolarizing line emission.  This might imply that the line emission region is still entrenched within the electron scattering photosphere, or that we do not have a good estimate of the ISP.  The change between day -15 and 17 does, however, still indicate that ASASSN-14il does have intrinsic polarization, even if it is difficult to tell whether it was instrinsically polarized at the earlier epoch, the later epoch, or perhaps both.

Previously published literature on SN~1997eg and SN~1998S also suggests an increase in the continuum polarization over time \citep{2000ApJ...536..239L,2008ApJ...688.1186H}.  In both cases, a diversity of scattering regions is evoked.  For SN~1997eg, \citet{2008ApJ...688.1186H}  proposed a dual-axis model with a toroidal CSM misaligned from the asymmetric underlying ejecta.  For SN~1998S, the change in polarization is described as interaction with a nearby CSM region that is then encompassed by the SN ejecta, that eventually runs into another disk of CSM at a later date that preserves interaction for a much longer period of time \citep{2000ApJ...536..239L}.  These are both consistent with the results from our sample of SNe IIn, which suggest that these objects have diverse CSM environments around them at the time of explosion which can result in a complicated series of increases and decreases in polarization, though rarely coupled to huge changes in the position angle.

\subsection{Light Echo from Dusty Distant CSM}
\label{sec:Dis:wavedepend}
In our measurements, the wavelength dependent slope of the polarization -- $S_{p}$ -- increases or decreases monotonically, or may change directions during its evolution (see \S~\ref{IIn:sec:Res:wavedep} for a summary of the measured evolution of the wavelength-dependent polarization). While we detect a tendency for SNe IIn to have a blue slope in the continuum, especially at early times, there is no clear trend in $S_{p}$ that unites all the SNe IIn in our sample.  Instead, we explore options that might allow for a wavelength-dependent polarization that produces a stronger polarization at blue wavelengths that can also occur at variable times.

\citet{2011MNRAS.415.3497D} showed that wavelength-dependent polarization can arise due to variation in albedo and continuum source function with wavelength and depth.  Primarily, bound-bound and bound-free transitions could cause this wavelength-dependent change in polarization.  However, the models in \citet{2011MNRAS.415.3497D} occur at relatively late times when the albedo is low and recombination is occurring.  Instead, a majority of our data show strong wavelength-dependent polarization at early times.  At these times, the SNe are expected to be hot and ionized, with a high albedo and negligible contribution from recombination.  Under these conditions, models considering the albedo show a mostly flat wavelength independent continuum polarization (L. Dessart, private communication), so we look to other options that can explain our early-time wavelength-dependence.

Light echoes included in the unresolved SN light have been observed for a variety of SNe \citep{1988A&A...198L...9G,2007ApJ...669..525W,2015MNRAS.451.1413A}. In the case of SNe IIn, it is plausible that distant CSM (whose presence is already likely for SNe IIn given their presumed progenitor history of eruptive mass-loss) causes such a light echo, which alters the polarization properties of the light significantly \citep{2018ApJ...861....1N}.  Light from this echo at the extrema tangential to our line of sight will be preferentially scattered at bluer wavelengths.  Additionally, because light is a transverse wave, the light from this echo that scatters orthogonally towards us will be highly polarized.  Given that the polarization signal we measure is on the order of a few percent, a light echo that does not significantly affect the overall luminosity of the SN could still significantly affect the polarization signal because its light is very strongly polarized.  This overall contribution to the light from tangential CSM dust may be why we see a strong predominance of a negative $S_{p}$ in most of our objects.

Additionally, the light from the CSM dust might arrive at variable times for the objects in our sample depending on the distance from the SN photosphere to the CSM.  This is supported by the diversity of the time evolution of $S_{p}$.  If the neutral and dusty CSM is close to the SN, we may see a light echo soon after explosion causing a wavelength-dependent shift in the overall polarization signal at early times, as is seen in SN~2015da and SN~2015bh.  In other cases where $S_{p}$ becomes more negative at later times (such as in the case of SN~2010jl), the CSM may be more distant and so the echo light arrives later.  Although it is beyond the scope of this paper, the time delay between the SN peak and the epoch with the most negative $S_{p}$ may provide a reasonable measure of the distance to the distant CSM that causes the wavelength-dependent polarization.  The magnitude of the wavelength-dependent polarization shift may also help inform the strength of the light echo and thus the scattering properties of the distant CSM.  With enough frequently-sampled spectropolarimetry, there might be a way to separate the polarization signal from the light echo and the CSM interaction region.  In the past, light echoes have been used as a powerful tool to explore the history of eruptive mass-loss and even separate such eruptions into different phases \citep{2018MNRAS.480.1466S,2018MNRAS.480.1457S}.

One of the most interesting objects from our sample that shows heavy wavelength-dependent polarization is SN~2015da.  In the 3rd epoch of spectropolarimetric data for SN~2015da, $S_{p}$ becomes significantly more negative at a time when the continuum polarization is also measured to be high and no line depolarization is seen.  Although the ISP constraint on SN~2015da from reddening is not very restrictive (ISP $<$8.73\%) because it is heavily reddened, the wavelength-dependent polarization is projected along the same position angle as the continuum polarization, suggesting that the external CSM dust shares a geometry with the continuum polarization region for SN~2015da.  It would be unlikely for the CSM dust that produces the wavelength-dependent polarization to be directly aligned with the ISP, so this suggests that the ISP for SN~2015da may actually be quite small.  This reinforces the idea that objects can indeed exist with reddening to ISP relations far below the upper limit set in \citet{1975ApJ...196..261S}.

Overall, we see a diversity of changes in $S_{p}$, which is consistent with a variable number of distant CSM shells producing light echoes with variable time delays depending upon their distance to the SN.  We find this to be the most plausible explanation for the general predominance of negative $S_{p}$ values, the diverse trends in $S_{p}$, and the numerous CSM-interaction-related conclusions that are already well-founded for SNe IIn.  

\subsection{Viewing Angle}
\label{sec:Dis:viewingangle}
Our sample of SNe IIn constitutes a diverse population with a variety of peak polarizations, rise times, and decay times, but it still has a few unifying factors like the steady drop in polarization at late times as H$\alpha$ equivalent widths and $V_{20}$ rise, and a general preference for showing a blueward slope in the continuum polarization level.  Here, we focus on the facts that 1. SNe~IIn show strong continuum polarization at early times, 2. a nonnegligble but significant fraction show little polarization at any time, and that 3. almost all SNe IIn have a negative $S_{p}$.  The combination of these properties suggests that a common axisymmetric geometry with a variety of viewing angles may play an important role in the spread of measured polarization signals for these objects, as already suggested in \citet{2020MNRAS.498.3835B} when studying SN~2014ab.

\begin{figure*}
\centering
\includegraphics[width=1\textwidth,height=1\textheight,keepaspectratio,clip=true,trim=0cm 0cm 0cm 0cm]{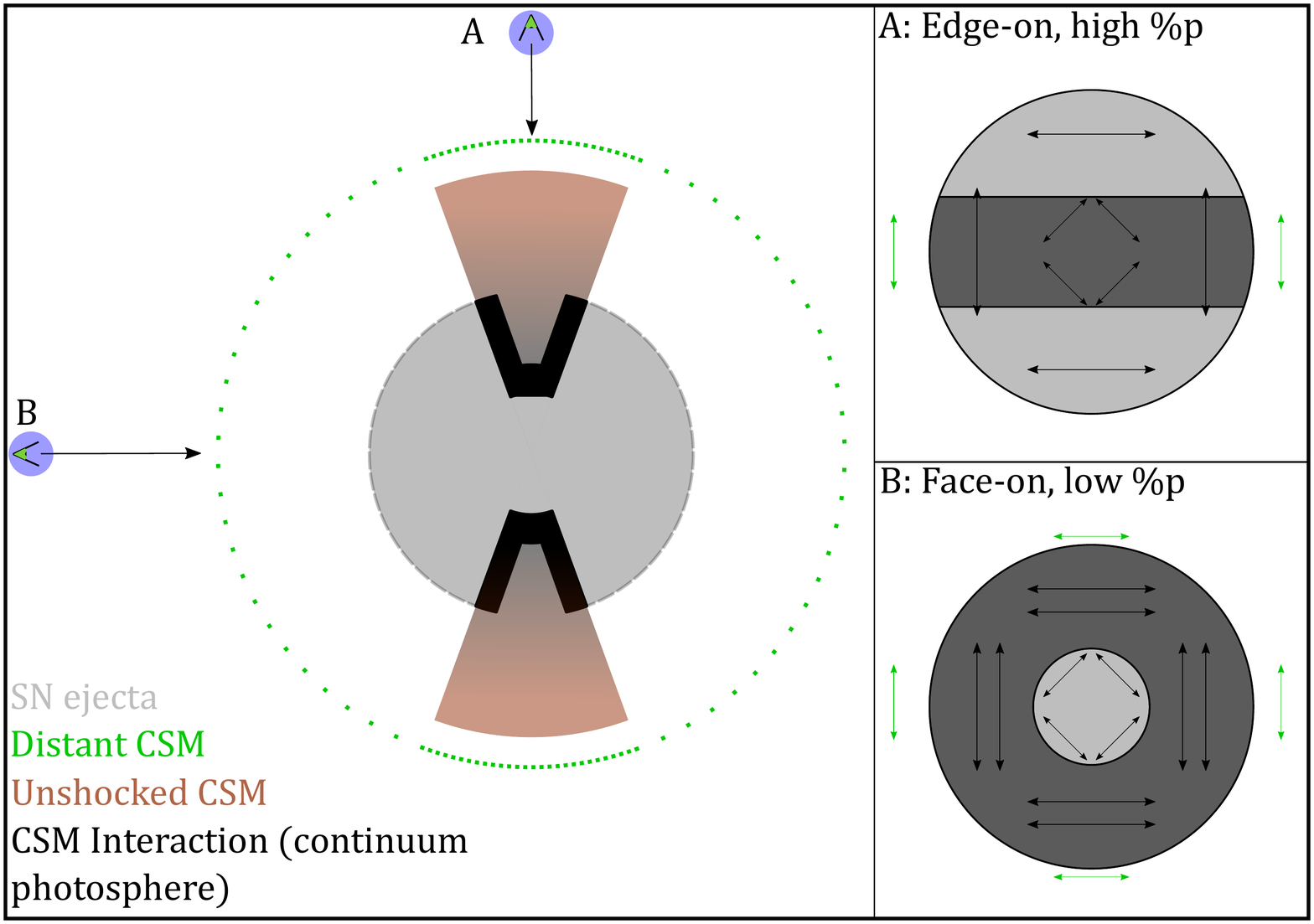}
\caption[Viewing angle schematic with dusty distant CSM]{An updated schematic from the previous work on SNe IIn viewing angles in \citet{2020MNRAS.498.3835B}.  We present key observable features for SNe IIn in this schematic as viewed from two orthogonal locations, viewing points A and B.  While we place A and B at two orthogonal extremes, the viewing angles at which we observe targets within our sample likely lie somewhere in between these two points.  There may be many CSM shells at a diversity of distances that produce narrow emission and absorption lines as well as potentially scatter light from the SN photosphere as an echo back into our line of sight.  These CSM shells may project anywhere in a 3-dimensional shell (into/out of the page).  Because the position angle does not typically change significantly when the polarization signal increases for most of our targets, we suggest that these CSM shells may be preferentially aligned with the equatorial CSM interaction regions.  The unshocked CSM shown in a brown color likely produces the majority of strong narrow Balmer-series emission and absorption.  The equatorial torus of CSM interaction with the interior SN ejecta (shown in black) is likely where the continuum photosphere resides in most of our observations.  The narrow/intermediate-width unpolarized line emission and broad Lorentzian polarized line emission likely originate in this CSM interaction region.  The SNe ejecta are shown in grey, though we do not actually constrain the geometry of the SN ejecta to be spherical.  The magnitude of the polarization from the distant CSM dust echoes may be larger than the polarization from the electron scattering geometry, even though the cartoon depicts it as having smaller polarization.}
\label{fig:IIn:cartoon}
\end{figure*}

In Figure \ref{fig:IIn:cartoon}, we show an updated schematic derived from the work in \citet{2020MNRAS.498.3835B}, with a predominantly disk-like or toroidal geometry for the densest CSM.  We have included the consideration that the distant CSM (shown in green) may be causing wavelength-dependent polarization due to CSM dust scattering.  At early times in most SNe IIn, the CSM is optically thick and the emitting photosphere is ahead of the forward shock \citep{2017hsn..book..403S}, and so we do not see the SN ejecta directly.  As such, we are unable to probe the geometry of the SN ejecta with spectropolarimetry unless we observed the SN prior to the start of CSM interaction (as was the case for SN~2009ip; \citealt{2014MNRAS.438.1191S,2014MNRAS.442.1166M}).  Instead, spectropolarimetry of SNe IIn probes the asymmetry of the SN environment and helps us learn more about the SN progenitor and its final years that otherwise would have remained hidden (though see \citealt{2016ApJ...818....3K} for a discussion on how flash spectroscopy may inform mass loss from core collapse events as well).

Suppose that all SNe IIn have a similar axisymmetric geometry (such as the disk-like/toroidal one shown in Figure \ref{fig:IIn:cartoon}).  In that case, when various SNe IIn are viewed from a random distribution of angles relative to the polar axis, we should see a distribution of polarizations from some upper threshold, corresponding to systems seen nearly edge-on, down to zero.  If the CSM is really axisymmetric, nearly pole-on views (like that of SN~2014ab, similar to that of panel B in Figure \ref{fig:IIn:cartoon}) will show very low polarization at all times, whereas viewing directions that are between mid-latitudes and the equatorial plane will see the highest polarization (similar to panel A in Figure \ref{fig:IIn:cartoon}).  If the distant dusty CSM has the same axisymmetric geometry as the inner CSM disk, we would expect to see a strong increase in the polarization for objects viewed from mid-latitudes at early times (explaining how something like SN~2017hcc might have reached a polarization degree of almost 6\%).  As optical depths fade at late times, the polarization degree will fall for all objects with an electron scattering component in the polarization.

According to Figure \ref{fig:maximalpol}, 7 of the 10 objects shown there display significant polarization, whereas only 3 maintain low polarization at all epochs (the other 4 objects in our sample were excluded because they did not meet the signal-to-noise ratio cutoffs we set in this analysis).  Although it would likely require sophisticated models to predict a detailed relation between viewing angle and polarization degree, we can attempt to estimate the fraction of SNe IIn that we would expect to see with low or high polarization based on a simple statistical model.  In our equatorial disk model, we assume that objects viewed between 0\degree\, and 45\degree\,(0\degree\, being edge-on with the disk) would be significantly polarized, while objects viewed nearer the polar axis, with viewing angles between 45\degree\, to 90\degree, would be only weakly polarized.  In this axisymmetric model, we would expect that $\sim$ 71\% of objects would be found within the mid-latitudes and $\sim$ 29\% within the polar latitudes.  Although our sample is small, our observations are in line with these predictions.  Nevertheless, because our sample of SNe IIn is quite small (even though it is the largest spectropolarimetric data set for SNe IIn yet), we do not attempt to empirically constrain the range of viewing angles that might produce low or high polarization signals.

\subsection{Implications for Pre-SN Mass Loss}
Despite the large diversity of polarization properties in the SN IIn sample, there are two key properties that most SNe IIn exhibit.  Every object with late-time data shows a steady drop to low levels of polarization at late times.  This does not necessarily mean that the more distant CSM hit by the shock at late times is more spherical.  Instead, this drop may result because at late times when the optical depth has dropped due to lower CSM densities at large radii, the electron scattering continuum is making a weaker contribution to the total light, even though CSM interaction continues.  This is consistent with the observation that the H$\alpha$ equivalent width rises at late times as the continuum fades away (Figure \ref{fig:halphaEQWandFW20}).  Additionally, most SNe IIn exhibit a wavelength-dependent polarization at some point in their evolution, but only in rare cases (SN~2009ip) do we see a change in the position angle.  This implies that the scattering by distant CSM dust that contributes the wavelength-dependent polarization is asymmetric, but mostly aligned with the inner CSM that gives rise to the electron scattering polarization signal.  This provides an important constraint for pre-SN eruptive mass loss: namely, sources of progenitor mass loss must be able to produce highly axisymmetric CSM {\it with a persistent and stable orientation} during these eruptive episodes.

Based on the requirement of extremely high mass-loss rates needed to power luminous SNe IIn through CSM interaction, progenitors for SNe IIn have mainly been suggested to be LBVs \citep{2005ASPC..332..302S,2006ApJ...645L..45S,2007ApJ...666.1116S,2008ApJ...686..467S,2010ApJ...709..856S,2007ApJ...656..372G,2011ApJ...741....7F,2013MNRAS.430.1801M,2014ApJ...797..118F,2014ARA&A..52..487S,2017MNRAS.471.4047A}, though extreme red supergiant progenitors have also been proposed \citep{2009AJ....137.3558S,2009ApJ...695.1334S}.  Other related clues, such as variable winds, have also pointed to LBVs as potential SN IIn progenitors \citep{2008A&A...483L..47T}. The mechanism by which LBVs undergo their episodes of eruptive mass loss is uncertain.  Explosions resulting in a strong blast wave \citep{2008Natur.455..201S} or super-Eddington winds \citep{2000ApJ...532L.137S,2004ApJ...616..525O,2006ApJ...645L..45S} have been suggested, but the source of energy for either mechanism is still unclear.  Pulsational pair instabilities \citep{2002RvMP...74.1015W,2007Natur.450..390W} and wave-driven mass loss \citep{2007ApJ...667..448M,2012MNRAS.423L..92Q,2014ApJ...780...96S} models might account for this additional needed energy in SN precursors, but these models are studied in 1-D and they do not depend on nor result in axisymmetric geometries.  Thus, they drive mass loss without necessarily creating an axisymmetric CSM with a persistent orientation as is suggested by our observations.  Instead, repeated binary interactions \citep{2011MNRAS.415.2009S,2014ApJ...785...82S,2018MNRAS.480.1466S} or pre-SN mergers \citep{2010MNRAS.406..840P,2014ApJ...785...82S} could supply the energy needed for eruptive mass loss, and they are also consistent with axisymmetric CSM with a persistent orientation.

Bipolar shapes and binary companions appear to be common around evolved massive stars with visible nebulae, including famous examples like $\eta$ Carinae \citep{1996ApJ...460L..49D,1997NewA....2..107D,2006ApJ...644.1151S,2018MNRAS.474.4988S} and the progenitor of SN~1987A \citep{1995ApJ...452L..45C}.  In particular, a bipolar nebula with rarified CSM along the pole and dense CSM along the equator, much like what is seen for $\eta$ Carinae \citep{2006ApJ...644.1151S,2018MNRAS.474.4988S} fits this picture well.  The multiple CSM shells at a range of distances from the SN that we predict from our observations are reminiscent of the multiple eruptions that $\eta$ Carinae has experienced in the recent past \citep{2004ApJ...605..854S,2016MNRAS.463..845K,2018MNRAS.480.1466S}.  \citet{2020MNRAS.499.3544S} show how this bipolar geometry with dense equatorial CSM can also explain the spectroscopic evolution of SN~2017hcc.  Since this bipolar nebula model is also consistent with SNe IIn that exhibit low polarization (such as SN~2014ab) if viewed along the polar axis, we suggest that this is the most promising unified picture for the environments of SNe IIn. Obviously, some individual objects may also deviate from this clean picture; for example, binary systems with eccentric orbits may act to disrupt the axisymmetry of the CSM \citep{2018MNRAS.475.1104B}.

\section{Summary}
For the first time, we present multi-epoch spectropolarimetric data for a sample of SNe IIn.  This sample includes 14 separate SNe IIn.  The continuum polarization measurements exhibit a diversity of trends, which is expected for this class of SNe that already exhibit tremendous heterogeneity \citep{2011MNRAS.412.1441L,2014AJ....147..118R}.  Below we summarise a few key unifying results discovered across our data set and from past published spectropolarimetric studies of SNe IIn:

\begin{itemize}
\item{Estimating the exact ISP contribution for SNe IIn can be difficult.  Reddening constraints (i.e. Na I D) only set upper limits on the ISP without placing it at a particular location in the $q-u$ plane, while depolarization of strong emission lines is uncertain because the lines themselves are often polarized to some extent (especially at early times).}

\item{SNe IIn can exhibit intrinsic polarization in the continuum as high as 5.76\%.  This is higher than the polarization degree level measured for any other type of SN, and is also beyond the expected polarization from models of SNe IIn that adopt electron scattering as the dominant source of continuum polarization, though modelling the polarization signals of interacting SNe is still in its infancy.}

\item{At late times, the gradual decline in the continuum polarization seen in many of our targets with multi-epoch spectropolarimetric data can be explained effectively by a drop in the optical depth of the CSM interaction region with time.  We generally see an increase in the equivalent width of H$\alpha$ as the continuum polarization fades and the H$\alpha$ line profile becomes broader.}

\item{At early times in some objects like SN~2017gas and SN~2017hcc, the continuum polarization drops rapidly as the equivalent width of H$\alpha$ decreases and the H$\alpha$ emission lines become narrower.  This different behavior at early times could be due to real geometrical changes in the photosphere, increased multiple scattering within the CSM interaction region, or a decreased contribution to the total luminosity from light scattered off CSM dust (perhaps because the dust was destroyed soon after the initial explosion).}

\item{Many SNe IIn show sudden increases in the continuum polarization or changes in their $S_{p}$.  These can generally be attributed to CSM regions existing at a diversity of distances from each SN IIn.  Some experience strong CSM interaction early on, while others experience a delay before the interaction begins.  Some SNe IIn even show evidence of multiple CSM shells, which are generally aligned with each other.}

\item{The majority of SNe IIn exhibit wavelength-dependent continuum polarization with a stronger polarization at blue wavelengths.  This is not expected for wavelength-independent electron scattering.  We are likely observing the combination of a polarization signal from the continuum electron scattering region found within CSM interaction and additional wavelength-dependent polarization from a light echo scattered towards us by CSM dust.  When these two geometries are aligned, they add constructively.}

\item{The  diversity of features seen in spectropolarimetric data for SNe IIn can potentially be explained by a combination of diverse environments with multiple CSM shells at various distances combined with a persistent axisymmetric geometry that is seen from a range of different viewing angles.}

\item{Most importantly, SNe IIn require an eruptive pre-SN mass loss mechanism that is highly asymmetric and maintains a persistent, perhaps axisymmetric, geometry.  Mass-loss mechanisms that lead to spherically symmetric ejections prior to death do not adequately match the observed CSM properties of SNe IIn.  Binary interactions and eruptive mass-loss focused within an equatorial disk may provide a plausible explanation for the polarization features we observe, whereas deep seated energy deposition in spherically symmetric stars would seem to be strongly disfavored overall.}

\end{itemize}
\section{Future Prospects}	
This is the first study of a sample of more than one SN IIn, and it has revealed a number of interesting trends regarding the evolution of the polarization and its wavelength dependence over time.  However, there are a few outstanding questions that could be answered with improved temporal coverage with spectropolarimetry at either early or late times, and with other types of observations that may help clarify some outstanding mysteries.  Specifically, future studies could benefit the understanding of SNe IIn explosion geometries and their environments in ways that we outline below.  

\subsubsection{High-cadence early spectropolarimetry}	
Additional high-cadence early-time spectropolarimetry would help examine the source of the high polarization signal sometimes seen near peak in more detail, and might clarify why other objects do not show this.  By following the early-time evolution of both the polarization and the intrinsic slope parameter, one may be able to estimate the distance to the nearest CSM and the extent of the contribution from dusty CSM light echoes.  Although we have early-time data for a few objects, only one (SN~2009ip) exhibits a rotation in the polarization consistent with a transition from a SN ejecta photosphere to one located in a CSM interaction region \citep{2014MNRAS.442.1166M}.  There may also be unusual spectropolarimetric signatures at early times due to pre-SN outbursts or double-peaked light curves.  For instance, SN~2009ip is unique in having spectropolarimetric data during its first peak in a double-peaked light curve.  If one could observe SNe IIn with spectropolarimetry at earlier times, one might be able to constrain the transition from explosion until the onset of CSM interaction.  If it is common for SNe IIn to experience a $\sim$ 90\degree\, rotation in their geometry between the time of explosion and first CSM interaction, this would support the bipolar nebula with an equatorial disk picture for SNe IIn.

\subsubsection{Improved late-time coverage with spectropolarimetry}	
Several SNe within our sample lack deep late-time spectropolarimetry, but the ones that do have such data all show a decline in polarization.  Acquiring more late-time spectropolarimetry would help confirm that the continuum polarization does fade for all SNe IIn at late times.  Additionally, late-time spectropolarimetry (especially using larger telescopes that can still detect a significant signal from the SN as it fades) could provide another estimate of the ISP when the intrinsic polarization of the SN has faded. Spectropolarimetry at late times requires relatively nearby SNe IIn and large telescopes.

\subsubsection{Higher resolution spectropolarimetry}	
SNe IIn are different from other core-collapse SNe in that they have strong narrow lines.  At all times, higher resolution spectropolarimetry would be useful in detecting specific differences across emission line features.  This would be particularly useful at early times when line emission shows significant polarization, so that estimates of the ISP from depolarization could isolate all polarized flux (both polarized continuum and polarized broad emission-line flux) from the unpolarized flux (the narrow-component of the emission line).

\subsubsection{Light curve comparisons and constraints}	
Although light curves are available for a number of the SNe IIn within our sample, many only have sporadic photometric information.  Well-sampled light curves would help confirm whether the general drop in the continuum polarization towards late times occurs alongside a similar drop in the continuum optical depth.  Current estimates of the time of peak for many of the SNe IIn within our sample are uncertain, especially when the time of peak is coincident with the time of discovery.  Additionally, a well-sampled early-time light curve could help estimate the explosion date for the SN, which would be useful in estimating the distance to the external CSM, especially if a wavelength-dependent light echo is observed.	

\subsubsection{X-ray, radio, and infrared observations}	
The spectropolarimetry we use is all observed at visual wavelengths.  However, studies of X-ray- or radio-wavelength emission could help corroborate the axisymmetric model we present.  X-rays generated in the shock interaction region may escape more easily from aspherical environments.  In particular, X-rays should escape more easily along the polar caps in the axisymmetric geometry we present.  Thus, we would expect to see greater X-ray emission at early times from objects with low continuum polarization like SN~2014ab, SN~2011ht, or ASASSN-14il.  Similarly, it would be useful to know if significant emission from dust is present at thermal-infrared wavelengths at the same times that we detect a strong wavelength dependence in the continuum polarization which we have attributed to scattering by CSM dust.

\section*{Acknowledgments}
The SNSPOL project is supported by the National Science Foundation under awards AST-1210599 to the University of Arizona, AST-1210372 and AST-2009996 to the University of Denver, and AST-1210311 and AST-2010001 to San Diego State University.  N.S. received additional support from NSF grants AST-1312221 and AST-1515559, and by a Scialog grant from the Research Corporation for Science Advancement.  Research by D.J.S. is supported by NSF grants AST-1821967, 1821987, 1813708, 1813466, 1908972, and by the Heising-Simons Foundation under grant \#2020-1864.  P.S. was supported by NASA/Fermi Guest
Investigator Program grants NNX09AU10G, NNX12AO93G, and NNX15AU81G. 
 D.C.L. acknowledges support from NSF grants AST-1009571 and AST-1210311, under which part of this research was carried out. J.L.H. acknowledges that the University of Denver resides on the ancestral territories of the Arapaho, Cheyenne, and Ute, and that its history is inextricably linked with the violent displacement of these indigenous peoples.  This paper made use of the NASA/IPAC Extragalactic Database (NED), which is operated by the Jet Propulsion Laboratory, California Institute of Technology, under contract with NASA.  This paper made use of data from Pan-STARRS1 acquired after May 2014.  Operation of the Pan-STARRS1 telescope is supported by the National Aeronautics and Space Administration under Grant No. NNX12AR65G and Grant No. NNX14AM74G issued through the NEO Observation Program.

We thank the staffs at the MMT, Bok, and Kuiper telescopes for their assistance with the observations.  Observations using Steward Observatory facilities were obtained as part of the large observing program AZTEC: Arizona Transient Exploration and Characterization. Some observations reported here were obtained at the MMT Observatory, a joint facility of the University of Arizona and the Smithsonian Institution.

\section*{Data Availability}
The data underlying this article will be shared on reasonable request to the corresponding author.

\clearpage
\appendix
\appendixpage

\section{$q-u$ Plots for the Entire Sample of SNe IIn}
\label{App:A}
In this appendix we show all 49 epochs of spectropolarimetric data within our sample as $q-u$ plots.  If an estimate of the ISP was made from line depolarization for the target, we also show the spectropolarimetric data after ISP correction.

\newcommand{\qucaption}[6]{\caption{$q-u$ plots as described at the start of Appendix \ref{App:A}.  {\it Top panel:} #1 #2 Day #3 {\it Bottom panel:} #4 #5 Day #6}}

\newcommand{\qufig}[8]{
\begin{figure*}
    \centering
    \includegraphics[width=1.0\textwidth,height=1.1\textheight,keepaspectratio,clip=true,trim=0cm 0cm 0cm 0cm]{figures/appendixQU/#4.pdf}
    \includegraphics[width=1.0\textwidth,height=1.1\textheight,keepaspectratio,clip=true,trim=0cm 0cm 0cm 0cm]{figures/appendixQU/#8.pdf}
    \qucaption{#1}{#2}{#3}{#5}{#6}{#7}
\end{figure*}
}

\qufig
{SN~2010jl}{Kuiper}{25}{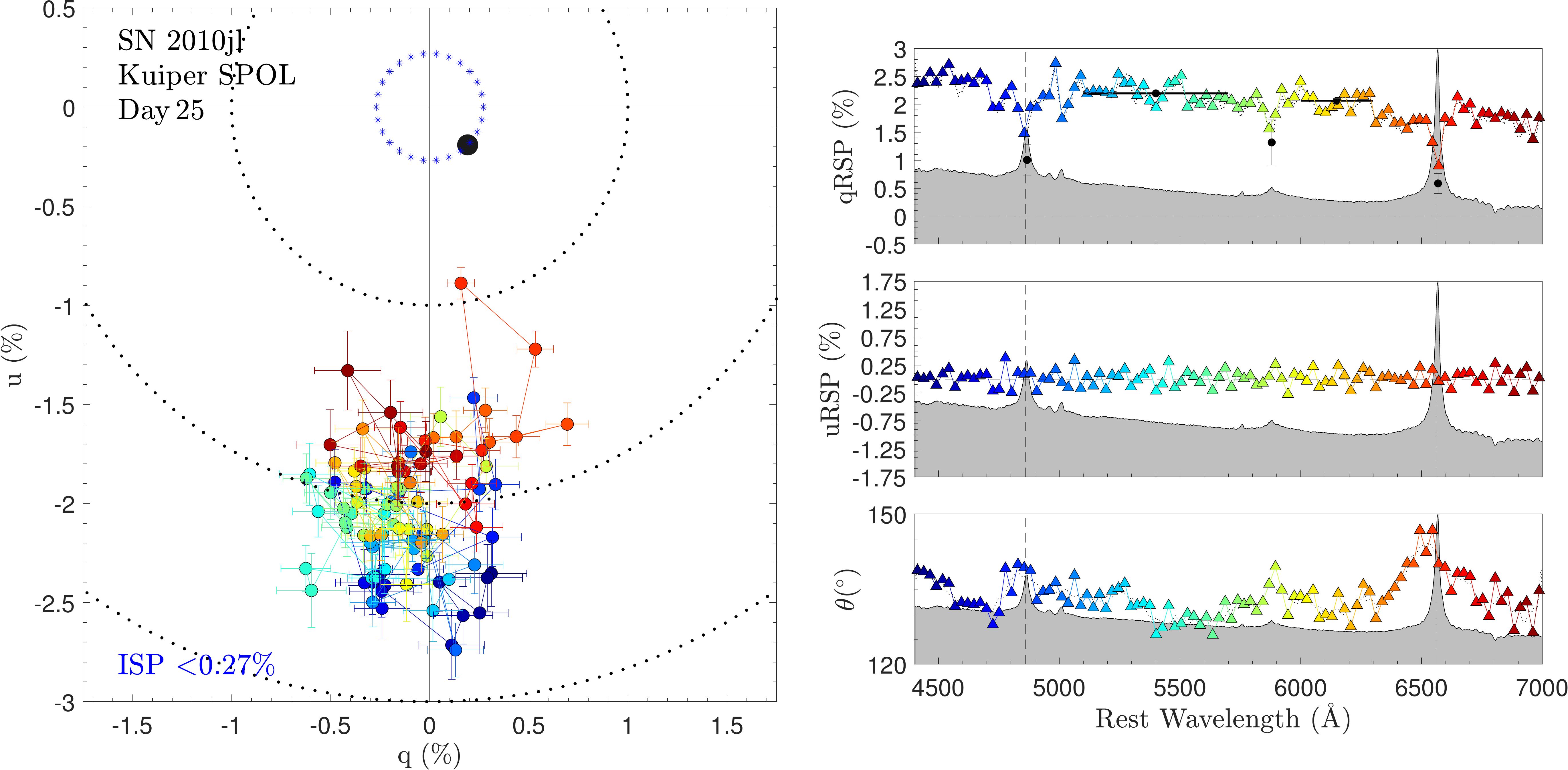}
{SN~2010jl}{Kuiper}{25, ISP-corrected}{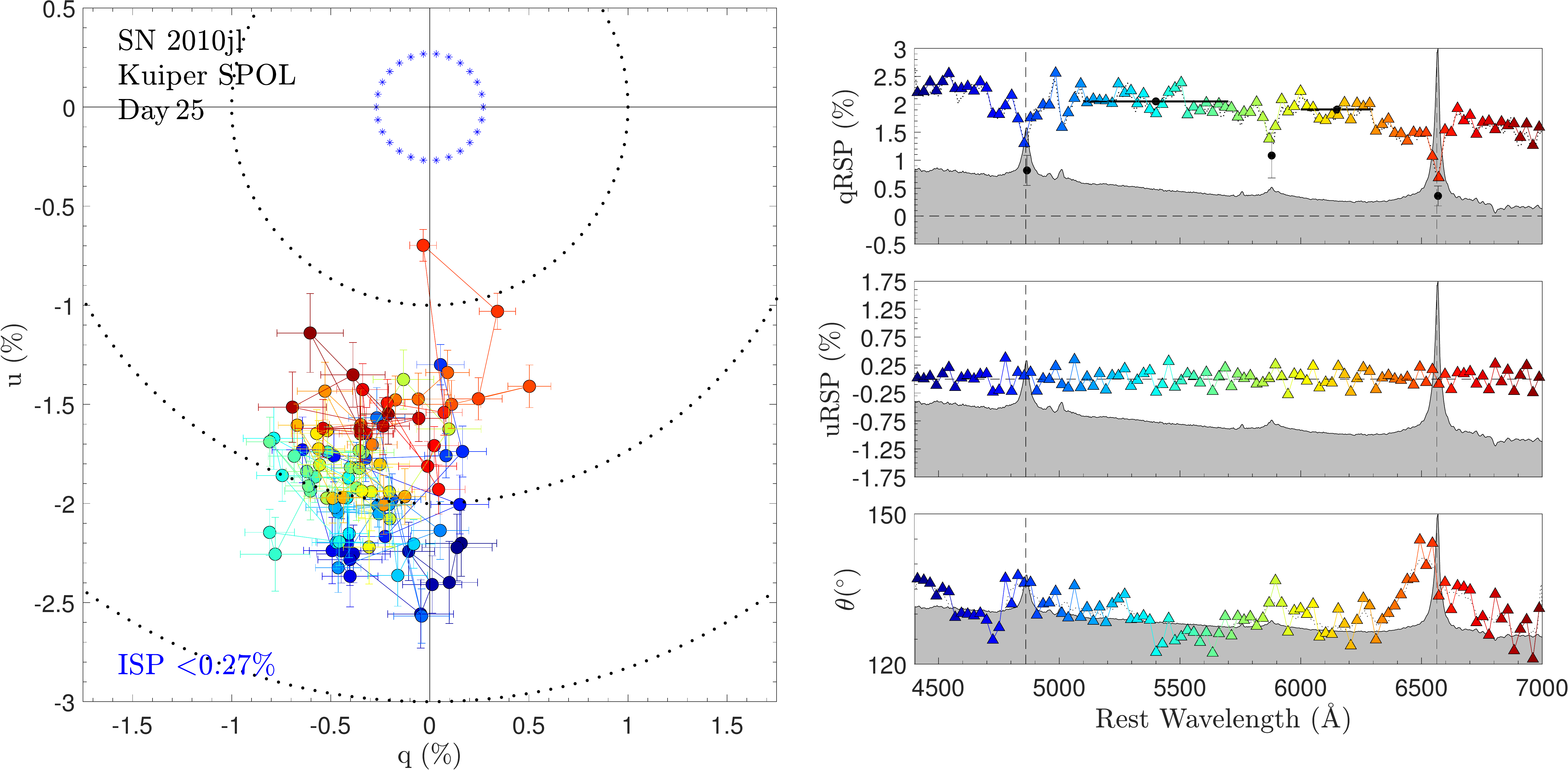}
\qufig
{SN~2010jl}{Kuiper}{45}{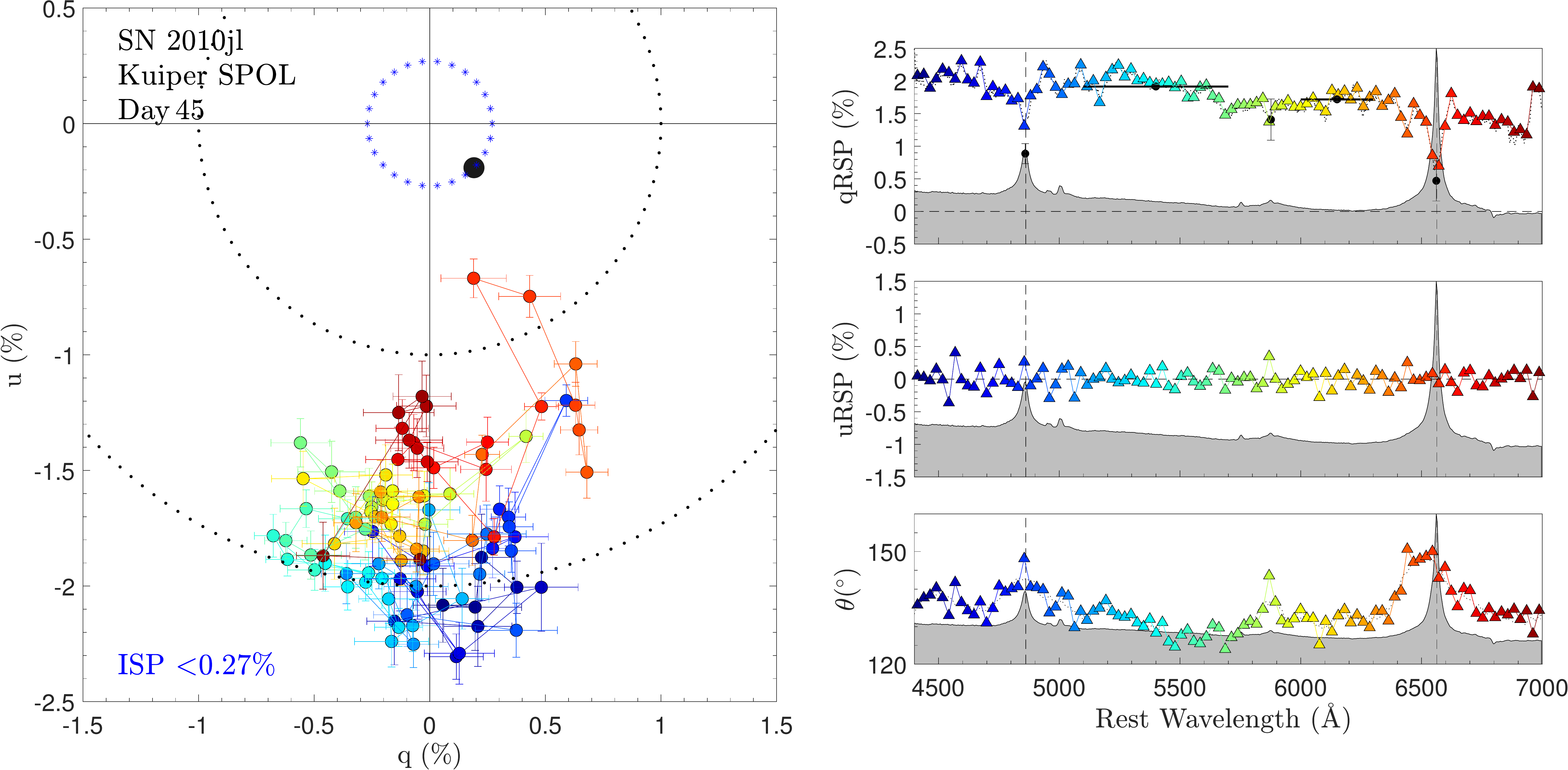}
{SN~2010jl}{Kuiper}{45, ISP-corrected}{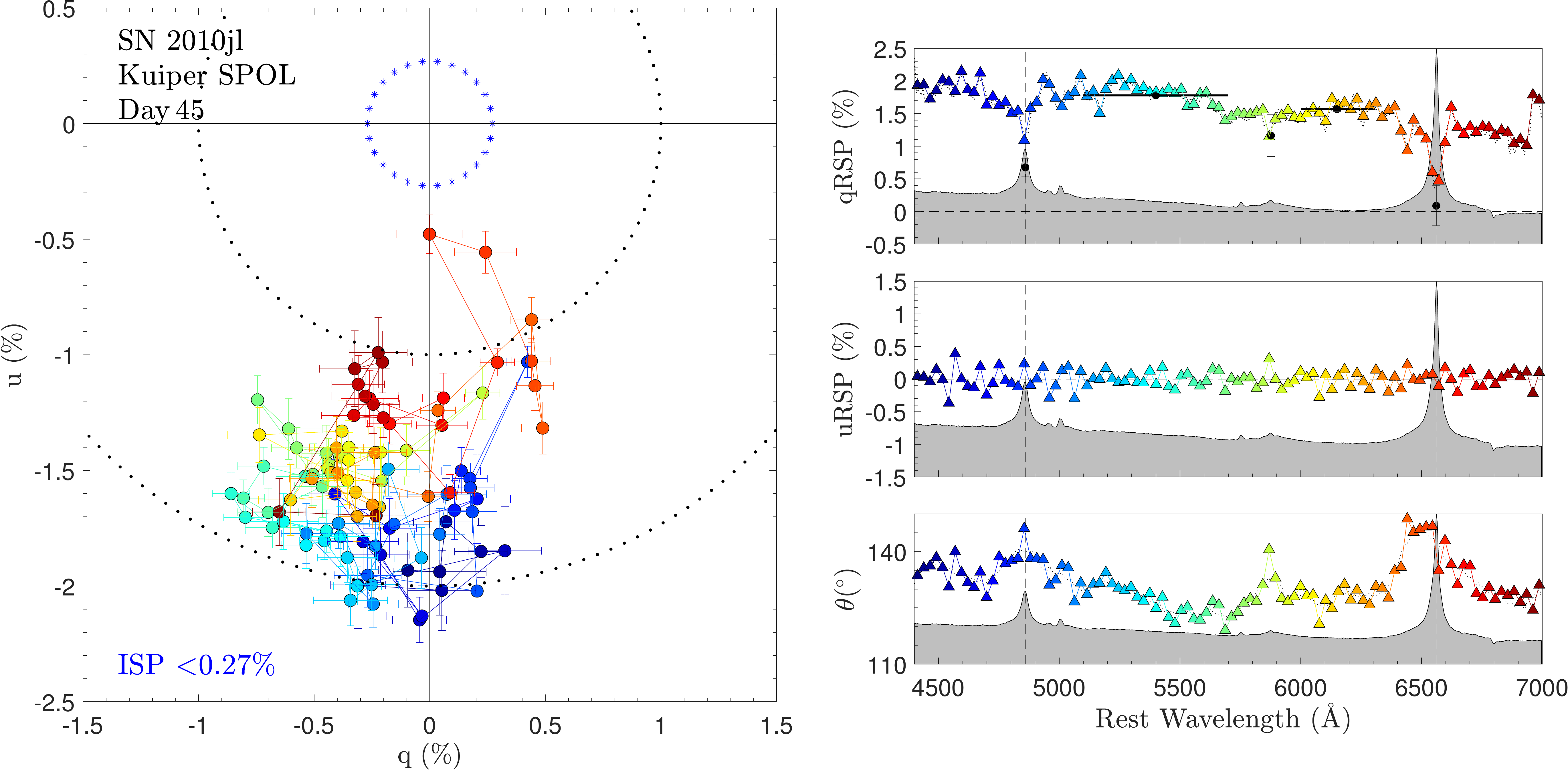}
\qufig
{SN~2010jl}{Bok}{76}{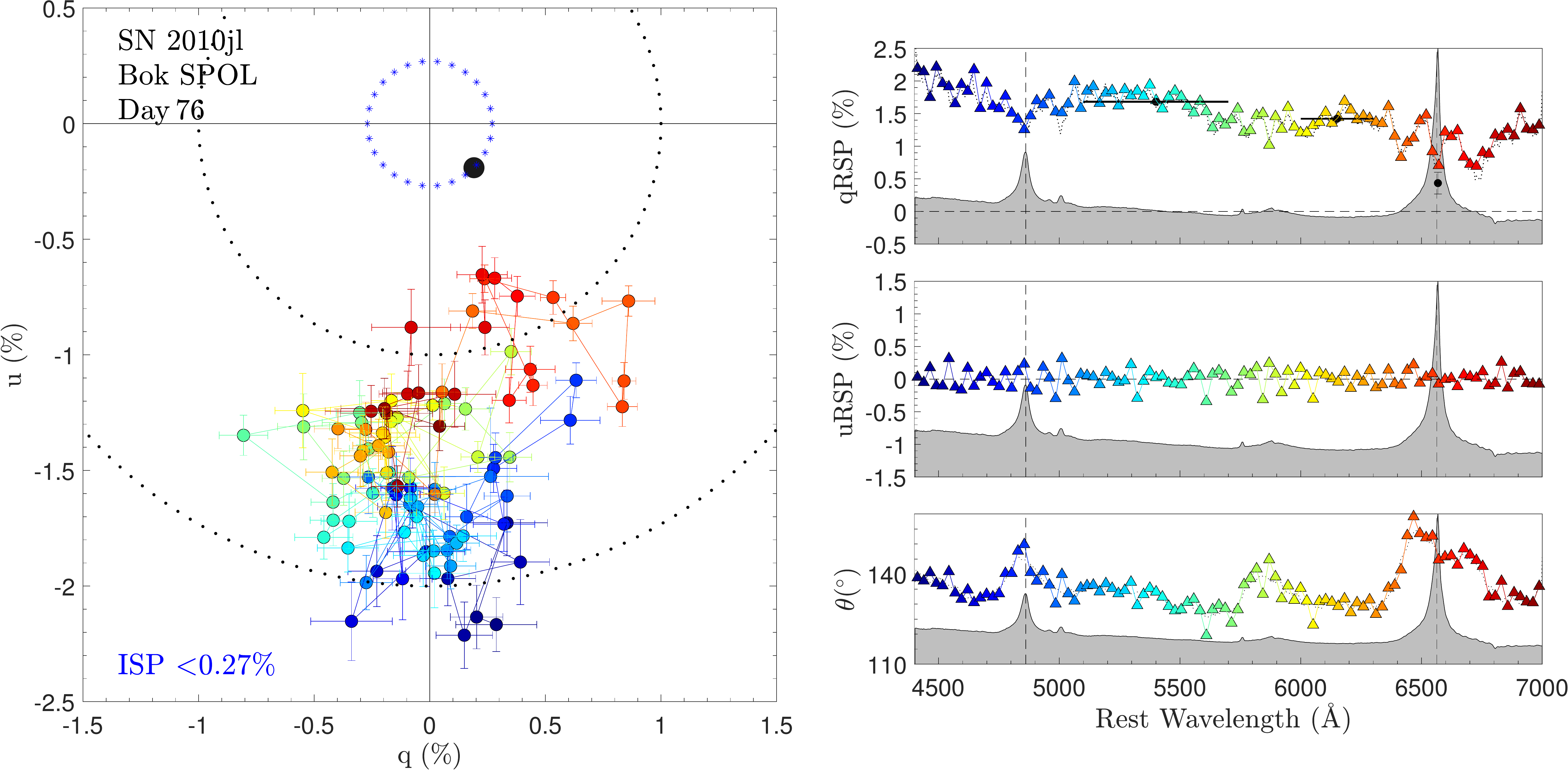}
{SN~2010jl}{Bok}{76, ISP-corrected}{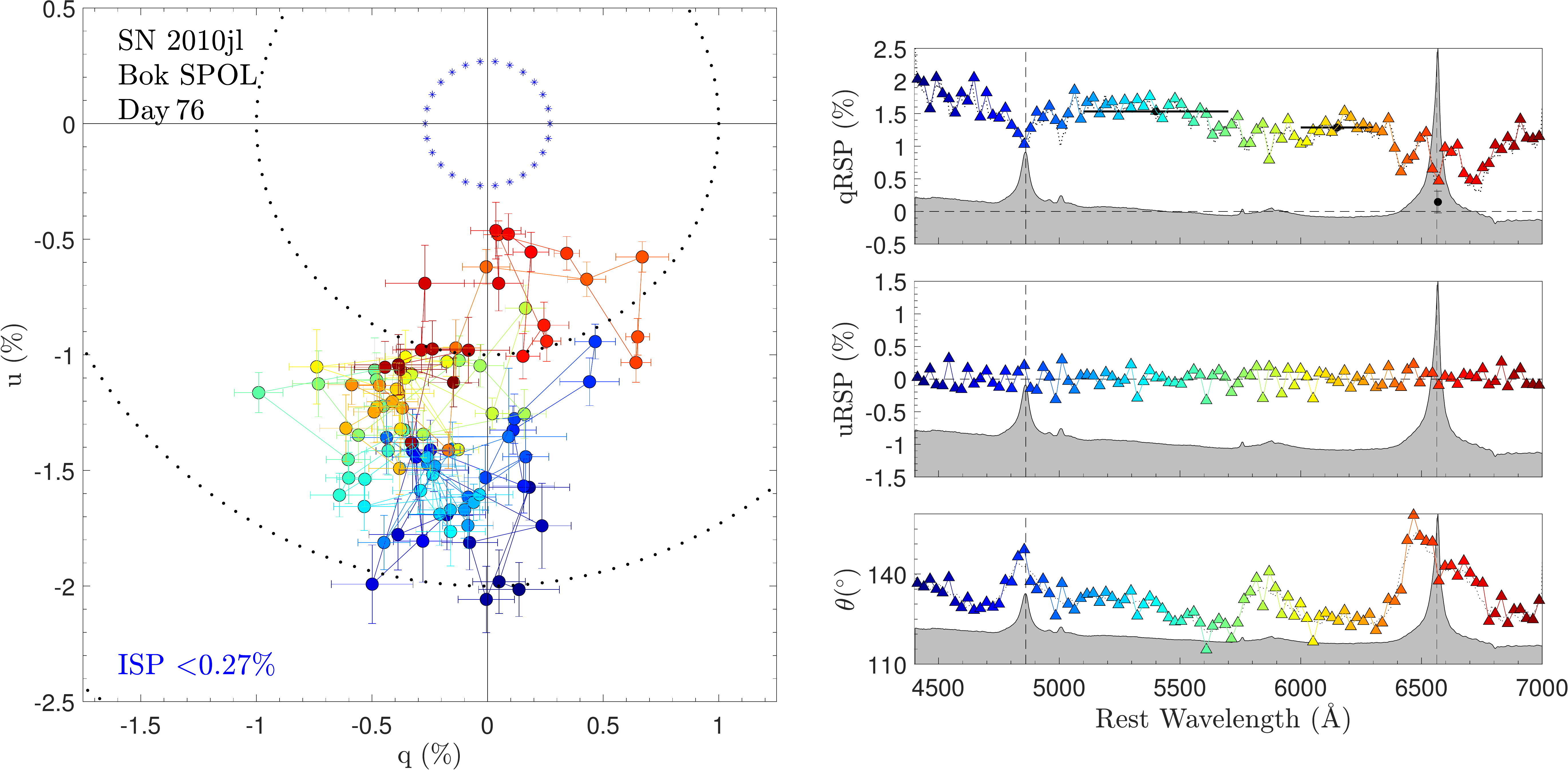}
\qufig
{SN~2010jl}{Bok}{109}{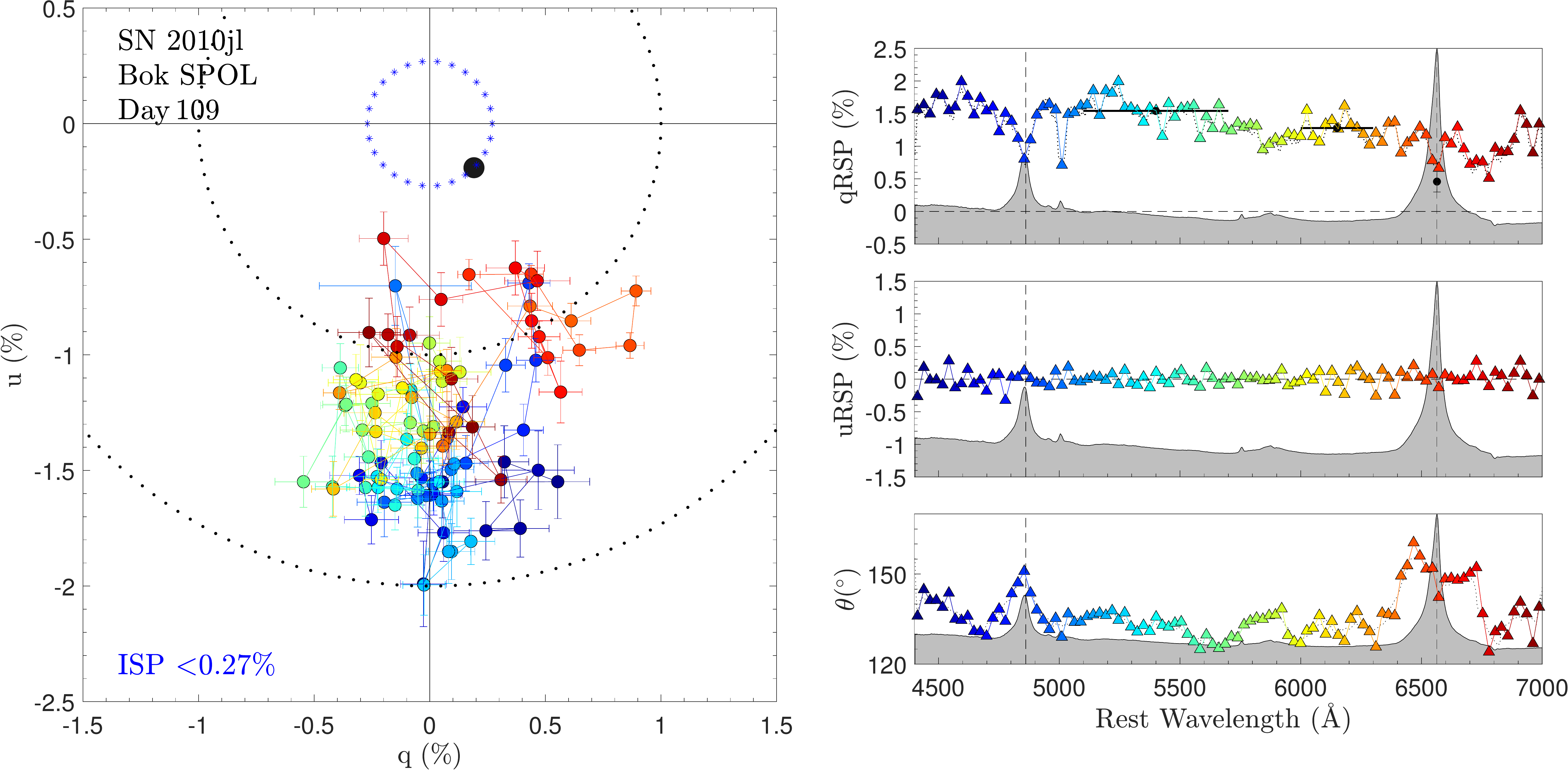}
{SN~2010jl}{Bok}{109, ISP-corrected}{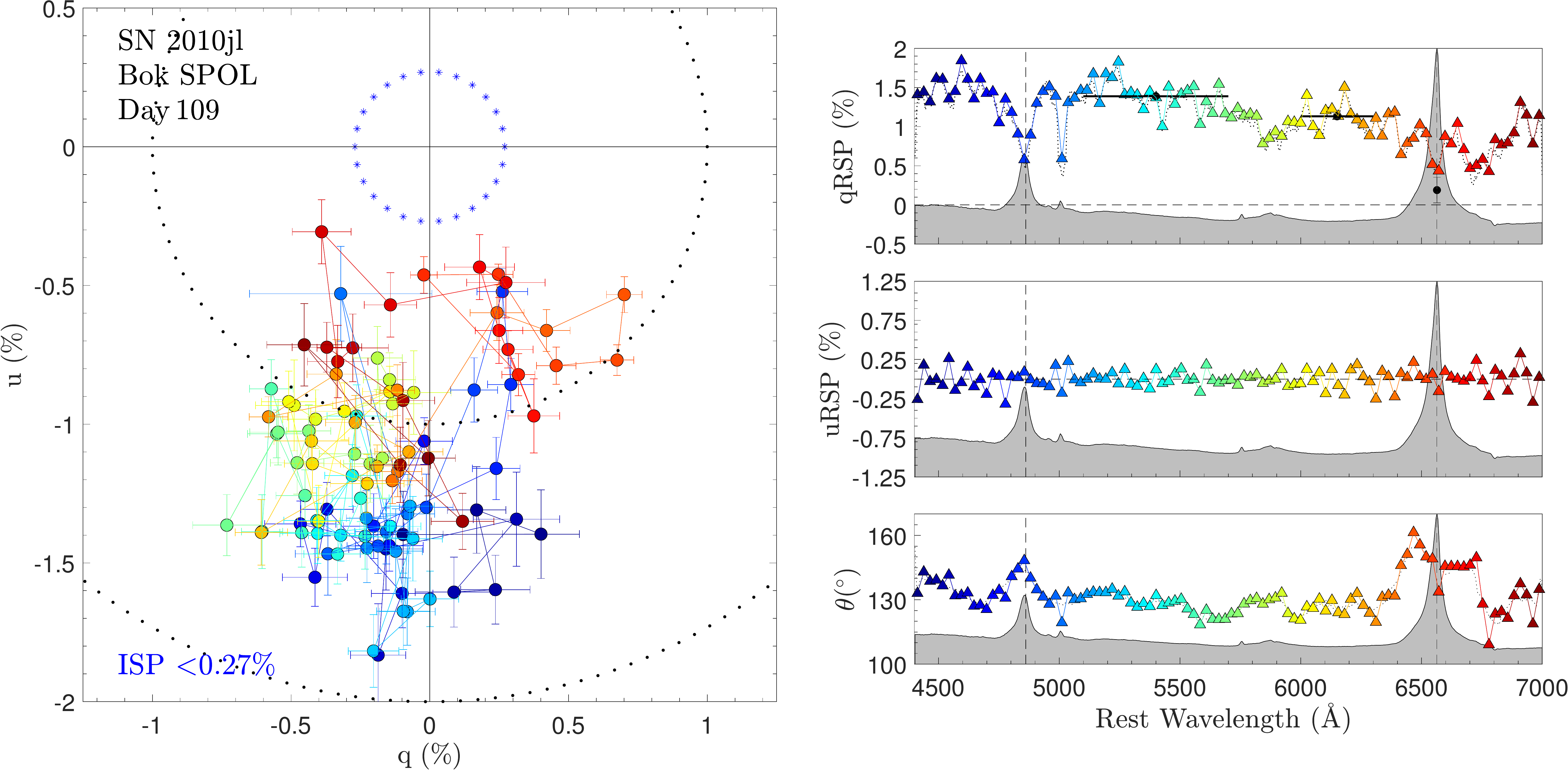}
\qufig
{SN~2010jl}{Bok}{137}{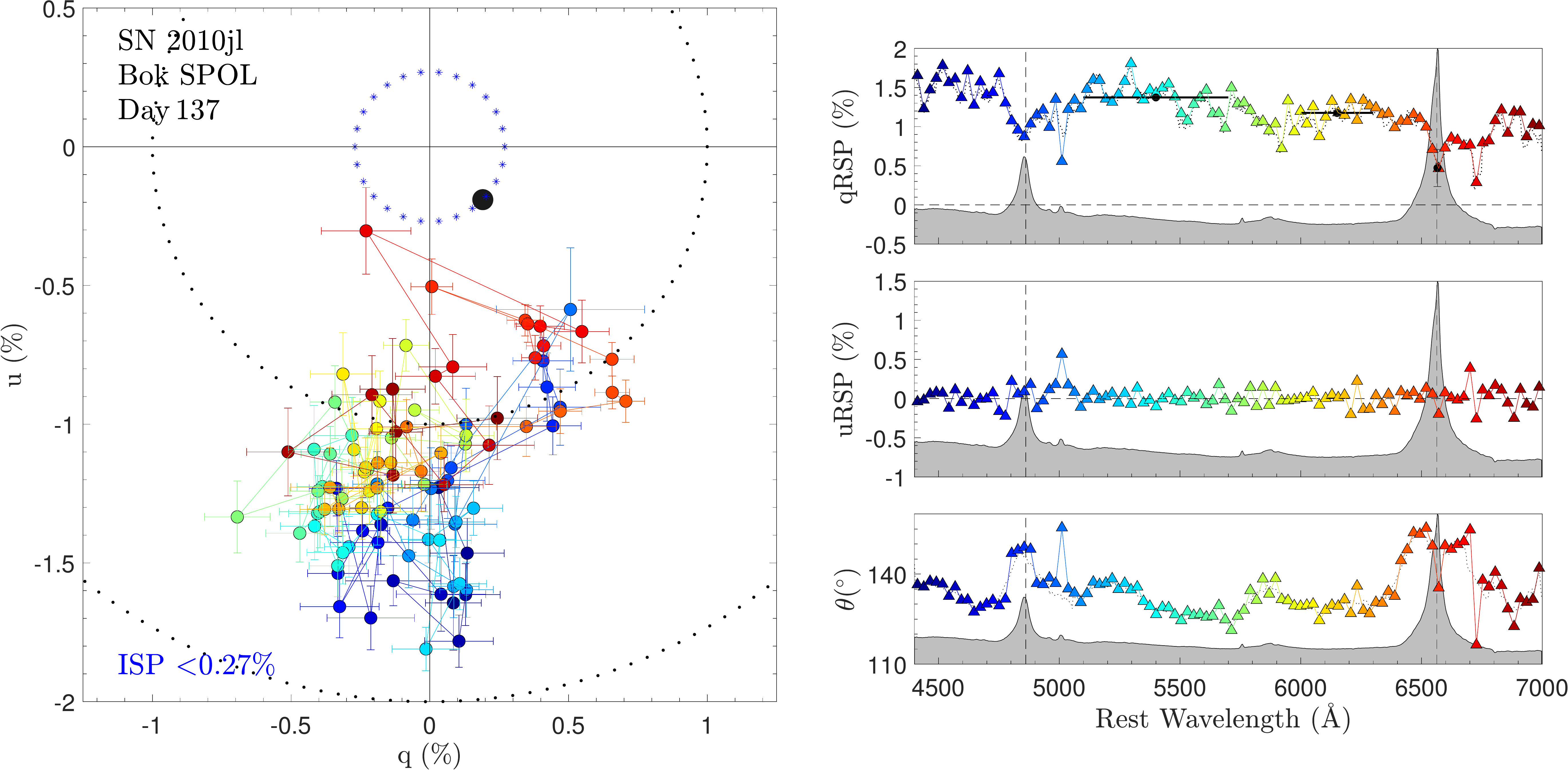}
{SN~2010jl}{Bok}{137, ISP-corrected}{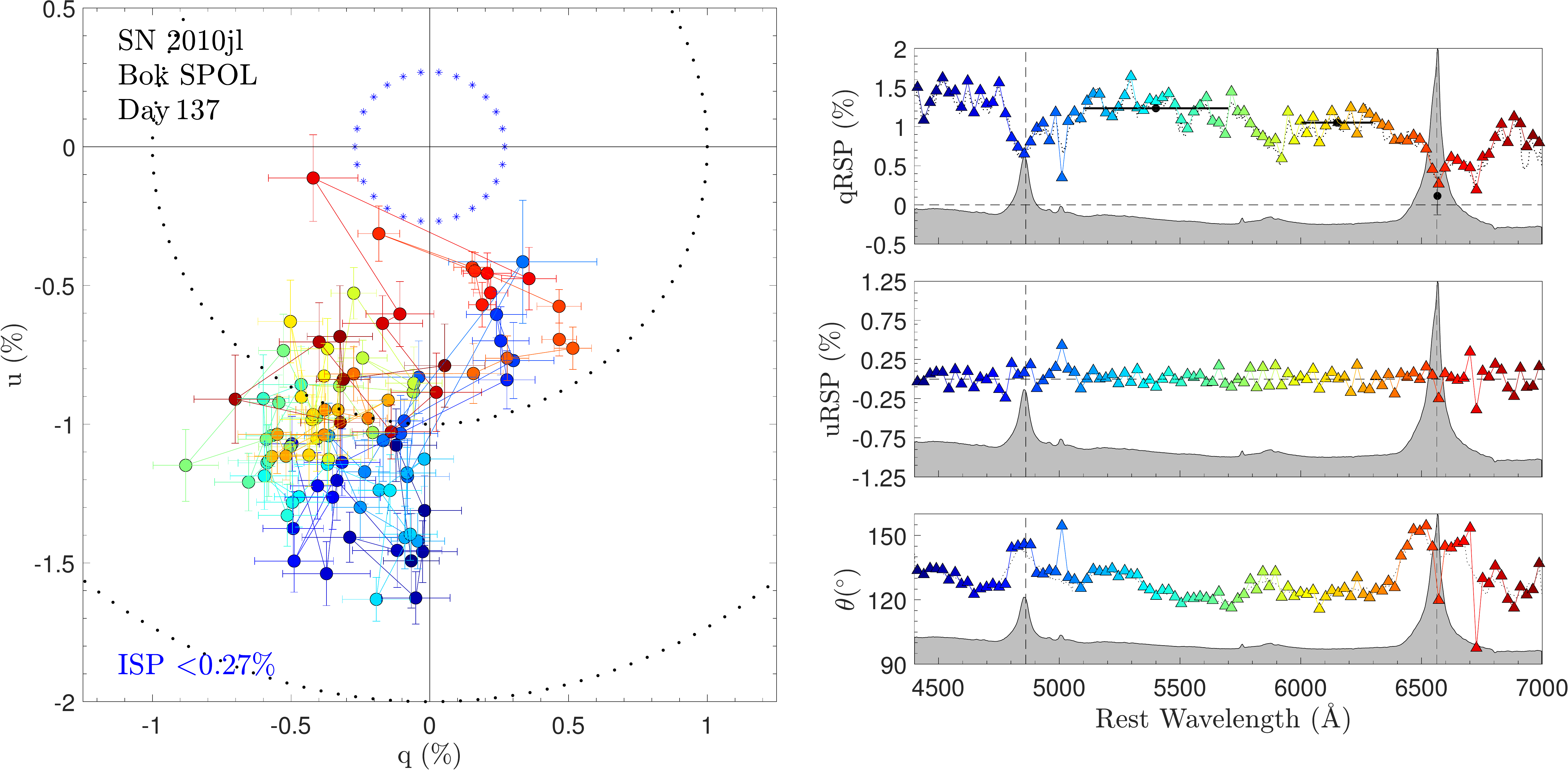}
\qufig
{SN~2010jl}{Kuiper}{168}{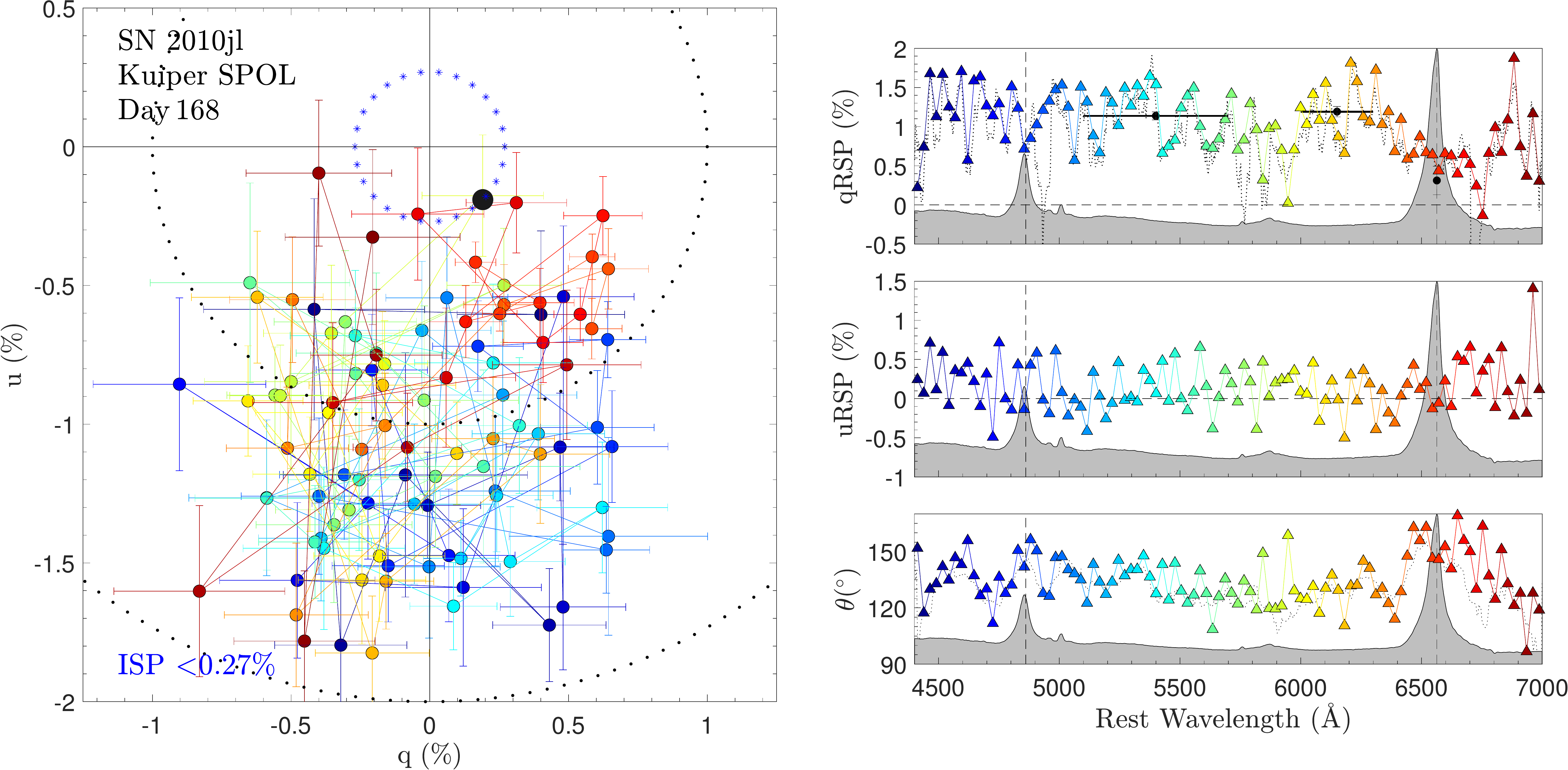}
{SN~2010jl}{Kuiper}{168, ISP-corrected}{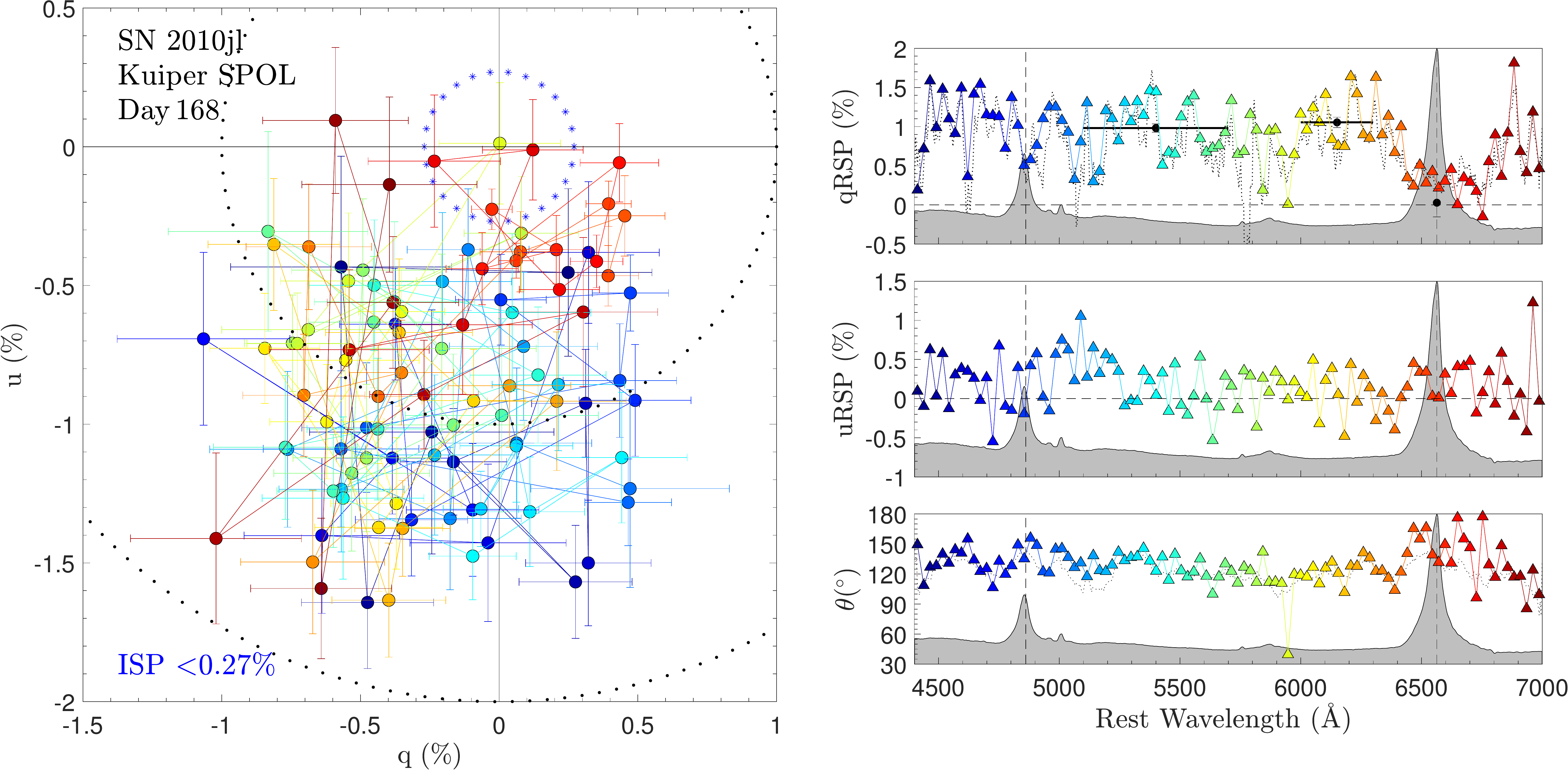}
\qufig
{SN~2010jl}{Bok}{221}{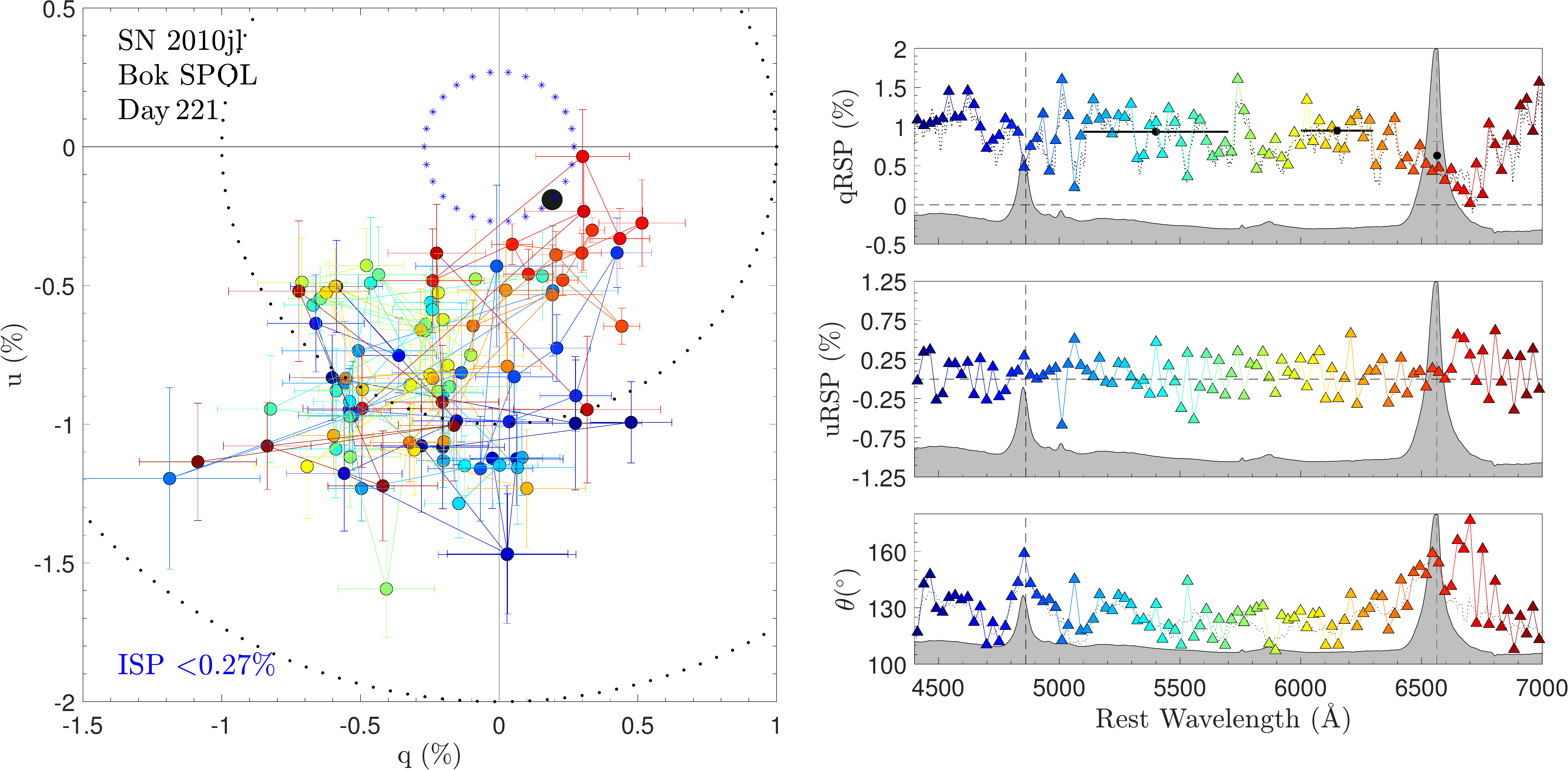}
{SN~2010jl}{Bok}{221, ISP-corrected}{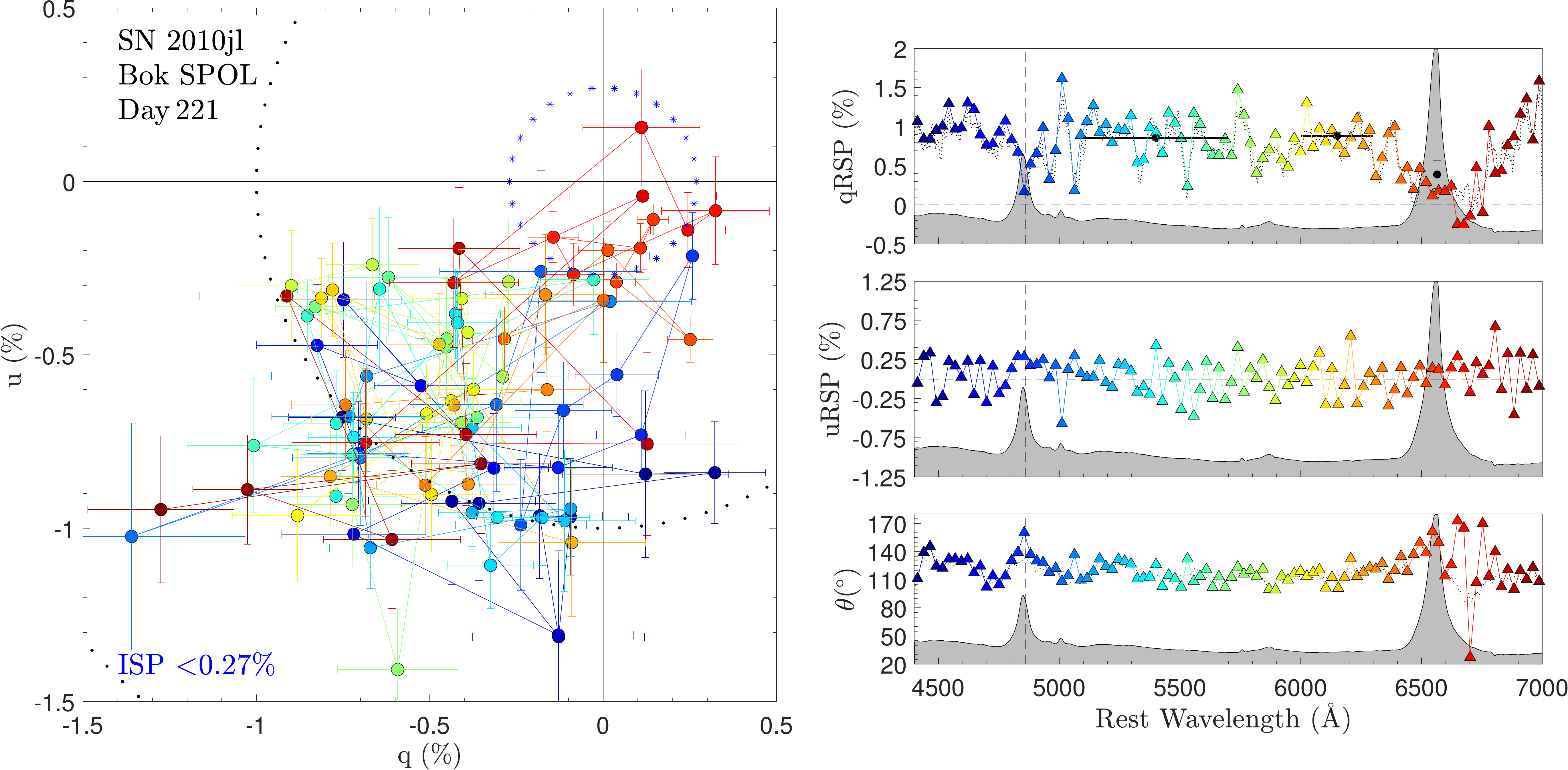}
\qufig
{SN~2010jl}{Bok}{239}{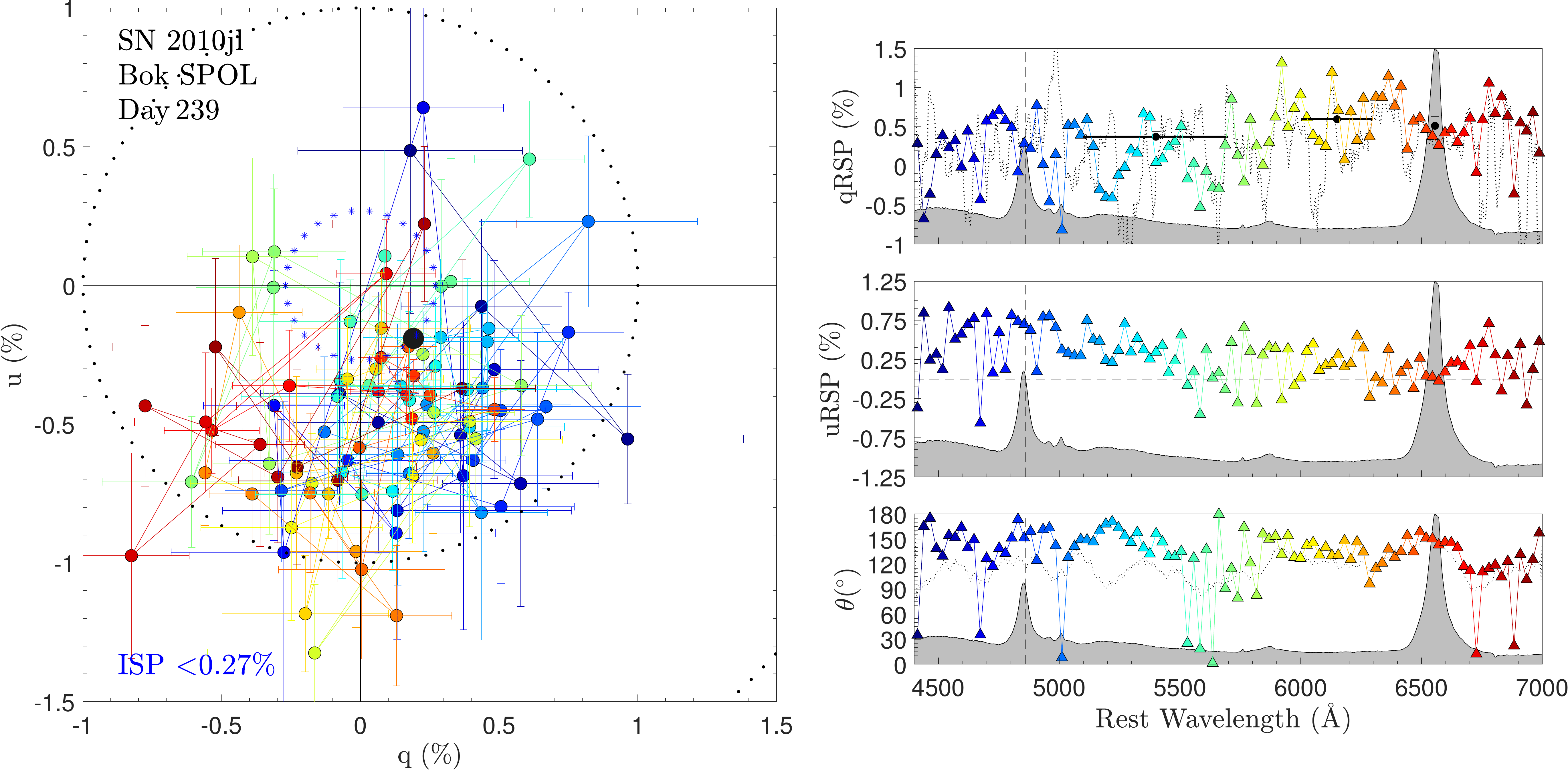}
{SN~2010jl}{Bok}{239, ISP-corrected}{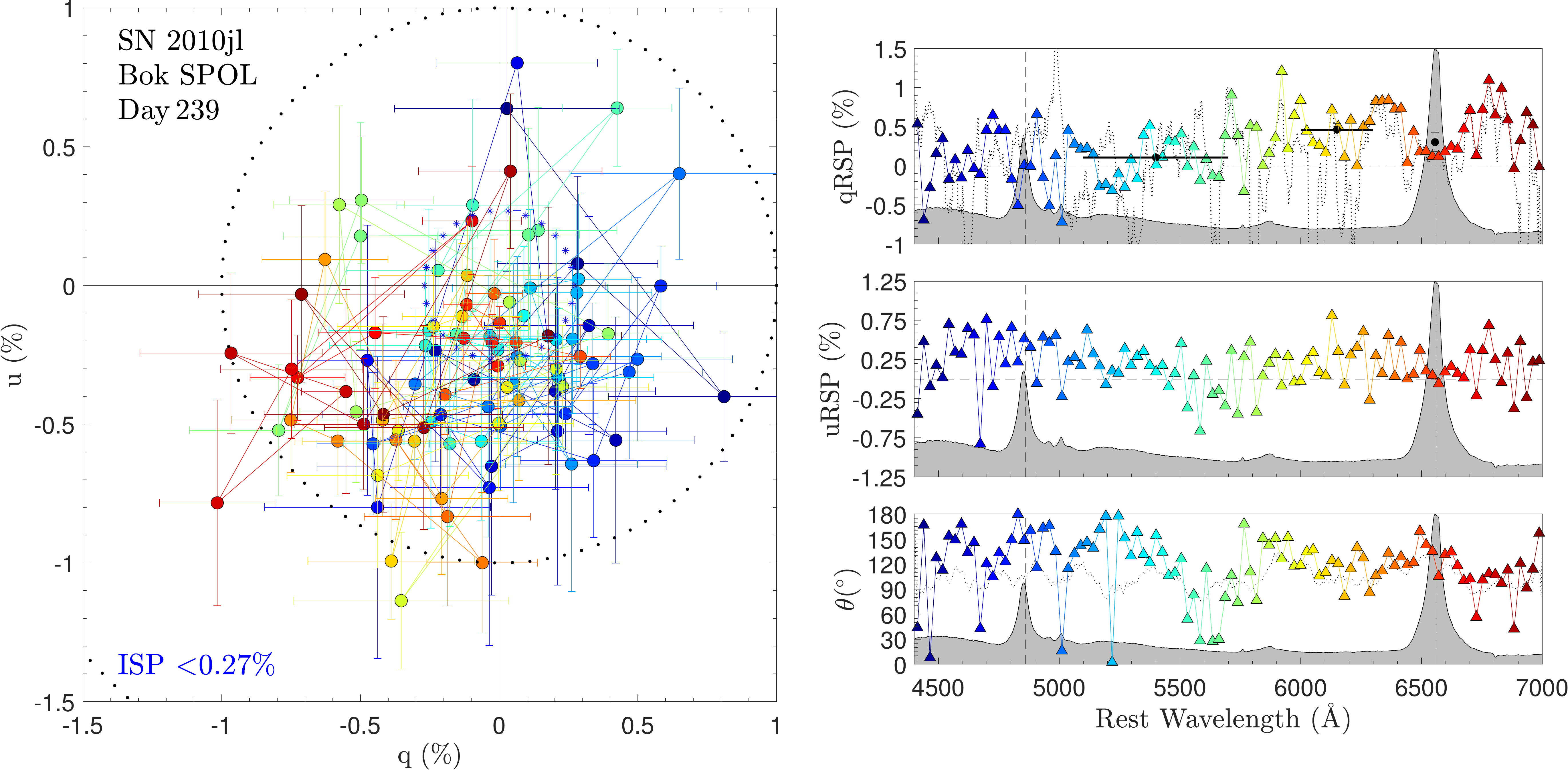}
\qufig
{SN~2010jl}{Bok}{465}{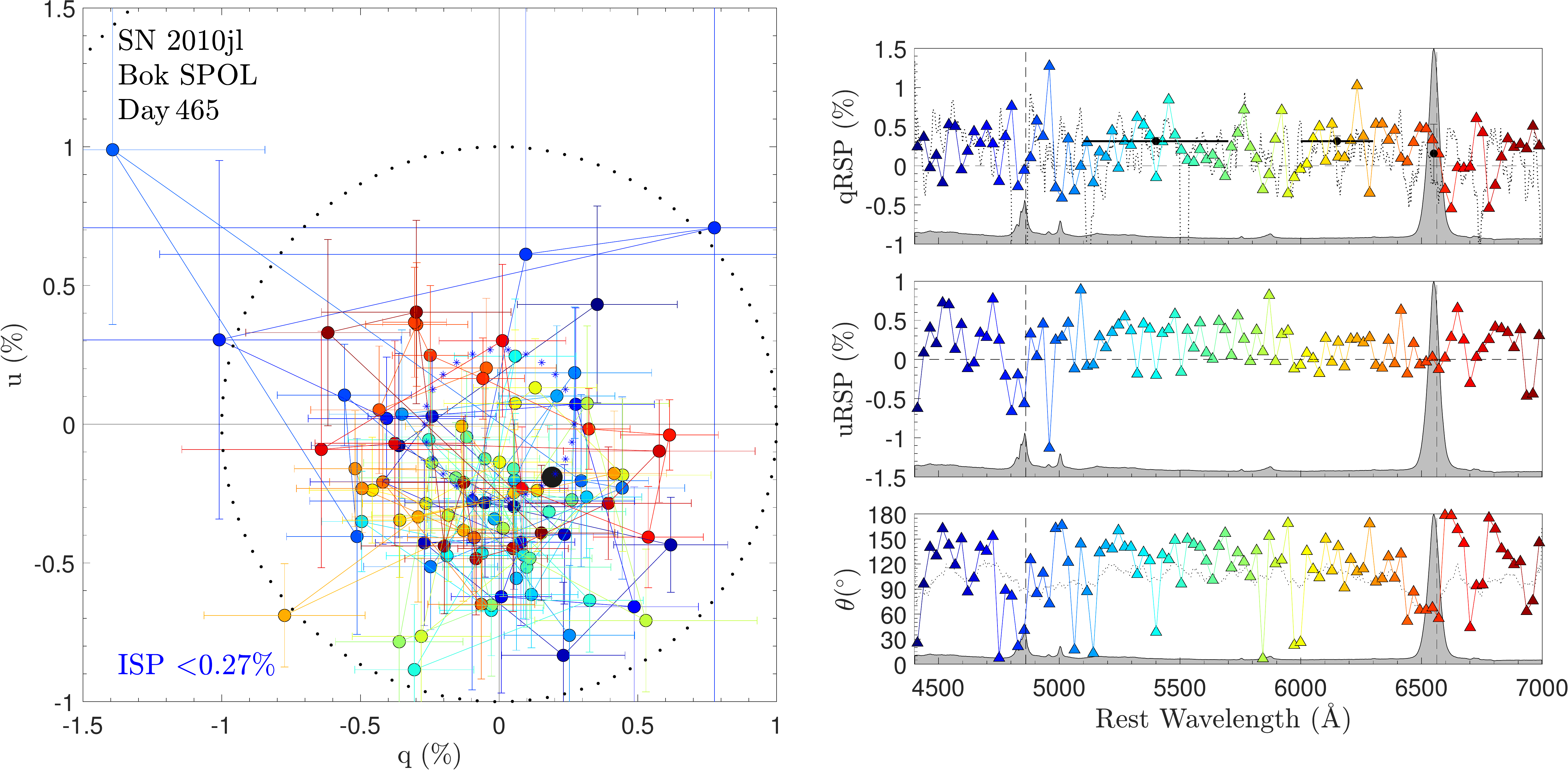}
{SN~2010jl}{Bok}{465, ISP-corrected}{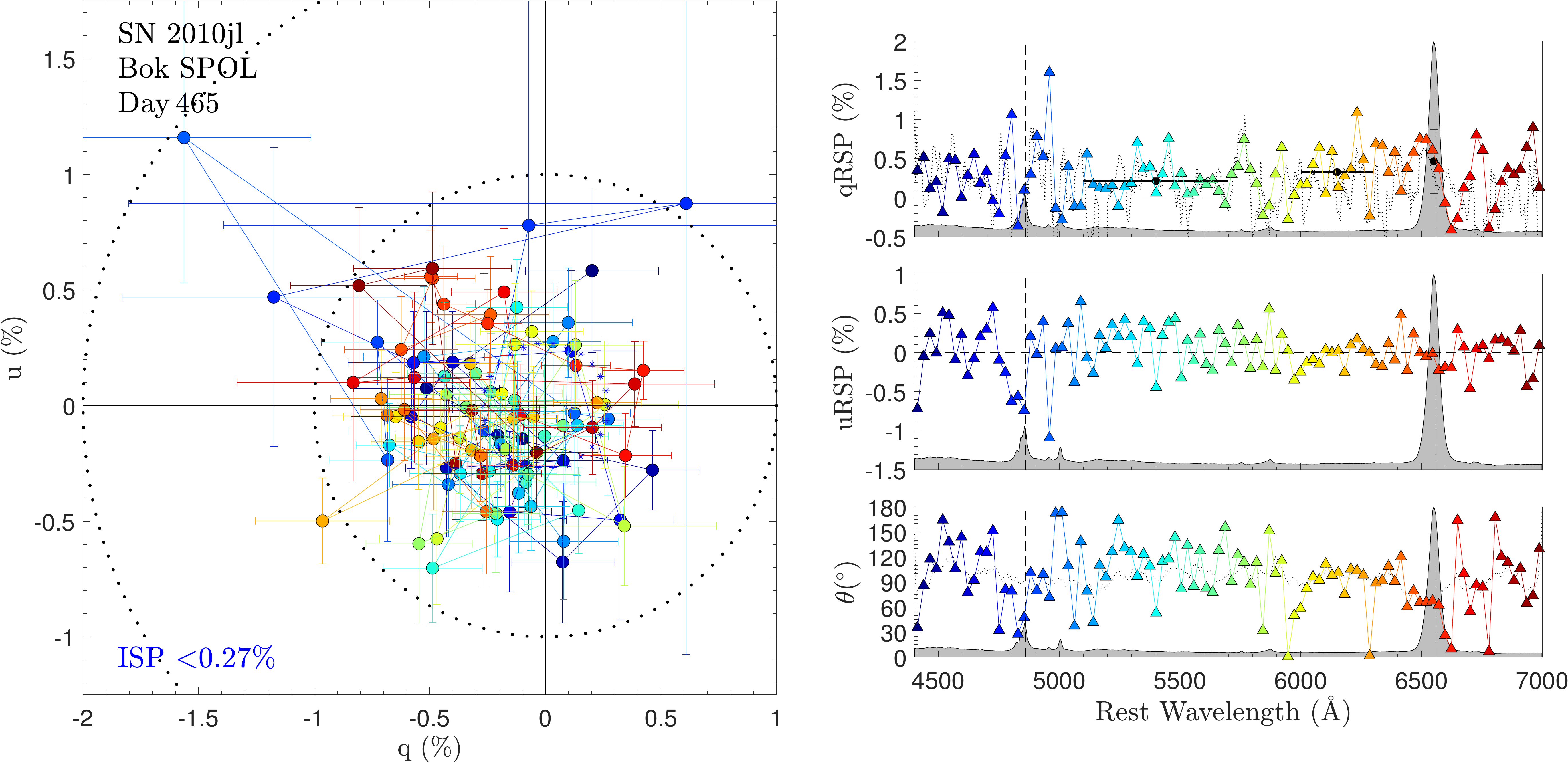}
\qufig
{SN~2010jl}{Bok}{488}{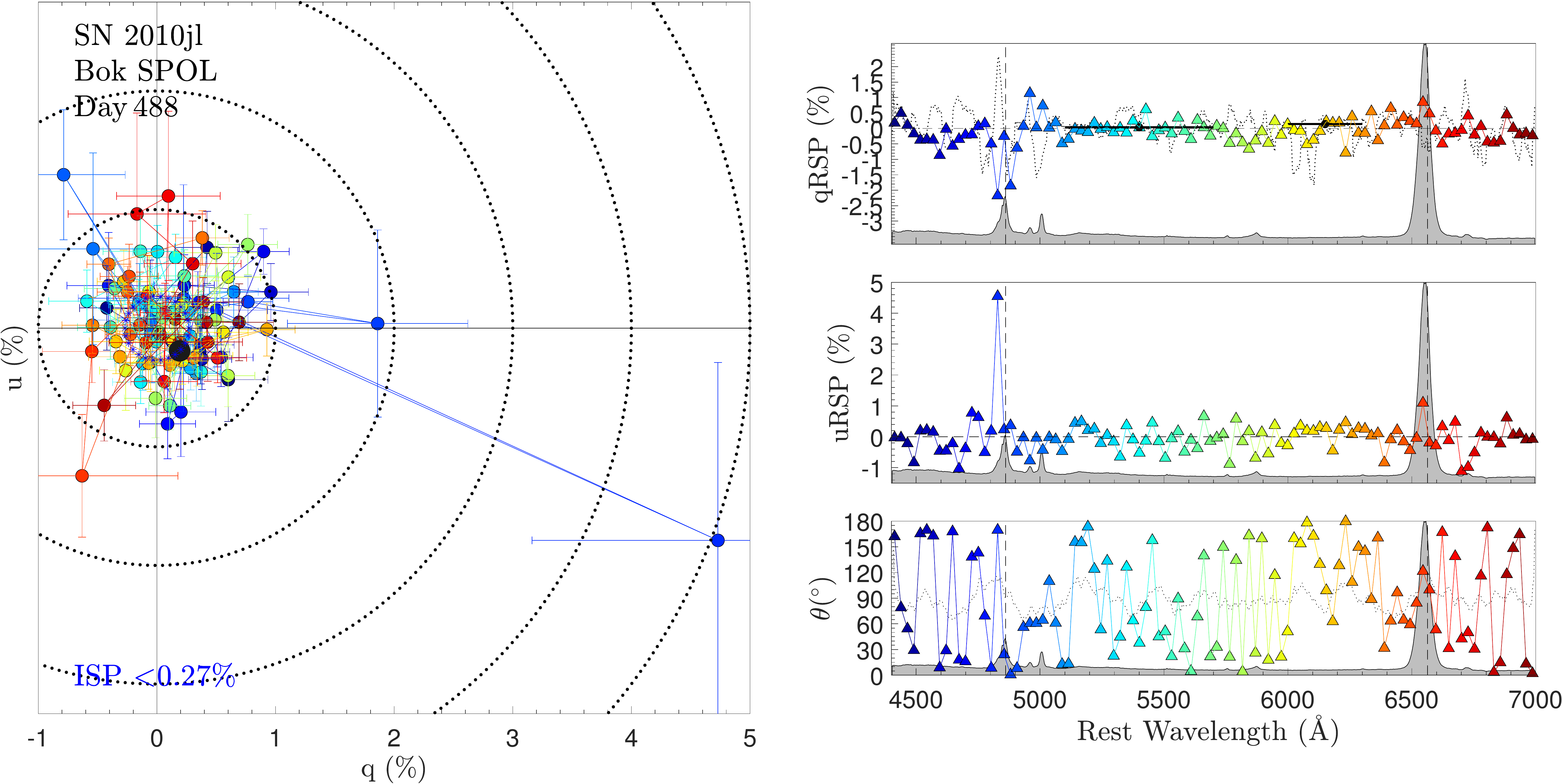}
{SN~2010jl}{Bok}{488, ISP-corrected}{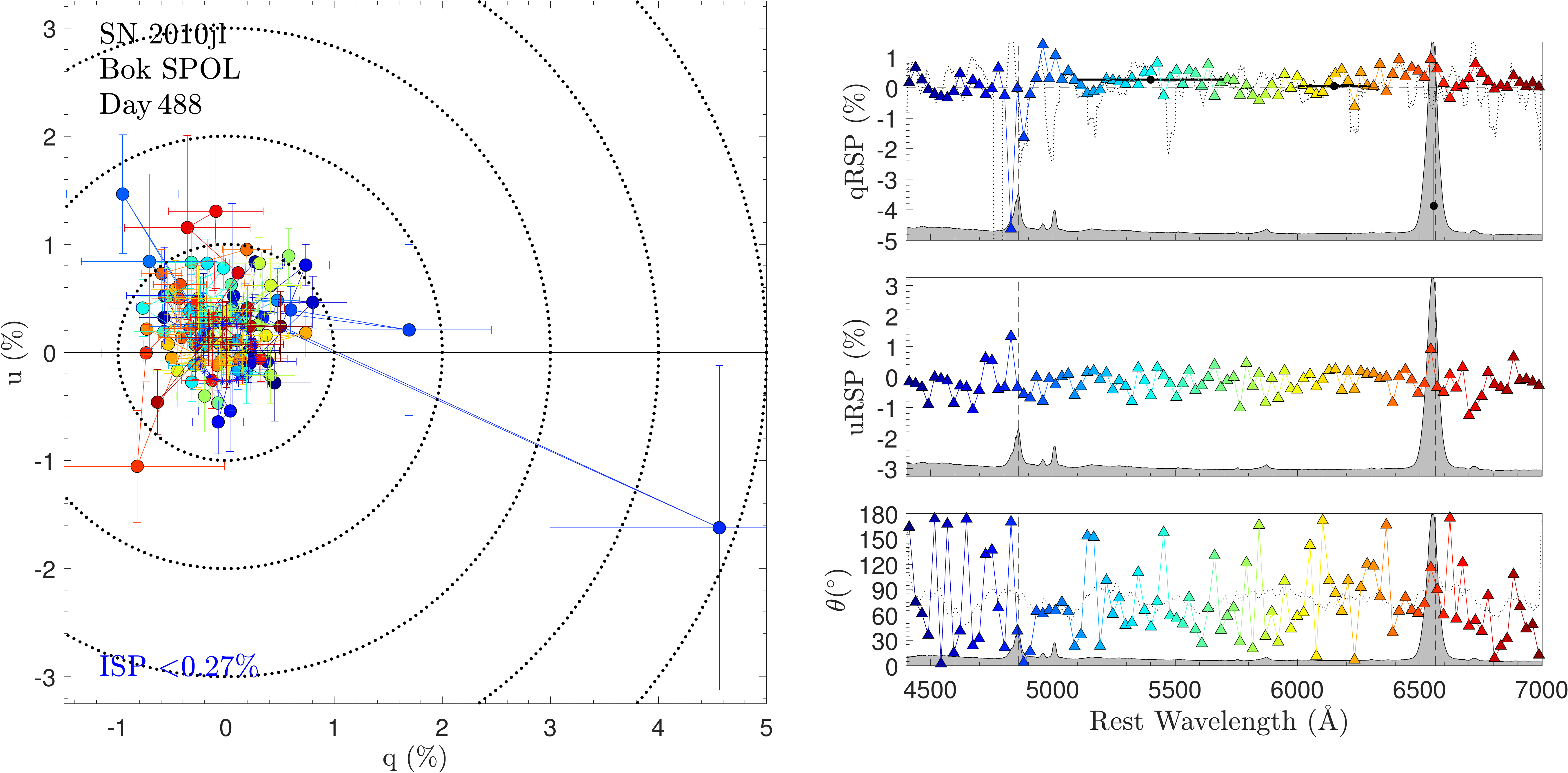}
\qufig
{SN~2010jl}{MMT}{546}{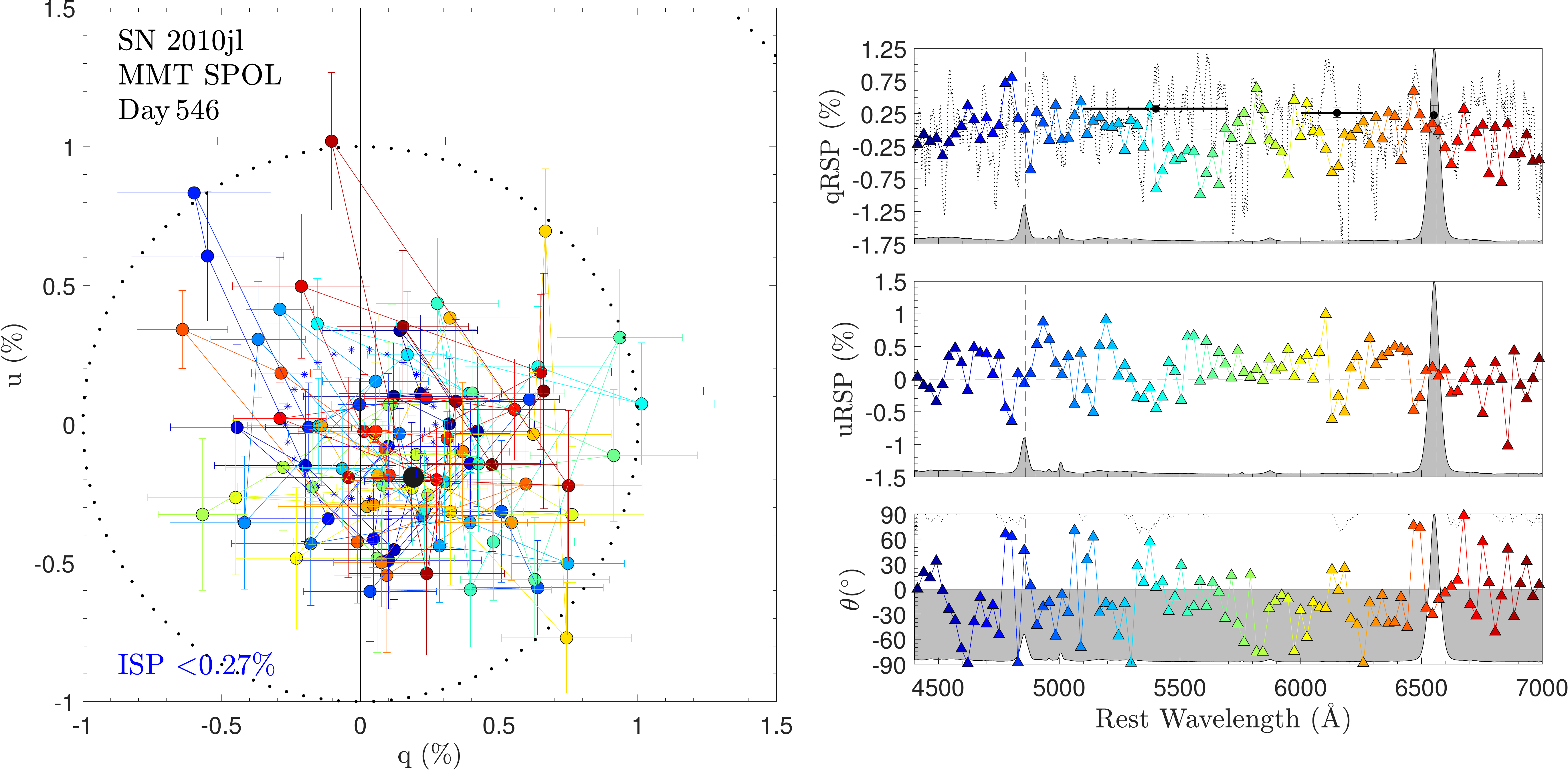}
{SN~2010jl}{MMT}{546, ISP-corrected}{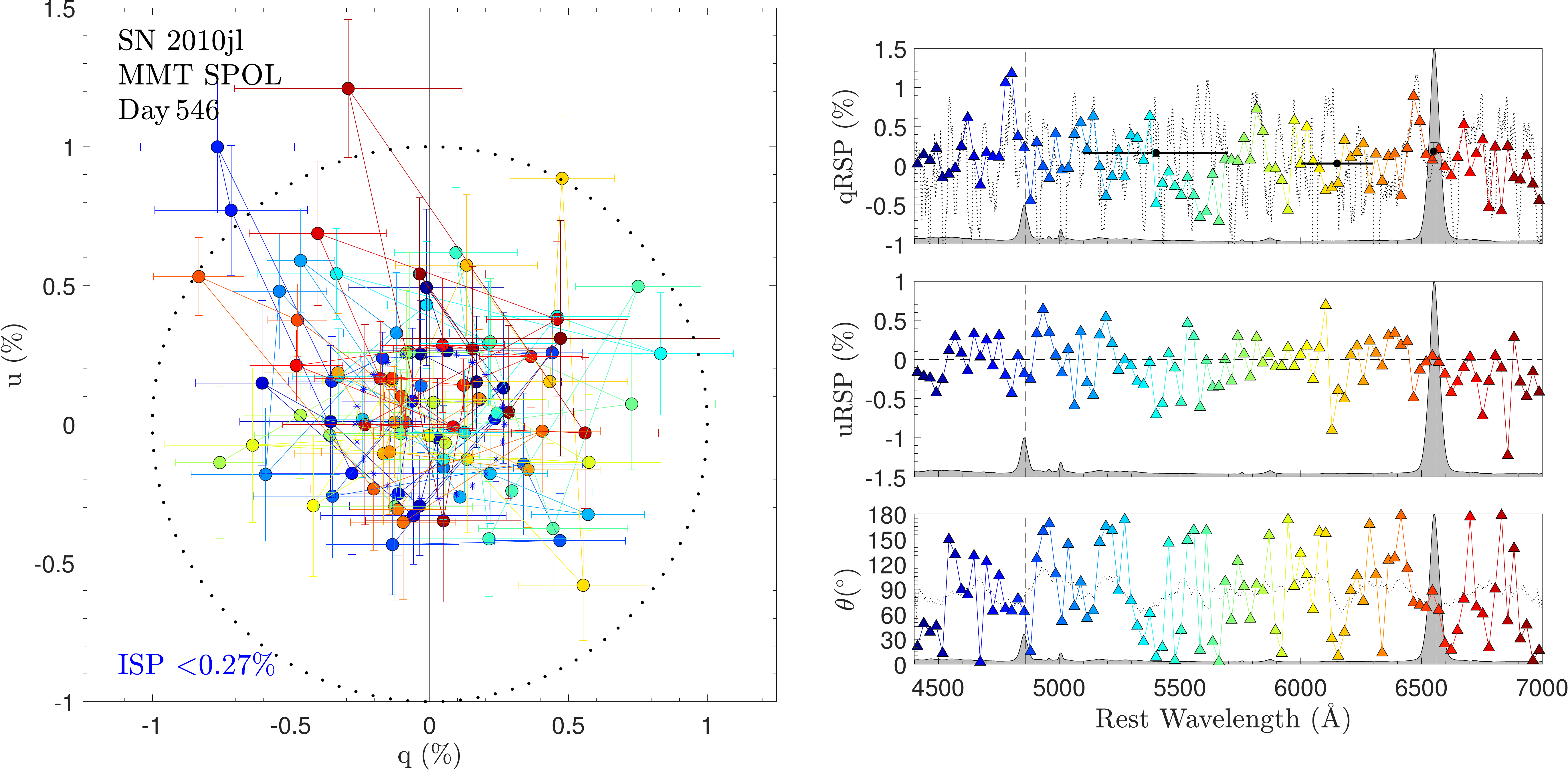}
\qufig
{SN~2011cc}{Bok}{63}{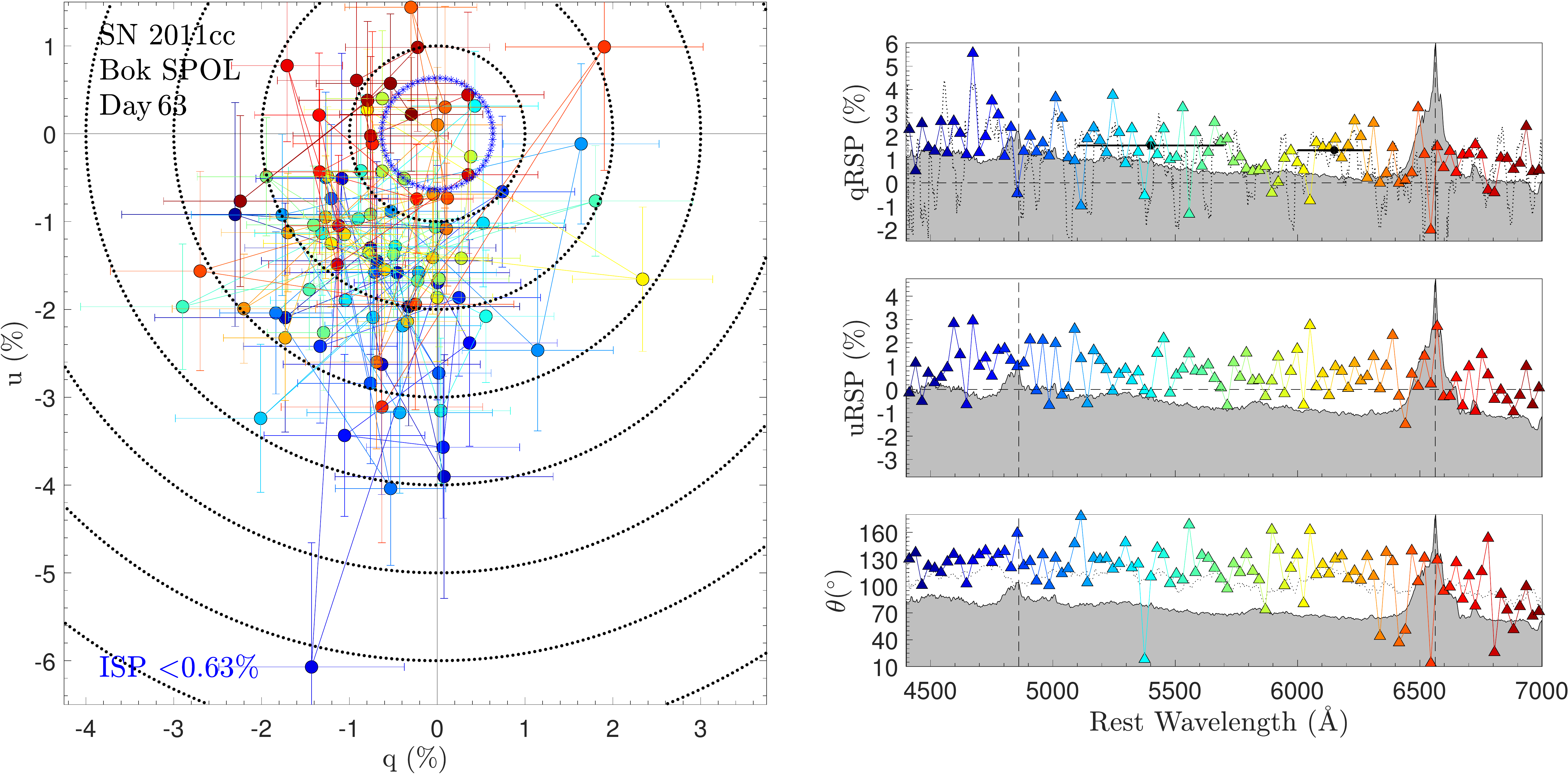}
{PTF11iqb}{Bok}{176}{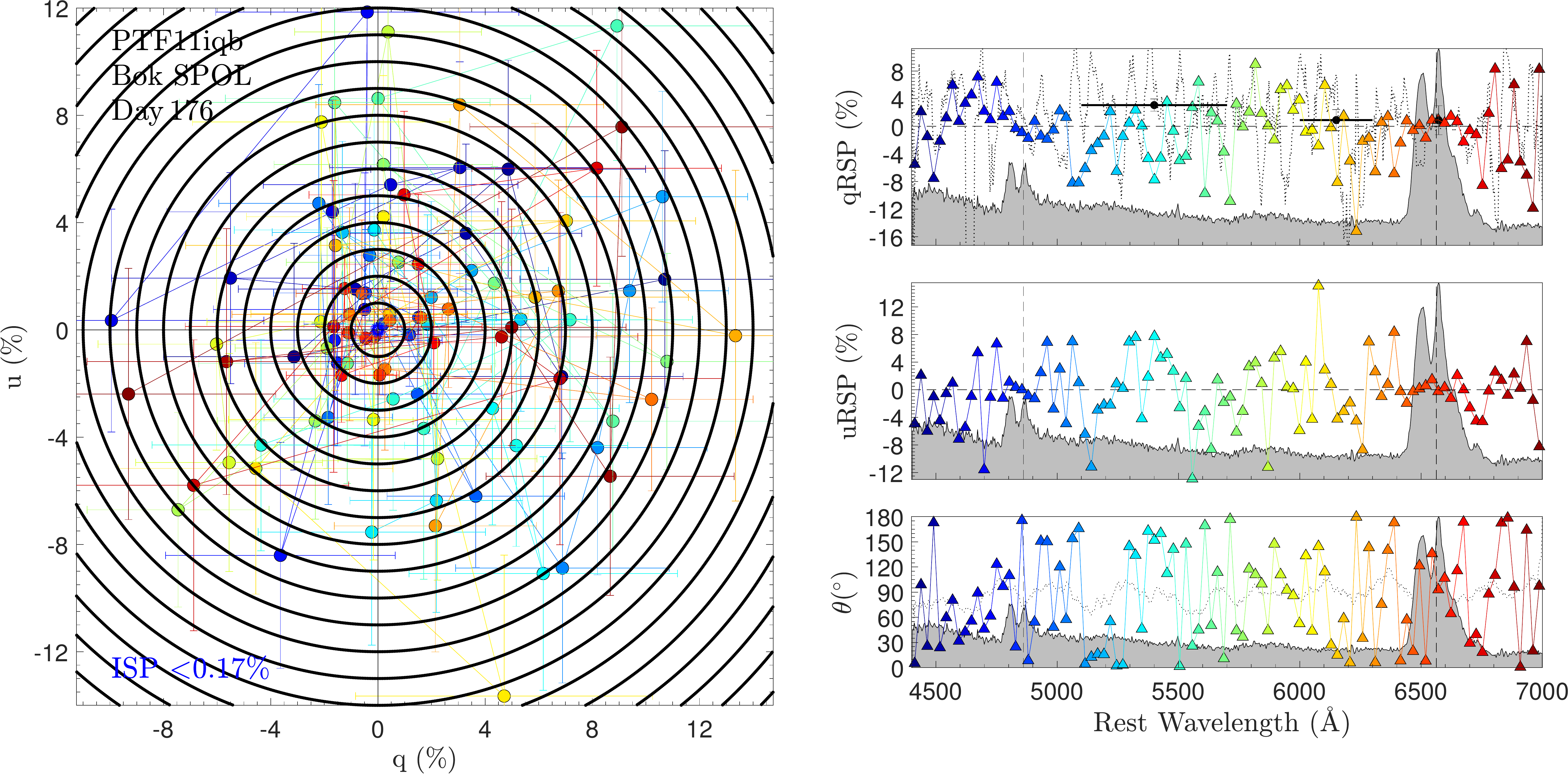}
\qufig
{SN~2011ht}{Bok}{64}{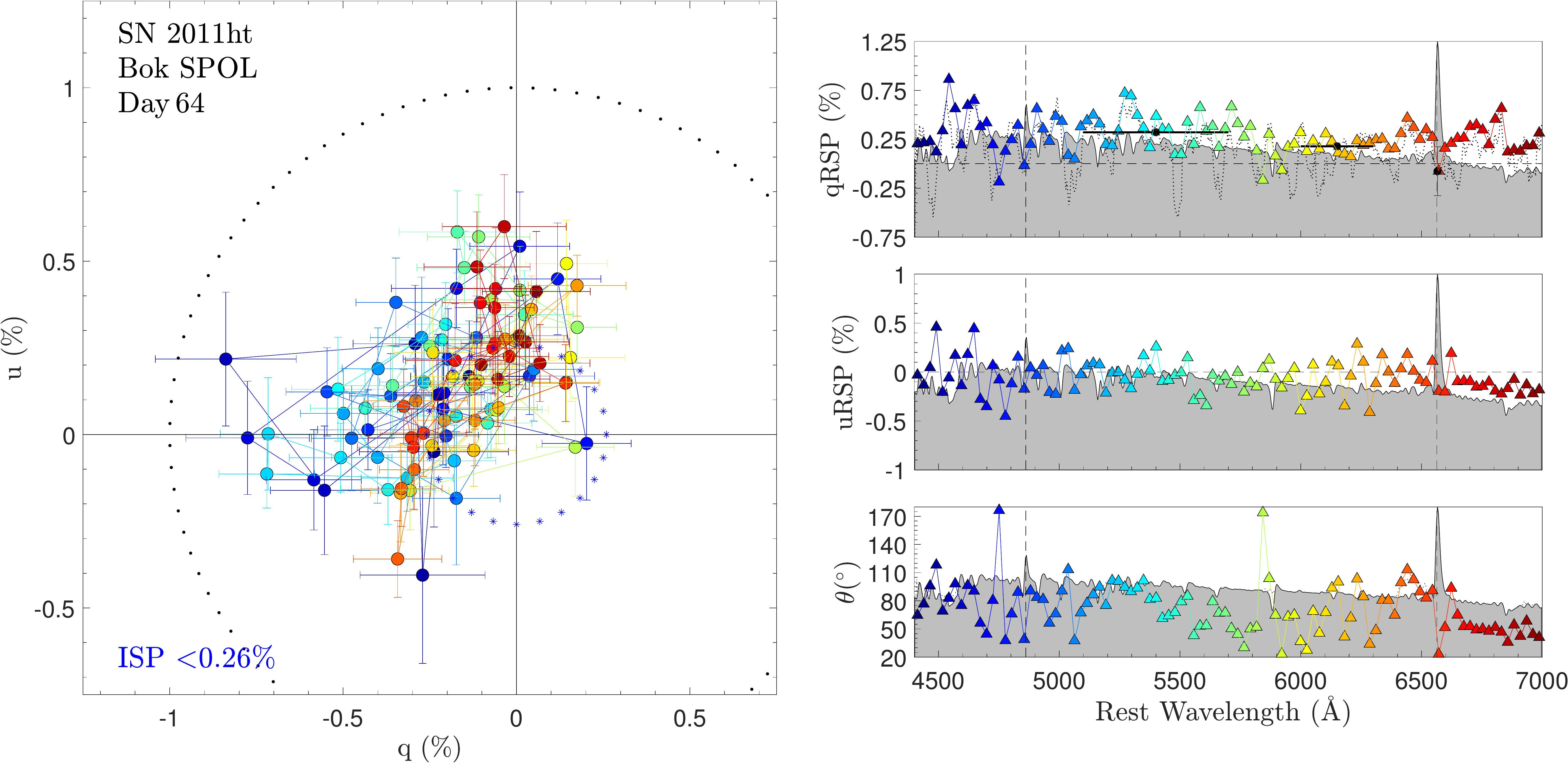}
{SN~2012ab}{Bok}{25}{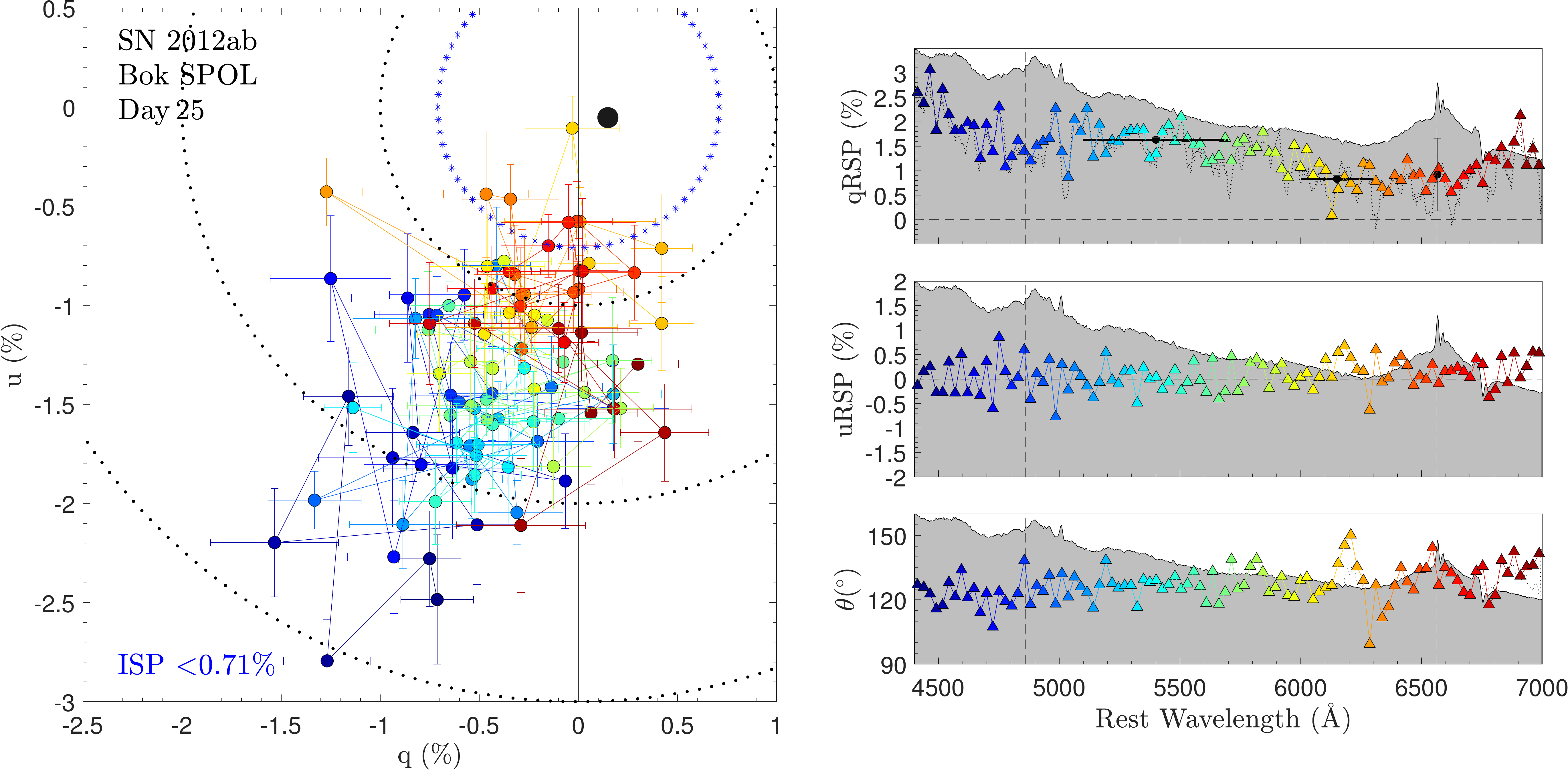}
\qufig
{SN~2012ab}{Bok}{25, ISP-corrected}{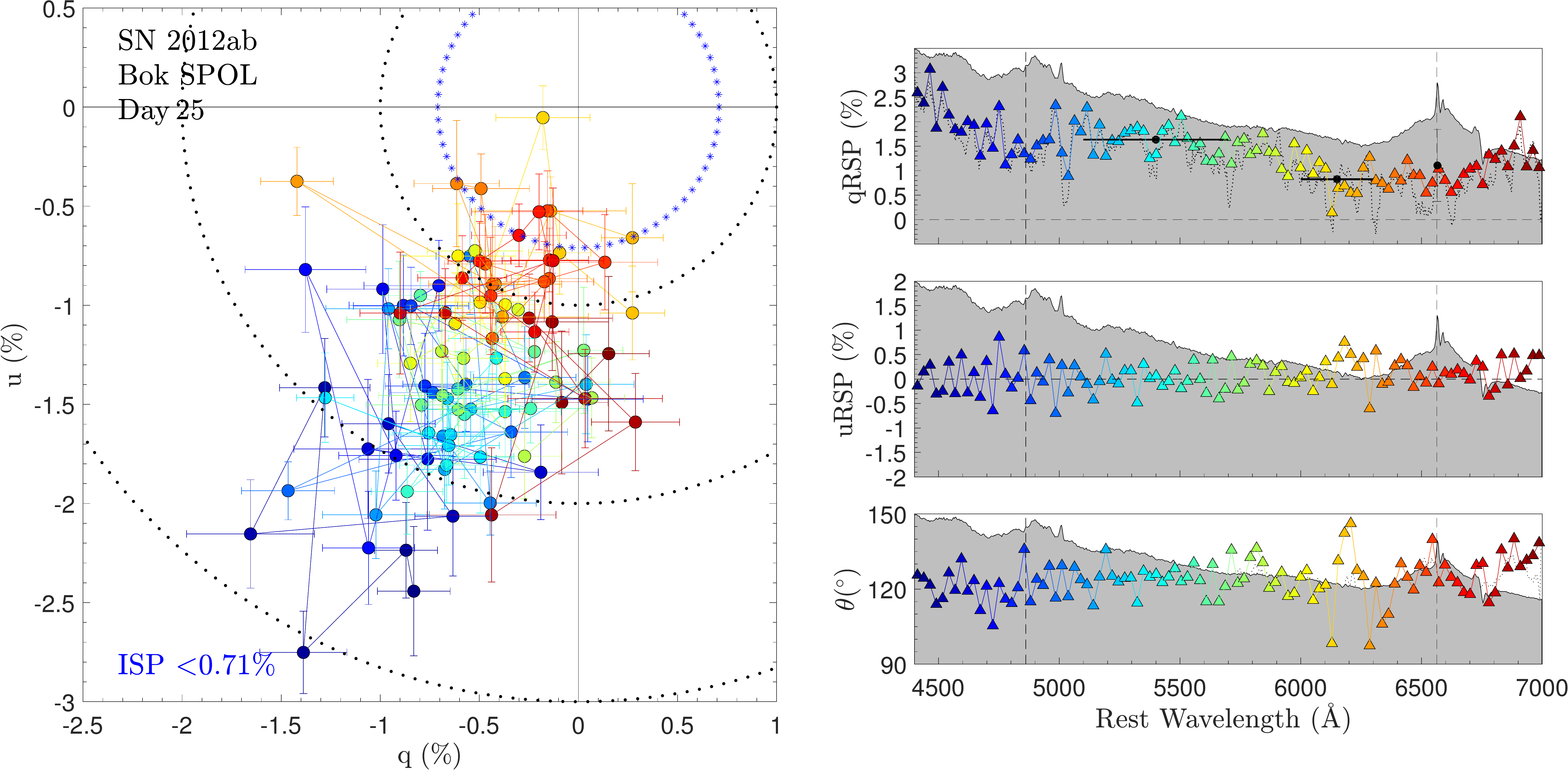}
{SN~2012ab}{Bok}{49}{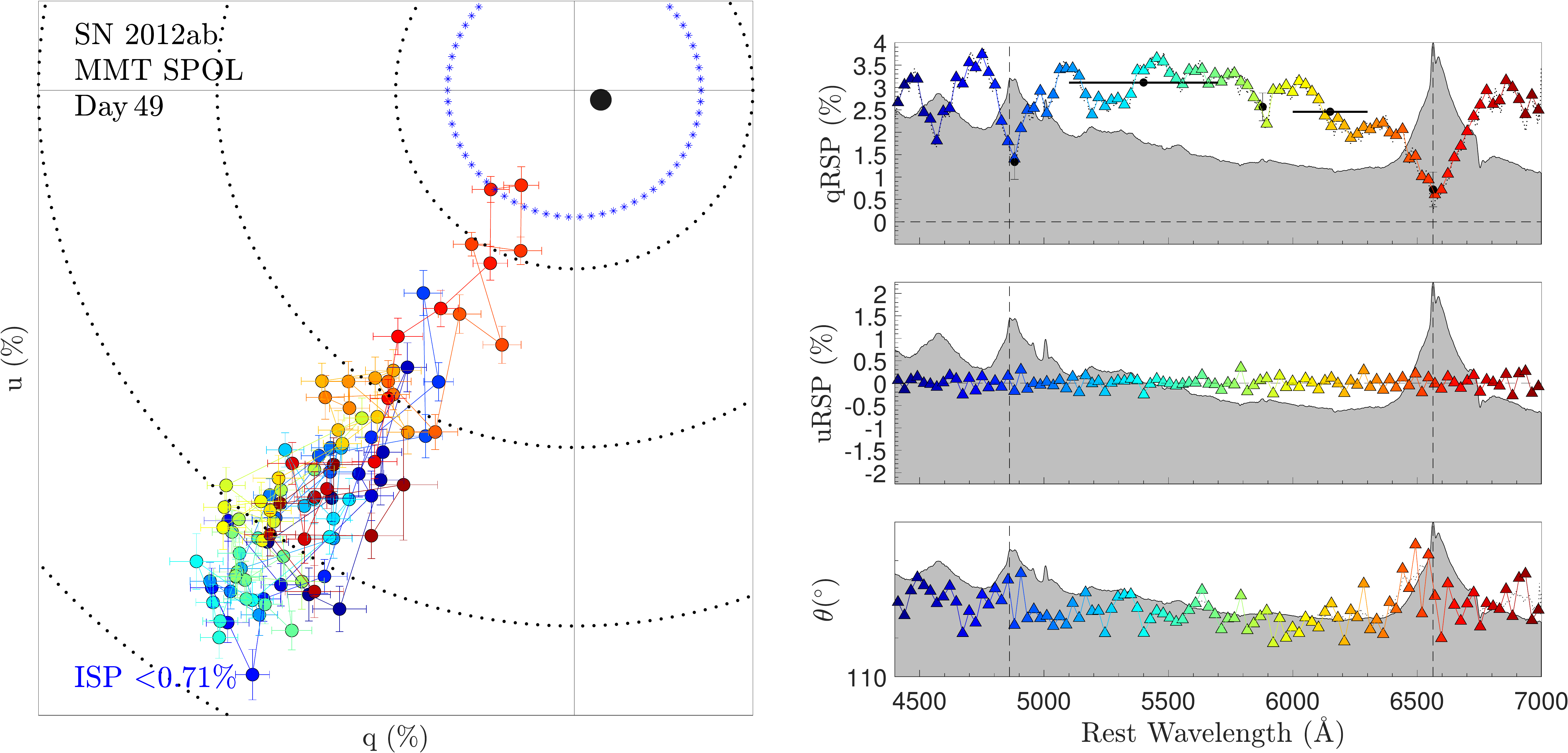}
\qufig
{SN~2012ab}{MMT}{49, ISP-corrected}{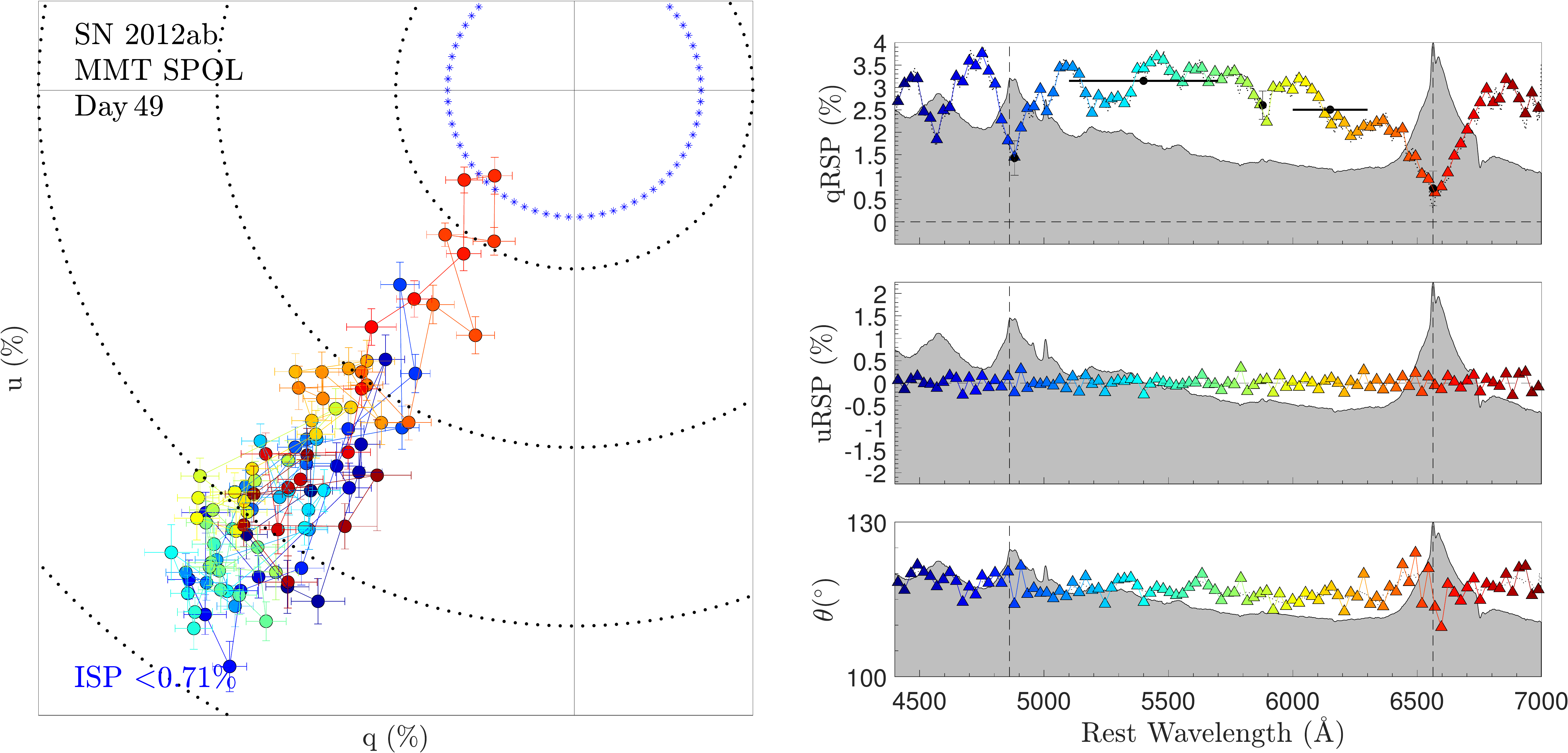}
{SN~2009ip}{MMT}{-14}{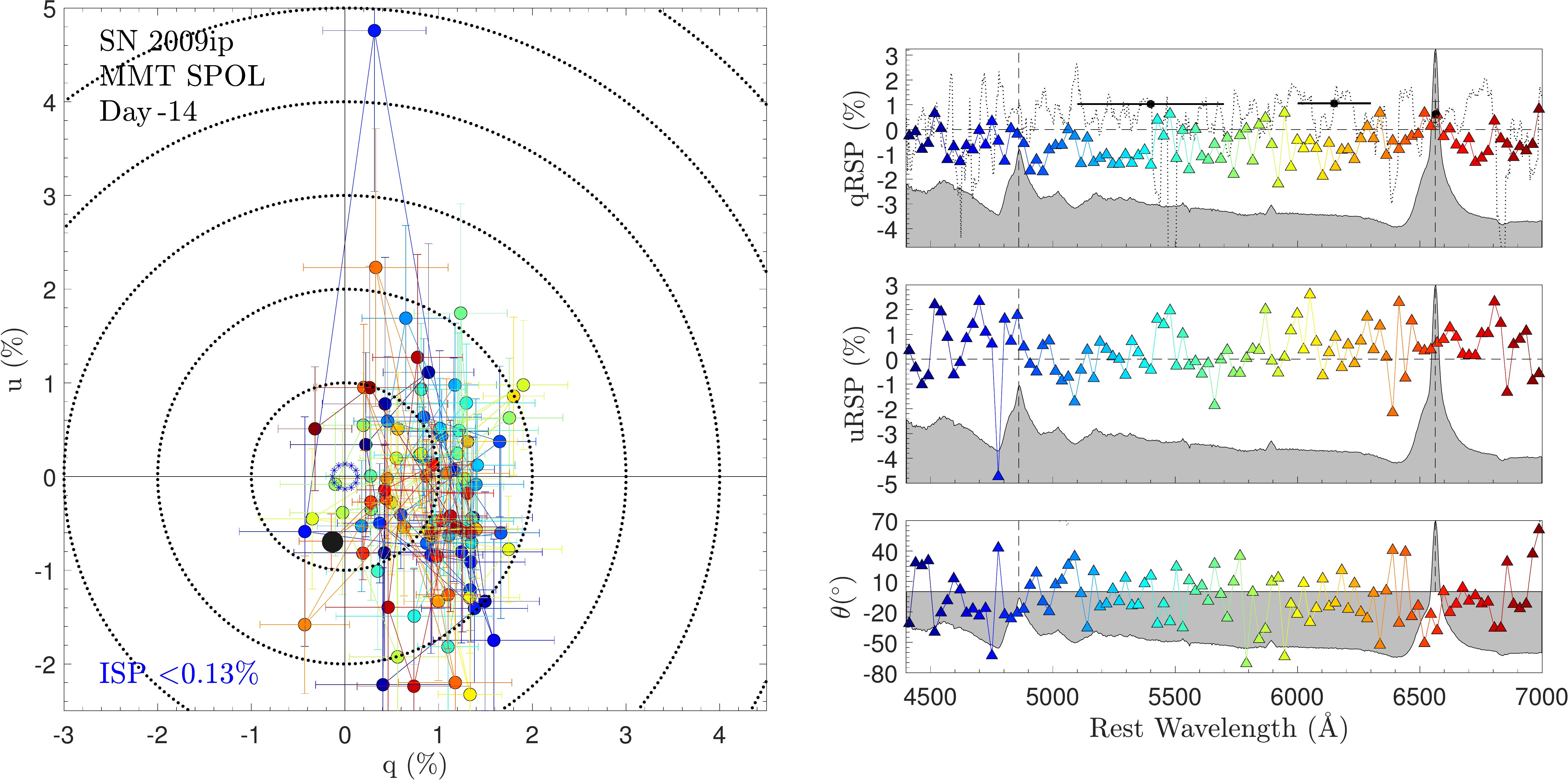}
\qufig
{SN~2009ip}{MMT}{-14, ISP-corrected}{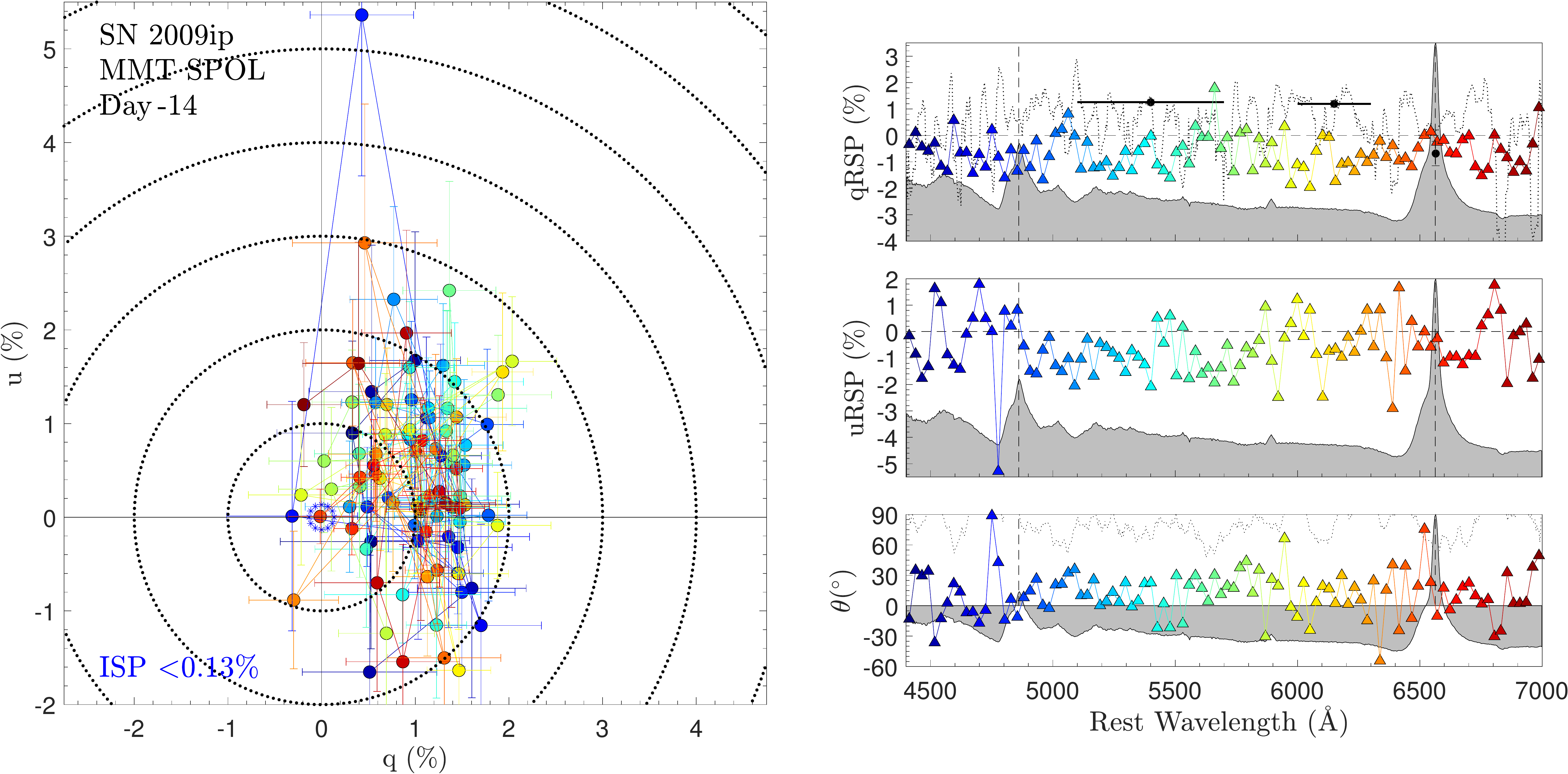}
{SN~2009ip}{Kuiper}{7}{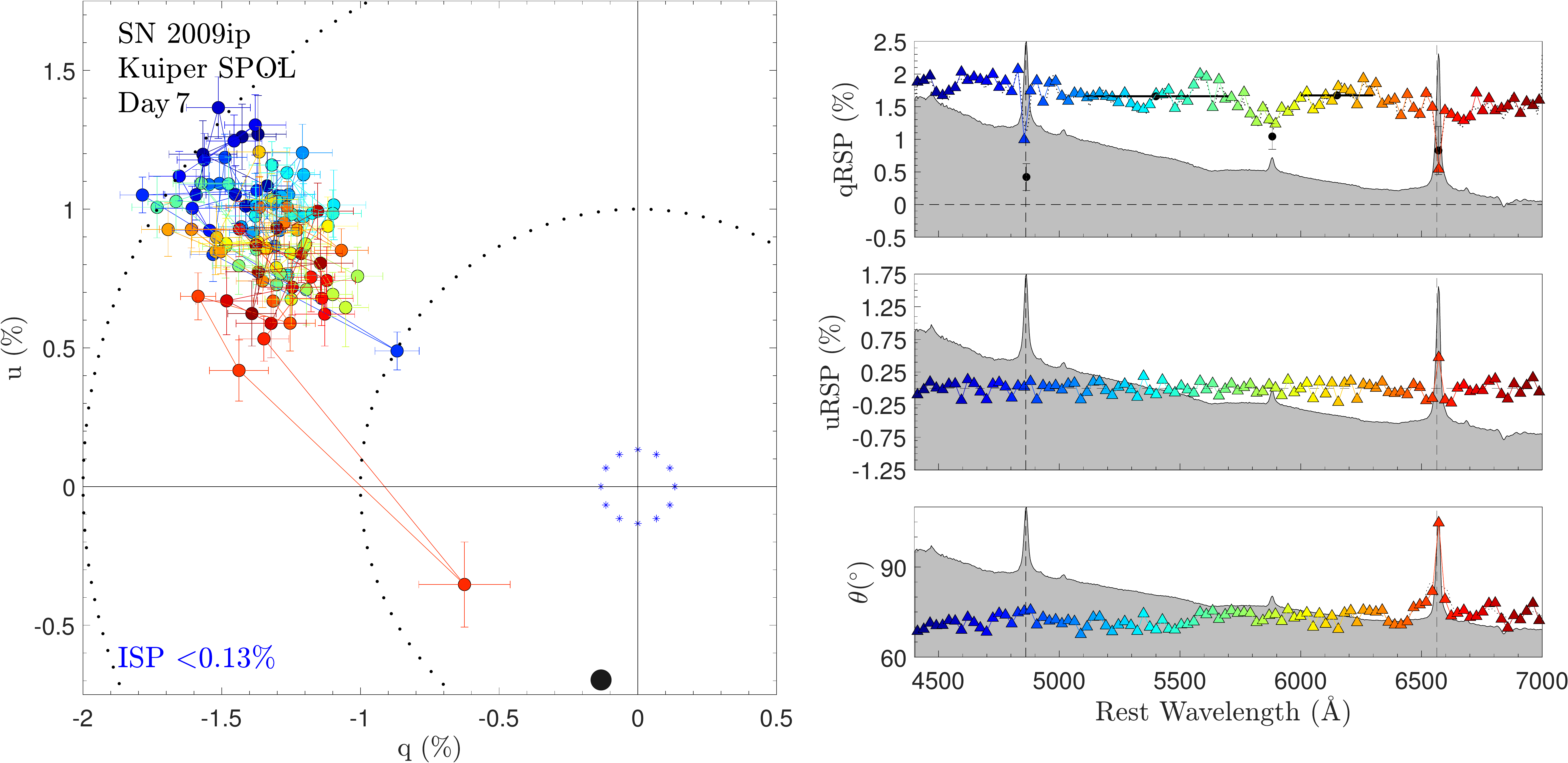}
\qufig
{SN~2009ip}{Kuiper}{7, ISP-corrected}{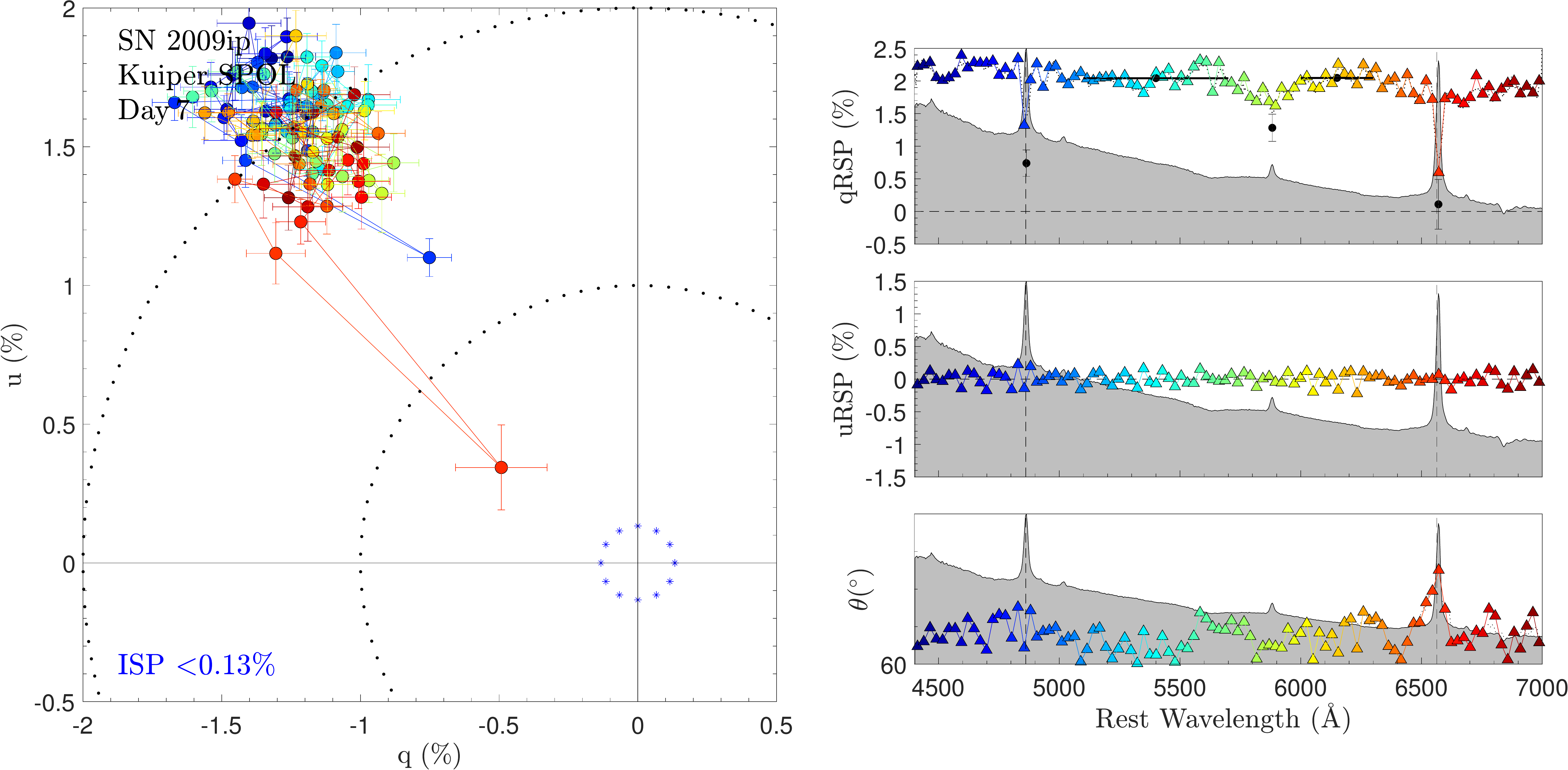}
{SN~2009ip}{Kuiper}{37}{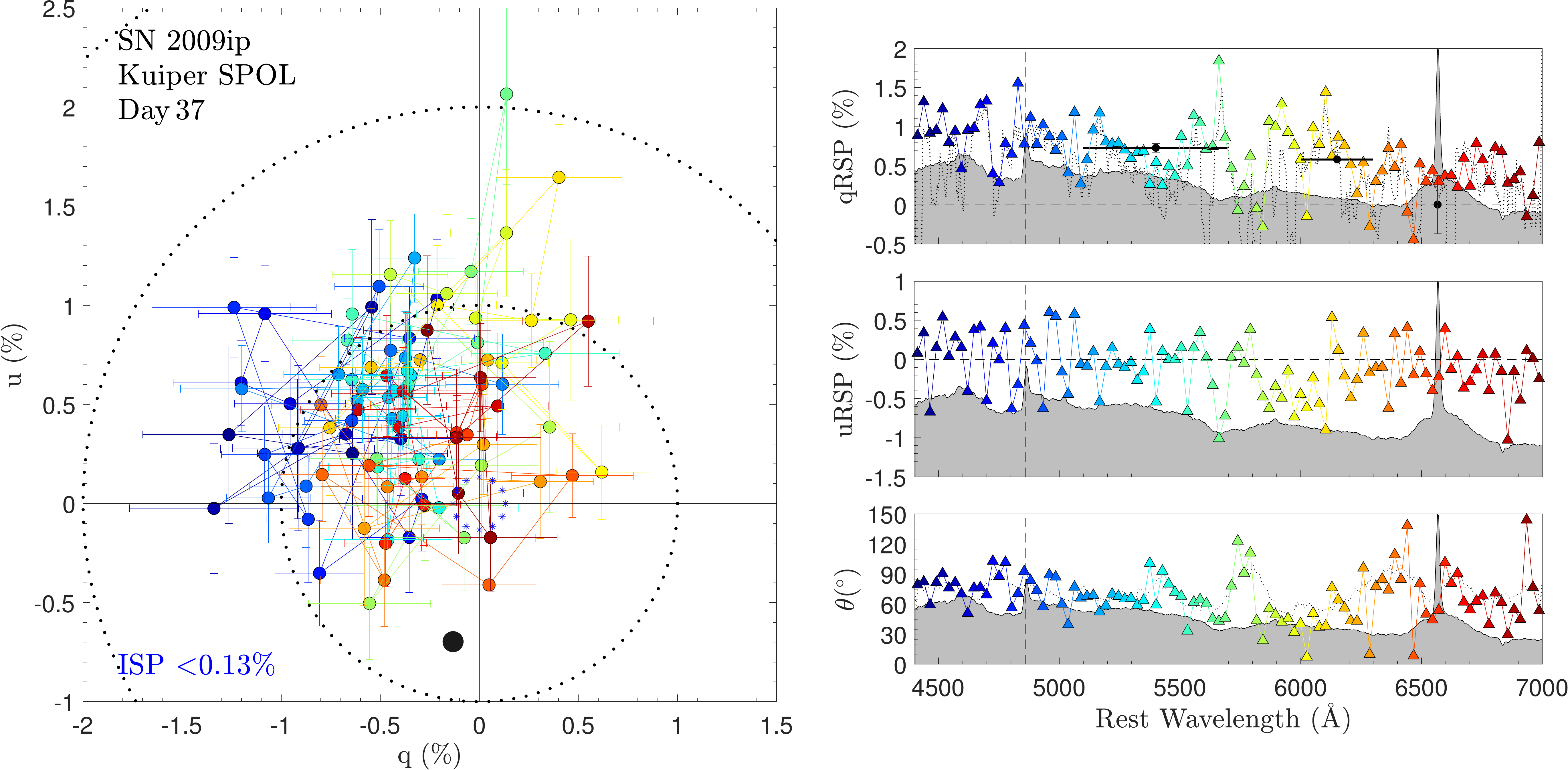}
\qufig
{SN~2009ip}{Kuiper}{37, ISP-corrected}{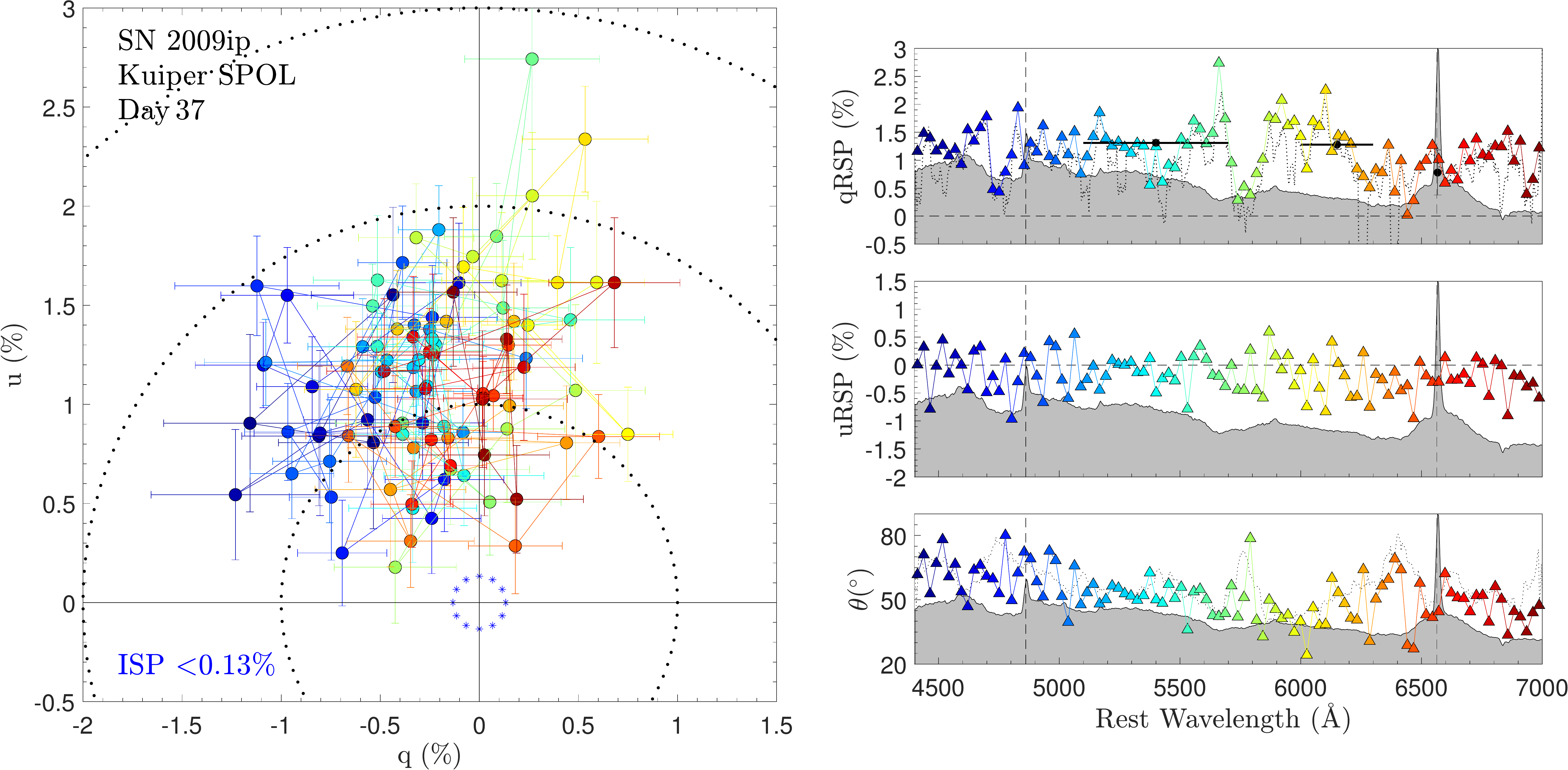}
{SN~2009ip}{Bok}{60}{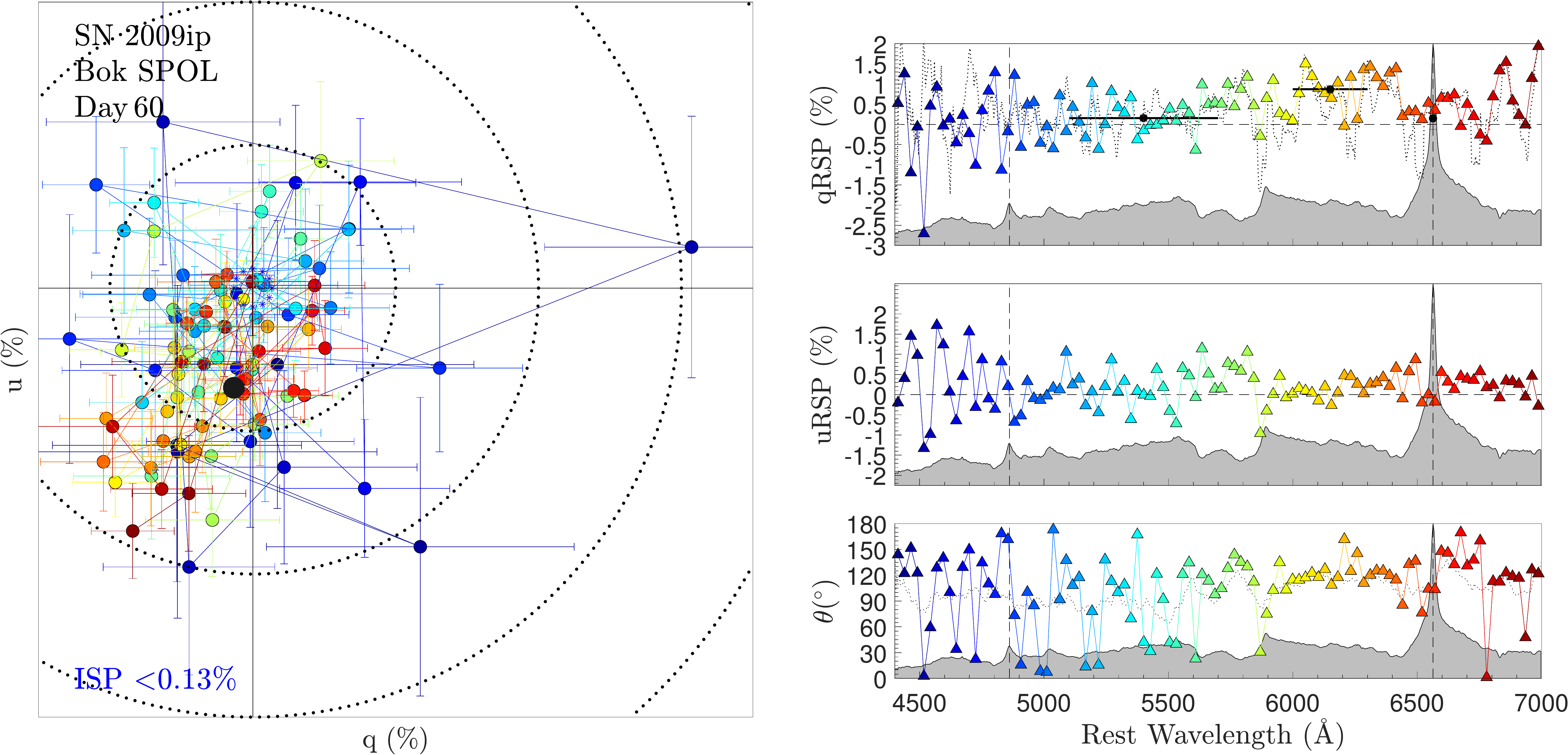}
\qufig
{SN~2009ip}{Bok}{60, ISP-corrected}{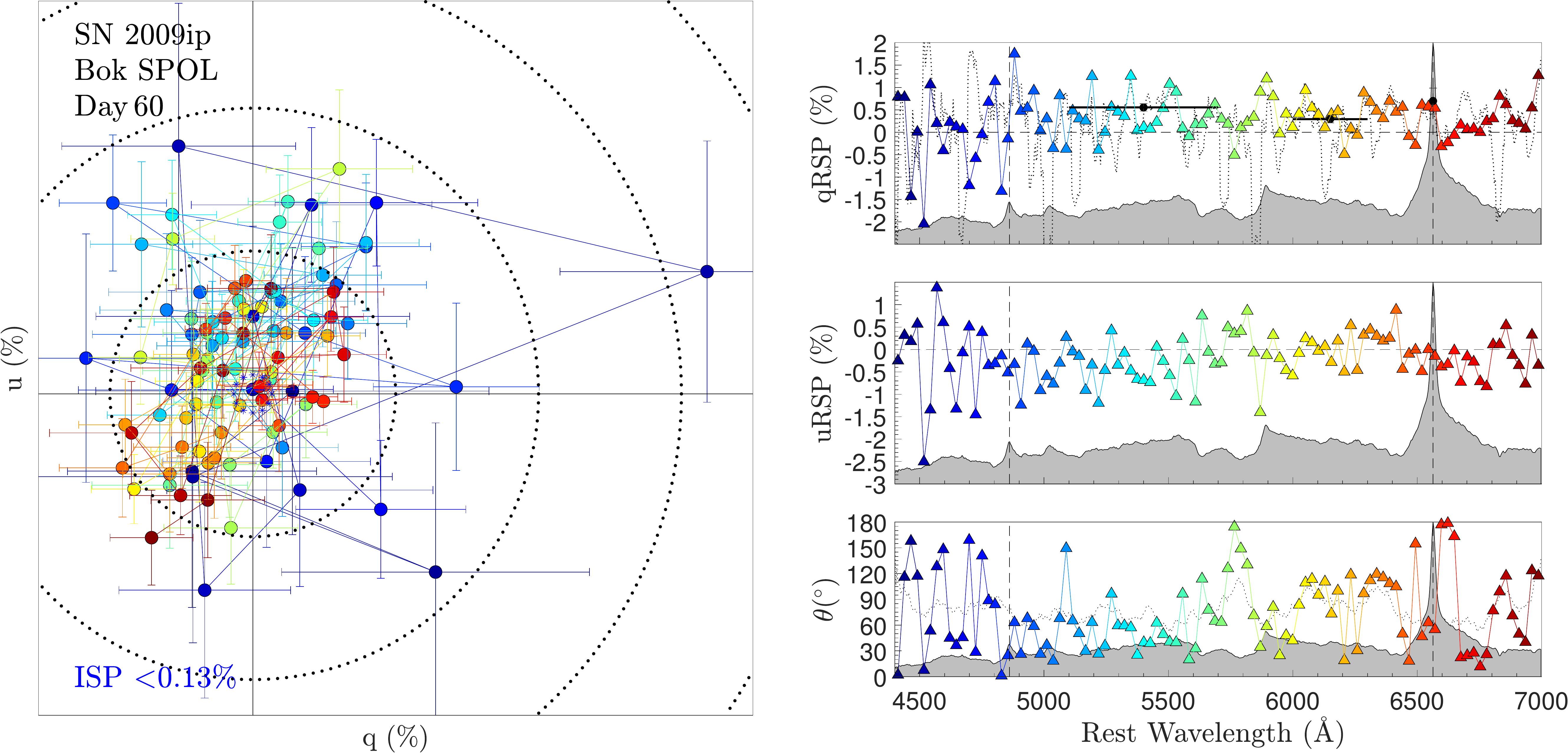}
{SN~2014ab}{Bok}{77}{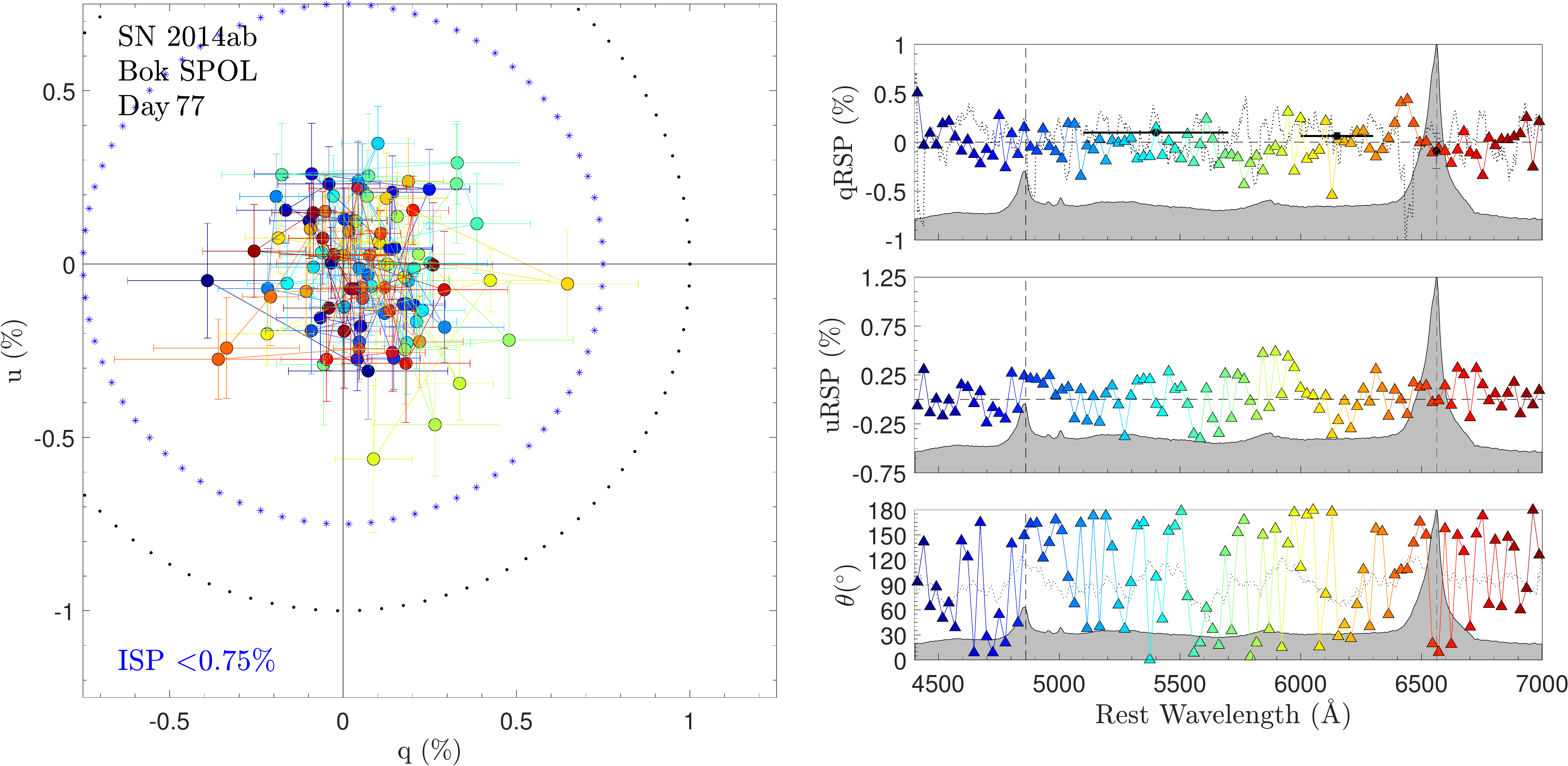}
\qufig
{SN~2014ab}{MMT}{99}{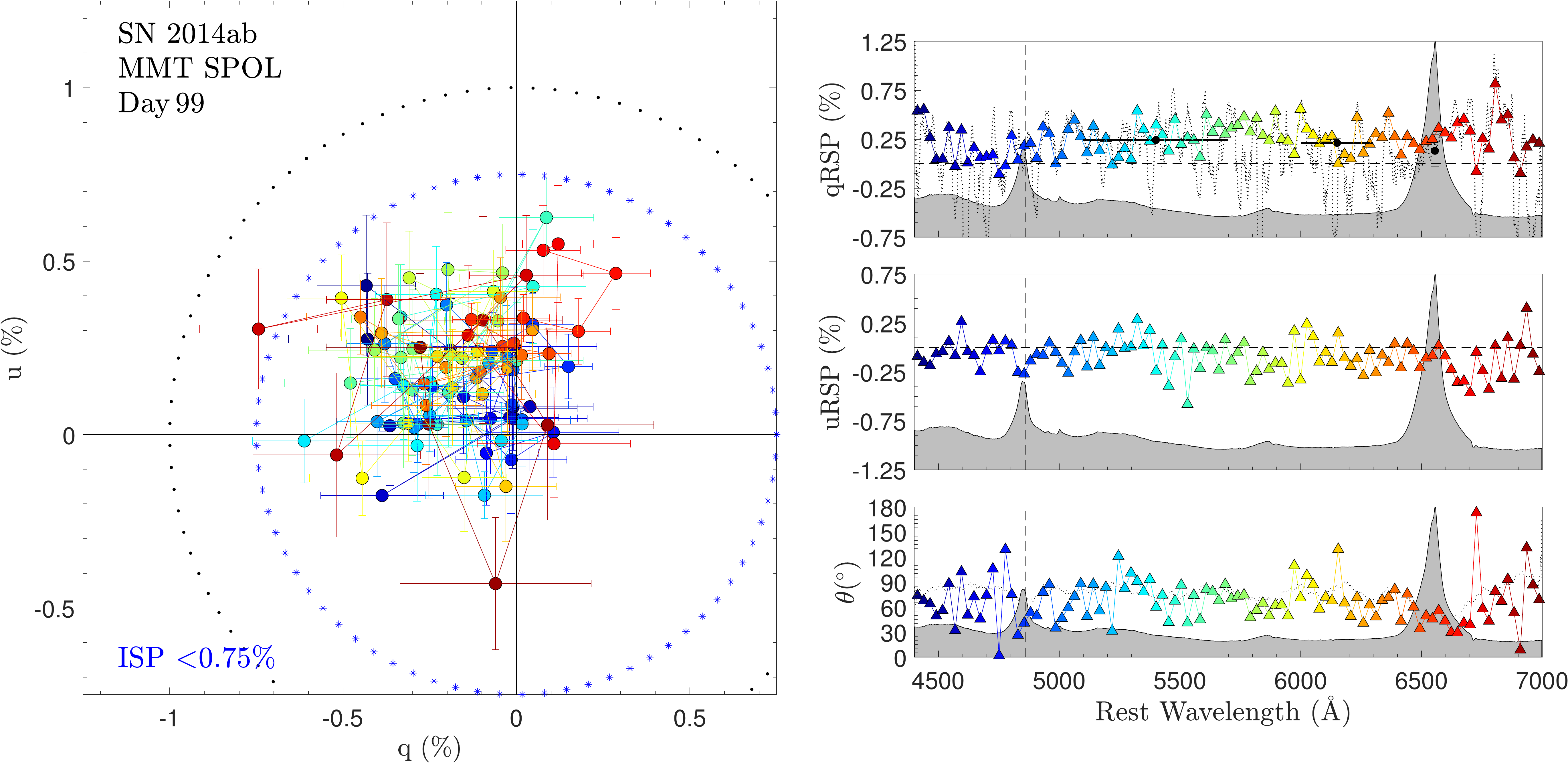}
{SN~2014ab}{Kuiper}{106}{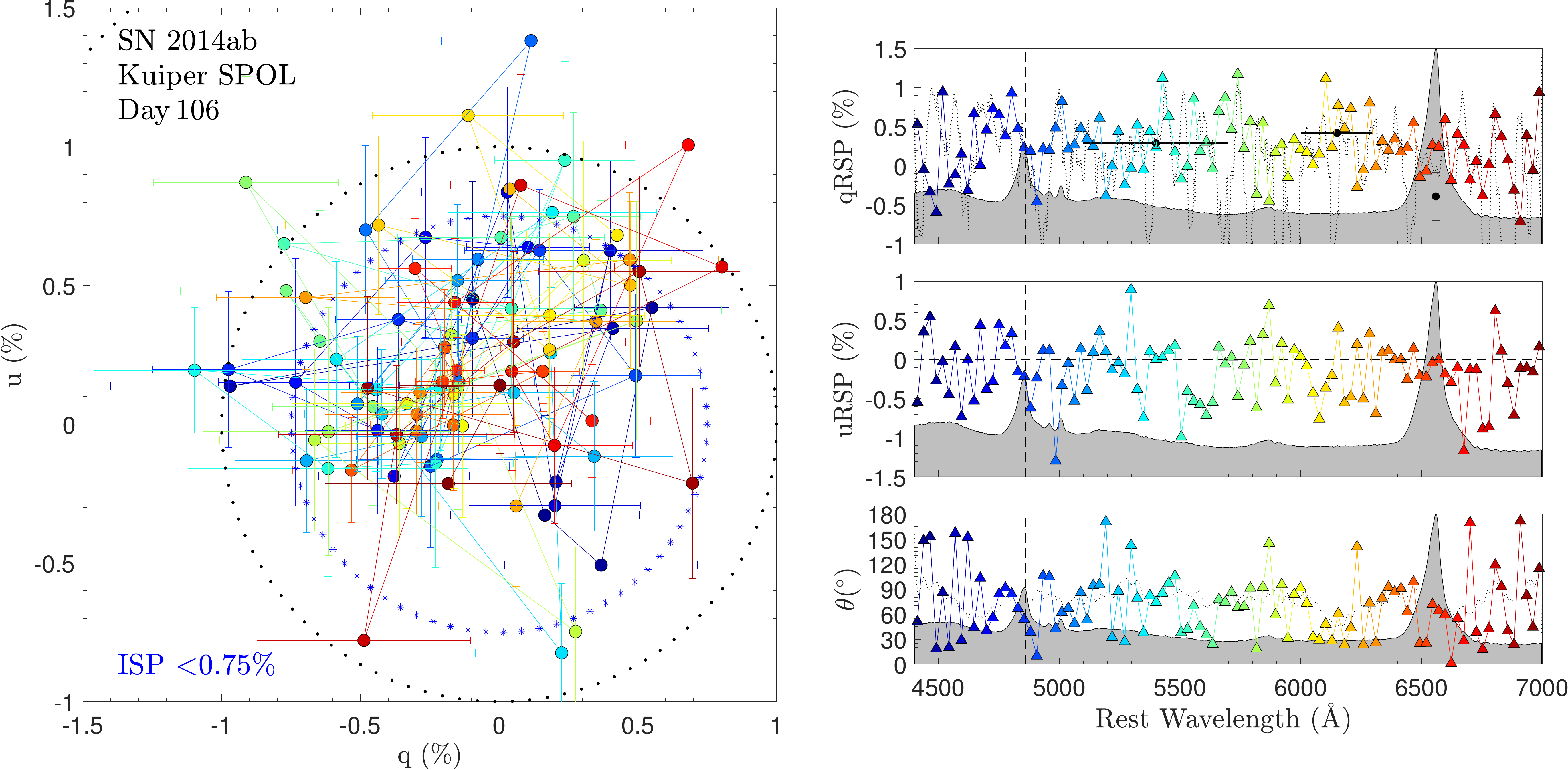}
\qufig
{SN~2014ab}{Bok}{132}{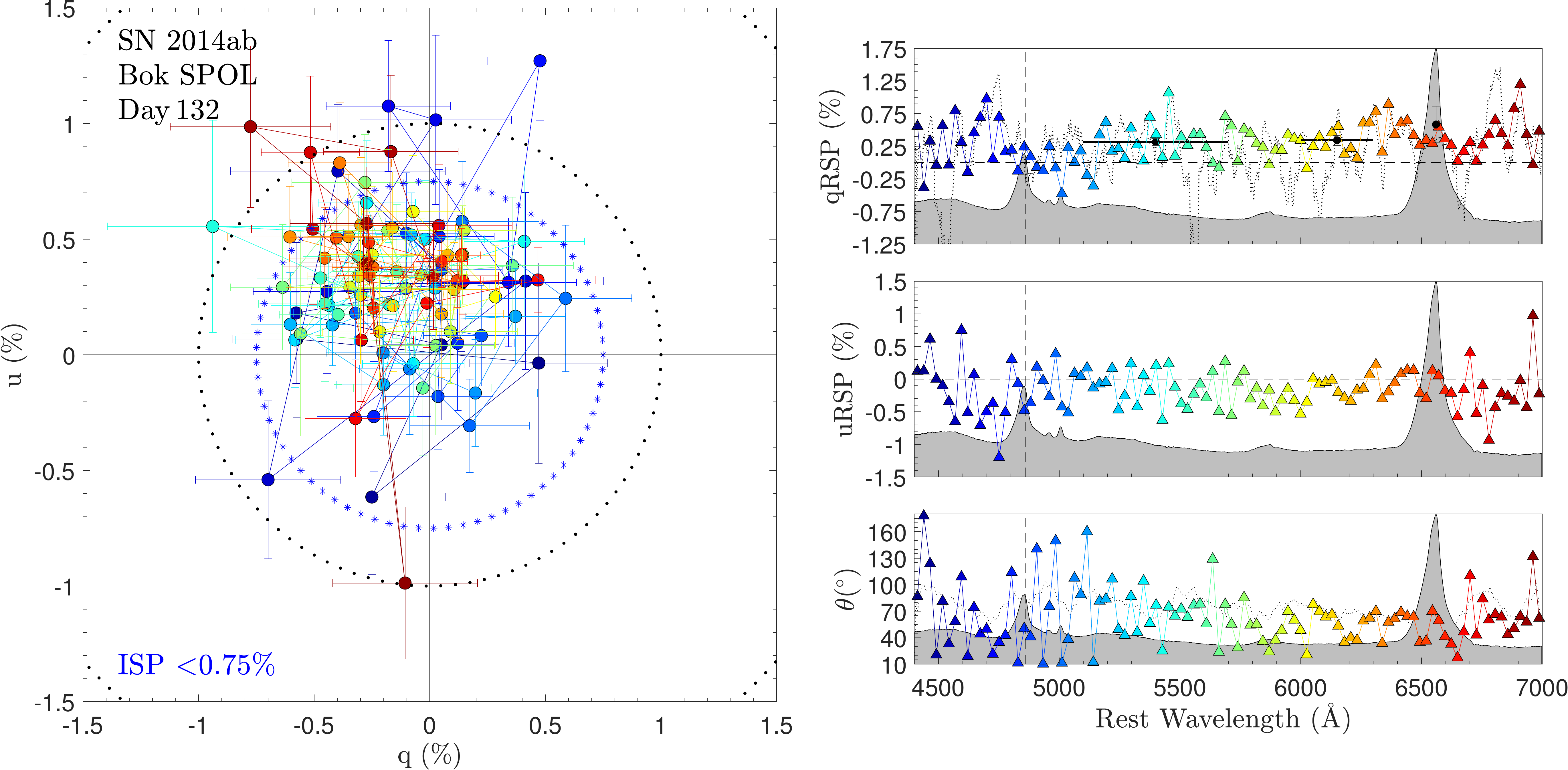}
{SN~2014ab}{Kuiper}{162}{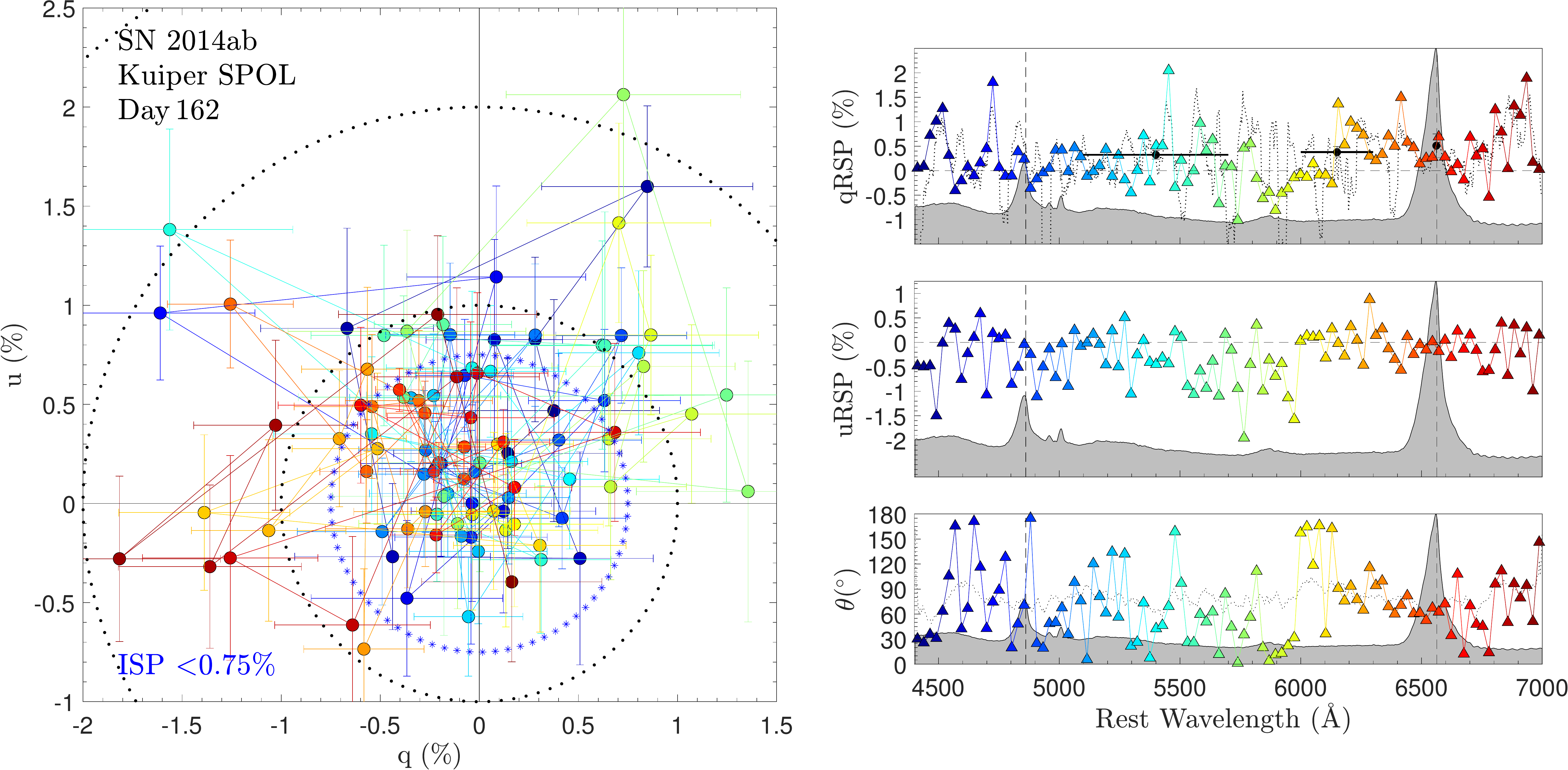}
\qufig
{M04421}{Kuiper}{118}{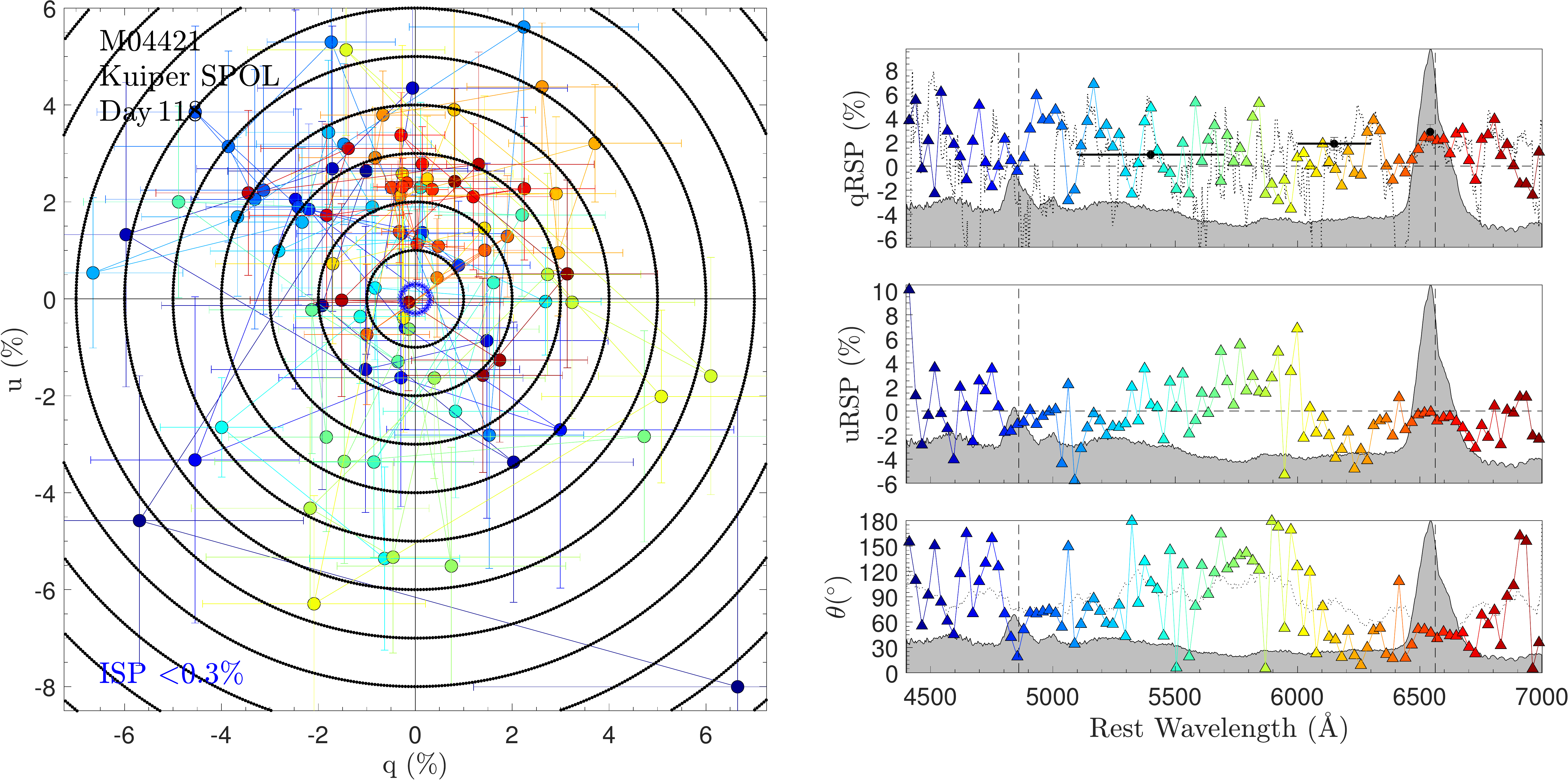}
{ASASSN-14il}{Kuiper}{-15}{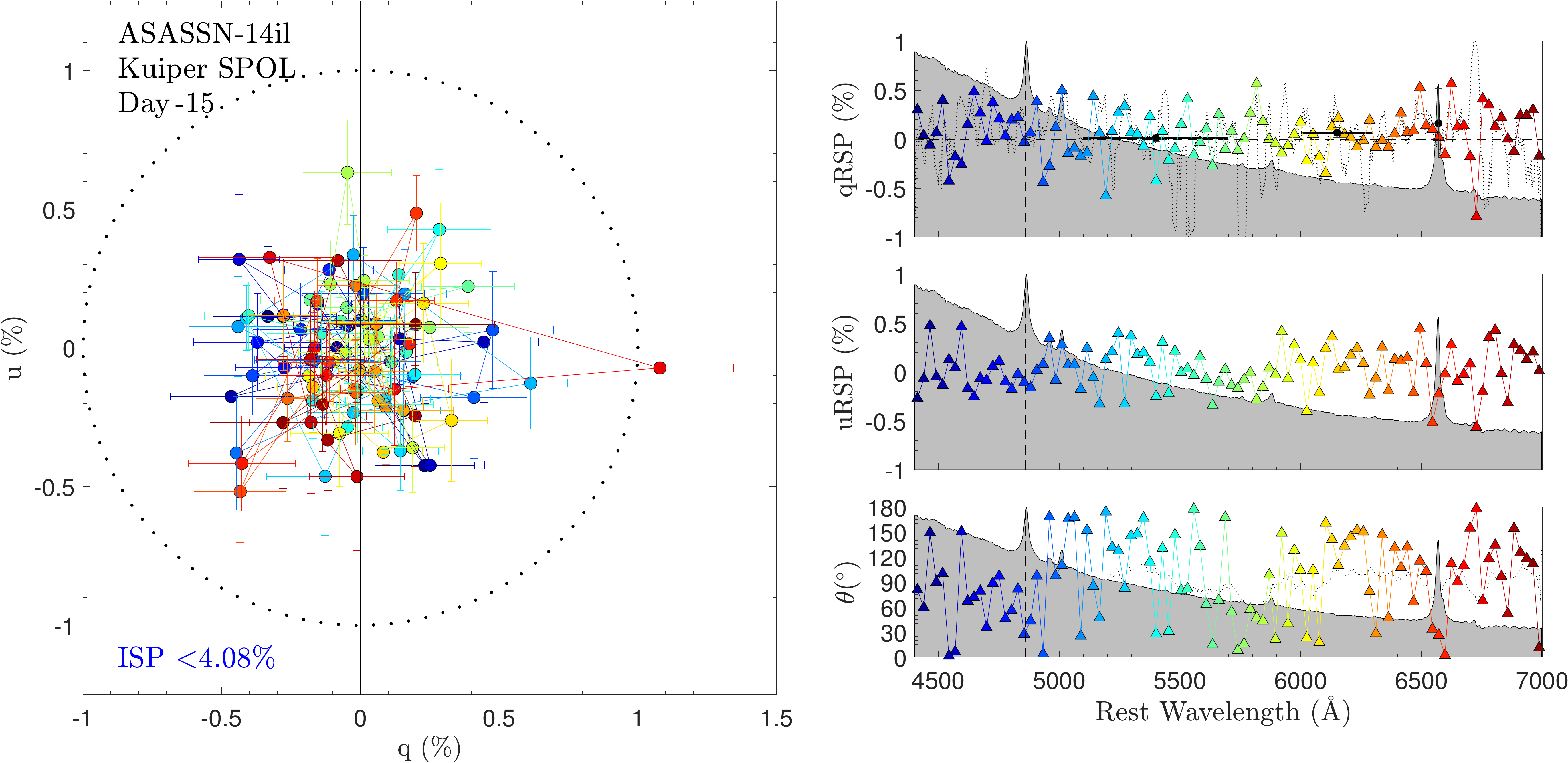}
\qufig
{ASASSN-14il}{Kuiper}{17}{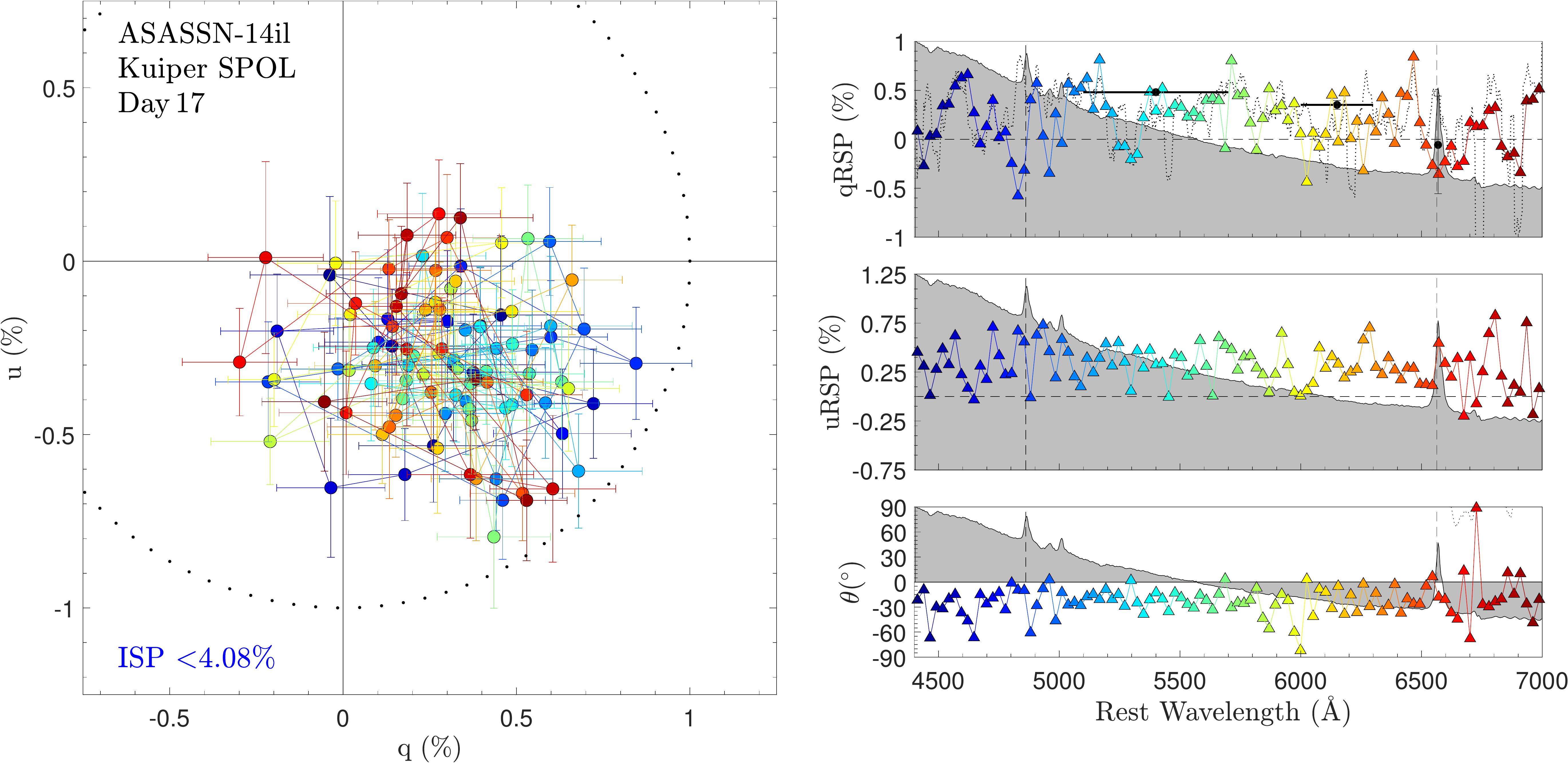}
{ASASSN-14il}{Bok}{73}{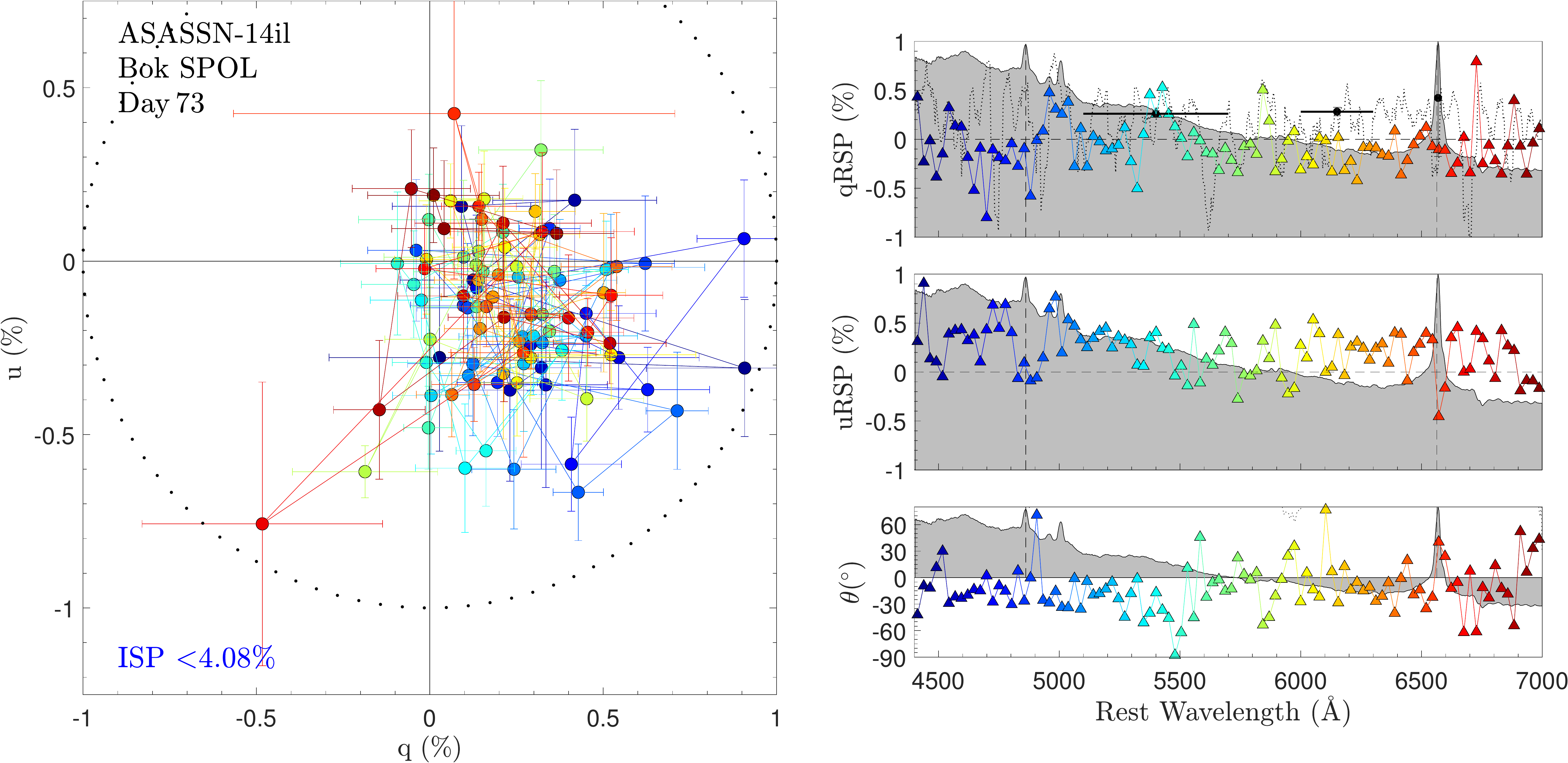}
\qufig
{SN~2015da}{Kuiper}{-55}{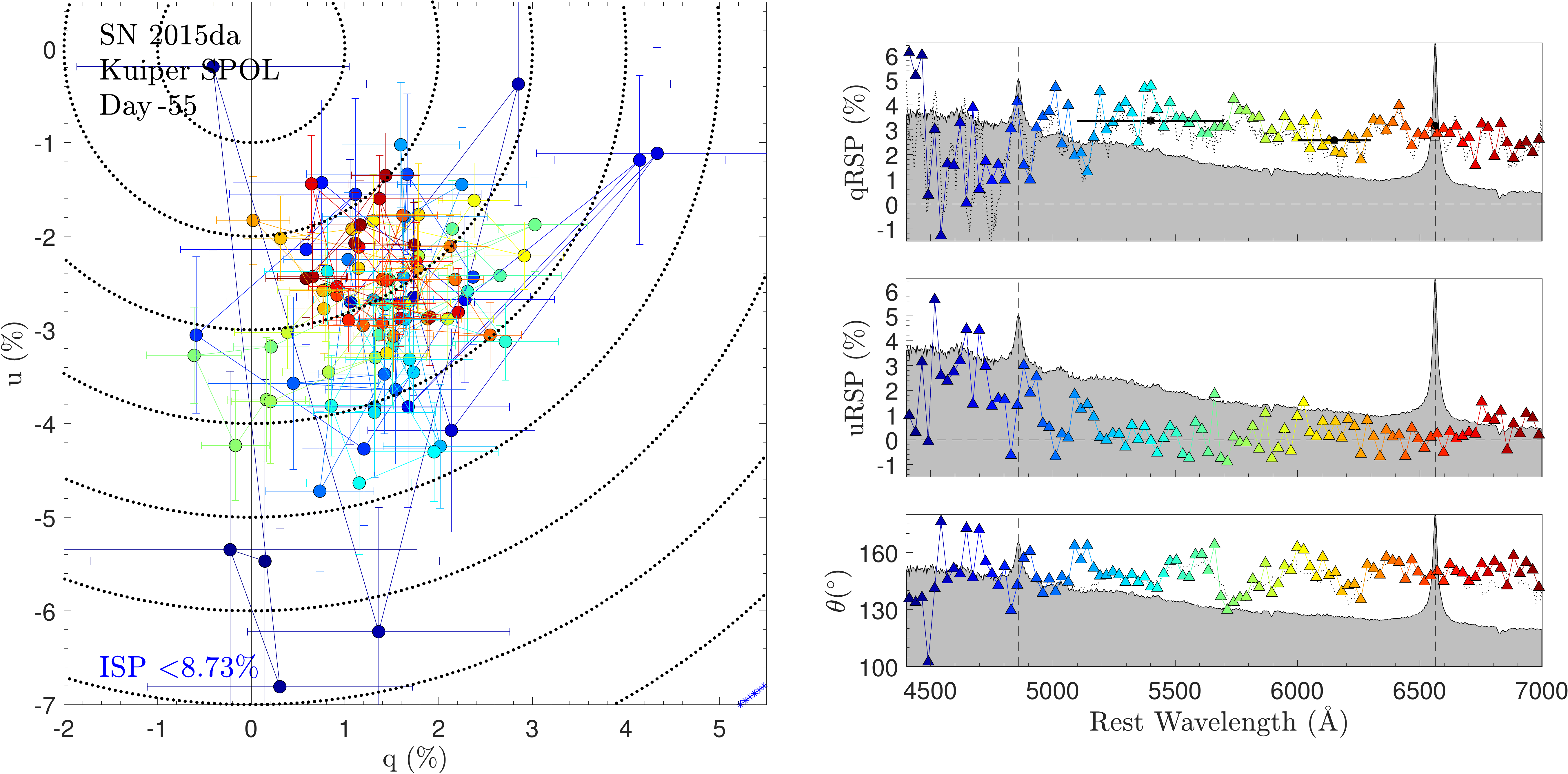}
{SN~2015da}{Kuiper}{-33}{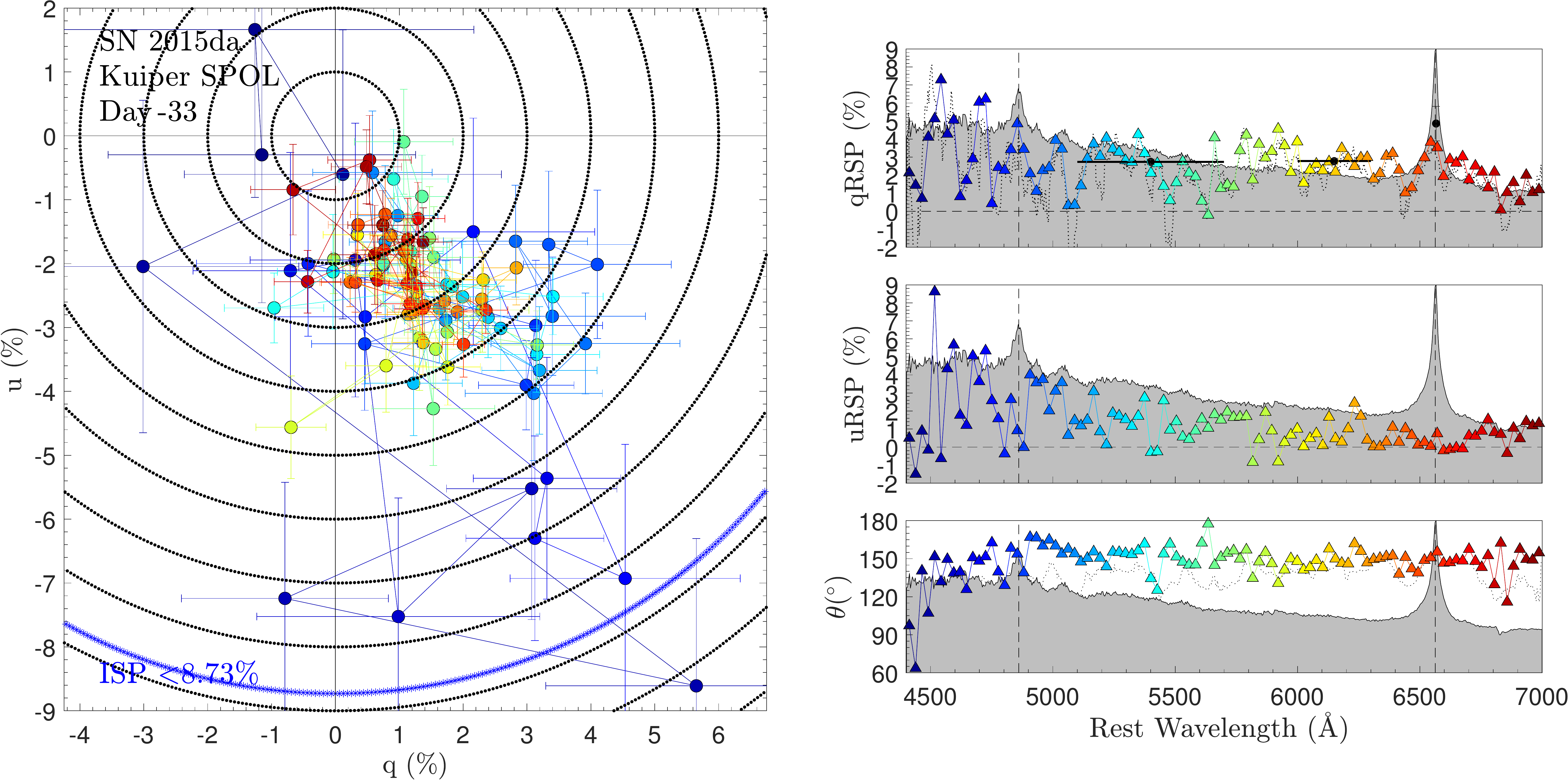}
\qufig
{SN~2015da}{MMT}{-22}{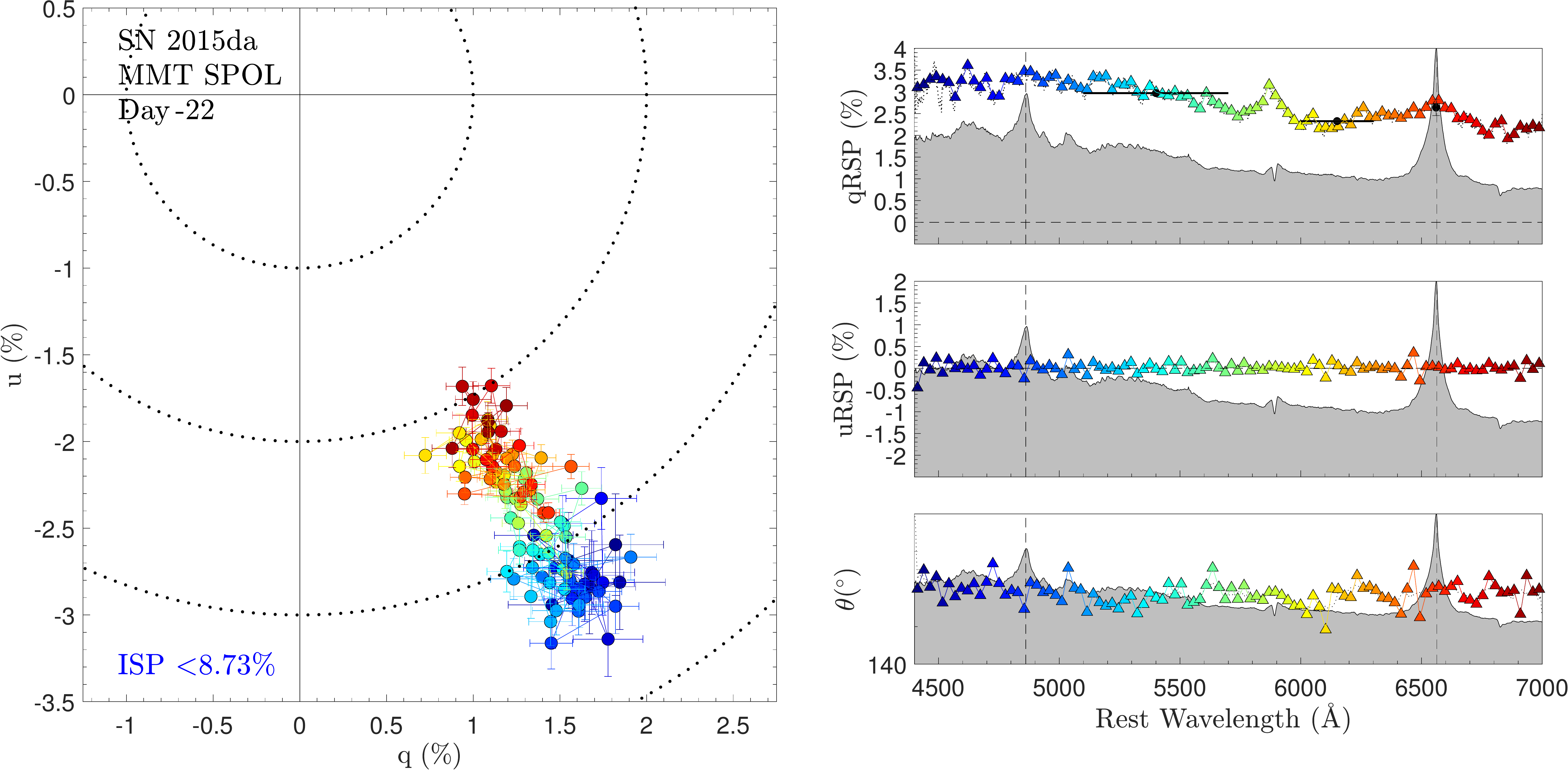}
{SN~2015da}{Bok}{4}{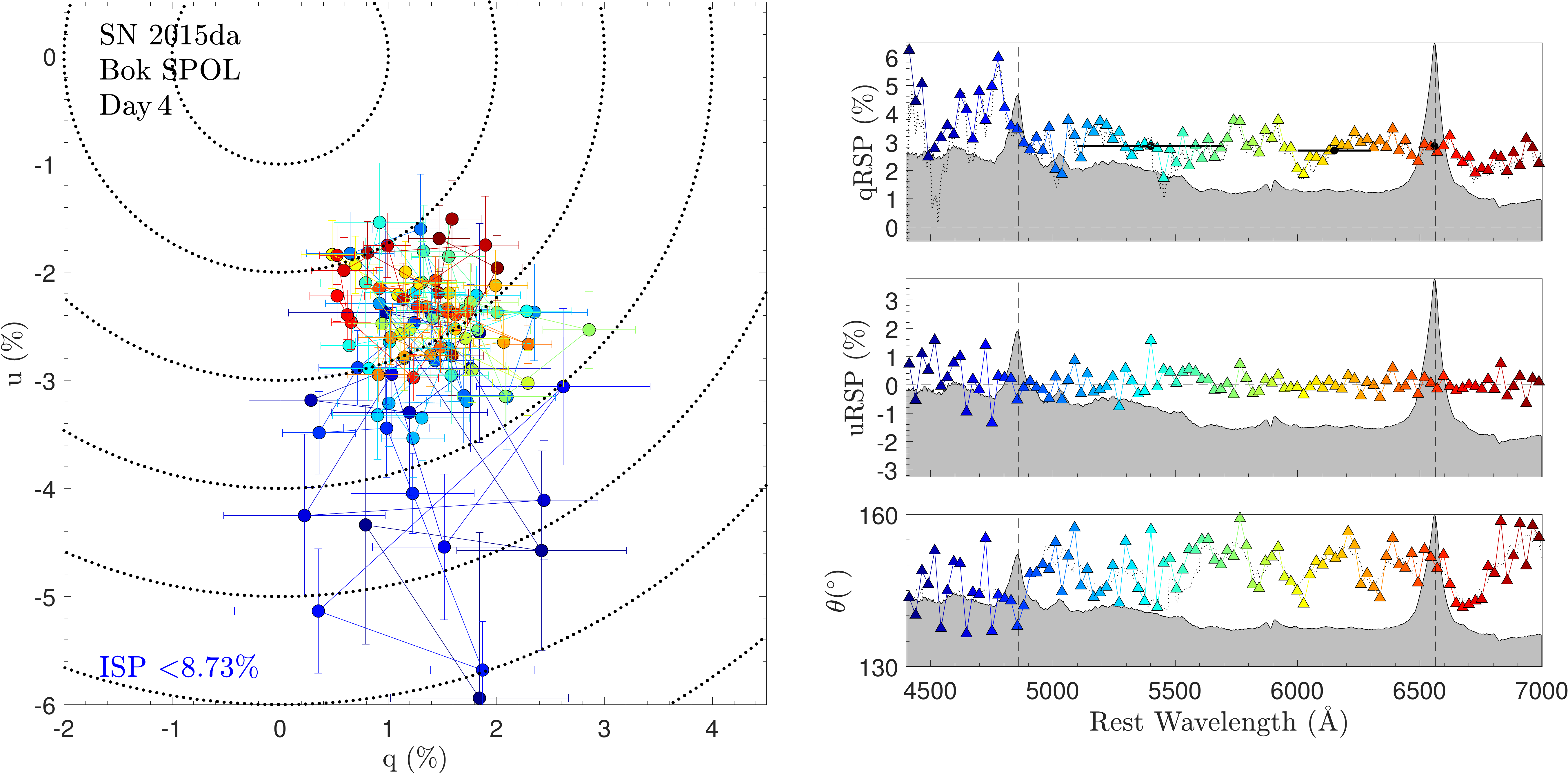}
\qufig
{SN~2015bh}{Bok}{-3}{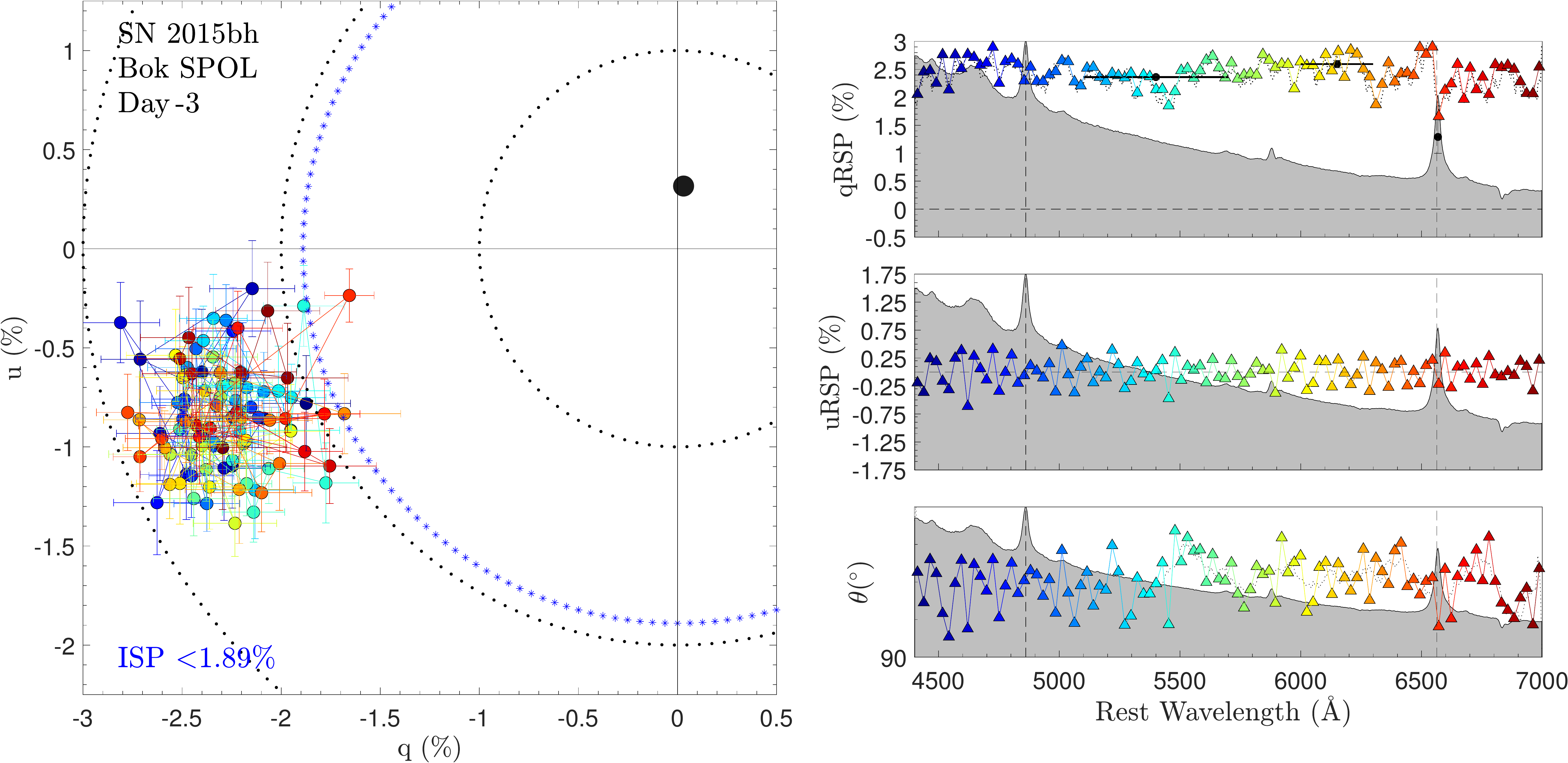}
{SN~2015bh}{Bok}{-3, ISP-corrected}{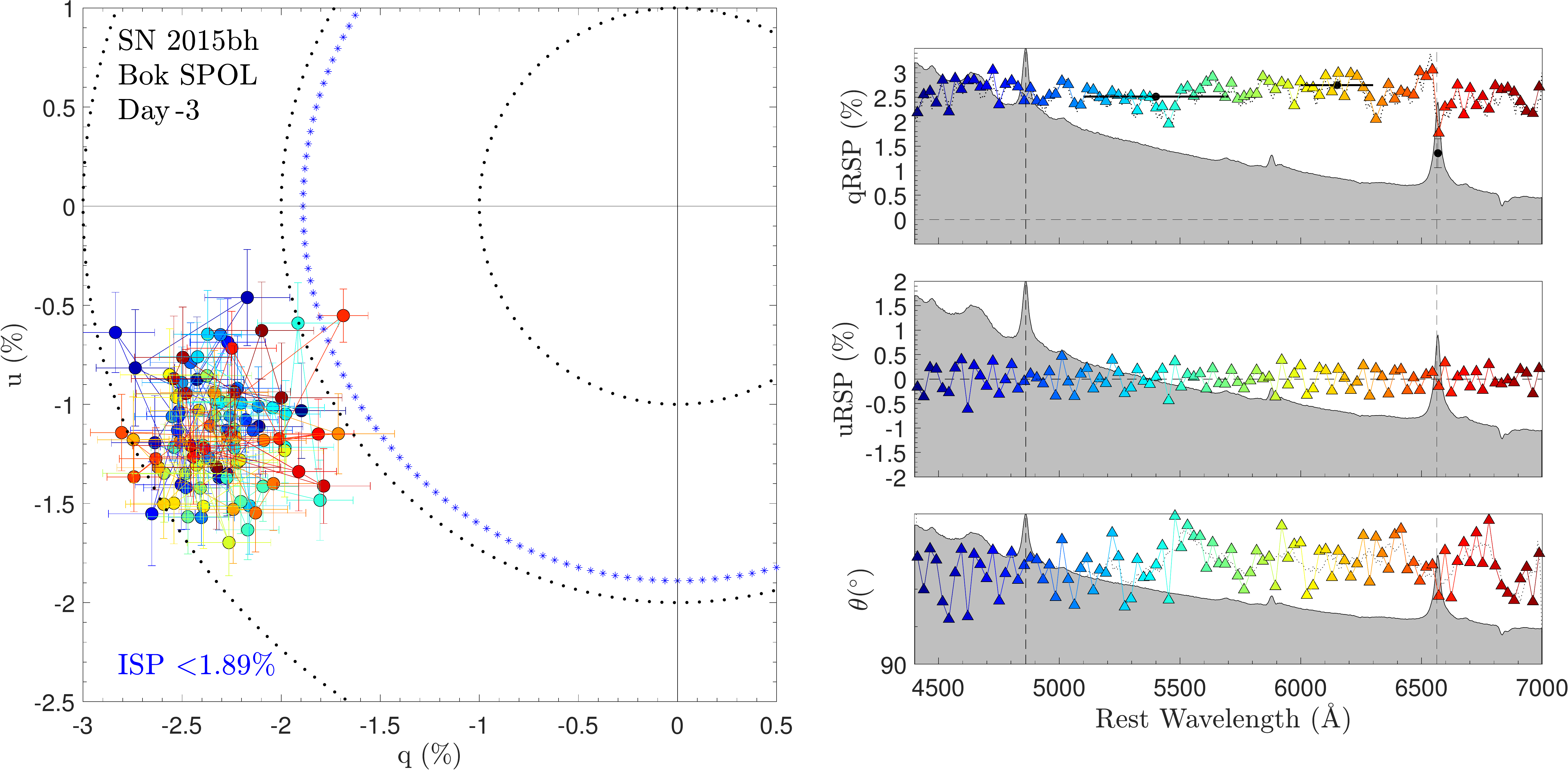}
\qufig
{SN~2015bh}{MMT}{18}{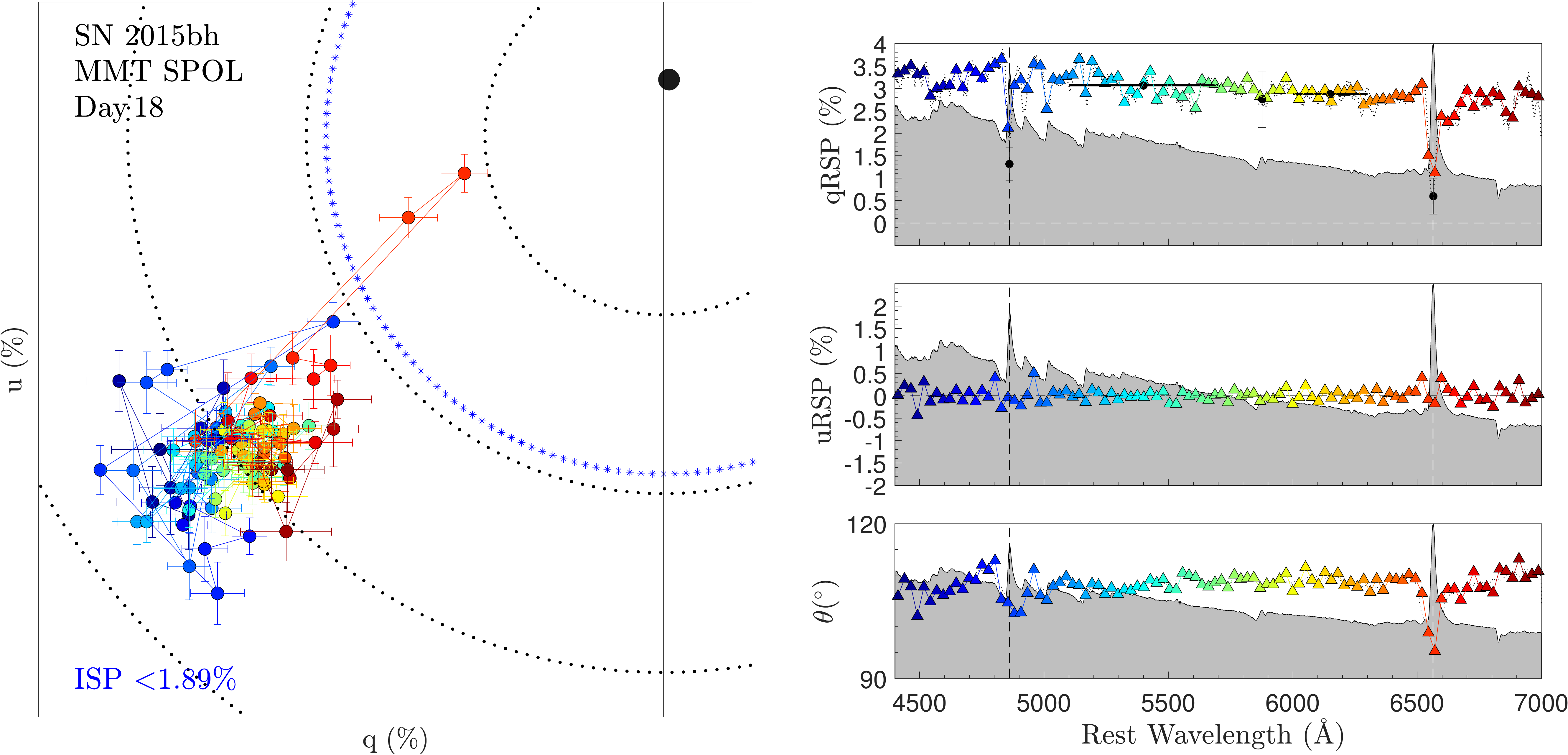}
{SN~2015bh}{MMT}{18, ISP-corrected}{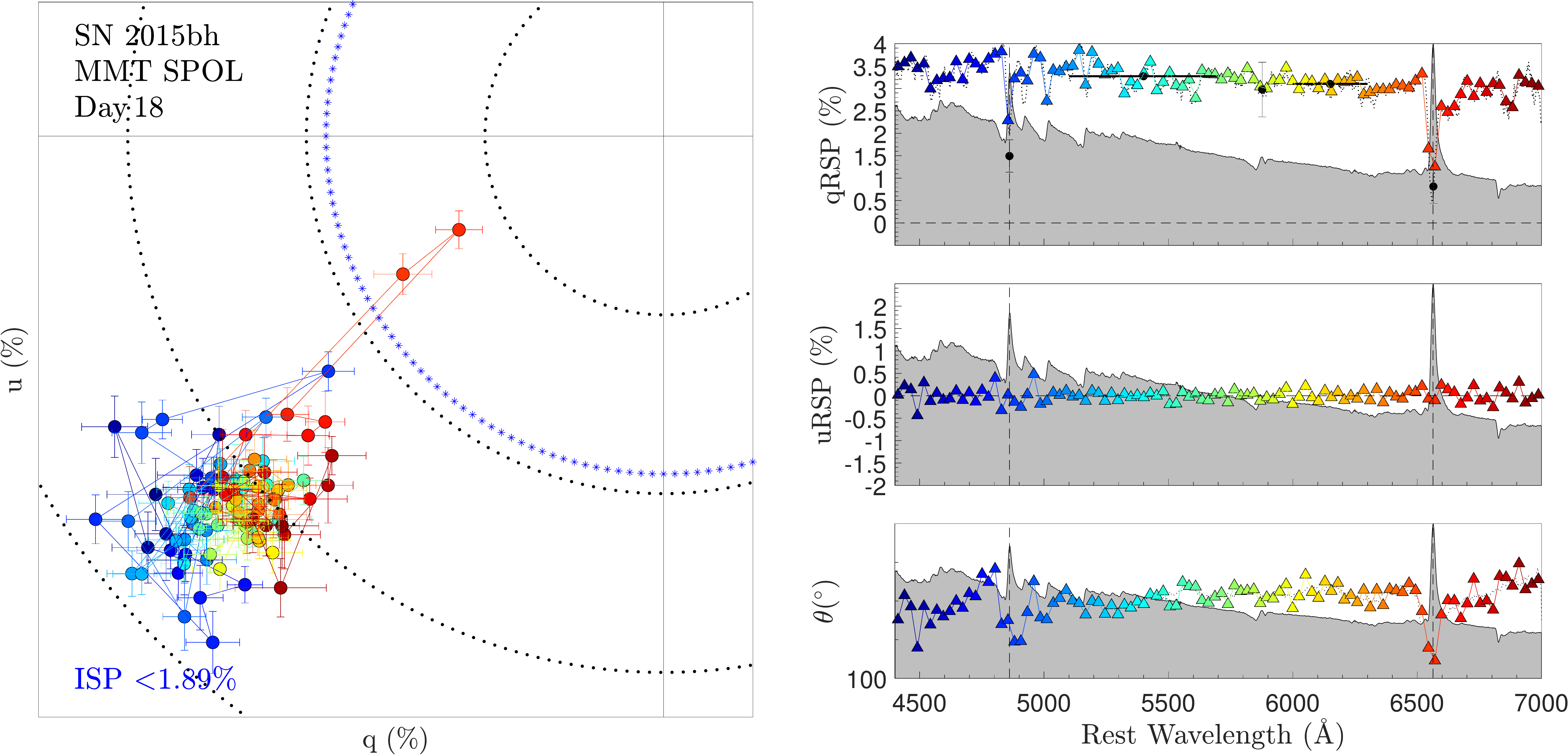}
\qufig
{SN~2015bh}{Kuiper}{23}{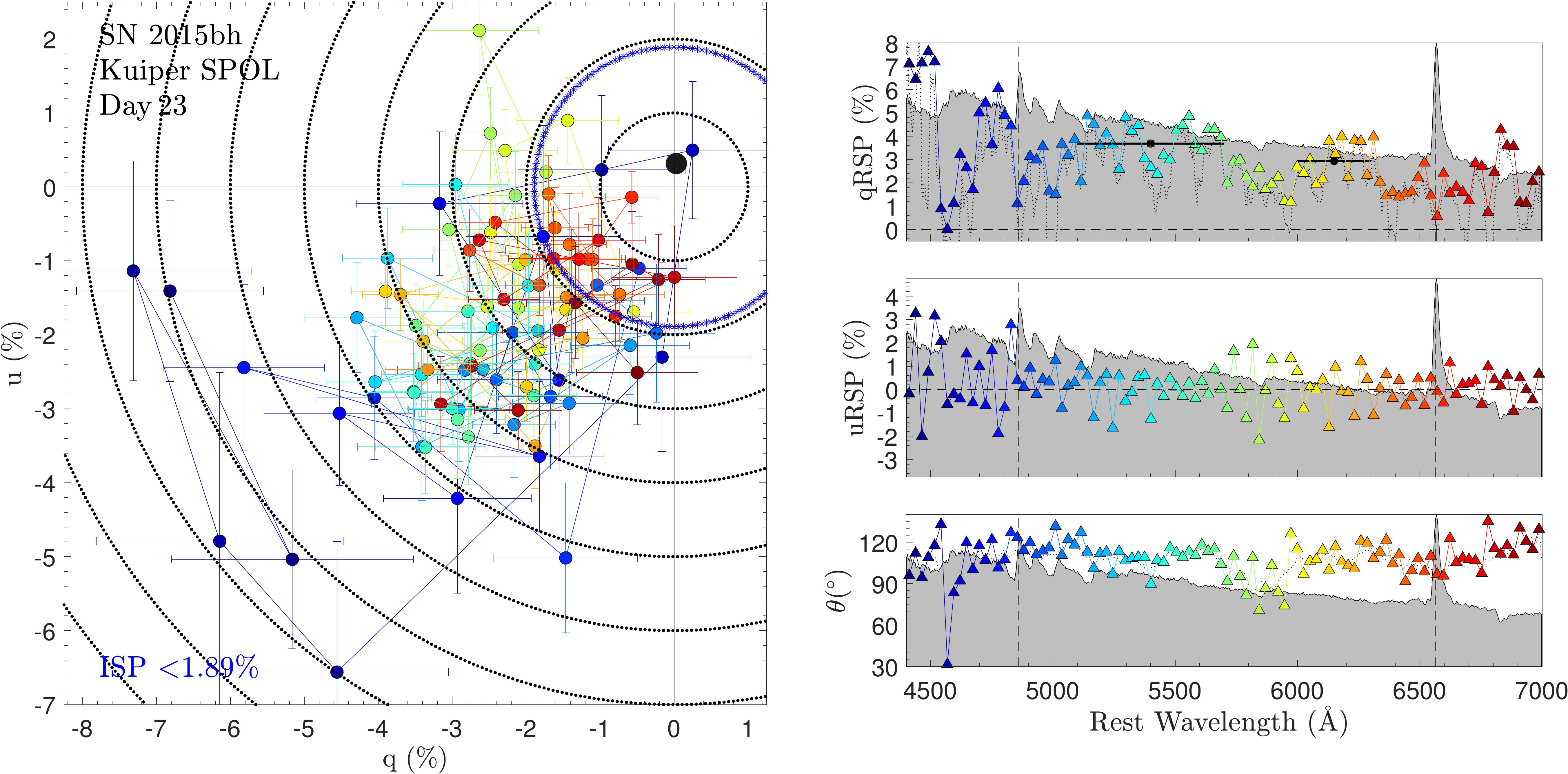}
{SN~2015bh}{Kuiper}{23, ISP-corrected}{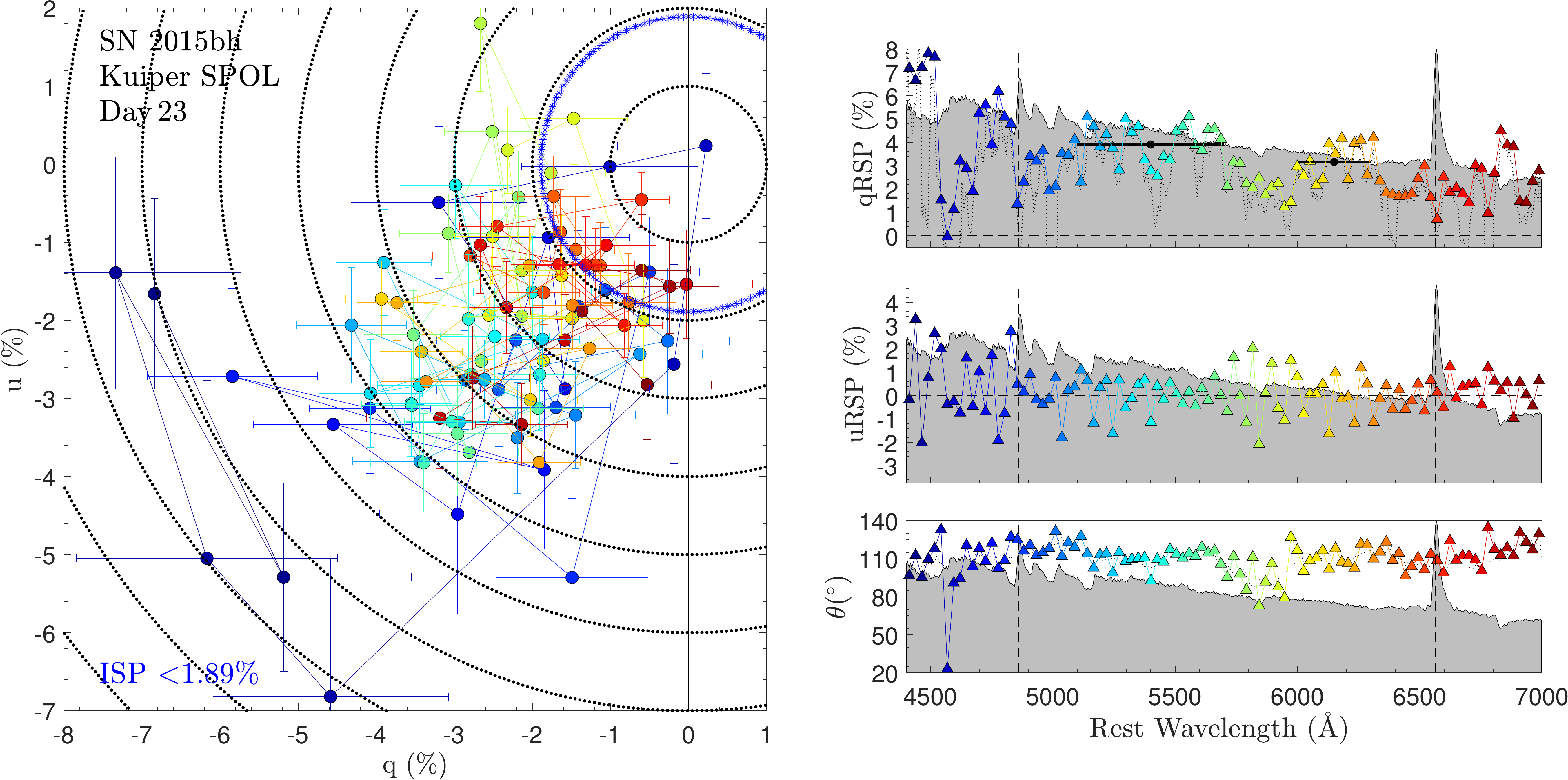}
\qufig
{PS15cwt}{Bok}{74}{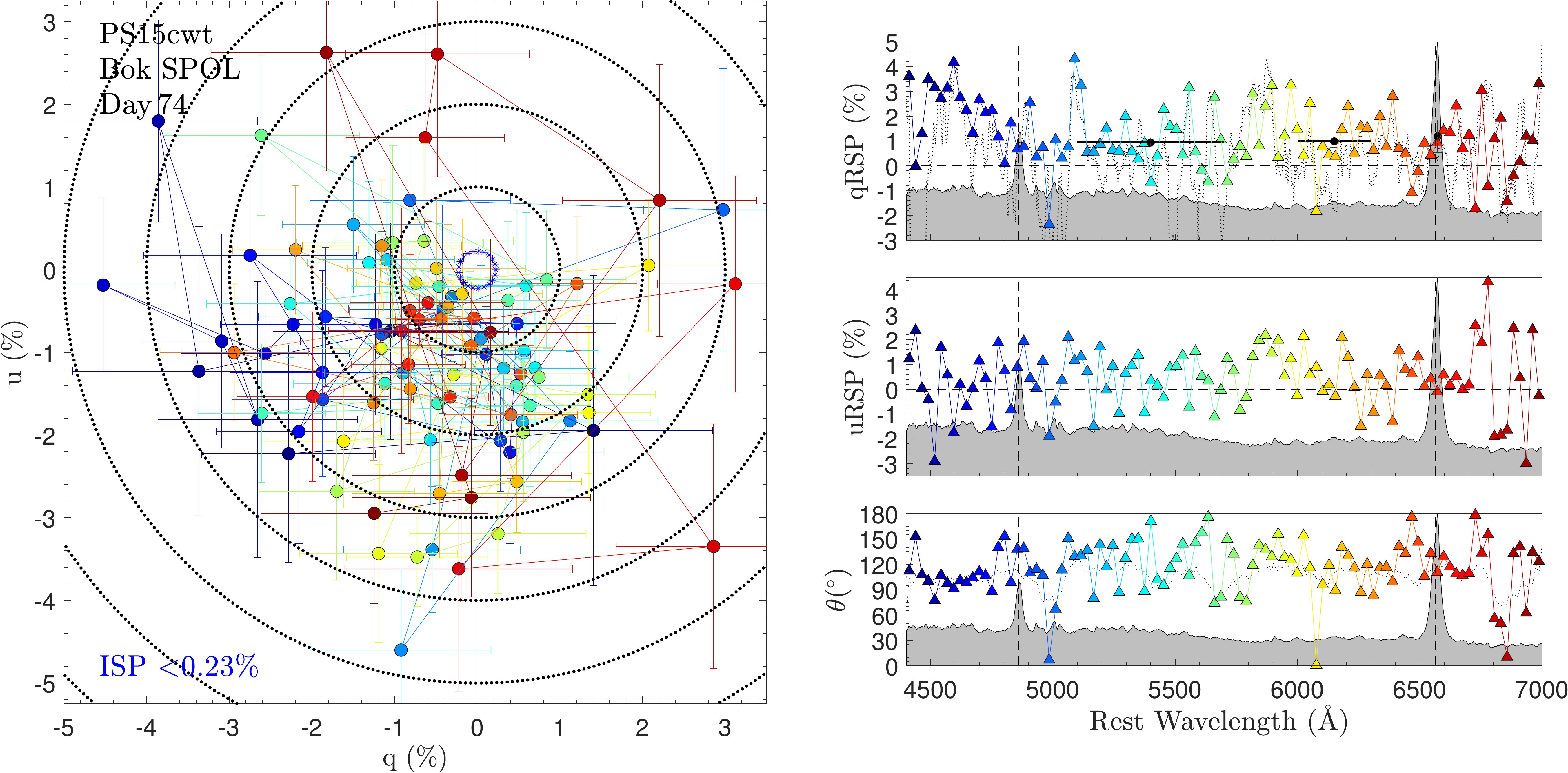}
{SN~2017gas}{MMT}{-3}{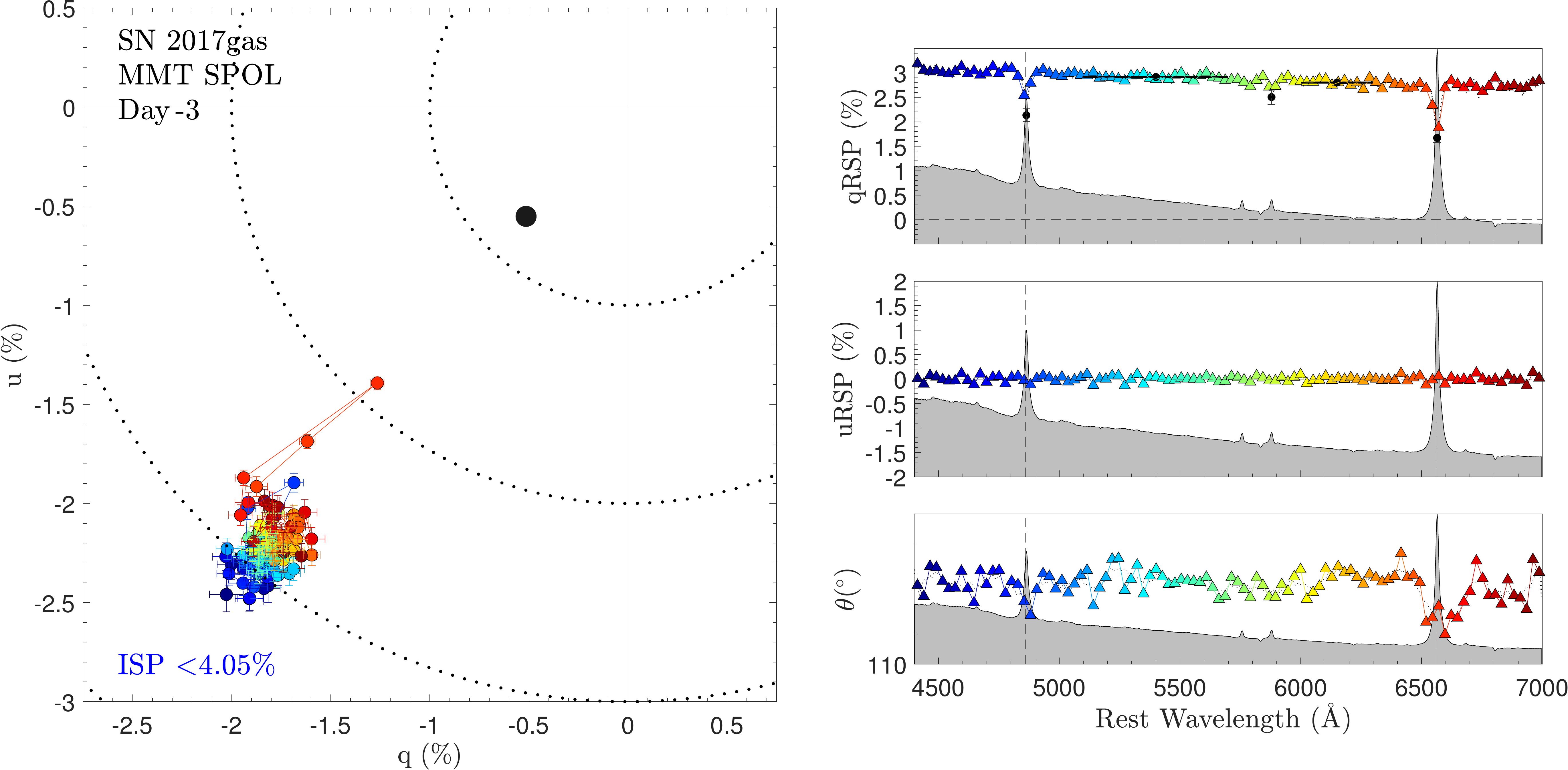}
\qufig
{SN~2017gas}{MMT}{-3, ISP-corrected}{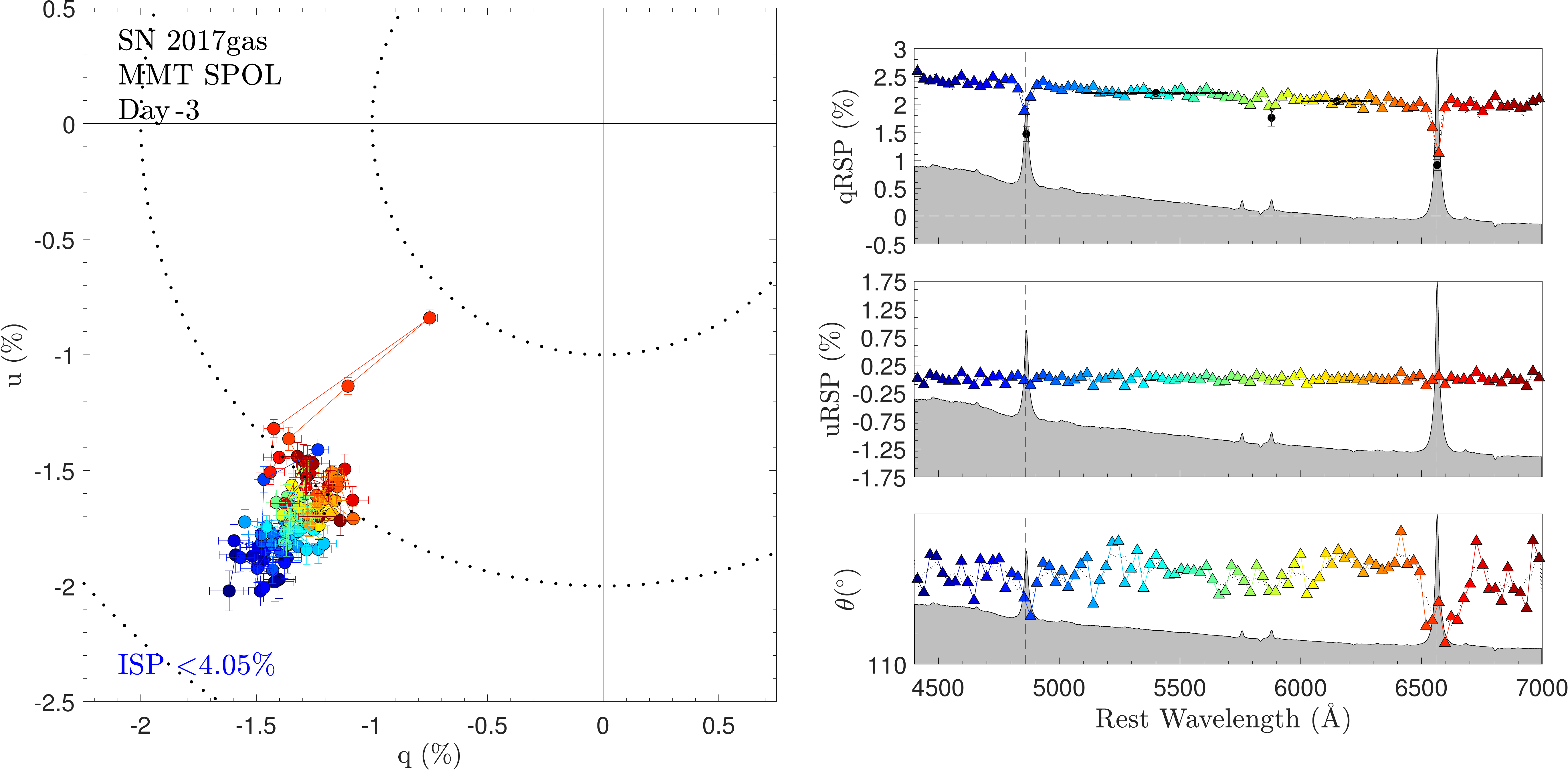}
{SN~2017gas}{Kuiper}{17}{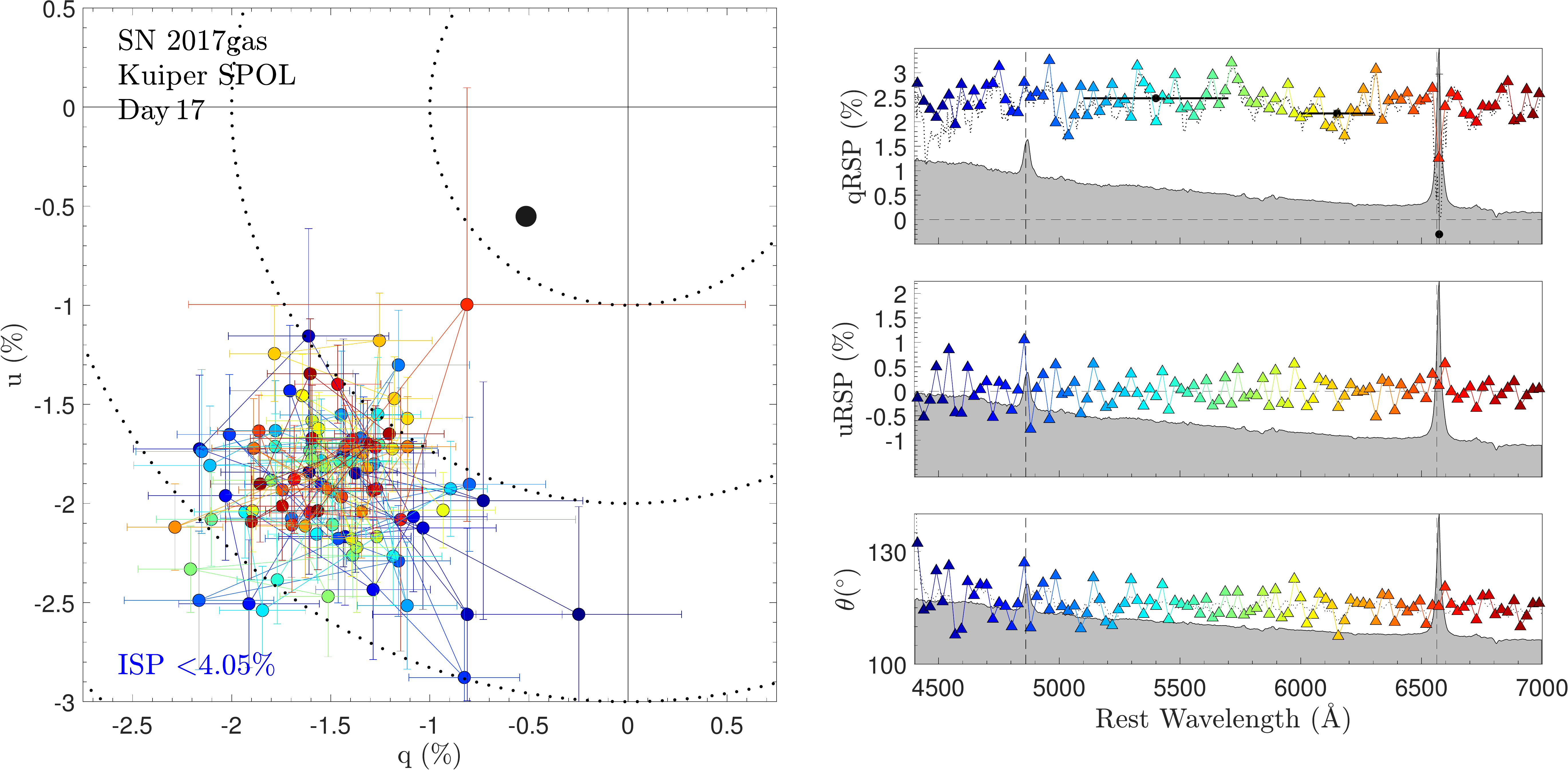}
\qufig
{SN~2017gas}{Kuiper}{17, ISP-corrected}{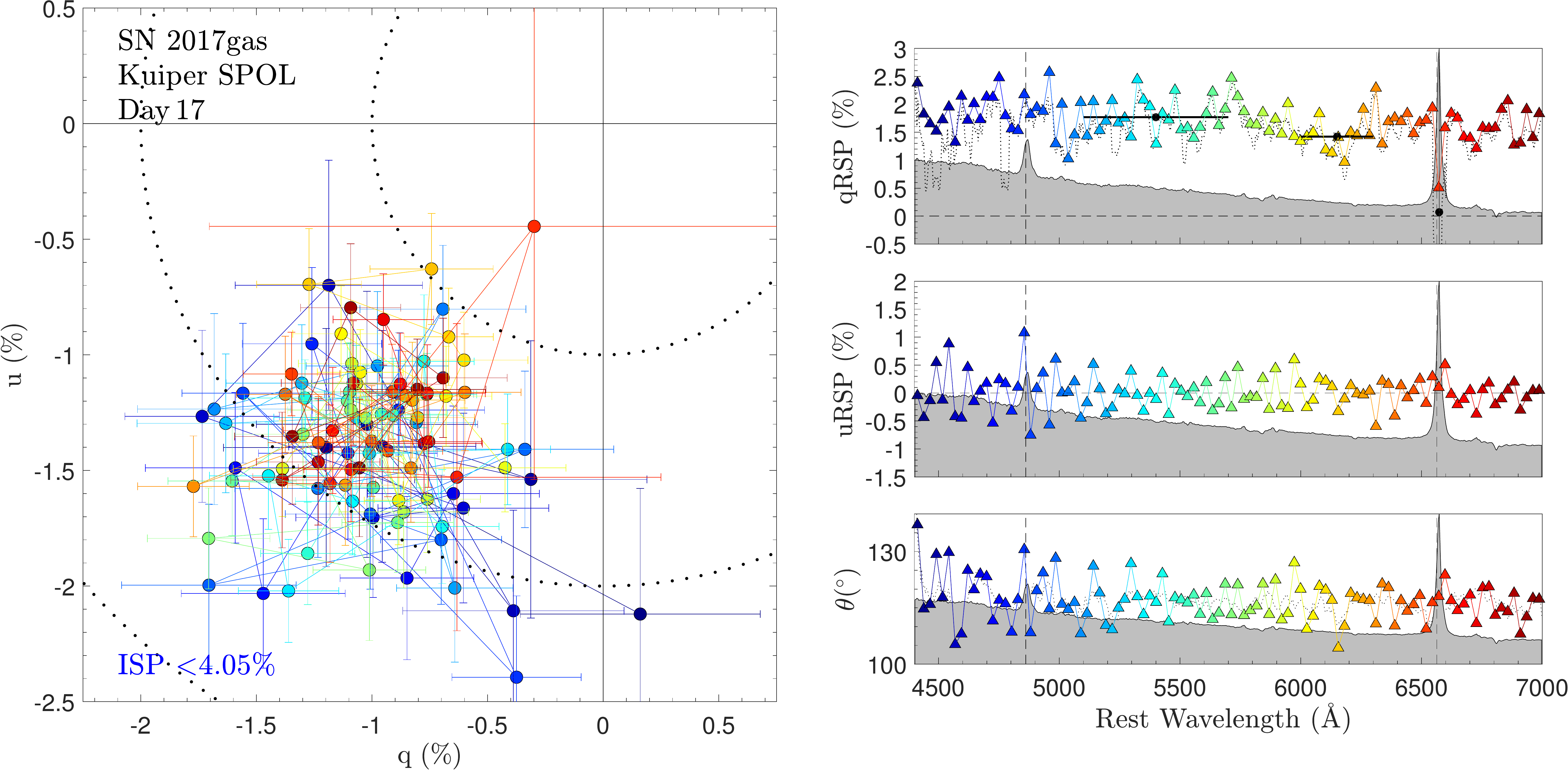}
{SN~2017gas}{Kuiper}{46}{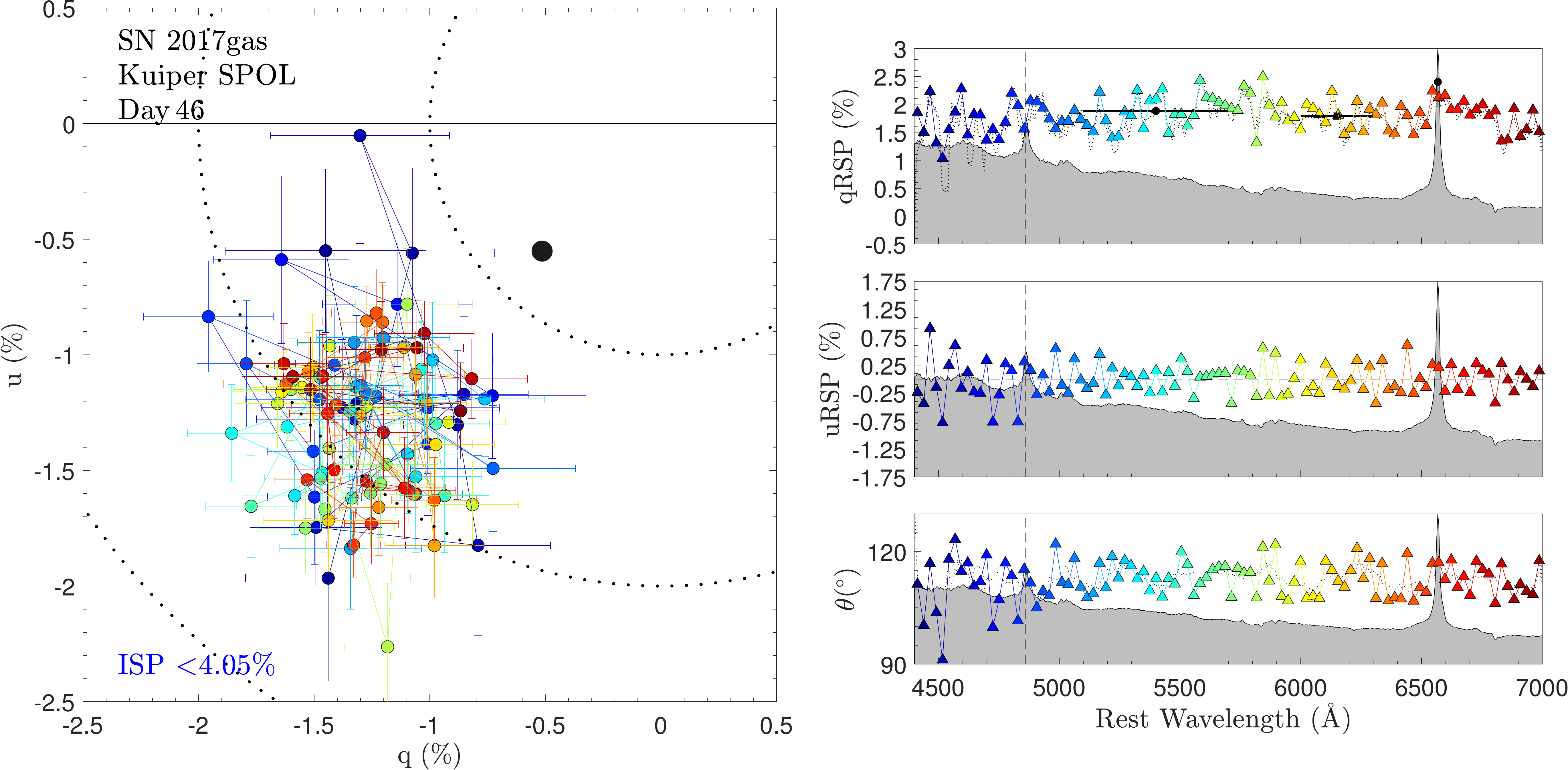}
\qufig
{SN~2017gas}{Kuiper}{46, ISP-corrected}{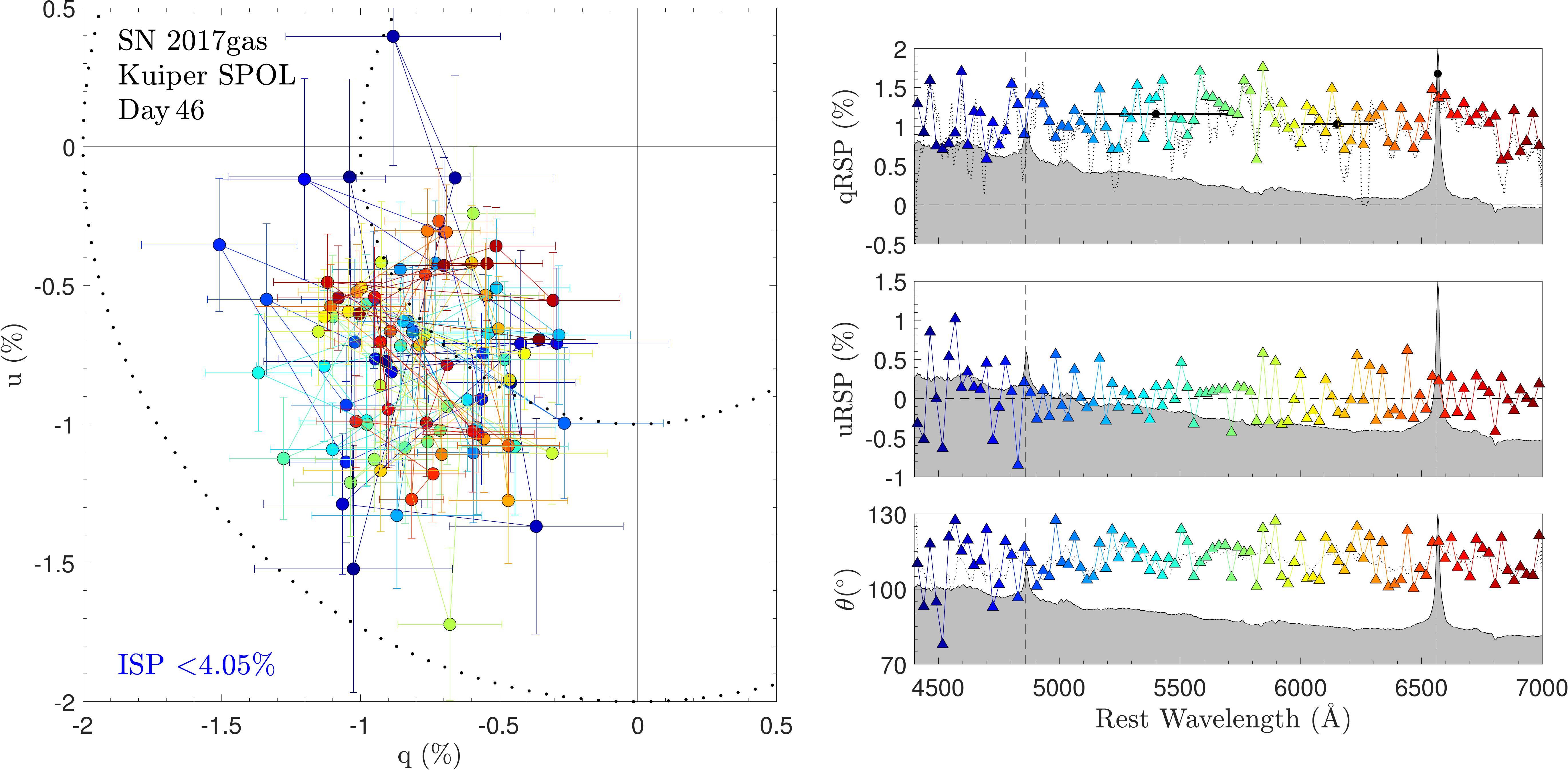}
{SN~2017gas}{Kuiper}{78}{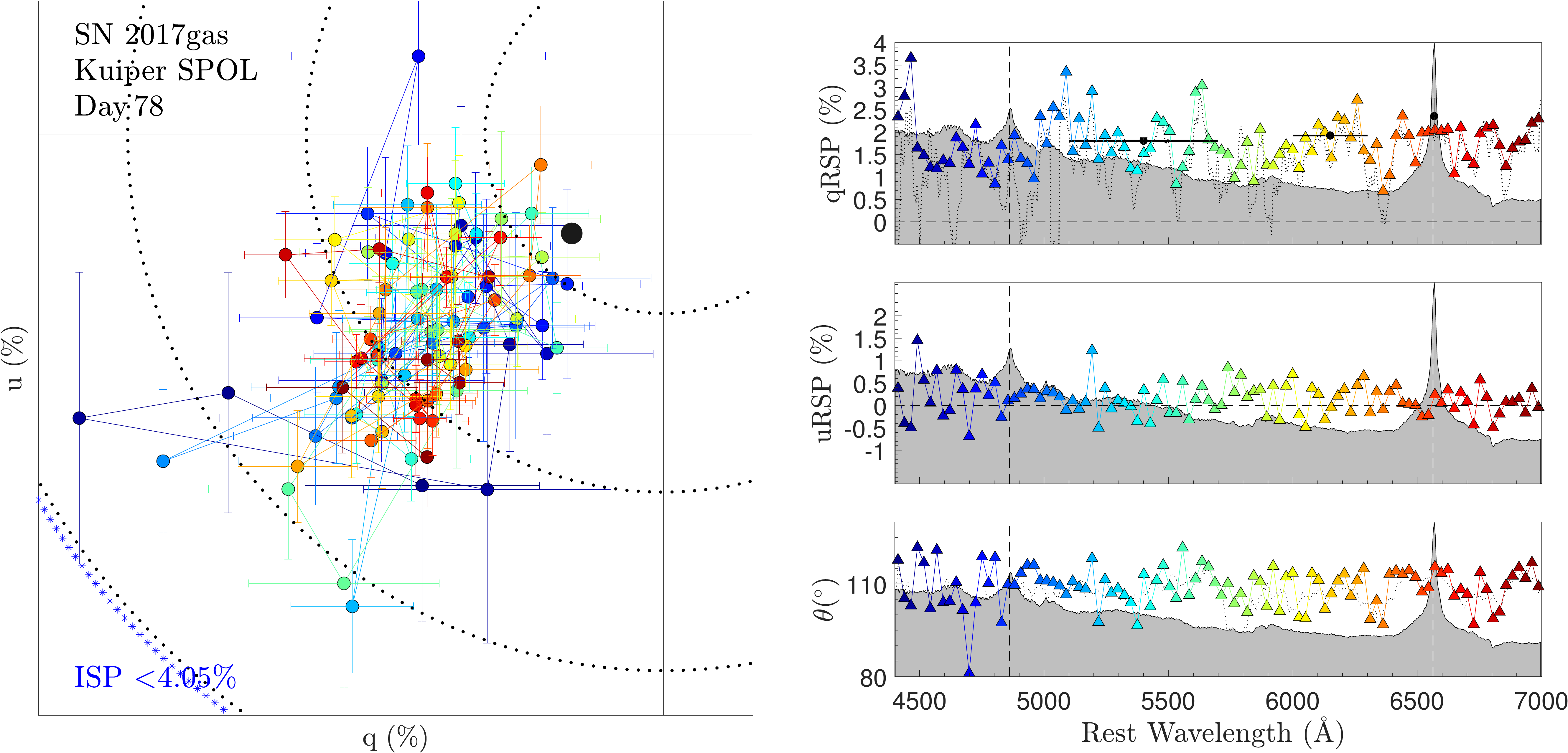}
\qufig
{SN~2017gas}{Kuiper}{78, ISP-corrected}{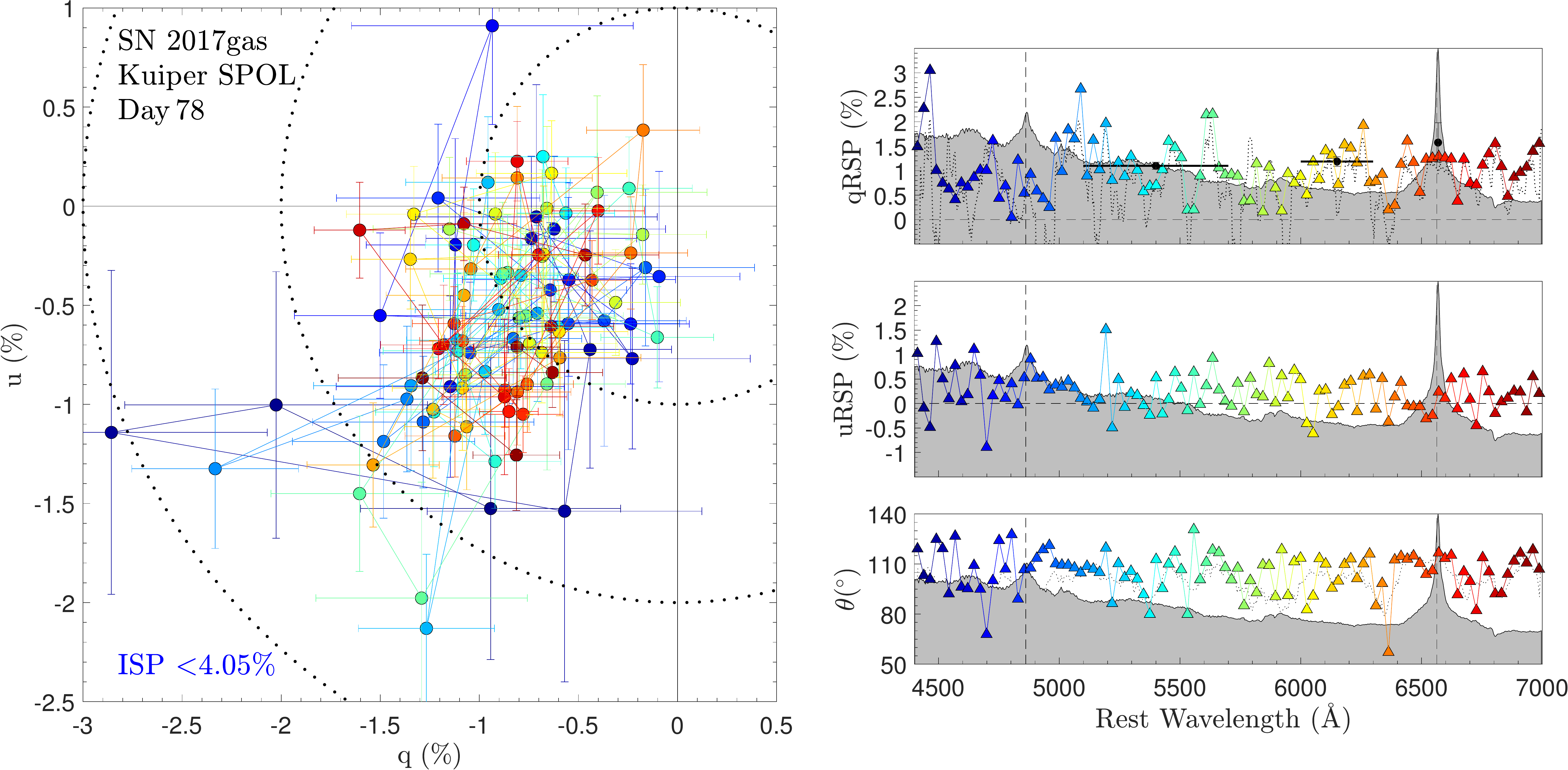}
{SN~2017gas}{MMT}{110}{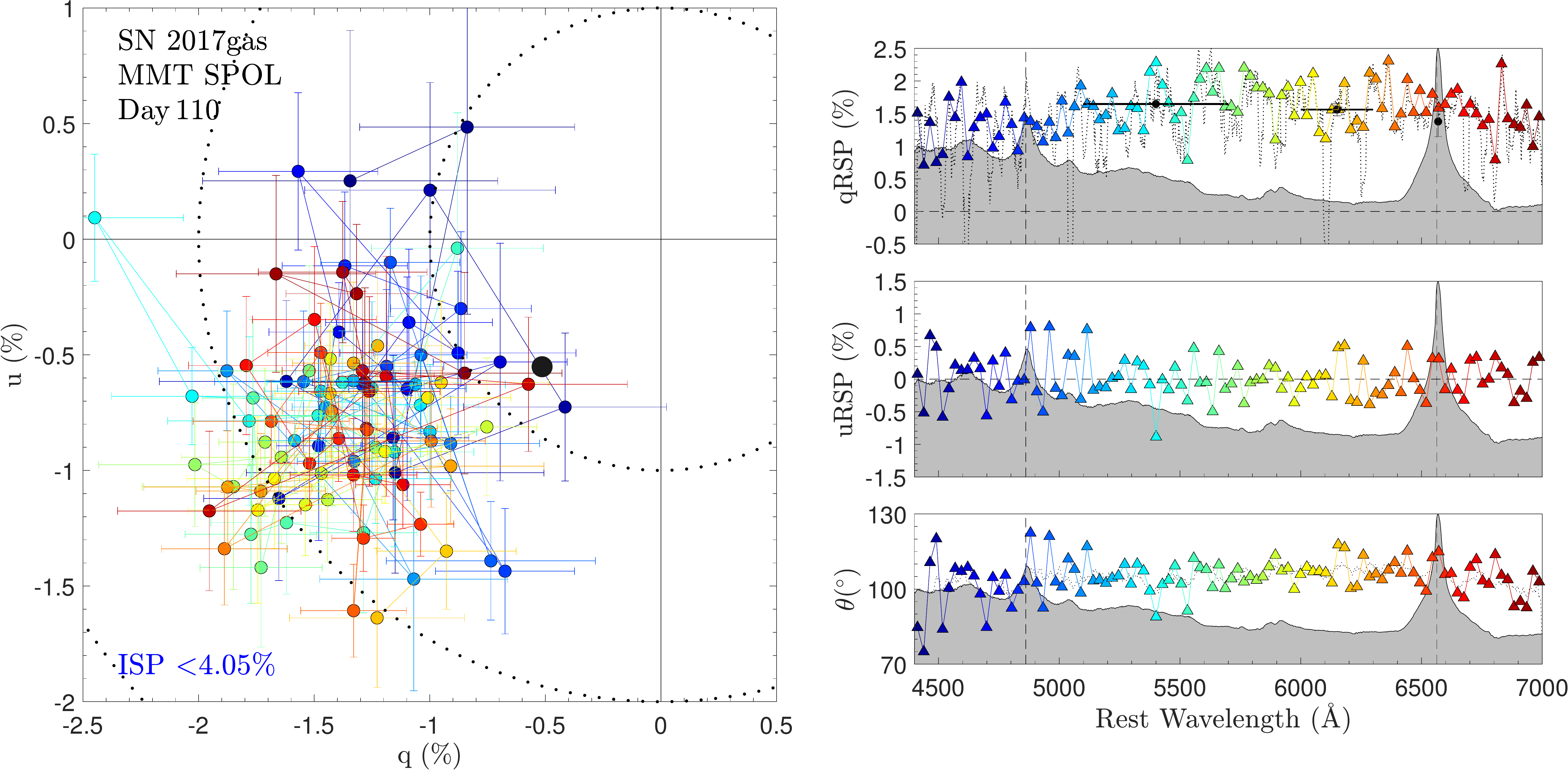}
\qufig
{SN~2017gas}{MMT}{110, ISP-corrected}{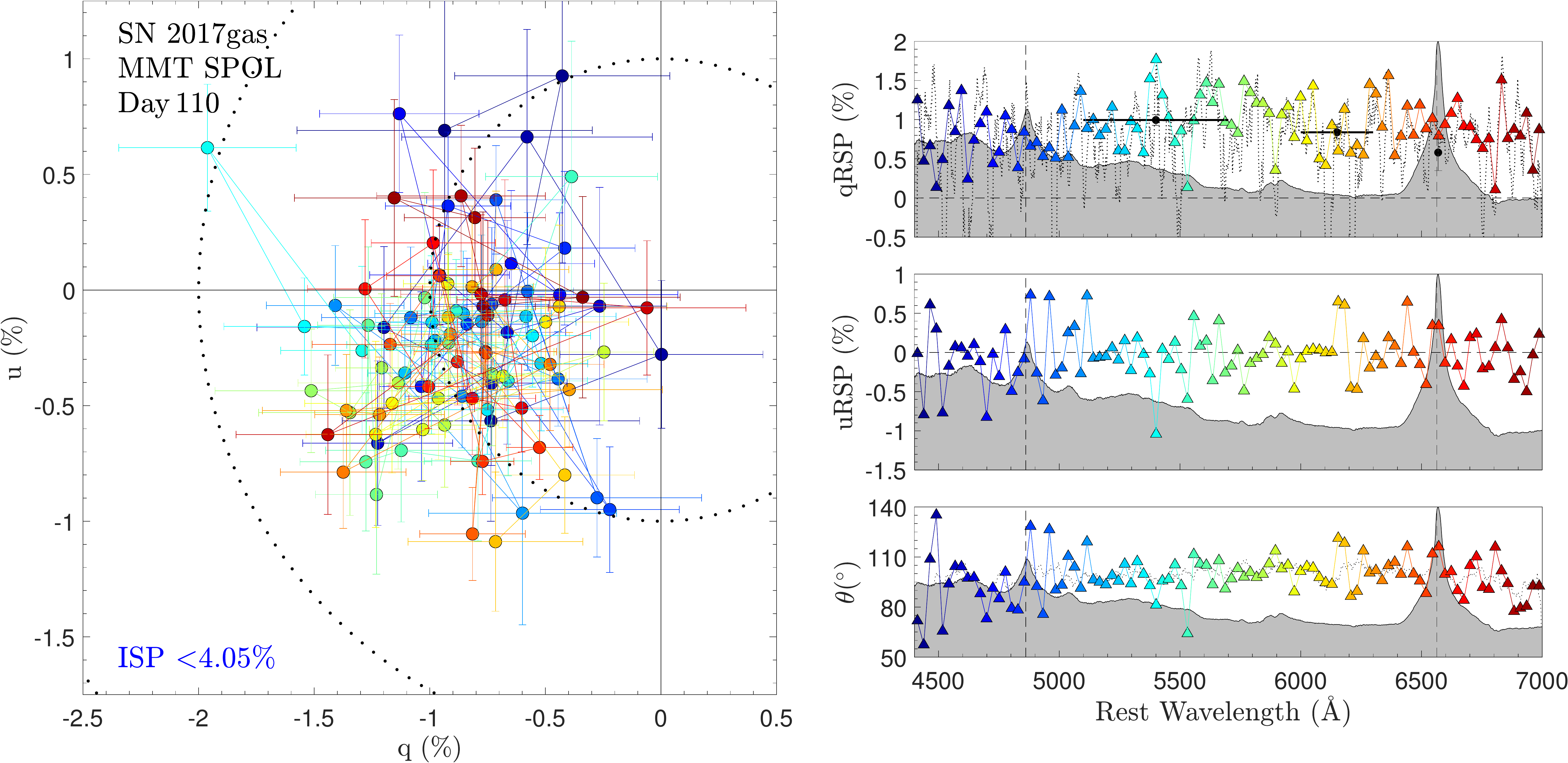}
{SN~2017hcc}{Kuiper}{-45}{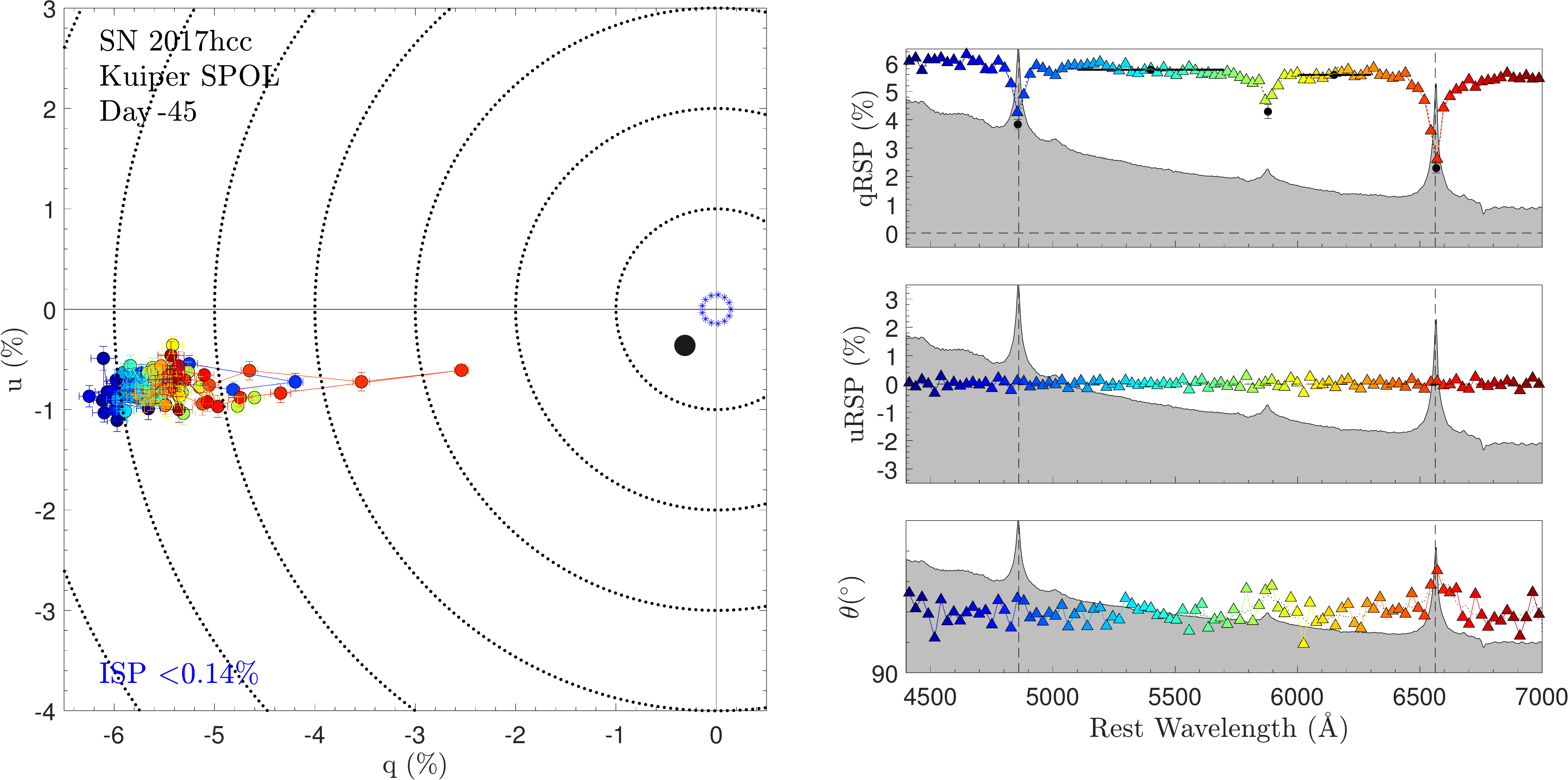}
\qufig
{SN~2017hcc}{Kuiper}{-45, ISP-corrected}{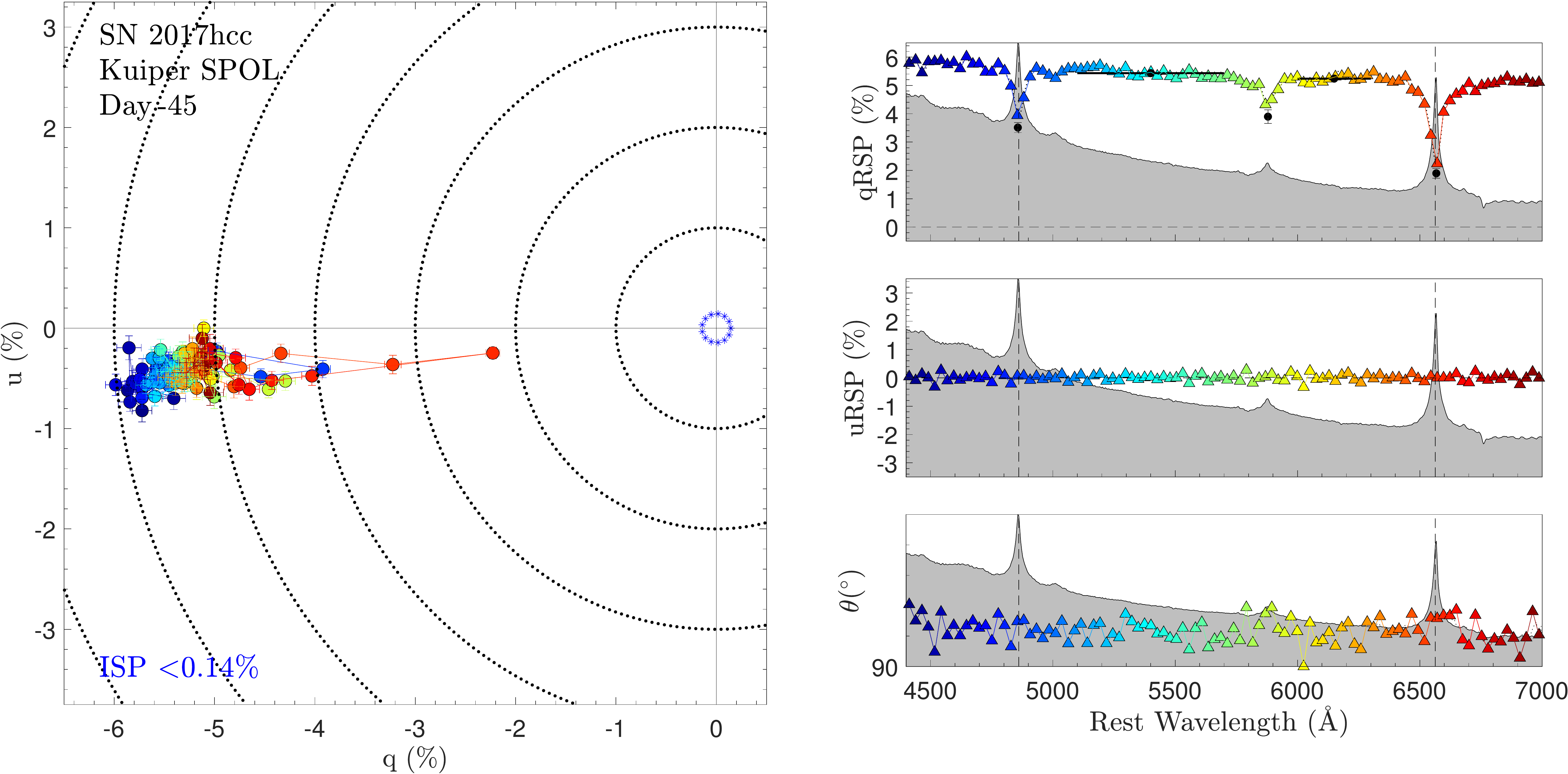}
{SN~2017hcc}{Kuiper}{-15}{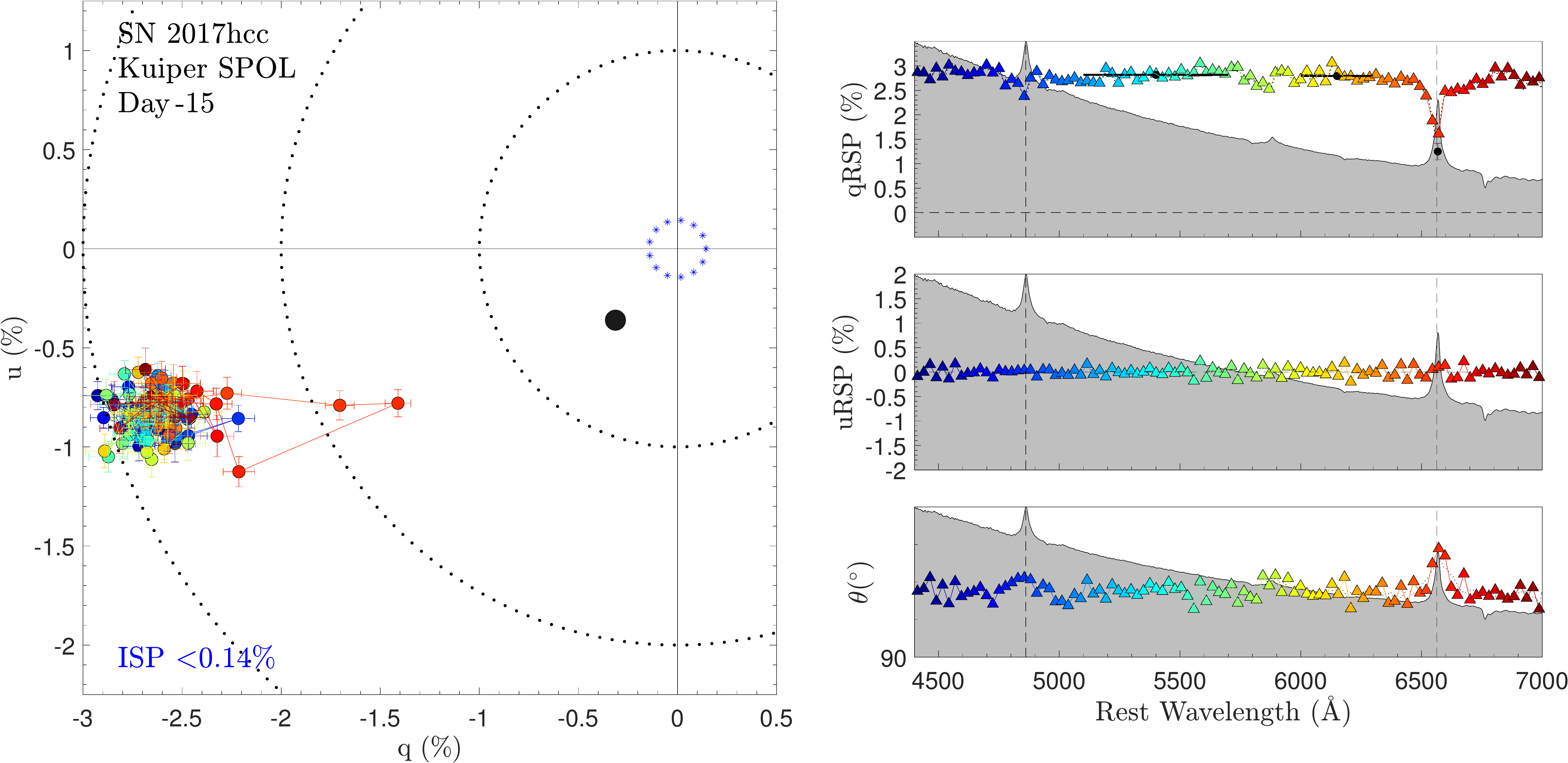}
\qufig
{SN~2017hcc}{Kuiper}{-15, ISP-corrected}{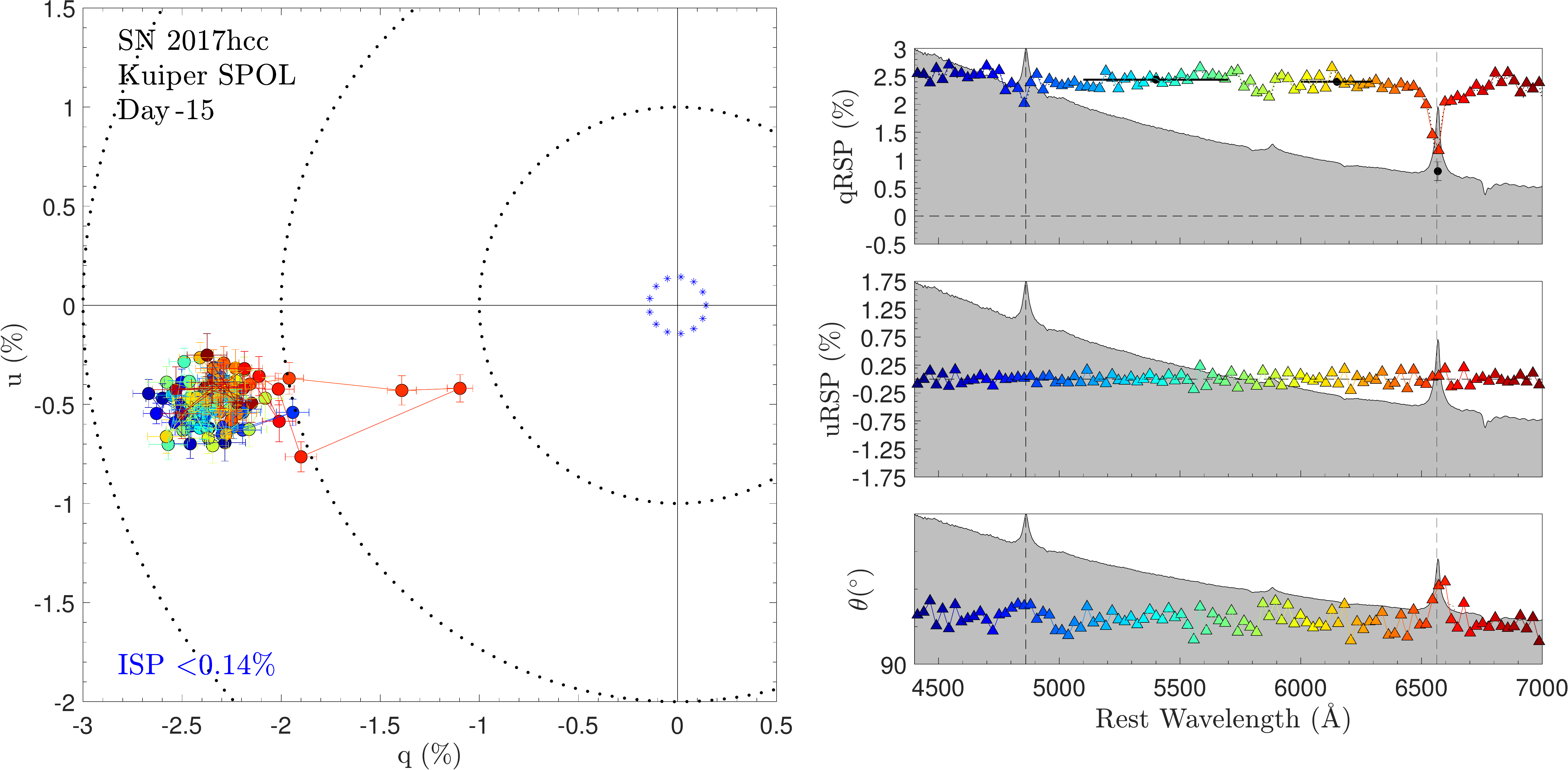}
{SN~2017hcc}{Kuiper}{9}{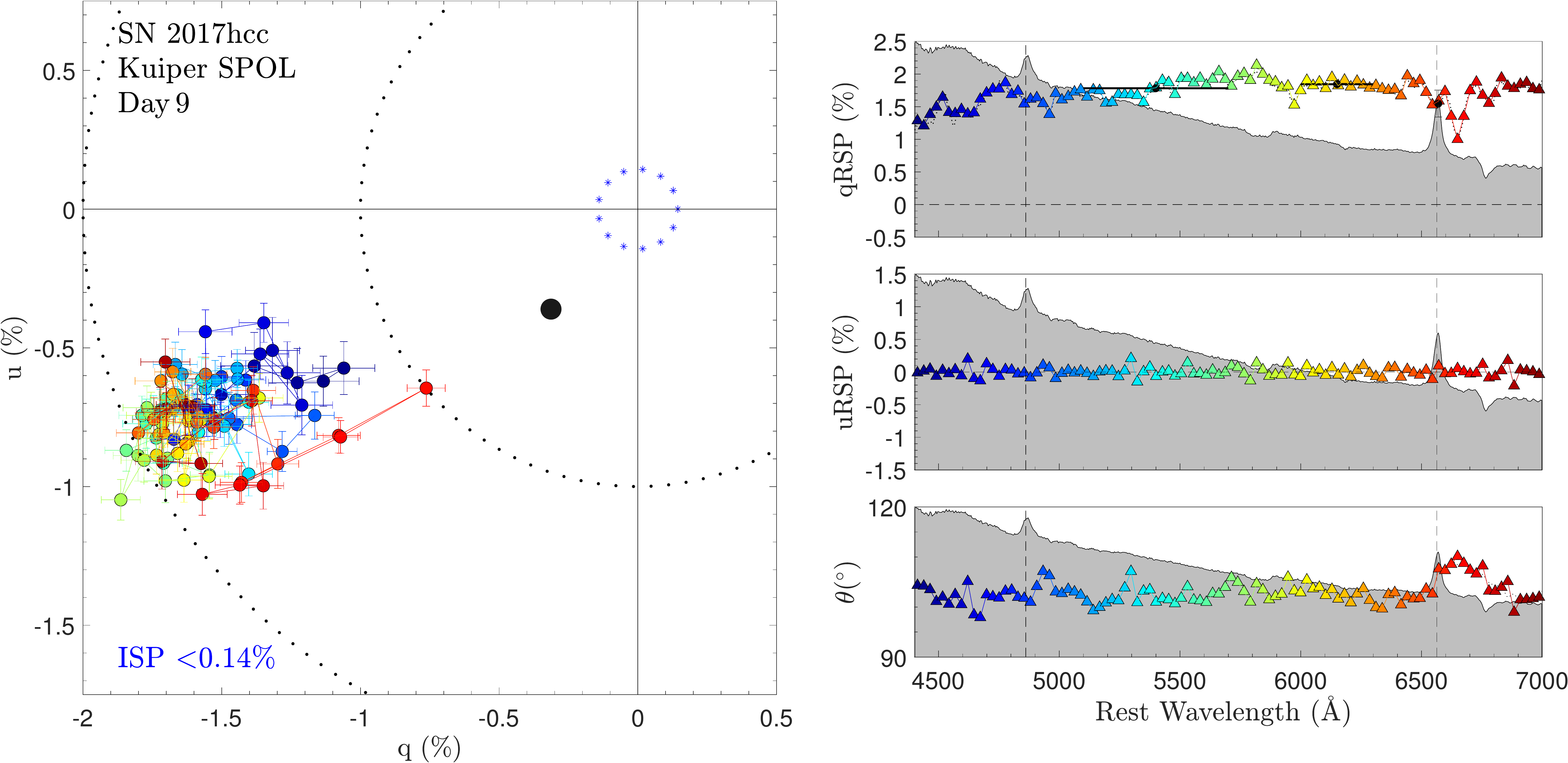}
\qufig
{SN~2017hcc}{Kuiper}{9, ISP-corrected}{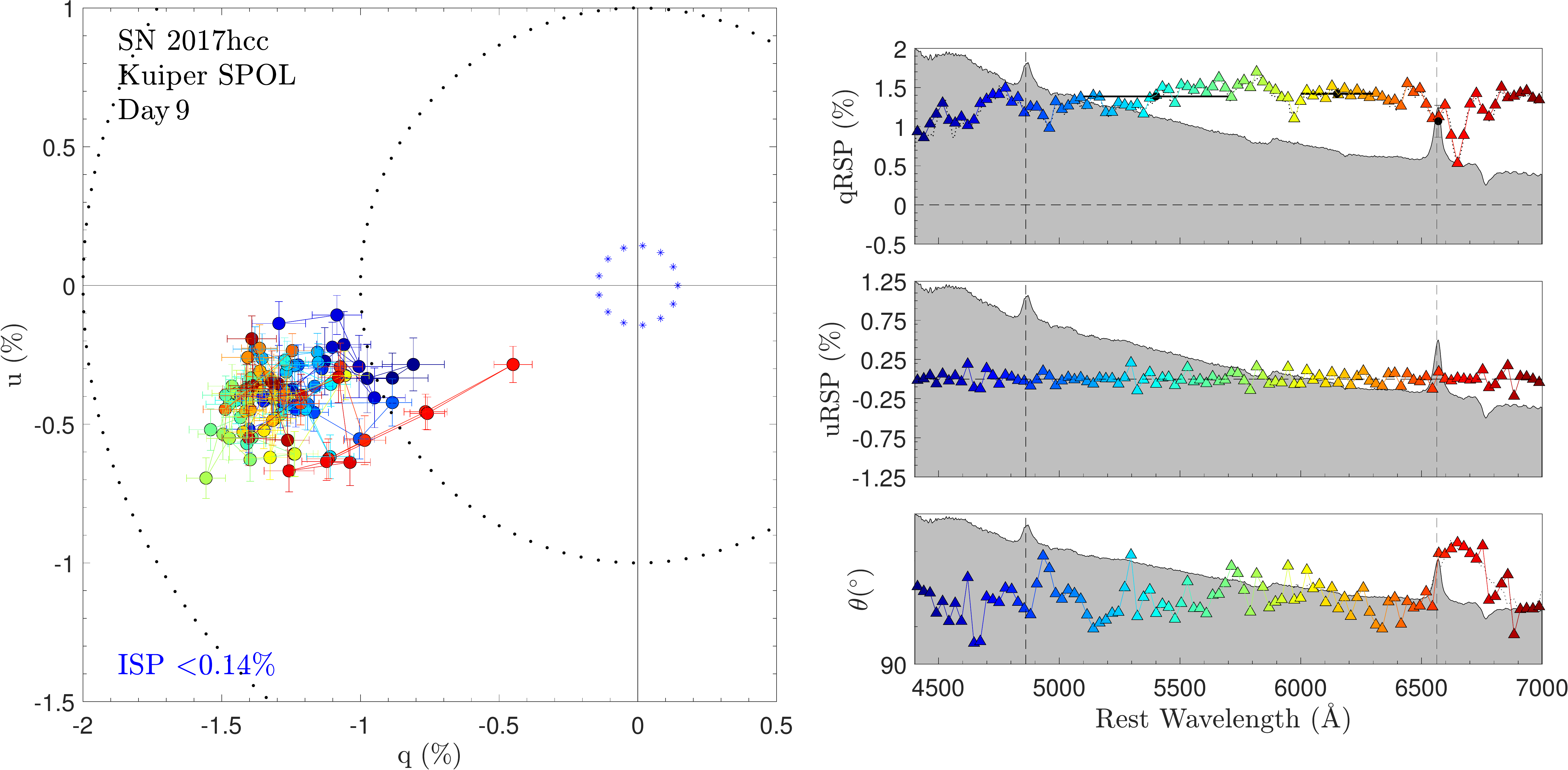}
{SN~2017hcc}{MMT}{17}{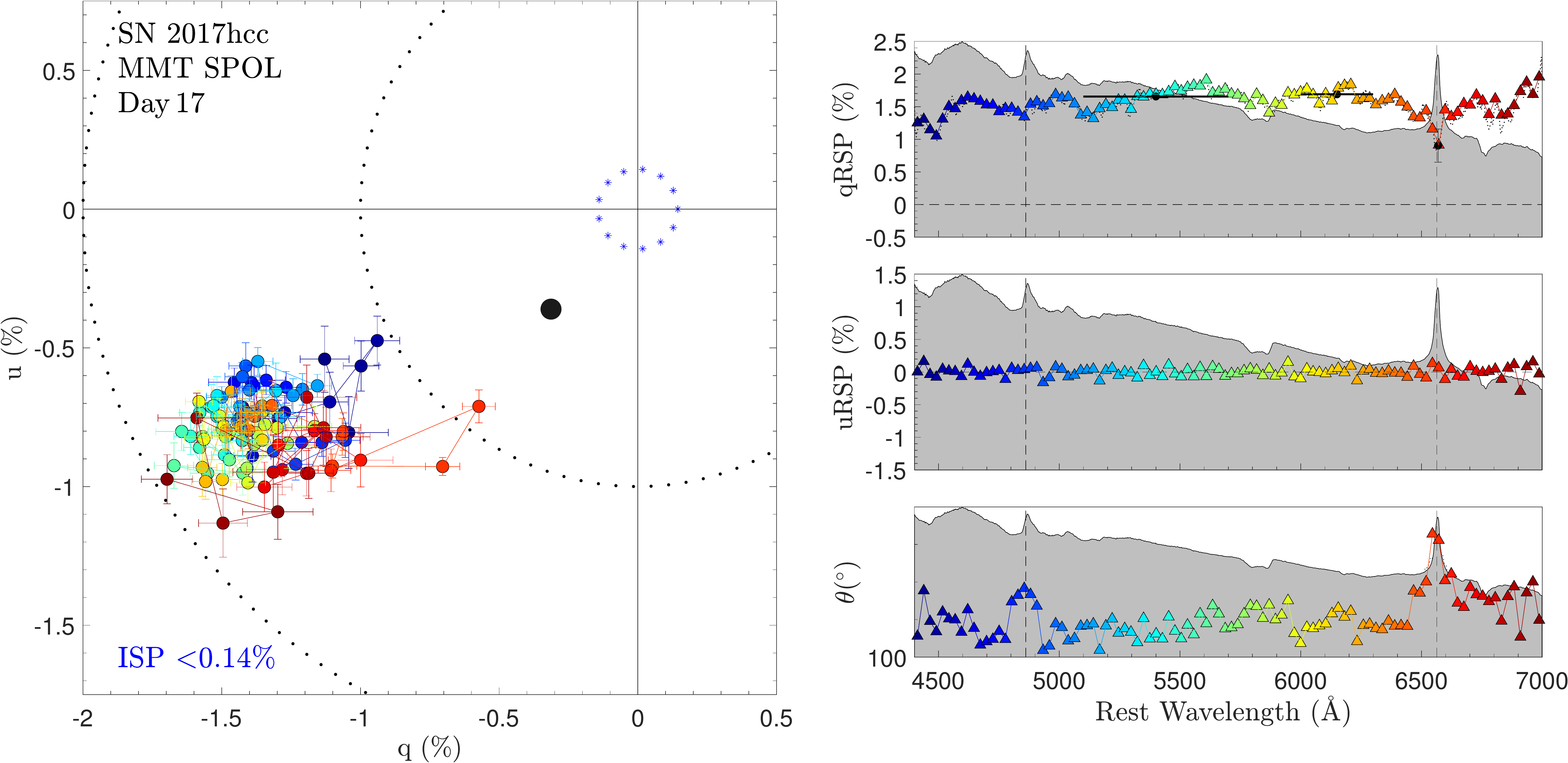}
\qufig
{SN~2017hcc}{MMT}{17, ISP-corrected}{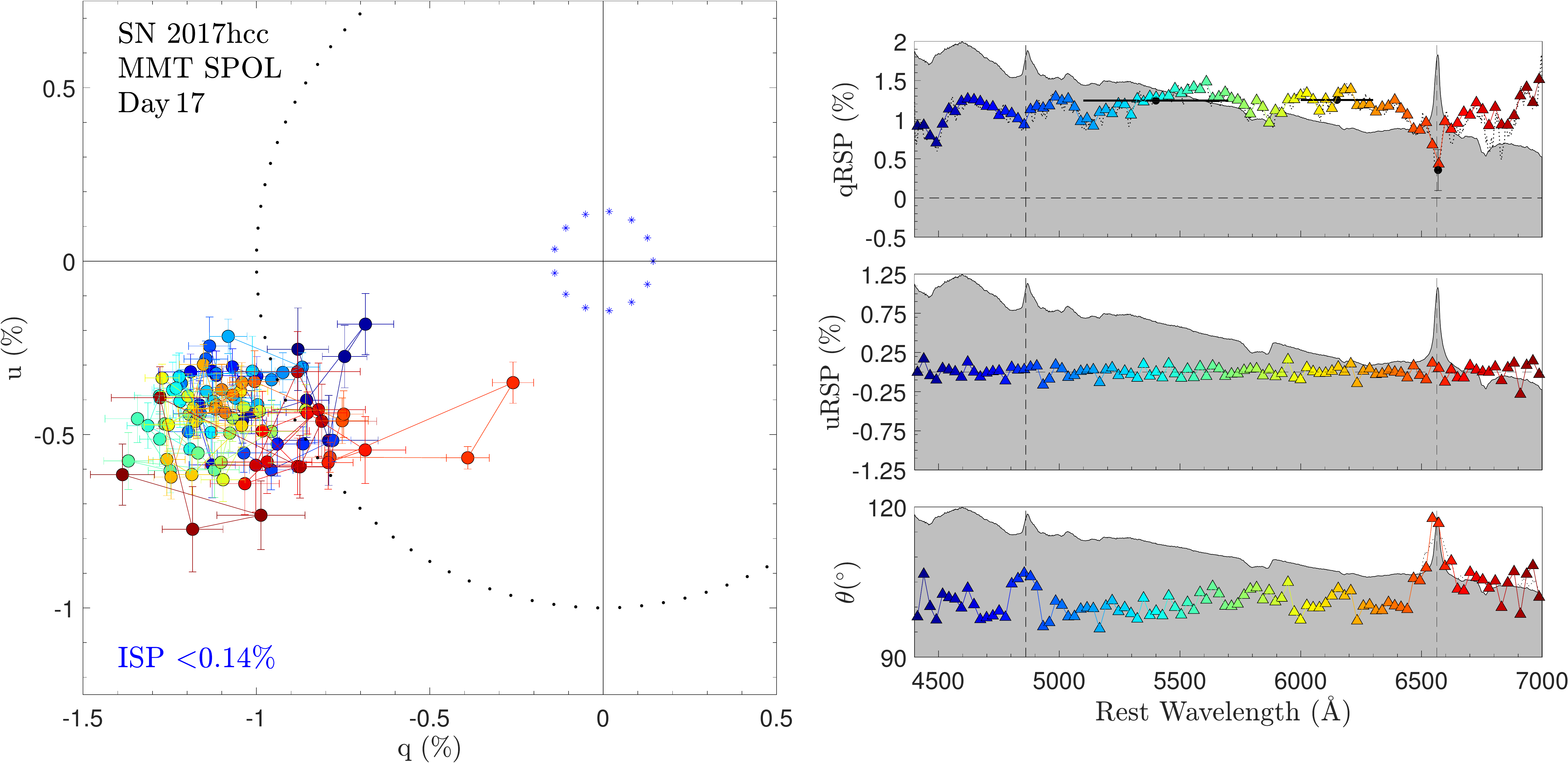}
{SN~2017hcc}{Kuiper}{41}{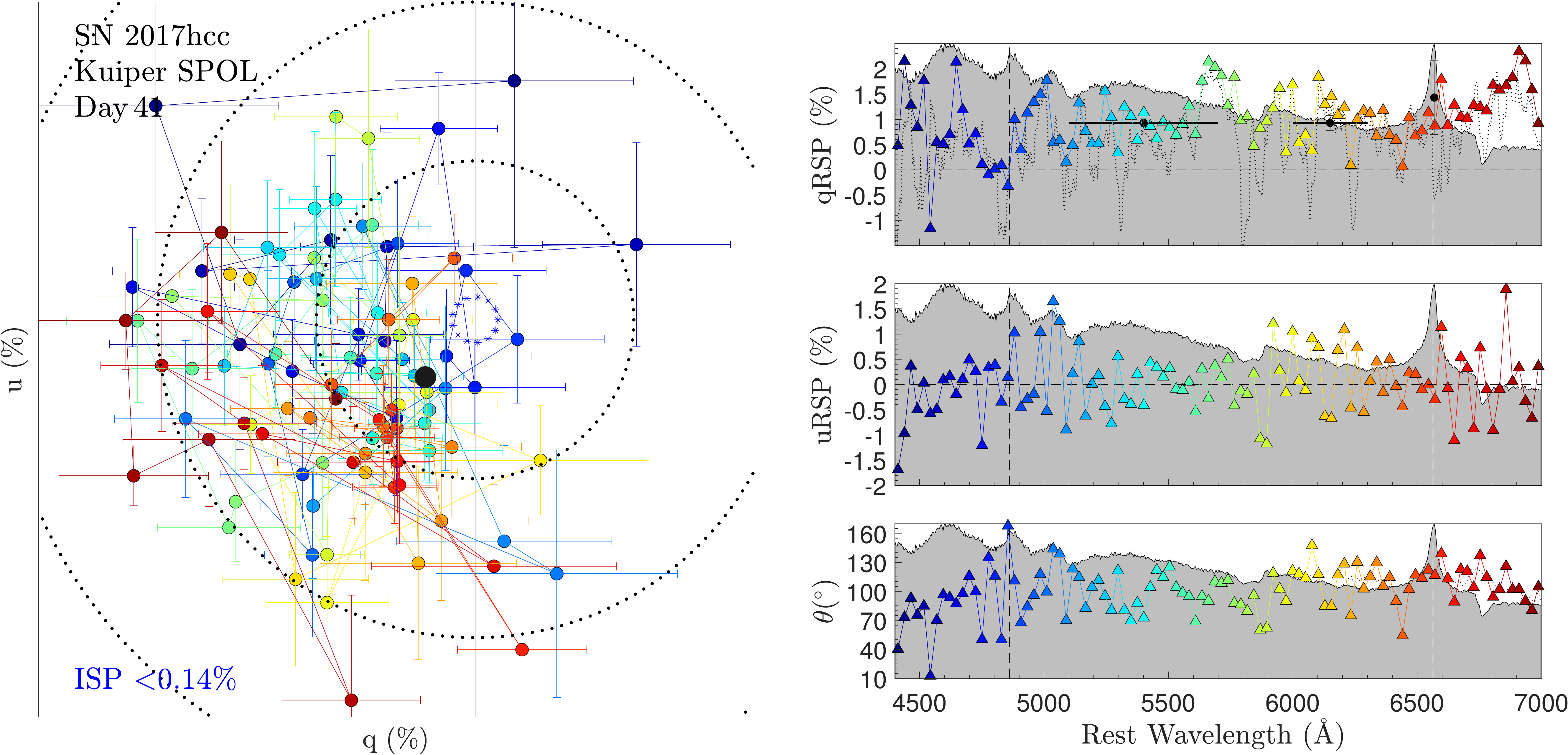}
\qufig
{SN~2017hcc}{Kuiper}{41, ISP-corrected}{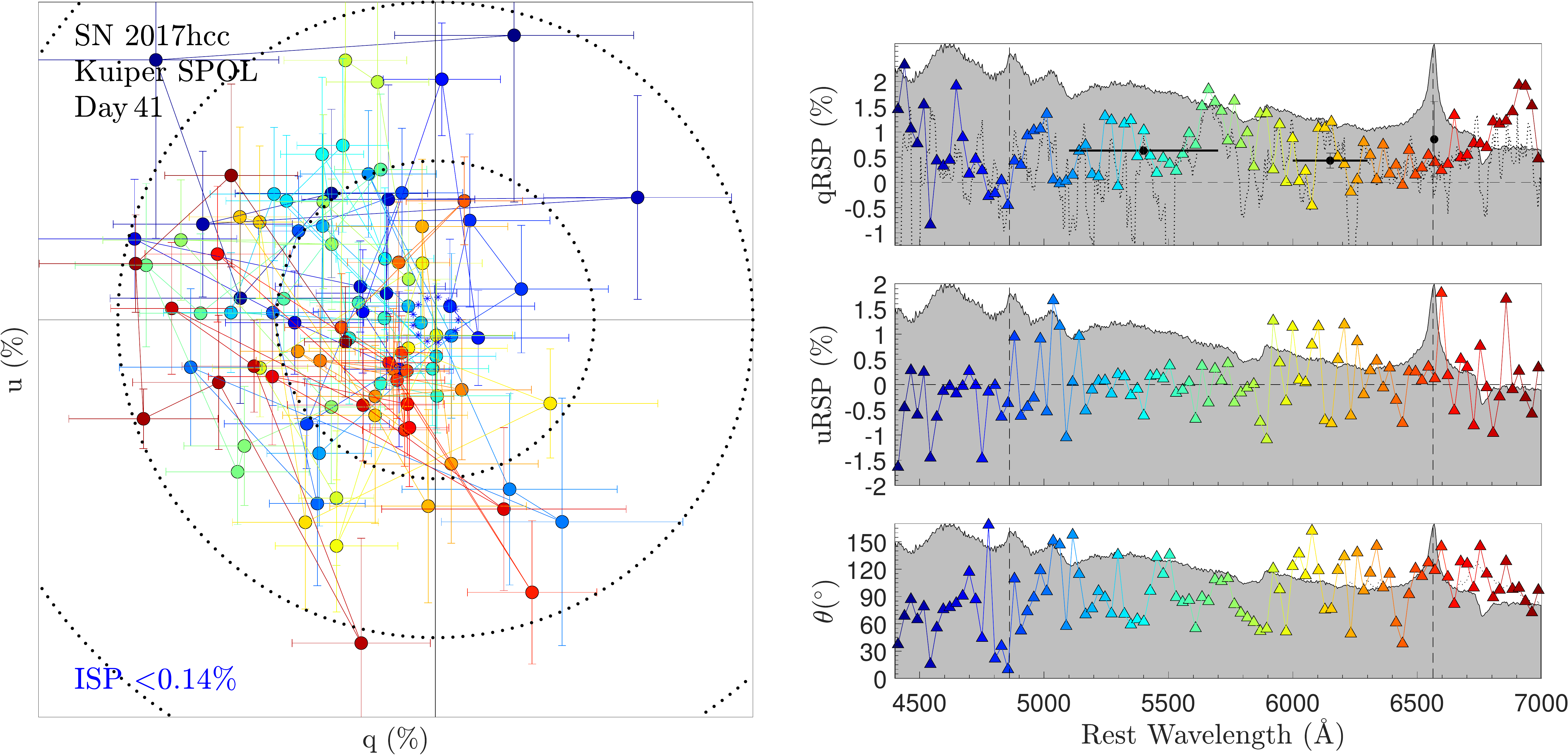}
{SN~2017hcc}{Bok}{48}{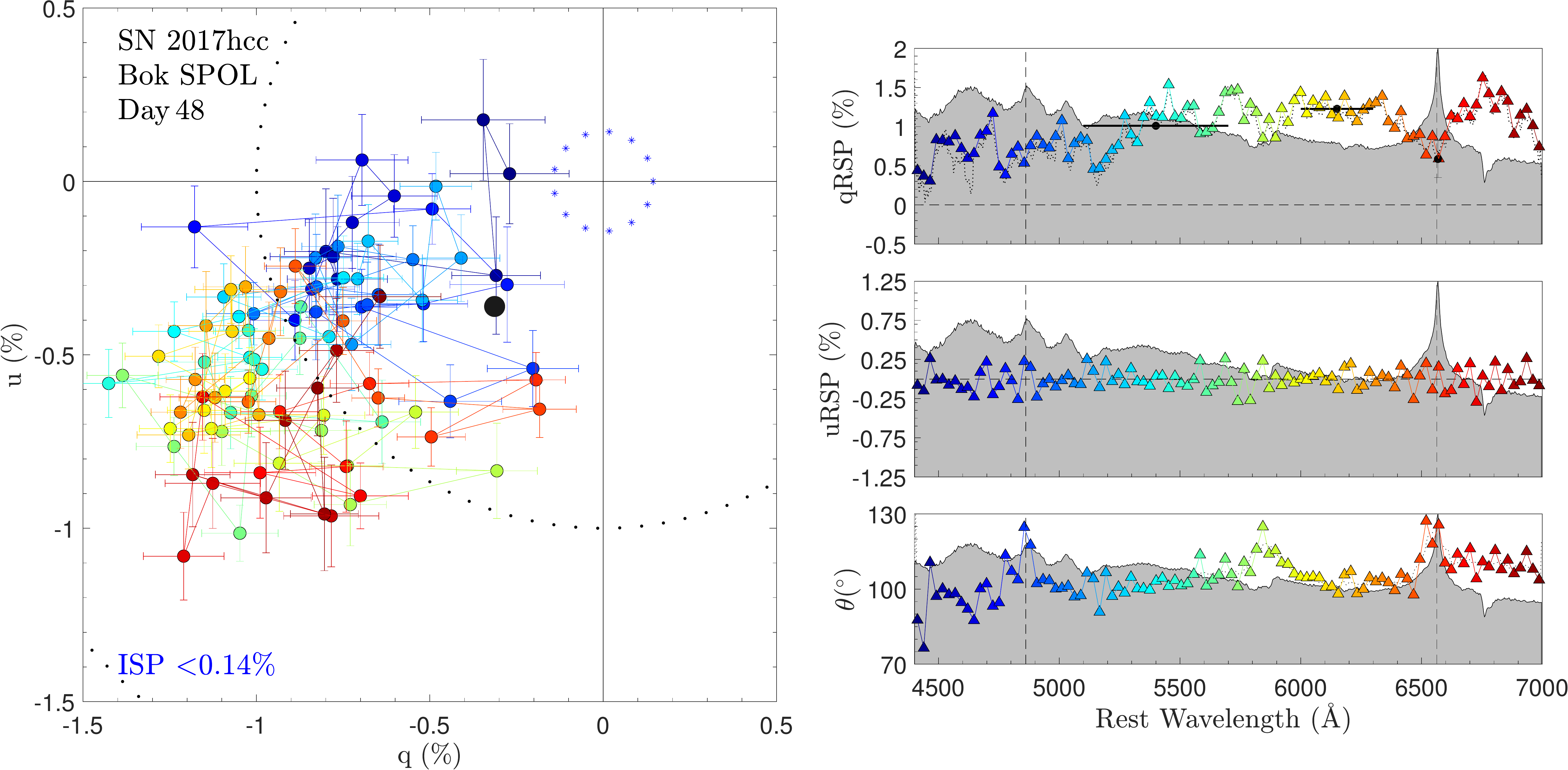}
\qufig
{SN~2017hcc}{Bok}{48, ISP-corrected}{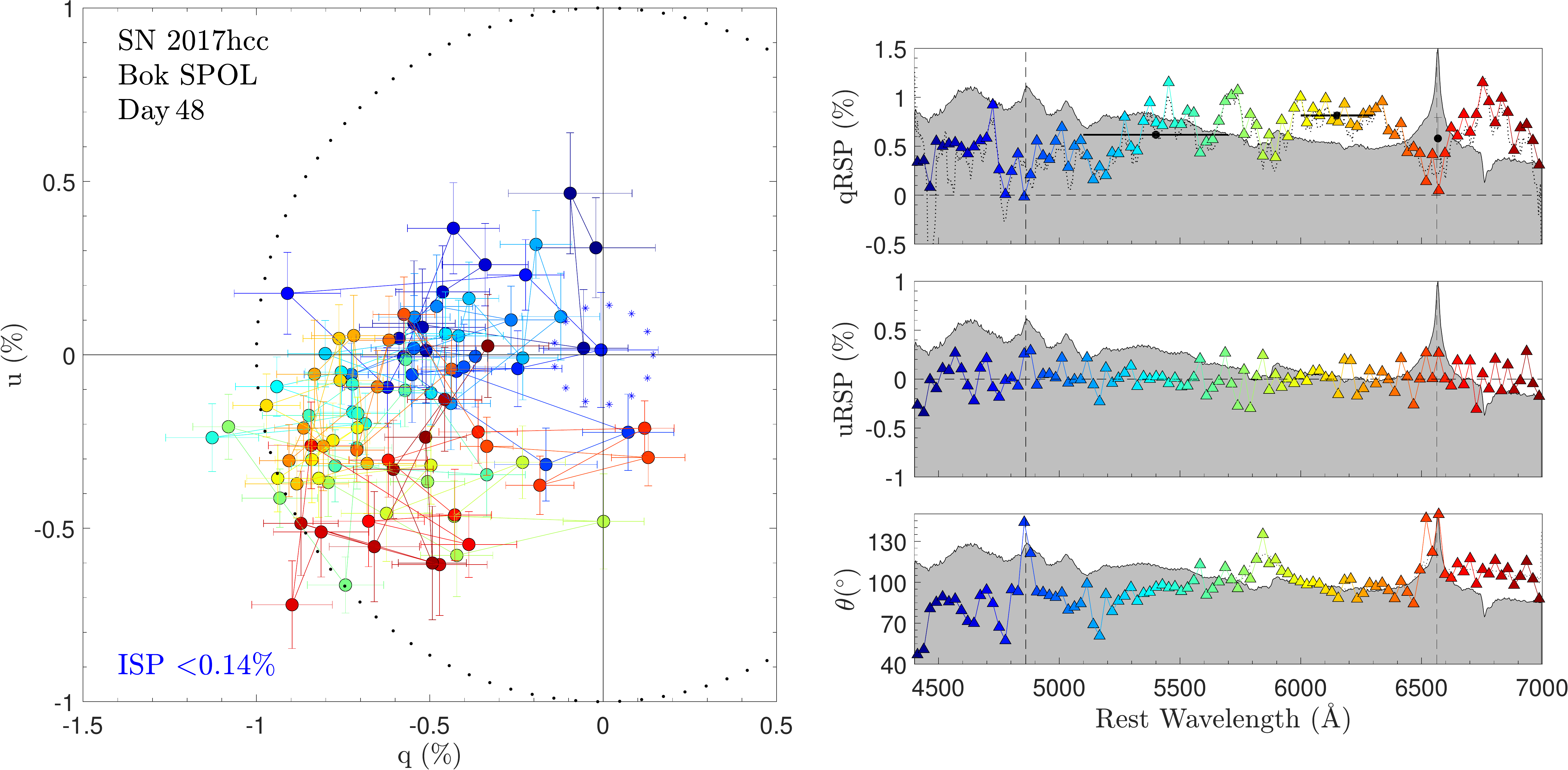}
{SN~2017hcc}{Bok}{328}{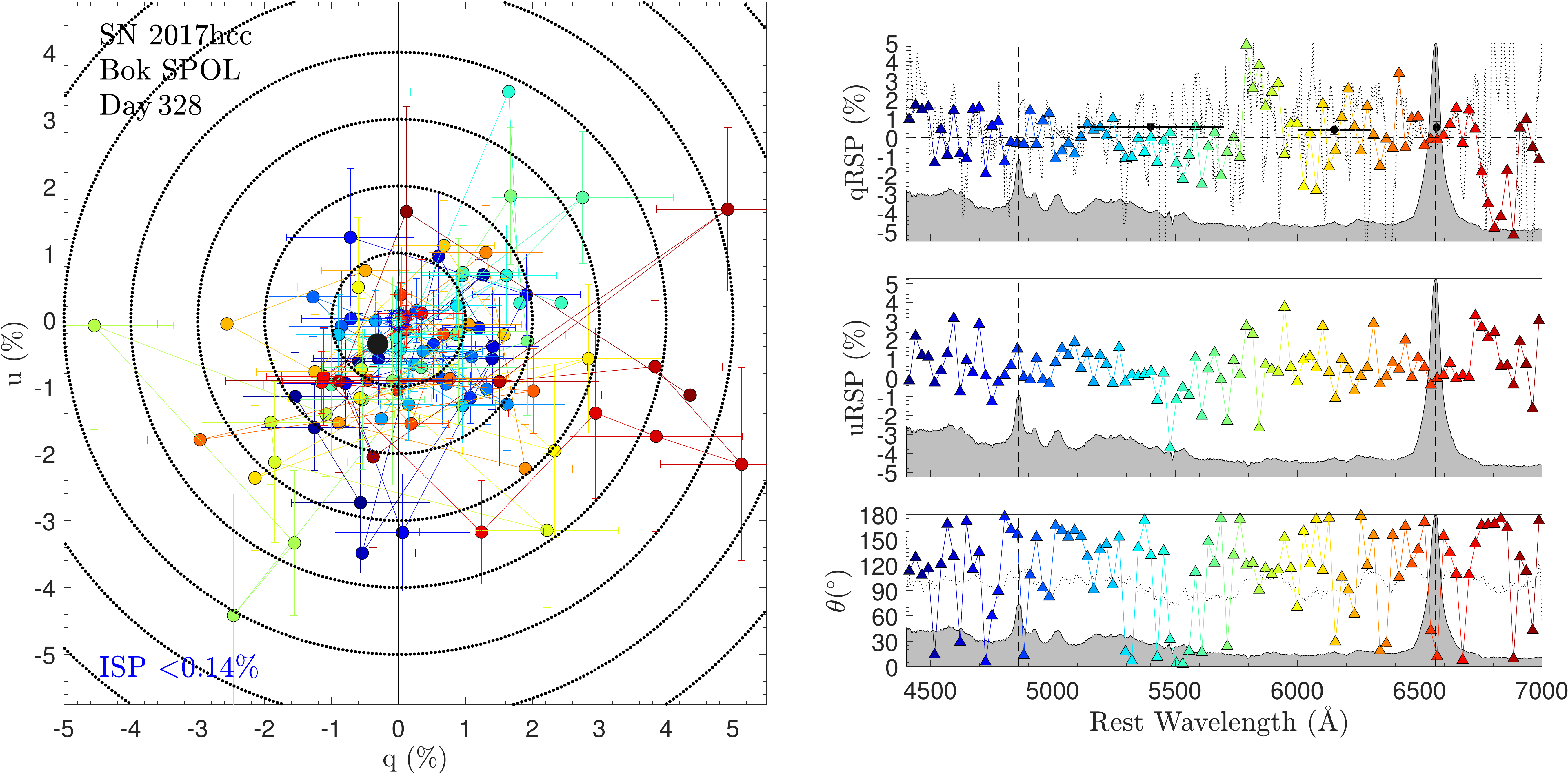}
\begin{figure*}
\centering
\includegraphics[width=1.0\textwidth,height=1.1\textheight,keepaspectratio,clip=true,trim=0cm 0cm 0cm 0cm]{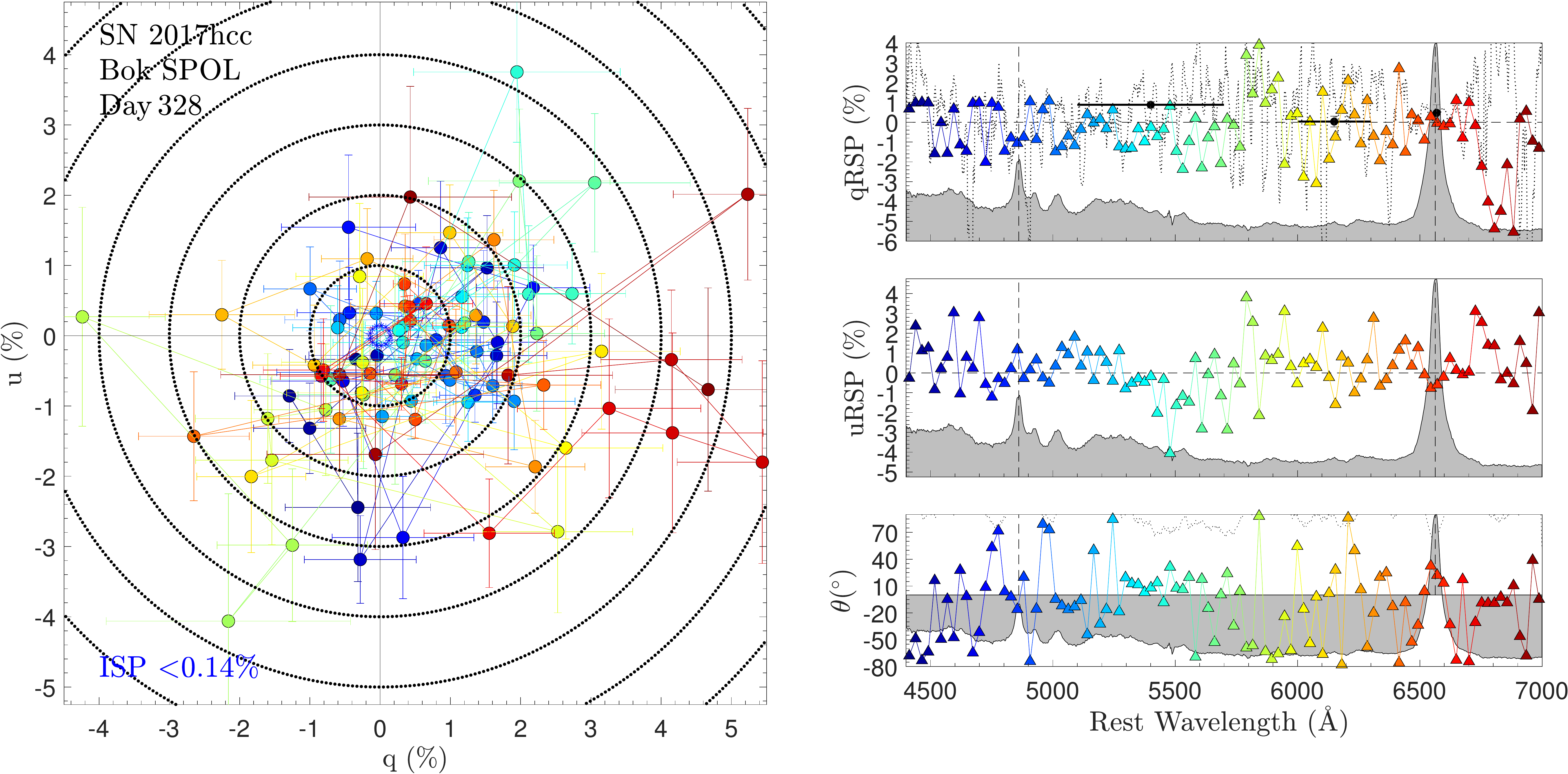}
\caption{$q-u$ plots as described at the start of Appendix \ref{App:A}.  SN~2017hcc Bok Day 328, ISP-corrected}
\end{figure*}


\label{lastpage}

\end{document}